\documentclass[article,twocolumn]{aastex62}

\newcommand{\cs}{G4Jy-3CRE}
\newcommand{\chn}{{\it Chandra}}
\newcommand{\swf}{{\it Swift}}
\newcommand{\xmm}{XMM-{\it Newton}}

\usepackage{graphicx}
\usepackage{longtable}
\usepackage{mdwlist}
\usepackage{color}
\usepackage{enumitem}
\usepackage{lipsum}
\shorttitle{Powerful Radio Sources in the Southern Sky I}
\shortauthors{F. Massaro et al.}

\begin{document}

\title{Powerful Radio Sources in the Southern Sky. I. Optical Identifications}

\author[0000-0002-1704-9850]{F. Massaro}
\affiliation{Dipartimento di Fisica, Universit\`a degli Studi di Torino, via Pietro Giuria 1, I-10125 Torino, Italy.}
\affiliation{Istituto Nazionale di Astrofisica (INAF) - Osservatorio Astrofisico di Torino, via Osservatorio 20, 10025 Pino Torinese, Italy.}
\affiliation{Istituto Nazionale di Fisica Nucleare (INFN) - Sezione di Torino, via Pietro Giuria 1, I-10125 Torino, Italy.}
\affiliation{Consorzio Interuniversitario per la Fisica Spaziale, via Pietro Giuria 1, I-10125 Torino, Italy.}

\author[0000-0002-2340-8303]{S. V. White}
\affiliation{Department of Physics and Electronics, Rhodes University, PO Box 94, Grahamstown, 6140, South Africa.}

\author[0000-0002-9896-6430]{A. Garc\'ia-P\'erez}
\affiliation{Dipartimento di Fisica, Universit\`a degli Studi di Torino, via Pietro Giuria 1, I-10125 Torino, Italy.}
\affiliation{Instituto Nacional de Astrof\'isica, \'Optica y Electr\'onica, Luis Enrique Erro 1, Tonantzintla, Puebla 72840, México.}

\author[0000-0003-4413-7722]{A. Jimenez-Gallardo}
\affiliation{Dipartimento di Fisica, Universit\`a degli Studi di Torino, via Pietro Giuria 1, I-10125 Torino, Italy.}
\affiliation{Istituto Nazionale di Astrofisica (INAF) - Osservatorio Astrofisico di Torino, via Osservatorio 20, 10025 Pino Torinese, Italy.}
\affiliation{Dipartimento di Fisica e Astronomia, Universit\`a di Bologna, via P. Gobetti 93/2, 40129 Bologna, Italy.}

\author[0000-0003-3684-4275]{A. Capetti}
\affiliation{Istituto Nazionale di Astrofisica (INAF) - Osservatorio Astrofisico di Torino, via Osservatorio 20, 10025 Pino Torinese, Italy.}

\author[0000-0002-4377-0174]{C.C. Cheung}
\affiliation{Space Science Division, Naval Research Laboratory, Washington, DC 20375, USA.}

\author[0000-0002-9478-1682]{W. R. Forman}
\affiliation{Center for Astrophysics | Harvard \& Smithsonian, 60 Garden Street, Cambridge, MA 02138, USA.}

\author[0000-0002-5941-5214]{C. Mazzucchelli}
\affiliation{European Southern Observatory, Alonso de C\'ordova 3107, Vitacura, Regi\'on Metropolitana, Chile.}
\affiliation{Instituto de Estudios Astrof\'{\i}sicos, Facultad de Ingenier\'{\i}a y Ciencias, Universidad Diego Portales, Avenida Ejercito Libertador 441, Santiago, Chile.}

\author[0000-0002-5646-2410]{A. Paggi}
\affiliation{Dipartimento di Fisica, Universit\`a degli Studi di Torino (UniTO), via Pietro Giuria 1, I-10125 Torino, Italy.}
\affiliation{Istituto Nazionale di Astrofisica - Osservatorio Astrofisico di Torino, via Osservatorio 20, 10025 Pino Torinese, Italy.}
\affiliation{Istituto Nazionale di Fisica Nucleare - Sezione di Torino, via Pietro Giuria 1, I-10125 Torino, Italy.}

\author[0000-0001-5783-6544]{N. P. H. Nesvadba}
\affiliation{Universit\'e de la C\^{o}te d'Azur, Observatoire de la C\^{o}te d'Azur, CNRS, Laboratoire Lagrange, Bd de l'Observatoire, CS 34229,06304 Nice cedex 4, France.} 

\author{J. P. Madrid}
\affiliation{Department of Physics and Astronomy, The University of Texas Rio Grande Valley, Brownsville, TX 78520, USA.}

\author[0000-0003-1562-5188]{I. Andruchow}
\affiliation{Instituto Argentino de Radioastronom\'{\i}a, CONICET-CICPBA-UNLP, CC5 (1897) Villa Elisa, Prov. de Buenos Aires, Argentina.}

\author[0000-0002-3866-2726]{S. Cellone}
\affiliation{Facultad de Ciencias Astron\'omicas y Geof\'isicas, Universidad Nacional de La Plata, Paseo del Bosque, B1900FWA La Plata, Argentina.} 
\affiliation{Complejo Astron\'omico El Leoncito (CASLEO), CONICET-UNLP-UNC-UNSJ, Av. Espa\~na 1512 (sur), J5402DSP San Juan, Argentina.}

\author[0000-0003-0032-9538]{H. A. Pe\~na-Herazo}
\affiliation{East Asian Observatory, 660 N. A'oh$\bar{o}$k$\bar{u}$ Place, Hilo, HI 96720, USA.}

\author[0000-0003-3471-7459]{R. Grossov\'a}
\affiliation{Department of Theoretical Physics and Astrophysics, Faculty of Science, Masaryk University, Kotl\'a\u{r}sk\'a 2, Brno, CZ-611 37, Czech Republic.}
\affiliation{Astronomical Institute of the Czech Academy of Sciences, Bocn\'i II 1401, 141 00, Prague, Czech Republic.}

\author[0000-0002-0690-0638]{B. Balmaverde}
\affiliation{Istituto Nazionale di Astrofisica - Osservatorio Astrofisico di Torino, via Osservatorio 20, 10025 Pino Torinese, Italy.}

\author[0000-0002-3140-4070]{E. Sani}
\affiliation{European Southern Observatory, Alonso de C\'ordova 3107, Vitacura, Regi\'on Metropolitana, Chile.}

\author[0000-0002-2558-0967]{V. Chavushyan}
\affiliation{Instituto Nacional de Astrof\'isica, \'Optica y Electr\'onica, Luis Enrique Erro 1, Tonantzintla, Puebla 72840, México.}
\affiliation{Center for Astrophysics | Harvard \& Smithsonian, 60 Garden Street, Cambridge, MA 02138, USA.}

\author[0000-0002-0765-0511]{R. P. Kraft}
\affiliation{Center for Astrophysics | Harvard \& Smithsonian, 60 Garden Street, Cambridge, MA 02138, USA.}

\author[0000-0002-6472-6711]{V. Reynaldi}
\affiliation{Facultad de Ciencias Astron\'omicas y Geof\'isicas, Universidad Nacional de La Plata, Paseo del Bosque, B1900FWA La Plata, Argentina.} 
\affiliation{Instituto de Astrof\'isica de La Plata (CCT La Plata-CONICET-UNLP), La Plata, Argentina.} 

\author{C. Leto}
\affiliation{ASI - Agenzia Spaziale Italiana, Via del Politecnico snc, 00133 Roma, Italy.}

\begin{abstract} Since the early sixties our view of radio galaxies and quasars has been drastically shaped by discoveries made thanks to observations of radio sources listed in the Third Cambridge catalog and its revised version (3CR). However the largest fraction of data collected to date on 3CR sources  was performed with relatively old instruments, rarely repeated and/or updated. Importantly, the 3CR contains only objects located in the Northern Hemisphere thus having limited access to new and innovative astronomical facilities. To mitigate these limitations we present a new catalog of powerful radio sources visible from the Southern Hemisphere, extracted from the GLEAM 4-Jy (G4Jy) catalog and based on equivalent selection criteria as the 3CR. This new catalog, named \cs, where the E stands for ``equivalent'', lists a total of 264 sources at declination below $-5^\circ$ and with 9\,Jy limiting sensitivity at $\sim$178 MHz. We explored archival radio maps obtained with different surveys and compared then with optical images available in the Pan-STARRS, DES and DSS databases to search for optical counterparts of their radio cores. We compared mid-infrared counterparts, originally associated in the G4Jy, with the optical ones identified here and we present results of a vast literature search carried out to collect redshift estimates for all \cs\ sources resulting in a total of 145 reliable $z$ measurements.
\end{abstract}


\section{Introduction}
\label{sec:intro} Since the early sixties, the Third Cambridge catalog of radio sources \citep[3C;][]{edge59}, together with all its revised editions \citep[3CR and 3CRR;][respectively]{bennett62,laing83}, has been considered a long-standing fundamental sample to understand the nature and evolution of powerful radio galaxies and quasars \citep{begelman84,urry95,harvaneck01,tremblay09,odea09,massaro11a}, and the relationships with their environments (i.e., feedback processes), at all scales and across cosmic time \citep{boehringer93,churazov00,churazov01,mcnamara07,mcnamara12,alexander12,fabian12,morganti17}. The classical classification scheme for radio galaxies, distinguishing between edge-darkened (i.e., FR\,I) and edge-brightened (i.e., FR\,II) sources, was developed on the basis of the 3C observations \citep{fanaroff74,leahy93}. 

The first revised version of the Third Cambridge catalog of radio sources \citep[3CR;][]{spinrad85} lists the most powerful radio sources detected in the Northern Hemisphere at 178\,MHz above a 9\,Jy flux density threshold listing 298 extragalactic sources, with only a small fraction (i.e., less than $\sim$8\%) still unidentified.
 
In the last decades a vast suite of multifrequency campaigns was dedicated to the 3CR radio sources and carried out at radio \citep[see e.g.,][]{clarke64,willis66,law95,hardcastle00,giovannini05}, infrared \citep[see e.g.,][]{simpson99,madrid06,baldi10,werner12,dicken14}, optical \citep[see e.g.,][]{wyndham66,sandage67,longair84,saunders89,longair95,mccarthy95,dekoff96,mccarthy96a,mccarthy97,martel99,lehnert99,chiaberge00} and X-ray frequencies \citep[see e.g.,][]{prieto96,evans06,hardcastle06,balmaverde12,wilkes13,kuraszkiewicz21}. Initial studies of the 3CR catalog were performed to estimate source redshifts and obtain optical classifications \citep[see e.g.,][]{smith76,kristian78,djorgovski88,hiltner91,martel98,buttiglione09,buttiglione10,capetti11,buttiglione11,baldi13}. Later work was carried out to dig deeper into the physical properties of active galactic nuclei (AGNs), their large-scale environments and intracluster medium (ICM) when harbored in galaxy clusters \citep[see e.g.,][]{baum88,blanton11,kotyla16,balmaverde18a}. 

The legacy value of 3CR follow up surveys can be also highlighted by results achieved thanks to: (i) the Hubble Space Telescope (HST) Snapshot Survey of 3CR Radio Source Counterparts\footnote{https://archive.stsci.edu/prepds/3cr/} \citep{privon08,chiaberge15,hilbert16}, (ii) the 3CR \chn\ Snapshot Survey \citep{massaro10,massaro12a,massaro13a,massaro18,stuardi18,jimenez20} and (iii) the MUse RAdio Loud Emission line Snapshot survey \citep{balmaverde18b,balmaverde19,balmaverde21,speranza21,balmaverde22}. The former campaign allowed us to obtain a full overview of optical properties of these powerful radio sources at $\sim$90\% level of completeness (even if not uniform in terms of instruments and filters adopted) while the latter one (i.e., the MURALES campaign), still ongoing, can be performed only on 3CR sources at $z<0.8$ and at Declination $<20^\circ$, being visibile from the Very Large Telescope (VLT) in Chile. In 2008 the 3CR \chn\ Snapshot survey also began aiming to (i) detect new jets, hot spots and lobes emitting in the X-ray band (ii) investigate nuclear emission of powerful radio sources and (iii) discover new galaxy clusters \citep[see also][]{hardcastle10,hardcastle12,dasadia16,madrid18,ricci18,jimenez21,jimenez22a,missaglia23}, covering all 3CR sources lacking X-ray observations. Despite a small number of 3CR sources that are still unidentified and unobserved in the X-rays \citep{maselli16,missaglia21}, more than 95\% of the 3CR catalog has high energy data already available in the \chn\ archive \citep[see e.g.,][]{massaro15a}. 

However, the 3CR has, unfortunately, the following drawbacks and limitations. It was created more than six decades ago and the largest fraction of 3CR radio, infrared and optical data collected to date was obtained with relatively old instruments, only rarely repeated and/or updated. It lists radio sources lying in the Northern Hemisphere, with limited access to observations that can be performed with state-of-the-art and upcoming astronomical facilities,  such as Multi Unit Spectroscopic Explorer \citep[MUSE;][]{bacon10} mounted at the Very Large Telescope (VLT), the Atacama Large Millimeter/submillimeter Array (ALMA) and in the near future the world's most powerful radio telescope: Square Kilometre Array\footnote{https://www.skao.int} \citep[SKA;][]{mcmullin20}, the Large Synoptic Survey Telescope \citep[LSST;][]{ivezic19} as well as the Extremely Large Telescope\footnote{elt.eso.org} (ELT). 

In the eighties the Molonglo Reference Catalog of Radio Sources \citep[MRC;][]{large81} containing nearly 12000 discrete sources with flux densities greater than 0.7\,Jy at 408\,MHz in the declination range between +18.5$^\circ$ and -85$^\circ$ and excluding regions within 3$^\circ$ of the Galactic equator was created. Several multifrequency campaigns were then dedicated to augment the information of MRC sources, eventually restricted to bright samples \citep[see e.g.][]{mccarthy96b,kapahi98}. 

A first attempt to create a complete sample similar to the 3CR, but selected at 408\,MHz, was performed by Best et al. (1999) using the Molonglo Reference Catalogue. They selected a sample listing 178 radio sources with flux density $S_{408}$ above 5\,Jy, in the range of declination between -30$^\circ$ and 10$^\circ$ and having Galactic latitudes $|b|>=10^\circ$. The equatorial location of all sources listed therein allowed them to achieve high spectroscopic completeness, and its footprint certainly mitigating one of the previously mentioned limitations of the 3CR: visibility from Southern Hemisphere telescopes. 

An additional attempt to build a catalog equivalent to the 3CRR was carried out by Burgess \& Hunstead in 2006, starting from the Molonglo Southern 4\,Jy sample \citep[MS4; see also][]{burgess06b}. The MS4 is a complete sample of 228 southern radio sources detected at 408\,MHz with integrated flux densities above 4.0\,Jy, Galactic latitude $|b|>$10$^\circ$ and declination in the range between -85$^\circ$ and +30$^\circ$, all imaged at 843\,MHz with the Molonglo Observatory Synthesis Telescope to obtain positions with an accuracy of $\sim$1\,\arcsec. Then radio spectra for the MS4 sources were compiled from the literature to estimate flux densities at 178\,MHz and the subset of SMS4, with $S_{178}>$9\,Jy was extracted. Some sources listed in the SMS4 were also recently observed in the soft X-rays \citep{maselli22}.

The recent GaLactic and Extragalactic All-sky Murchison Widefield Array (MWA) survey \citep[GLEAM;][]{wayth15,hurley17} offers today a unique opportunity to create a southern sample of powerful radio sources matching the selection criterion of the 3CR catalog. Thanks to the MWA observations available for the whole southern sky at declinations $\delta<30^\circ$ in the frequency range between 72 and 231 MHz, White et al. (2018, 2020a, 2020b) built a complete sample of radio sources with flux density above 4\,Jy at 151\,MHz, namely the GLEAM 4-Jy sample (hereinafter labelled as G4Jy). The majority of radio sources included therein are extragalactic, mainly AGNs with extended structures detected at low radio frequencies (i.e., at hundreds of MHz). This sets the basis for extracting, from the G4Jy catalog, a new sample of radio sources equivalent to the 3CR (hereinafter \cs) but including only those located in the Southern Hemisphere. 

Here we introduce the \cs\ catalog, listing the 264 very brightest radio sources, selected from the larger parent G4Jy catalog, above a flux density threshold of 9\,Jy at $\sim$178 MHz, as the nominal value of the 3C, at declinations below -5$^\circ$. There are several differences with respect to the previous analysis carried out on the G4Jy catalog as listed below. In the present work we restricted our investigation to a small fraction of sources, 264 with respect to $\sim$1900 listed in the G4Jy catalog, with the potential advantage of performing deeper analyses with more astronomical facilities and instruments, but with the disadvantage of a limited number of high $z$ sources implying less robust claims from a statistical perspective. On the other hand, radio sources listed in the \cs\ sample are potentially primary targets for SKA, being the brightest ones and preparing the sample before its advent allows us to start collecting multifrequency observations that will be crucial to investigate their nature and that of their environments.

This first paper is mainly devoted to the comparison between radio images, at higher resolution than that achievable with previously available radio maps, with mid-IR and optical archival observations to confirm counterparts previously assigned to each radio source and/or determine potential incorrect associations. The analysis presented here is based on archival radio maps with higher angular resolution than those used for the G4Jy associations, such as those retrieved from the Very Large Array (VLA) Sky Survey \citep[VLASS;][]{lacy20} and the National Radio Astronomy Observatory (NRAO) VLA Archive Survey (NVAS)\footnote{http://www.vla.nrao.edu/astro/nvas/} databases, available for at least 60\% of the \cs\ catalog. This counterpart search is crucial to identify targets for future spectroscopic campaigns that are necessary to obtain source redshifts and their optical classification. We also present results of an extensive literature search carried out to obtain redshift estimates for \cs\ sources.  
 
The paper is structured as follows. In \S~\ref{sec:sample} we present the criteria underlying the \cs\ sample selection. In \S~\ref{sec:counterparts} we present the results of our search for optical counterparts and a comparison with the mid-IR sources associated with the G4Jy catalog, while \S~\ref{sec:literature} is dedicated to a brief description of a literature search for redshift estimates. Summary, conclusions and future perspectives are given in \S~\ref{sec:summary}. Appendix A is dedicated to a brief description of individual sources while Appendix B is devoted to the cross-identifications obtained comparing the \cs\ sample with other radio catalogs based on observations carried out in the Southern Hemisphere. Finally, Appendix C is dedicated to a statistical test for the radio and mid-infrared crossmatches carried out to search for counterparts of G4Jy sources, in comparison with the optical analysis presented here.

We adopt cgs units for numerical results and we assume a flat cosmology with $H_0=69.6$ km s$^{-1}$ Mpc$^{-1}$, $\Omega_\mathrm{M}=0.286$ and $\Omega_\mathrm{\Lambda}=0.714$ \citep{bennett14}. For optical photometric data we used the catalog obtained from the Panoramic Survey Telescope \& Rapid Response System \citep[Pan-STARRS;][]{flewelling20} survey, where magnitudes are reported in the AB system \citep{oke74,oke83}. The same applies for the Dark Energy Survey \citep[DES;][]{abbott18} for which observations are performed in optical filters similar to those of Pan-STARRS and Sloan Digital Sky Survey \citep[SDSS;][]{ahn12}. Limiting sensitivity for both the Pan-STARRS and DES optical survey reaches $\sim$23 mag in the $r$ band. For optical magnitudes we did not apply the correction for Galactic extinction but we report the total extinction $A_V$, extracted from the Galactic Dust Reddening and Extinction database\footnote{https://irsa.ipac.caltech.edu/applications/DUST/} \citep{schlegel98,schlafly11}. {Spectral indices, $\alpha$, are defined by flux density, S$_{\nu}\propto\nu^{-\alpha}$.} Finally, given the large number of acronyms used in the paper, these are summarized in Table \ref{tab:acronyms}.
\begin{table}[h]
\tiny
\caption{Table of acronyms used across the paper.}           
\label{tab:acronyms}      
\begin{tabular}{ll}
 \hline
ACRONYM & MEANING \\
\hline
& Generic \\
\hline
AGN & Active Galactic nucleus \\ BCG & Brightest Cluster Galaxy \\ BLRG & Broad emission Line Radio Galaxy \\ CSS & Compact Steep Spectrum radio source \\ FSRQ & Flat Spectrum Radio Quasar \\ GPS & Gigahertz Peaked-Spectrum \\ HyMoR & Hybrid Morphology Radio Source \\ ICM & Intracluster Medium \\ HERG & High Excitation Radio Galaxy \\ LERG & Low Excitation Radio Galaxy \\ NLRG & Narrow emission Line Radio Galaxy \\ QSO & Quasi Stellar Object \\ USS & Ultra Steep Spectrum radio source\\ WAT & Wide-Angle Tail radio galaxy \\ \hline
& Institutes \\
\hline
NRAO & National Radio Astronomy Observatory  \\ TIFR & Tata Institute of Fundamental Research  \\ \hline
& Telescopes \& Instruments \\
\hline
ALMA & Atacama Large Millimeter/submillimeter Array \\ ATCA & Australia Telescope Compact Array \\ CASLEO & Complejo Astron\'omico El Leoncito \\ELT & Extremely Large Telescope \\ GMRT & Giant Metrewave Radio Telescope  \\ HST & Hubble Space Telescope \\ LSST &  Large Synoptic Survey Telescope \\ MUSE & Multi Unit Spectroscopic Explorer  \\ MWA & Murchison Widefield Array \\ SKA & Square Kilometre Array \\ VLT & Very Large Telescope \\ WISE &  Wide Infrared Survey Explorer  \\ \hline
& Catalogs \& Surveys \\
\hline
2Jy & Bright extragalactic radio sources at 2.7 GHz \\ 3C & Third Cambridge Catalog  \\ 3CR & Third Cambridge Catalog Revised \\ 6dFGS & Six-degree Field Galaxy Survey \\ AT20G & Australia Telescope 20-GHz Survey Catalog \\ CRATES & Combined Radio All-Sky Targeted Eight GHz Survey \\ DES & Dark Energy Survey \\ DSS & Digital Sky Survey \\ GLEAM & GaLactic and Extragalactic All-sky MWA  \\ G4Jy & GLEAM 4-Jy sample \\ MRC & Molonglo Reference Catalog of Radio Sources \\ MS4 & Molonglo Southern 4\,Jy sample \\ MURALES & MUse RAdio Loud Emission line Snapshot \\ NED & NASA Extragalactic Database \\ NVAS & NRAO VLA Archive Survey \\ NVSS & NRAO VLA Sky Survey \\ Pan-STARRS & Panoramic Survey Telescope \& Rapid Response System \\ PMN & Parkes-MIT-NRAO Surveys \\ PKSCAT90 & Parkes radio catalog \\ SDSS & Sloan Digital Sky Survey \\ SUMSS & Sydney University Molonglo Sky Survey \\ TGSS & TIFR GMRT Sky Survey \\ TXS & Texas Survey of Radio Sources at 365 MHz \\ VLA & Very Large Array \\ VLASS & VLA Sky Survey \\ VLSSr & VLA Low-Frequency Sky Survey Redux \\ \hline
\end{tabular}\\
\end{table}

\section{Sample selection}
\label{sec:sample} The MWA, operating since 2013 and being the SKA precursor at low radio frequencies \citep{tingay13}, performed the GLEAM survey\footnote{https://www.mwatelescope.org/gleam} \citep{wayth15}. The extragalactic GLEAM catalog \citep{hurley17} covers $\sim$25000 square degrees, at declinations south of +30$^\circ$ and Galactic latitudes $|b|>10^\circ$, and excluding some regions such as the Magellanic Clouds. It lists $\sim$3$\times$10$^5$ radio sources with 20 separate flux density measurements across the frequency range 72-231 MHz, selected from a time- and frequency-integrated image centered at 200 MHz, with an angular  resolution of $\sim$2 arcmin. Based on the GLEAM survey and selecting all radio sources with $S_{151}$ above 4\,Jy, White et al. (2020a,b) built the G4Jy catalog \citep[see also][]{wayth15}, a flux limited sample listing nearly 2000 sources over an area of $\sim$ 25000 square degrees. The selection criterion on the Galactic latitudes (i.e., Galactic latitudes $|b|>10^\circ$) was used to lower the contamination of Galactic sources and focus on the extragalactic sky. This sets the basis to extract a southern catalog of extragalactic radio sources: the \cs, fully equivalent, in terms of radio flux density selection, to the northern 3CR extragalactic sample, that lists powerful radio sources at declinations above -5$^\circ$ and having a flux density higher than 9\,Jy at 178\,MHz \citep{bennett62,spinrad85}. We remark that the brightest sources at declinations below +30$^\circ$ and Galactic latitudes $|b|>10^\circ$, including the Orion Nebula, were all masked in the GLEAM extragalactic catalog and thus are not listed in the G4Jy Sample \citep[see][and references therein for a list of them and additional details]{white20a}.

Sources listed in the \cs\ are selected to have (i) $Dec.<-5^\circ$ and (ii) flux densities at 174 MHz and at 181 MHz, integrated over the GLEAM bands, above the following thresholds: $S_{174}>$8.13 Jy and  $S_{181}>$7.85 Jy, respectively\footnote{These thresholds are computed assuming a power-law description for the radio spectrum of the G4Jy sources and adopting the spectral index reported in the G4Jy catalog}. The flux density thresholds adopted here correspond to the 9\,Jy limiting sensitivity at $\sim$178 MHz, the nominal value adopted to prepare the 3CR. Sources close to the Galactic plane (i.e., at Galactic latitudes $|b|<$10$^\circ$) are excluded since they were not originally listed in the G4Jy catalog \citep{white20a}. All sources listed in the \cs\ catalog are visible from the Southern Hemisphere and lie in a footprint not covered by the 3CR catalog, whereas the handful of targets in common were removed. The final \cs\ catalog lists 264 radio sources. At declinations below +20$^\circ$ there are 106 out of 298 3CR objects thus the new \cs\ sample includes about three times the number of powerful radio sources visible with southern telescopes. 

After the release of the 3CR, several radio analyses \citep[see e.g.,][]{roger73} were carried out to refine the sample, some of them led to the 3CRR release \citep{laing83}. The flux density threshold adopted to create the G4Jy-3CRE, even if similar to the nominal value of the 3CR, allowed us to create a comparable statistical sample for radio sources mainly visible from the Southern Hemisphere, with access to the state-of-the-art observing facilities there. For reference the 3CR lists 298 radio sources while the 9\,Jy cut used for the G4Jy-3CRE results in a list of 264 objects. However, to make a more precise comparison with the 3CR, we added a flag to all 181 radio sources out of 264 filtered for having radio flux densities above 9.8\,Jy\footnote{This flux density threshold was computed at 178\,MHz  extrapolating that between 174\,MHz  and 181\,MHz reported in the G4Jy catalog using the radio spectral index between these two frequencies. It corresponds to that used to create the original 3C catalog taking into account refined intercalibrations.} at 178\,MHz, this threshold has been computed to account calibration differences with respect to original 3C observations \citep[see e.g.,][and references therein]{roger73}.

The analysis of White et al. (2020a, 2020b) revealed that in the G4Jy catalog 86\% of the radio sources appear to be associated with a mid-infrared (mid-IR) counterpart, detected in the all-sky survey performed with the Wide Infrared Survey Explorer \citep[WISE;][]{wright10}. In the present analysis we only report WISE images at 3.4$\mu$m but we refer to mid-IR counterparts of G4Jy sources considering those listed in the allWISE data release and thus based on the detection in all filters \citep{cutri12,cutri13}. In our selected sample, 225 \cs\ sources (i.e., $\sim$85\%) are associated with a mid-IR counterpart detected in the WISE all-sky survey \citep{white20b}. 

Radio positions reported in the G4Jy were then computed using NRAO VLA Sky Survey \citep[NVSS;][]{condon98} and/or Sydney University Molonglo Sky Survey \citep[SUMSS;][]{mauch03} images while data available from the Tata Institute of Fundamental Research (TIFR) Giant Metrewave Radio Telescope (GMRT) Sky Survey \citep[TGSS;][]{intema17} were inspected to obtain a better overview of the radio structure. Thus for sources at declinations above -39.5$^\circ$, the position reported in the NVSS catalog was used, while for the remaining ones the one reported in the SUMSS catalog was adopted. Thus, given their relatively low angular resolution, there are sources, having an unusual radio morphology and/or being asymmetric, for which the location of the radio core, and consequently that of their host galaxy, was not clearly identified. For these reasons, as we described below, we expended considerable effort to investigate higher resolution radio observations.  

\begin{figure}[]
\begin{center}
\includegraphics[height=6.8cm,width=9.6cm,angle=0]{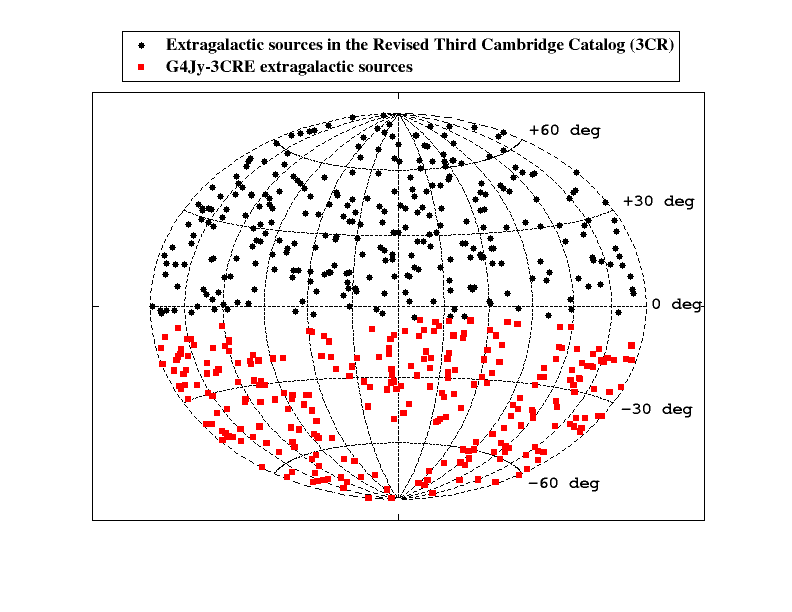}
\caption{The sky distribution, shown in the Hammer-Aitoff projection, of the 3CR sources located in the Northern Hemisphere (black circles) in comparison with those belonging to the G4Jy-3CRE catalog (red squares), equivalent to the 3CR but located in the Southern Hemisphere. It is clear how footprints of both surveys are complementary and the region of the Galactic plane was excluded to select the \cs\ catalog.}
\label{fig:aitoff}
\end{center}
\end{figure}

In Figure~\ref{fig:aitoff} we show the sky distribution, plotted adopting the Hammer-Aitoff projection, of the 3CR sources in the Northern Hemisphere alongside their equivalent sample of southern celestial objects, \cs, to highlight their complementary footprints. Sources at Galactic latitudes $|b|<$10$^\circ$ are excluded since they were not originally listed in the G4Jy catalog \citep{white20a}. 

\section{Searching for optical counterparts}
\label{sec:counterparts} \begin{table*}[!ht]
\caption{Details on the identification flags (IDF) adopted in our analysis and list of examples.}
\label{tab:flags}
\begin{center}
\begin{tabular}{clcc}
\hline
IDF & criteria & Example & Cases\\
\noalign{\smallskip}
\hline 
\noalign{\smallskip}
1.0 & optical counterpart $\equiv$ mid-IR counterpart & G4Jy\,312 & 184 (70\%)\\
2.1 & optical counterpart $\neq$ mid-IR counterpart  & G4Jy\,524 & 5 (2\%)\\
2.2 & optical counterpart but no mid-IR counterpart  & G4Jy\,593 & 21 (8\%)\\
3.0 & confused (not possible to assign either an optical and/or a mid-IR counterpart) & G4Jy\,934 & 14 (5\%)\\
4.1 & no optical counterpart but mid-IR counterpart & G4Jy\,1846 & 19 (7\%)\\
4.2 & no optical counterpart \& incorrect mid-IR counterpart & G4Jy\,1057 & 4 (2\%)\\
4.3 & no optical counterpart \& no mid-IR counterpart & G4Jy\,1587 & 17 (6\%)\\
\noalign{\smallskip}
\hline
\end{tabular}\\
\end{center}
\end{table*}

\subsection{Identification flags}
We identified optical counterparts of radio nuclei overlaying radio contours on optical images.  The same comparison with mid-IR images of the WISE all-sky survey\footnote{https://wise2.ipac.caltech.edu/docs/release/allwise/} was then crucial to verify both (i) the presence of a mid-IR counterpart and (ii) the robustness of all associations reported in the original G4Jy catalog. As previously stated, we only report in our analysis a comparison between radio maps and mid-IR images at 3.4$\mu$m and the mid-IR associated counterpart as in the G4Jy based on detections of the AllWISE survey \citep{cutri12}.

We initially explored several radio databases to search for radio images of sources listed in the \cs\ sample to identify the correct position of their radio cores. At radio frequencies we compared GLEAM images with those available in the databases of the VLA Low-Frequency Sky Survey Redux\footnote{http://cade.irap.omp.eu/dokuwiki/doku.php?id=vlssr} \citep[VLSSr;][]{cohen07}, TGSS\footnote{https://tgssadr.strw.leidenuniv.nl/doku.php}, NVSS\footnote{https://www.cv.nrao.edu/nvss/} and VLASS\footnote{https://public.nrao.edu/vlass/} \citep{lacy20} for sources visible from the Northern Hemisphere while we used mainly the SUMSS\footnote{http://www.astrop.physics.usyd.edu.au/sumss/} for those lying out of footprints of previous surveys. High resolution radio surveys, such as the VLASS, allowed us to obtain a precise measurement of the radio core location. 

To achieve our goals we additionally checked the NVAS radio archive, mainly searching for radio images at 1.4\,GHz, 5\,GHz and 8\,GHz, however these were useful only for a handful of sources, allowing us to clearly detect the position of their radio cores. NVAS radio maps, with angular resolution in the range between 0\arcsec.5 and $\sim$10\,\arcsec, are available for 124 out of 264 sources listed in the \cs. Here, host galaxy search and identification are partially biased, being more efficient, towards sources located above declinations of $\sim$-40$^\circ$ where JVLA data are available.

For optical images, we retrieved optical images for 178 \cs\ sources from the Pan-STARRS\footnote{https://panstarrs.stsci.edu} and the DES\footnote{https://www.darkenergysurvey.org} databases, while for the remaining 86 objects, lying outside the footprints of these surveys, we inspected observations/images provided in the red filter of the Digital Sky Survey\footnote{https://archive.eso.org/dss/dss} (DSS). 

\begin{figure*}[!ht]
\begin{center}
\includegraphics[height=6.cm,width=16.2cm,angle=0]{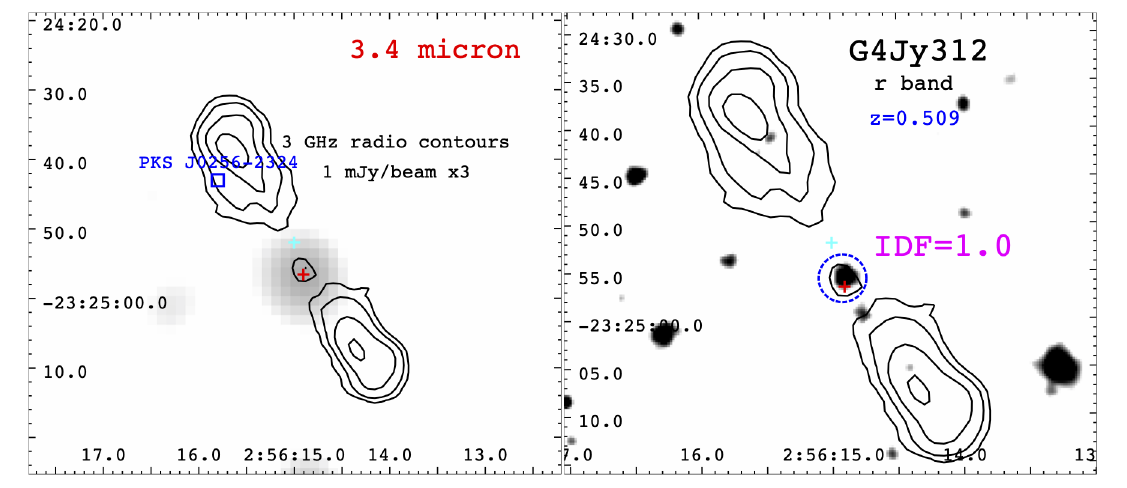}
\end{center}
\caption{(Left panel) The 3.4$\mu$m image available thanks to the WISE
    all-sky survey, with radio contours overlaid in black. The frequency of
    the radio map from which the radio contours were drawn is reported in
    the figure together with the intensity of the first level and the
    binning factor. The symbol x3, reported in the image indicates that
    radio contours, starting at 1 mJy/beam level, increase by a factor of
    three. Radio maps from VLASS, NVSS and SUMSS archives correspond to a
    nominal frequency of 3\,GHz, 1.4\,GHz and 843\,MHz, respectively. (Right
    panel) The optical image collected from one of the optical surveys used
    in our analysis. If, below the source name, the label ``red filter''  is
    reported then the optical image is collected from the DSS archive while
    when it is written ``$r$ band'', as in this case, optical images were
    retrieved from  Pan-STARRS or DES databases. Radio contours are also
    overlaid on the optical image. The red cross, if present, marks the
    position of the mid-IR counterpart associated in the G4Jy catalog while
    the cyan cross corresponds to the position of the brightness-weighted
    radio centroid of the G4Jy catalog \citep{white20a,white20b}. The blue
    dashed circle, if present, indicates the position of the optical
    counterpart identified from our analysis. Both these images are an
    example of a source, i.e., G4Jy\,312, having an IDF=1.0, for which our
    analysis revealed that the mid-IR counterpart associated in the G4Jy
    catalog corresponds to the optical source lying at the same position of
    the radio nucleus in the high angular resolution radio map used here.
    The blue open square or the blue X symbol, if present in the left panel,
    mark the location of the closest radio source belonging to the Parkes
    radio catalog \citep[PKSCAT90,][]{bolton79} or to the Molonglo Reference
    Catalog of Radio Sources \citep[MRC,][]{large81}, respectively (see also
    Appendix C for additional details about radio cross identifications).
    \textbf{The complete figure set (264 images), showing finding charts for all 
    sources in the G4Jy catalog, is available in the online journal.}}
\label{fig:example1}
\end{figure*}

\begin{figure*}[!ht]
\begin{center}
\includegraphics[height=6.cm,width=16.2cm,angle=0]{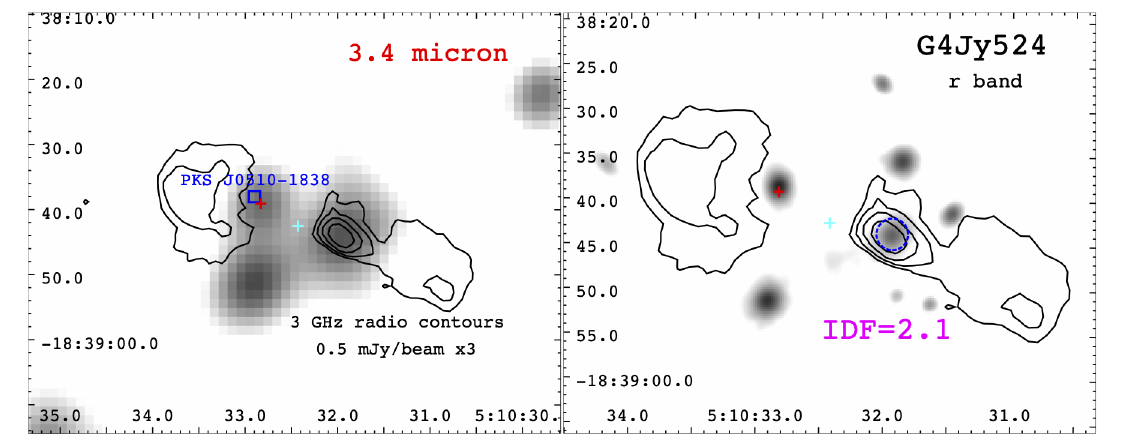}
\includegraphics[height=6.cm,width=16.2cm,angle=0]{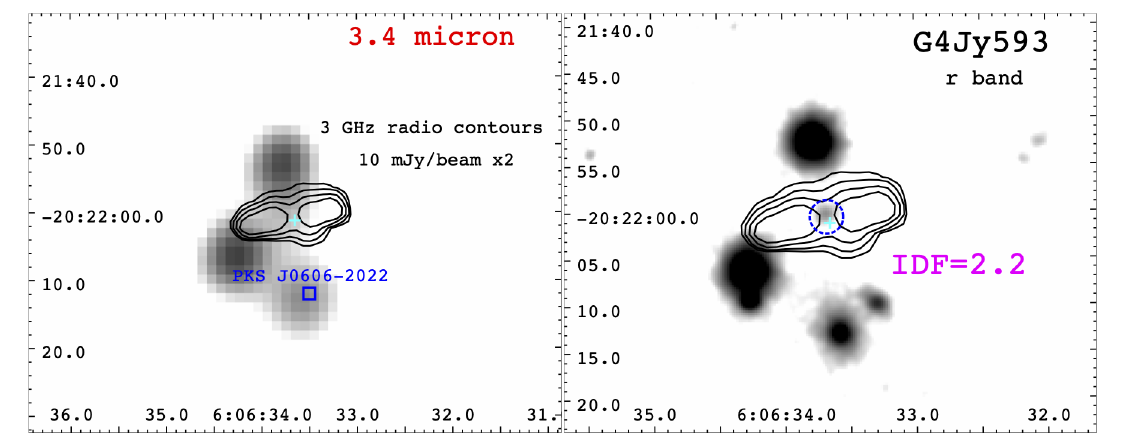}
\end{center}
\caption{(Top panels) Same as Figure~\ref{fig:example1} but for the case of G4Jy\,524, a radio source for which the mid-IR counterpart identified in the G4Jy catalog is different from the optical one that corresponds to the position of the radio core detected in the VLASS radio map at 3\,GHz. Sources showing the same behavior were marked with an IDF=2.1 in our analysis. (Bottom panels) Same as Figure~\ref{fig:example1} but for G4Jy\,593. This radio source, having IDF=2.2, is an example of those cases for which there is a clear optical counterpart but they lack an assigned mid-IR source in the G4Jy catalog.}
\label{fig:example2}
\end{figure*}

\begin{figure*}[!ht]
\begin{center}
\includegraphics[height=6.cm,width=16.2cm,angle=0]{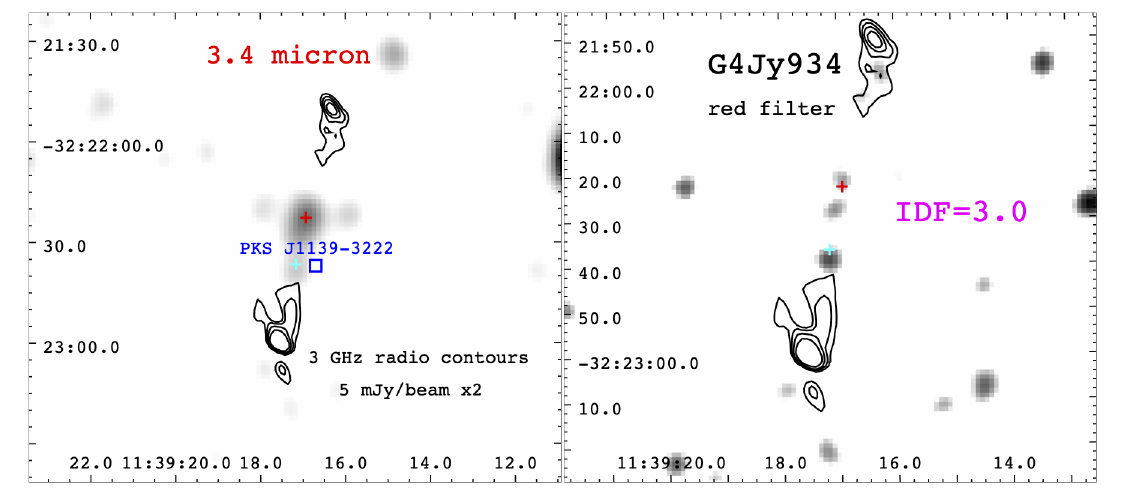}
\end{center}
\caption{Same as Figure~\ref{fig:example1} for G4Jy\,934. The lack of a high angular resolution radio map, in this case the one from which radio contours are computed was collected from the SUMSS archive, prevented us to clearly identify the host galaxy of the radio source. These cases are flagged as ``confused'' in our analysis and have IDF=3.0}
\label{fig:example3}
\end{figure*}

\begin{figure*}[!ht]
\begin{center}
\includegraphics[height=6.cm,width=16.2cm,angle=0]{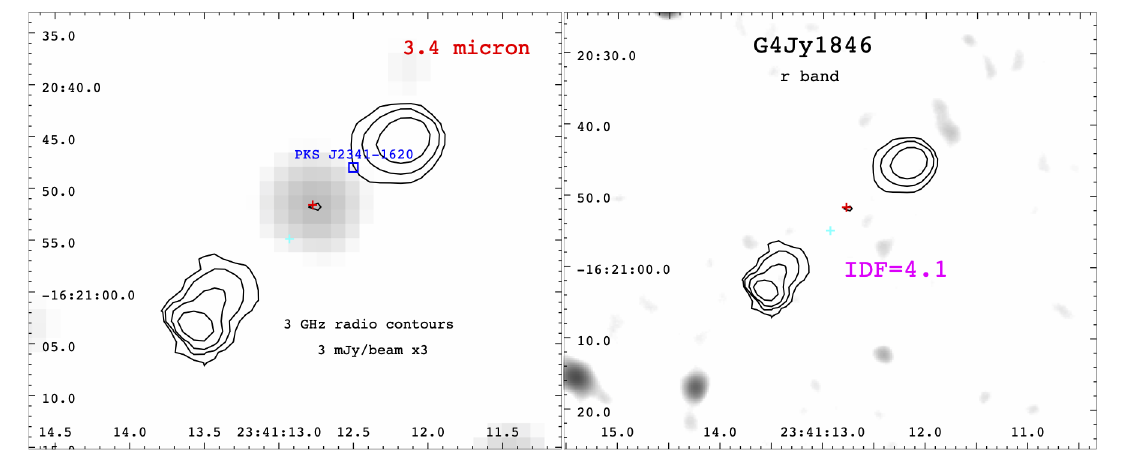}
\includegraphics[height=6.cm,width=16.2cm,angle=0]{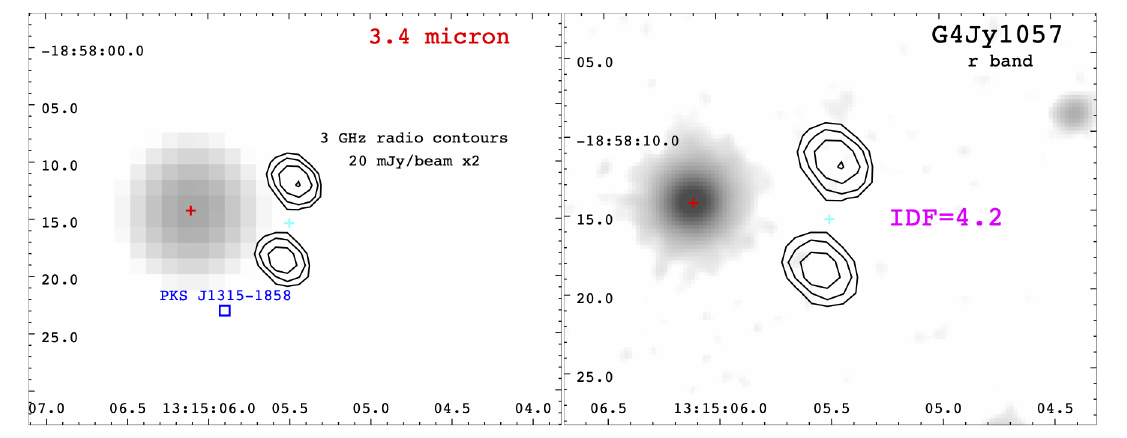}
\includegraphics[height=6.cm,width=16.2cm,angle=0]{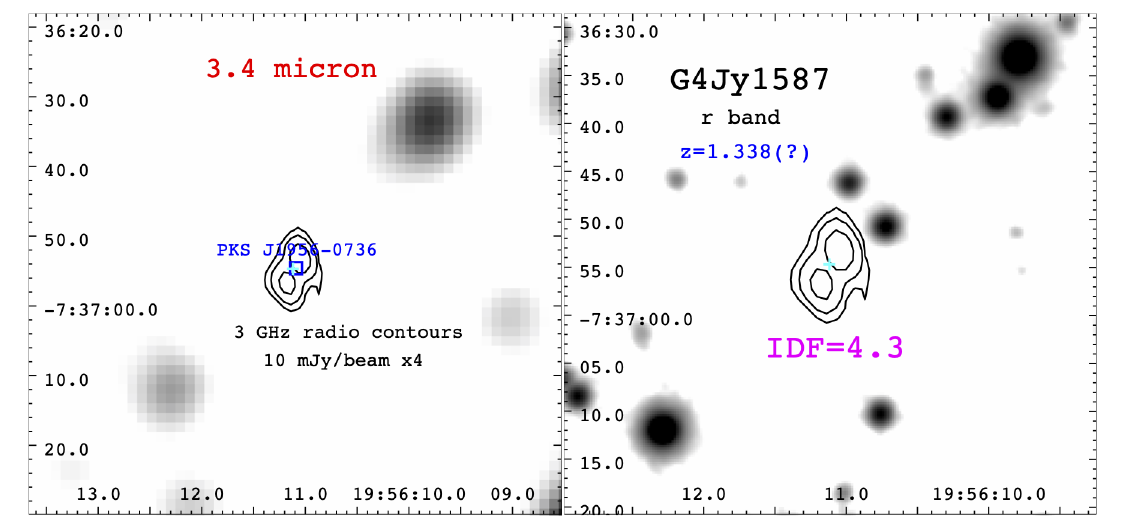}
\end{center}
\caption{(Top panels) Same as Figure~\ref{fig:example2} for G4Jy\,1846. As shown in the right panel, the optical counterpart of the radio nucleus is too faint to be detected in the images available from Pan-STARRS, DES and the DSS databases, however it is clear from the left panel that the radio source has an infrared counterpart. These cases are indicated with IDF=4.1. (Central panels) Same images for G4Jy\,1057 but in this case, despite the lack of a plausible optical counterpart the radio map available indicates that the assigned mid-IR counterpart appears to be incorrect, thus marked in our analysis with IDF=4.2. (Bottom panels) Same as above panels but for sources as G4Jy\,1587 that lack both a mid-IR and an optical counterpart of its radio core. G4Jy\,1587 has IDF=4.3. These are radio sources for which we also visually inspected other optical images available in the Pan-STARRS and in the DES databases in the $g$, $i$, $z$ and $y$ bands searching for signatures of their host galaxies.}
\label{fig:example4}
\end{figure*}

We adopted the following identification flags (IDFs) with a summary of all criteria described below. Association results based on these IDFs are all summarized in Table~\ref{tab:flags}, together with several examples. In Table~\ref{tab:main} we report all IDFs assigned to each source listed in the \cs, together with other parameters. For the IDFs we distinguished the following cases:

\begin{itemize}
\item \underline{IDF=1.0}: sources for which the position of the radio core is coincident with the mid-IR source associated in the original G4Jy catalog and it also corresponds to a unique optical source detected in the $r$-band of the Pan-STARRS/DES catalogs or in the red filter image obtained with the DSS, as shown for G4Jy\,312 in Figure~\ref{fig:example1};
\item \underline{IDF=2.0}: sources having an optical counterpart of their radio core. However, on the basis of the mid-IR counterpart, we distinguished:
\begin{itemize}
\item \underline{IDF=2.1}: those sources having the radio core associated with a unique optical counterpart, different from the mid-IR one listed in the original G4Jy catalog (IDF=2.1, as reported for G4Jy\,524 in Figure~\ref{fig:example2});
\item \underline{IDF=2.2}: those cases for which there is an optical counterpart but lacking an associated mid-IR source (i.e., IDF=2.2, as shown for G4Jy\,593 in Figure~\ref{fig:example2}).
\end{itemize}
According to the analysis performed by White et al. (2020a,2020b) the latter ones are cases for which the identification of the host galaxy was partially or completely limited by the mid-IR data, being either undetected in the AllWISE survey or having its emission affected by a relatively bright nearby object; 

\item \underline{IDF=3.0}: sources for which the low angular resolution of radio observations did not allow us to uniquely identify the optical counterpart (IDF=3.0, as reported for G4Jy\,934 in Figure~\ref{fig:example3}), despite the presence of an associated mid-IR source. These are simply ``confused'' cases;

\item \underline{IDF=4.0}: sources for which there is no optical counterpart, being undetected at the sensitivity limit of the survey data we used to search for it. We also distinguished here several subcategories:
\begin{itemize}
\item \underline{IDF=4.1}: those lacking an optical counterpart but with an assigned mid-IR counterpart in the original G4Jy catalog (IDF=4.1, as shown for G4Jy\,1846 in Figure~\ref{fig:example4}); 
\item \underline{IDF=4.2}: those for which there is no optical counterpart but the radio map used allowed us to verify that the previously assigned mid-IR one is incorrect (IDF=4.2, as shown for G4Jy\,1057 in Figure~\ref{fig:example4});
\item \underline{IDF=4.3}: those lacking counterparts at both mid-IR and optical frequencies (IDF=4.3, as reported for G4Jy\,1587 in Figure~\ref{fig:example4}).
\end{itemize}

\end{itemize}

Here we assumed that a source is detected only if it is reported in the catalog corresponding to each survey used in the current analysis, thus having a detection threshold equal to the level of confidence of the survey itself. No associations between radio and mid-IR/optical counterparts were considered reliable if the angular separation between their positions is greater than 5.\arcsec4. This is also supported by the seeing of optical surveys being $\sim$2\,\arcsec\ and by the statistical analysis reported in \S~\ref{sec:counterparts}.

In Figures~\ref{fig:example1}, \ref{fig:example2}, \ref{fig:example3} and \ref{fig:example4}, we show WISE images collected at 3.4$\mu$m with radio contours overlaid. The frequency of the radio map from which radio contours were drawn is reported in the figure together with the intensity of the first level. All radio contours increase in level by a binning factor also reported in the figure. Radio maps obtained through the VLASS, NVSS and SUMSS archives correspond to a nominal frequency of 3\,GHz, 1.4\,GHz and 843\,MHz, respectively. We also show optical images collected from one of the surveys used in our analysis. If the optical image has a label ``red filter'' it was obtained from the DSS archive, while ``$r$ band'' marks those retrieved from Pan-STARRS or DES archives. Radio contours are also overlaid on optical images. The red cross, if present, marks the position of the associated mid-IR counterpart according to the G4Jy catalog, while the cyan cross corresponds to the position of brightness-weighted radio centroids reported in the G4Jy \citep{white20a, white20b}. The blue dashed circle, whenever present, indicates the position of the optical counterpart identified in our analysis. We remark that the scale of the mid-IR and the optical images are different on purpose. The underlying reasons are: (i) the angular resolution is different and (ii) mid-IR images, reported with a larger field of view, allow us to highlight the large-scale radio structure, while optical images were mainly used to identify counterparts as finding charts.

\begin{figure}[]
\includegraphics[height=5.2cm,width=9.cm,angle=0]{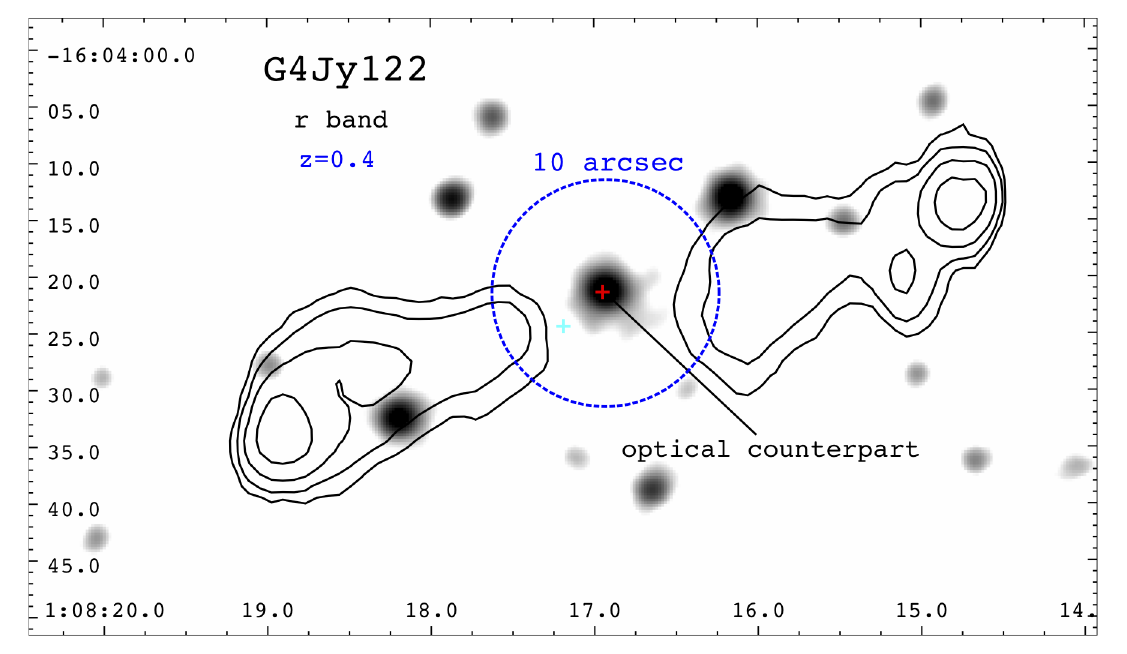}
\caption{Same as right panel of Figure~\ref{fig:example2} for G4Jy\,122. The dashed blue circle, having 10\,\arcsec\ radius, is centered on the location of the potential optical counterpart identified here since (i) it lies between the two radio lobes of G4Jy\,122 showing a classical FR\,II radio morphology and (ii) there are no other optical sources within this area.}
\label{fig:example5}
\end{figure}

Sources having IDF=1.0 (namely, those for which the resolution of radio maps allows us to firmly establish the position of the radio core) include also cases for which we do not clearly detect the radio nucleus but the brightness-weighted radio centroid of all radio maps used is coincident with a unique optical counterpart with no nearby companions, within an angular separation of 5\,\arcsec--10\,\arcsec. An example is shown in Figure~\ref{fig:example5}, where we report the case of G4Jy\,122, a classical FR\,II radio galaxy at $z$=0.4, for which the radio morphology and the lack of optical sources within an angular separation of 10\,\arcsec\ from the potential optical counterpart allowed us to assigned it an IDF=1.0. We emphasize that for several sources we also used archival radio maps, collected from the NVAS database, that allowed us to unequivocally assign an optical counterpart (i.e., IDF=1.0  rather than 3.0). 
\begin{table*}[]
\caption{Full list of the G4Jy-3CRE catalog}
\label{tab:main}
\begin{center}
\begin{tabular}{rlllllclclc}
\hline
G4Jy & WISE & R.A.$^{(r)}$ & Dec.$^{(r)}$ & R.A.$^{(o)}$ & Dec.$^{(o)}$ & IDF & $z$ & r & $A_V$ & sub.\\
name & name & (J2000) & (J2000) & (J2000) & (J2000) & & & (mag) & (mag) & \\
\noalign{\smallskip}
\hline 
\noalign{\smallskip}
4 & J000322.14-172714.1 & 00:03:22.001 & -17:27:11.412 & 00:03:21.9430 & -17:27:11.663 & 2.1 & 1.465(?) & 19.24 & 0.079 & \\
9 & J000557.94-562831.0 & 00:05:57.650 & -56:28:31.512 & 00:05:57.8611 & -56:28:30.930 & 1.0 & 0.2912 & 17.05 & 0.034 & \\
12 & J000616.50-830608.0 & 00:06:11.989 & -83:05:56.911 & 00:06:16.5559 & -83:06:07.280 & 1.0 & -- &  & 0.578 & \checkmark \\
20 & J001030.14-442257.1 & 00:10:30.550 & -44:22:57.000 & -- & -- & 3.0 & -- &  & 0.03 & \checkmark \\
26 & J001524.35-380438.6 & 00:15:24.283 & -38:04:35.375 & -- & -- & 3.0 & -- &  & 0.045 & \\
27 & J001602.93-631003.8 & 00:16:02.681 & -63:10:07.212 & 00:16:02.8827 & -63:10:04.436 & 1.0 & -- & 22.8 & 0.053 & \checkmark \\
33 & J001851.42-124233.7 & 00:18:51.379 & -12:42:33.516 & 00:18:51.3821 & -12:42:34.041 & 1.0 & 1.589 & 22.96 & 0.11 & \checkmark \\
43 & J002308.86-250229.6 & 00:23:09.341 & -25:02:35.362 & 00:23:08.8739 & -25:02:29.183 & 1.0 & 0.35 & 20.41 & 0.043 & \\
45 & J002430.15-292854.3 & 00:24:30.120 & -29:28:48.900 & 00:24:30.1273 & -29:28:54.426 & 1.0 & 0.40645 & 17.5 & 0.073 & \checkmark \\
48 & J002549.22-260212.3 & 00:25:49.169 & -26:02:12.804 & 00:25:49.1681 & -26:02:12.786 & 1.0 & 0.32188 & 19.85 & 0.049 & \checkmark \\
\noalign{\smallskip}
\hline
\end{tabular}\\
\end{center}
Note: col. (1) the name reported in the G4Jy catalog; col. (2) the name of the assigned mid-IR counterpart, detected in WISE, as in the G4Jy catalog; col. (3,4) right ascension (R.A.) and declination (Dec.), in J2000  Equinox, of the brightness-weighted radio centroid collected from the G4Jy catalog; col. (5,6) same as previous two columns but measured from the centroid of the optical counterpart in the Pan-STARRS, DES and DSS images; col. (7) identification flag (IDF) adopted in our analysis (see \S~\ref{sec:counterparts} for all details); col. (8) the redshift value reported in the literature, where question marks highlight those with uncertain estimates; col. (9) $r$-band magnitude from the Pan-STARRS and DES counterparts; col. (10) Galactic extinction; col. (11) The check mark indicates if the source belong to the subsample selected with radio flux density above 9.8\,Jy at $\sim$178\,MHz (see \S~\ref{sec:sample} for more details). The entire table is reported in the on-line version of the journal while only the first 10 lines are shown here for guidance regarding its form and content. \\
\end{table*}

Once we identified the optical counterpart we also searched both the latest releases of the Pan-STARRS and the DES catalogs to obtain an estimate of its $r$-band magnitude. In Table~\ref{tab:main} we summarize our main results indicating: (i) the G4Jy name; (ii) the name of the mid-IR counterpart associated in the G4Jy catalog using WISE images; both (iii) radio and (iv) optical positions; (v) the counterpart IDF and (vi) the $r$ magnitude of the optical counterpart, as previously mentioned, together with the optical magnitude and the Galactic extinction. Then, we added to the table the estimate of the Galactic extinction\footnote{https://irsa.ipac.caltech.edu/applications/DUST/} $A_V$ \citep{schlegel98, schlafly11}. The entire Table~\ref{tab:main} is reported in the on-line version of the journal while only the first 10 lines are shown here for guidance regarding its form and content. Optical positions are reported from the Pan-STARRS and/or DES catalogs, when available, having all uncertainties lower than $\sim$0\arcsec.5. We did not measure DSS magnitudes and positional uncertainties since we are already carrying out a photometric survey, mainly using telescopes at Complejo Astron\'omico El Leoncito (CASLEO), to estimate them for those sources lying out of the Pan-STARRS and DES footprints.

We also inspected other optical images available in the Pan-STARRS/DES searching in the $g$, $i$, $z$ and $y$ bands to detect the host galaxy of sources lacking an optical counterpart of their radio core but having an associated mid-IR object. This analysis confirmed previous results with the unique exception of G4Jy\,818, for which we identified a marginal detection of an optical counterpart in $y$-band only.

\subsection{Optical identifications}
\begin{figure}[!ht]
\includegraphics[height=6.5cm,width=9.2cm,angle=0]{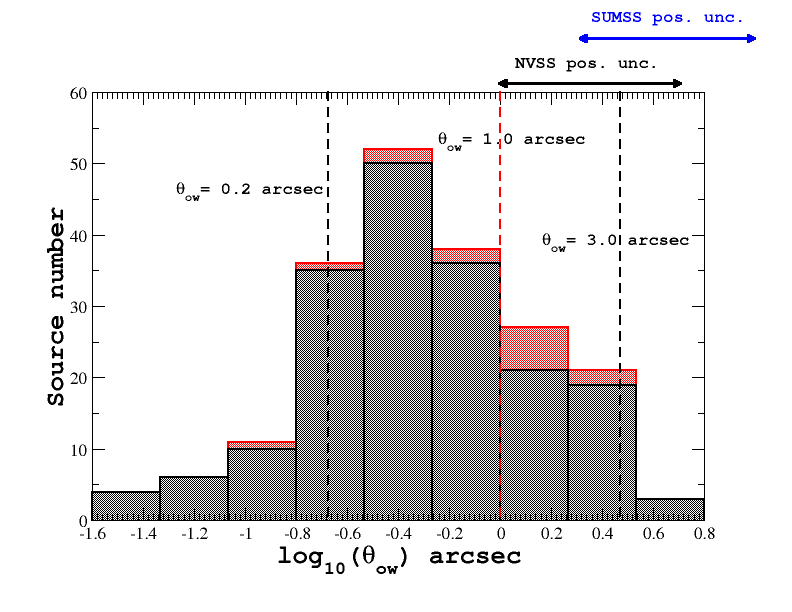}
\caption{The black histogram shows the distribution of the angular separation $\theta_{ow}$ between the mid-IR counterpart, assigned in the G4Jy catalog, and the optical one for all those G4Jy-3CRE radio sources having IDF=1.0, that, according to our analysis, implies that both counterparts are coincident. Two dashed black lines mark the location for $\theta_{ow}$ equal to 0\arcsec.2, and 3\,\arcsec, respectively, while the red one corresponds to 1\,\arcsec. The red histogram includes also those radio sources with IDF=2.2 for which our optical analysis helped to identify the mid-IR counterpart (see \S~\ref{sec:counterparts}.2 for more details). As a comparison, we also report here the typical range of positional uncertainties of the NVSS and the SUMSS catalogs \citep{vollmer05} mainly used in the G4Jy catalog to compute brightness-weighted radio centroids.}
\label{fig:angsep}
\end{figure}

We found that for 184 out of a total of 264 sources listed in the \cs\ catalog (i.e., 70\%) their optical counterpart is spatially coincident with the mid-IR match reported in the G4Jy catalog \citep{white20a,white20b}, thus confirming previous results. These are all marked with an IDF=1.0 in our Table~\ref{tab:main}. In Figure~\ref{fig:angsep} we report the distribution of the angular separation $\theta_{ow}$ between the optical and the mid-IR counterparts. 

Only 5 out of 264 sources (i.e., $\sim$2\%) show an incorrect association
between the optical source we assigned and the mid-IR counterpart reported in
the G4Jy catalog, namely: G4Jy\,4, G4Jy\,524, G4Jy\,1197, G4Jy\,1302,
G4Jy\,1854. For these five cases an IDF=2.1 was assigned. Then we identified
new optical counterparts for 21 sources out of 264 (i.e., $\sim$8\%,
corresponding to IDF=2.2) that did not have a mid-IR counterpart assigned in
the original G4Jy catalog. This was mainly possible thanks to the use of high
resolution radio maps as those retrieved from the VLASS and the NVAS archives.
In particular, we found that 4 sources with IDF=2.2, namely: G4Jy\,162,
G4Jy\,730, G4Jy\,1301 and G4Jy\,1330, have no mid-IR counterparts since they
lie close to bright WISE sources and their detection could be contaminated by
artifacts. In the case of G4Jy\,593, again with IDF=2.2, the detection of its
mid-IR counterpart is indeed compromised by the poor angular resolution of WISE
images. An additional 12 radio sources (out of 21) with IDF=2.2, marked with a
'u' host flag in the G4Jy catalog due to possible ambiguities related to the
complexity of their radio structure and/or the spatial distribution of nearby
mid-IR sources \citep[see][for more details]{white20a,white20b}, have mid-IR
counterpart, identified thanks to the refined optical analysis presented here.
These radio sources are: G4Jy\,113, G4Jy\,350, G4Jy\,530, G4Jy\,611, G4Jy\,672,
G4Jy\,680, G4Jy\,939, G4Jy\,1262, G4Jy\,1365, G4Jy\,1401, G4Jy\,1518,
G4Jy\,1740, and all their mid-IR associated counterparts are now reported in
Table~\ref{tab:main}. In addition for the two cases of G4Jy\,837 and
G4Jy\,1590, both having IDF=2.2 and both previously labelled with a 'm' host
flag in the G4Jy catalog \citep[i.e., identification of their host galaxy
limited by the mid-infrared data][]{white20b}, the optical analysis presented
here allowed us to recognize their mid-IR counterpart. In
Table~\ref{tab:newwise} we report the WISE name of these 14 mid-IR counterparts
identified by refined optical analysis performed here. These are also included
in Table~\ref{tab:main}. We marked the location of these 14 newly
associated mid-IR counterparts using a red circle on the WISE image at
3.4$\mu$m in the finding charts to distinguish them from
those associated in the G4Jy catalog labelled with a red cross. The only two
remaining sources: G4Jy\,1498 and G4Jy\,1532 are those clearly lacking a mid-IR
counterpart, being undetected in the WISE images. In Figure~\ref{fig:angsep} we
also report the distribution of the angular separation $\theta_{ow}$ between
the optical and the mid-IR counterparts including those with IDF=2.2 that have
been assigned thanks to the optical analysis presented here.
\begin{table*}[]
\caption{Newly assigned mid-IR counterparts of sources with IDF=2.2}
\label{tab:newwise}
\begin{center}
\begin{tabular}{rcc}
\hline
G4Jy & WISE &  $\theta_{ow}$ \\
name & name & (arcsec)\\
\noalign{\smallskip}
\hline 
\noalign{\smallskip}
113 & J010241.76-215254.2 & 2.65\\
350 & J032314.09-881605.2 & 1.84\\
530 & J051247.41-482416.5 & 0.63\\
611 & J062620.46-534135.1 & 1.31\\
672 & J074331.61-672625.5 & 1.78\\
680 & J080236.28-095739.9 & 0.38\\
837 & J102003.93-425130.0 & 1.48\\
939 & J114134.22-285048.0 & 0.32\\
1262 & J153014.29-423151.7 & 1.9\\
1365 & J164604.85-222804.6 & 1.05\\
1401 & J172011.01-070132.2 & 0.14\\
1518 & J191548.68-265257.4 & 0.28\\
1590 & J195816.66-550934.9 & 0.59\\
1740 & J215407.02-515012.8 & 1.18\\
\noalign{\smallskip}
\hline
\end{tabular}\\
\end{center}
Note: col. (1) the name reported in the G4Jy catalog; col. (2) the name of the mid-IR counterpart, detected in WISE, assigned thanks to the optical analysis presented here; col. (3) the angular separation $\theta_{ow}$ between the position of the optical and mid-IR counterpart, assigned thanks to the analysis performed here.
\end{table*}

We found that 14 out of the total 264 sources (i.e., $\sim$6\%) visually inspected are flagged as ``confused'' since we were not able to identify a unique optical counterpart, and thus require a deeper investigation and eventually additional follow up observations. { If these sources are high excitation radio galaxies \citep[HERGs;][]{hine79} or quasars, due to their relatively low source density (i.e., number of sources per square degree in the sky), X-ray or optical spectroscopic observations could reveal the position of their counterpart and thus identify their host galaxies. However the precise location of their radio core, necessary to identify the host galaxy position, can be only achieved using higher resolution radio maps, in particular when they are hosted in elliptical galaxies with weak optical emission lines, as often occurs in low excitation radio galaxies \citep[LERGs;][]{hine79}.} For all of them the value of IDF=3.0 is reported in Table~\ref{tab:main}. 

There are 19 out of 264 sources that lack an optical counterpart of the radio core but have at least a mid-IR source associated with it (i.e., $\sim$7\% of the whole \cs\ catalog), thus being marked with IDF=4.1, as for G4Jy\,1846 shown in Figure~\ref{fig:example4}. For 4 more objects out of 264 (i.e., $\sim$2\%), the associated mid-IR source does not appear to be correct thus having IDF=4.2 (see G4Jy\,1057 in Figure~\ref{fig:example4}, as well as G4Jy\,183, G4Jy\,1551, G4Jy\,1782). Lastly, 17 remaining sources ($\sim$6\% of the total) have no optical and no mid-IR counterpart associated with their radio core and are indicated with IDF=4.3. Two of those radio sources labelled with IDF=4.3 are: G4Jy\,77 and G4Jy\,1605, the former is a radio phoenix of the galaxy cluster Abell\,85 \citep[see e.g.,][]{bagchi98,kempner04,ichinohe15} while the latter is the radio relic of Abell\,3667 \citep[see e.g.,][]{johnston08,owers09}, as also discussed, with more details, in Appendix A.

All sources with IDF=4.1 or IDF=4.2 or IDF=4.3 have no detection of their radio cores in radio maps due to their poor angular resolution, with the only exceptions of four objects, namely: G4Jy\,854, G4Jy\,1010, G4Jy\,1136 and G4Jy\,1830. In these four cases the detection of the mid-IR counterpart could be contaminated by artifacts, due to the presence of bright WISE sources, for G4Jy\,854 and G4Jy\,1136, while for G4Jy\,1010 and G4Jy\,1830 no clear mid-IR emission is reported in the WISE all-sky catalog.

All numbers reported in this section are also summarized in Table~\ref{tab:flags}. 

Finally, there are 203 out of 264 sources listed in the G4Jy-3CRE sample (i.e.,$\sim$77\%) (i) having the optical counterpart coincident with the mid-IR one (i.e., IDF=1.0) or (ii) lacking optical counterpart of their radio core but having a mid-IR counterpart  (i.e., IDF=4.1). Thus in Appendix C we computed the probability of spurious associations between mid-IR sources listed in the WISE all-sky survey and the full G4Jy radio catalog, as originally performed to assign mid-IR counterparts, and we found a good agreement with the refined analysis presented here.

\section{Literature results}
\label{sec:literature} Once we identified optical counterparts of radio sources we performed a literature search to investigate the availability of redshifts. Results of this literature search, including uncertain $z$ estimates are also reported in each figure when available and in the finding charts. We also compared our final catalog with the several radio catalogs obtained from radio surveys (see Appendix B for a brief overview of all radio catalogs used here). All these identifications derived from these radio crossmatches are reported in Appendix B to simplify searches in astronomical databases as NED and SIMBAD. 

We found that for a total of 157 sources out of 264 (i.e., 59\% of the whole sample) there is a spectroscopic redshift measurement already available in the literature, twelve of them considered uncertain and thus labelled with a question mark for a total of 145 radio sources with firm $z$ estimates. Our search was carried out also using the NASA Extragalactic Database (NED)\footnote{http://ned.ipac.caltech.edu} and the SIMBAD Astronomical Database\footnote{http://simbad.u-strasbg.fr/simbad/}. According to previous analyses carried out during past follow up spectroscopic campaigns \citep{massaro16,herazo20,herazo22,kosiba22}, we adopted the same conservative criteria and we only considered confident redshift measurements reported in the literature those for which we could verify (i) a published image of the optical spectrum, or (ii) there is a description of it with emission and/or absorption lines clearly reported in a table or in the manuscript. In this way, we marked redshifts $z$ we could not verify with a question mark for 12 sources out 157.

Moreover we did not consider photometric redshifts since we are currently carrying out optical spectroscopic observations, based on the analysis presented here and more redshifts will be presented in a forthcoming paper \citep{perez22}. In some cases, and only for sources belonging to the Molonglo Southern 4\,Jy sample \citep[MS4][]{burgess06a,burgess06b}, we reported values of photometric redshift estimates in Appendix A.

A large fraction of the spectra we found are available in the Data Release 3 of the Six-degree Field Galaxy Survey (6dFGS) Database\footnote{http://www-wfau.roe.ac.uk/6dFGS/index.html} \citep{jones04, jones09} in addition to the spectroscopic observations of the equatorial sample of powerful radio galaxies \citep{best99}. In a few cases we also used other databases such as the On-Line Inventory of Extragalactic X-ray Jets\footnote{https://hea-www.harvard.edu/XJET/} \citep[XJET;][]{massaro11a} and references reported therein. 
 
\begin{figure}[]
\includegraphics[height=6.8cm,width=9.4cm,angle=0]{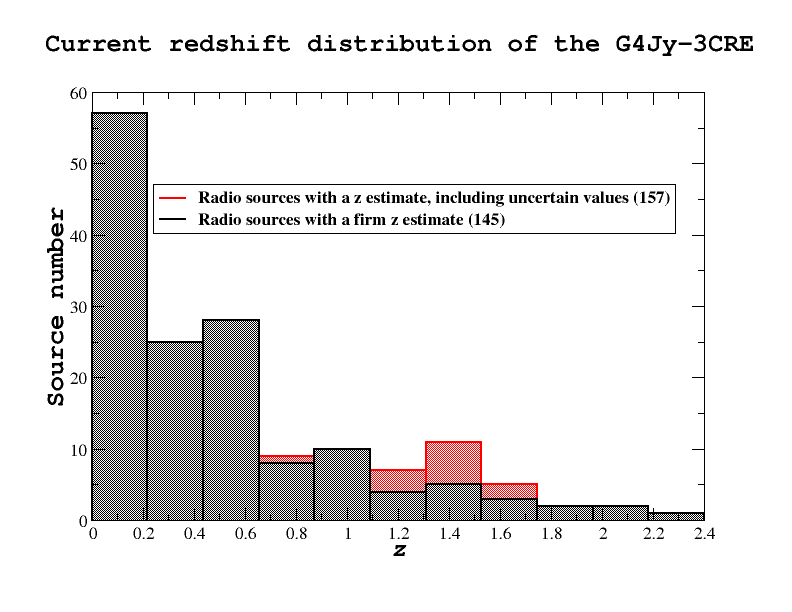}
\caption{The redshift distribution obtained from our literature search for those 157 radio sources listed in the \cs\ catalog with a $z$ estimate. This red histogram includes also 12 radio sources with uncertain values in comparison with that of radio sources with firm $z$ measurements shown in black.}
\label{fig:redshifts}
\end{figure}

According to our literature search we found that 131 sources out of 157 having IDF=1.0 already have an available redshift measurement. For all these cases, with only two exceptions, namely: G4Jy\,1158 and G4Jy\,1225, we found these estimates reliable (i.e., no question mark reported). The underlying choice of considering these estimates, at a first look correct, is also based on the relatively low sky density (i.e., number of sources per square degree) of radio galaxies and QSOs, thus the probability of having more than a single radio-loud AGN close to the radio position within a circle of a few arcseconds radius is extremely low. This also proves the importance of collecting spectroscopic information that could confirm optical counterparts associated in the present analysis \citep[see e.g., results of spectroscopic campaigns carried out on radio loud AGNs lying within the positional uncertainty regions of gamma-ray sources][]{massaro15c,landoni15,ricci15,crespo16}. In Figure~\ref{fig:redshifts} we show the comparison between the redshift distribution for 157 radio sources, out of 264 listed in the \cs\ catalog, having a $z$ estimates, and that excluding those 12 with uncertain $z$.

For those sources with IDF=2.1 or IDF=2.2 we found a redshift estimate for 17 cases out of 26, namely 3 objects with IDF=2.1 and 14 with IDF=2.2. In 2 out of the 3 cases with IDF=2.1, we report $z$ estimates in Table~\ref{tab:main} adding the question mark to indicate that these measurements are uncertain given the lack of positional data about the target, while this situation does not occur for all radio sources with IDF=2.2. 

For only one case out of the ``confused'' sources (i.e., those labelled with IDF=3.0), namely: G4Jy\,1626, that could potentially reside in galaxy-rich large-scale environments (i.e., groups or cluster of galaxies), we also report a $z$ estimate with a question mark while the remaining 13 sources all lack a $z$ measurement. 

Finally, we found that for 6 sources (out of 19) marked with IDF=4.1, for which we were not able to identify the optical counterpart in the archival images, there is a $z$ estimate in the literature, but since we could not verify which sources were targeted, they are all flagged as uncertain, with the only exception: G4Jy\,417 for which we found the finding chart in the literature since it belongs to the 2\,Jy sample\citep{wall85,morganti97}, even if we could not detect it in the DSS archival images used here. Then all 4 radio sources with IDF=4.2 have no $z$ measurement and the same occurs for 15 out of 17 of those labelled with IDF=4.3. Both remaining 2 radio sources with IDF=4.3 have the $z$ estimate marked with a question mark thus considered uncertain.

\section{Summary, conclusions and future perspectives}
\label{sec:summary} We present the \cs\ catalog extracted from the G4Jy catalog \citep{white20a, white20b} and based on the low radio frequency observations of the MWA as part of the GLEAM survey. The \cs\ catalog lists 264 sources with 9\,Jy limiting sensitivity at $\sim$178 MHz, which is the same as the nominal threshold for the 3CR, but including only targets at $Dec.<-5^\circ$, having Galactic latitudes $|b|>10^\circ$, all lying in the footprint not covered by the 3CR.

Thanks to a huge amount of effort carried out on the G4Jy catalog, a large fraction of radio sources listed therein (i.e., $\sim$85\%) are associated with a mid-IR counterpart \citep[being their host galaxy][]{white20b} detected in the WISE all-sky survey \citep{wright10}. Here we present a refined analysis, restricted only to the \cs\ sample, aimed at locating optical counterparts of their host galaxies. Thanks to recent high angular resolution radio observations, such as those available in the VLASS, we can obtain a precise estimate of the position of their radio cores, for a significant fraction of the \cs\ sources (i.e., 207 out of 264, nearly 78\%). This allowed us to (i) improve the localization of the host galaxy in the optical images of Pan-STARRS, DES and DSS archives and then (ii) search the literature for sources for which a redshift estimate is already available.

Results achieved by our inspection of archival radio, infrared and optical images can be summarized as follows:
\begin{enumerate} 
\item We found that for 184 out of 264 \cs\ sources the optical counterpart associated in the present analysis is coincident with the mid-IR counterpart reported in the G4Jy catalog, confirming the robustness of the previous analysis \citep{white20a,white20b}. 

\item There are 26 \cs\ sources for which the optical counterpart we identified is different from the, previously assigned, mid-IR one. In particular for 21 of them there is not a mid-IR source assigned/associated in the G4Jy catalog. For 14 out of these 21 radio sources we have been able to assign a mid-IR counterpart on the basis of our optical analysis.

\item For 14 sources the poor angular resolution of radio maps available and the presence of several optical sources around the position of their radio cores did not allow us to confirm their host galaxies. 

\item There are an additional 40 sources with no optical counterpart of their radio core. For 4 cases the mid-IR counterpart associated in the G4Jy catalog does not appear to be correct (IDF=4.2), while for 17 sources there is no mid-IR counterpart detected at the location of the radio cores, as reported in the G4Jy catalog (i.e., IDF=4.3). The remaining 19 sources are only detected in the mid-IR images (IDF=4.1). \end{enumerate}

According to our analysis, radio sources having the identification flags: IDF=1.0 or IDF=4.1 are $\sim$77\% (i.e., 203 out of 264) of the whole \cs\ sample and correspond to the more reliable mid-IR associations. This is in agreement with the statistical test we used to compute the expected number of spurious associations when matching the G4Jy with the AllWISE catalog.

Given the identified location of the host galaxies for a large fraction of the \cs\ sources, we also checked the literature to search for possibile redshift estimates. Adopting a conservative criterion we found a total of 157 spectra, and 145 of them appear to provide a firm $z$ estimate. Moreover, 129 are reliable since their optical counterpart coincides with that associated at mid-IR frequencies and are not labelled as possible sources having an uncertain $z$ measurement. 

Finally, we conclude by highlighting future perspectives on the potential use of the \cs\ sample. Several proposals were already submitted to collect optical spectroscopic information for all sources listed in the \cs\ catalog, and in a forthcoming paper of this series, part of these observations, already acquired, will be presented \citep{perez22}. A dedicated paper presenting the X-ray analysis, based on \swf\ data collected for $\sim$80 \cs\ sources, is also in preparation, to highlight the potential use of X-ray snapshot observations to refine the search for optical counterparts and host galaxies \citep{massaro23}.

\begin{acknowledgments}
We thank the anonymous referee for useful and valuable comments that led to improvements in the paper.
We wish to dedicate this paper to D. E. Harris and R. W. Hunstead. Their insight, passion and contributions to radio astronomy are an inspiration.

A. J. acknowledges the financial support (MASF-CONTR-FIN-18-01) from the Italian National Institute of Astrophysics under the agreement with the Instituto de Astrofisica de Canarias for the ``Becas Internacionales para Licenciados y/o Graduados Convocatoria de 2017’’. S.V.W. acknowledges financial assistance of the South African Radio Astronomy Observatory (SARAO)\footnote{https://www.sarao.ac.za}. W.F. and R.K. acknowledge support from the Smithsonian Institution and the \chn\ High Resolution Camera Project through NASA contract NAS8-03060.  W.F. also acknowledges support from NASA Grants 80NSSC19K0116, GO1-22132X, and GO9-20109X. This investigation is supported by the National Aeronautics and Space Administration (NASA) grants GO9-20083X, GO0-21110X and GO1-22087X. I.A., S.A.C. and V.R. are partially supported by grant PIP 1220200102169CO, Argentine Research Council (CONICET). A. G.-P. acknowledges support from the CONACyT program for their Ph.D. studies. V. C. acknowledges support from the Fulbright  - García Robles scholarship. This work was partially supported by CONACyT (Consejo Nacional de Ciencia y Tecnología) research grant 280789. 

This research has made use of the NASA/IPAC Extragalactic Database (NED), which is operated by the Jet Propulsion Laboratory, California Institute of Technology, under contract with the National Aeronautics and Space Administration.
This research has made use of the SIMBAD database, operated at CDS, Strasbourg, France \citep{wenger00}.

This research has made use of the CIRADA cutout service at URL cutouts.cirada.ca, operated by the Canadian Initiative for Radio Astronomy Data Analysis (CIRADA). CIRADA is funded by a grant from the Canada Foundation for Innovation 2017 Innovation Fund (Project 35999), as well as by the Provinces of Ontario, British Columbia, Alberta, Manitoba and Quebec, in collaboration with the National Research Council of Canada, the US National Radio Astronomy Observatory and Australia’s Commonwealth Scientific and Industrial Research Organisation.
The National Radio Astronomy Observatory is a facility of the National Science Foundation operated under cooperative agreement by Associated Universities, Inc.
Part of this work is based on the NVSS (NRAO VLA Sky Survey): The National Radio Astronomy Observatory is operated by Associated Universities, Inc., under contract with the National Science Foundation and on the VLA low-frequency Sky Survey (VLSS). 
We thank the staff of the GMRT that made these observations possible. GMRT is run by the National Centre for Radio Astrophysics of the Tata Institute of Fundamental Research.
The Molonglo Observatory site manager, Duncan Campbell-Wilson, and the staff, Jeff Webb, Michael White and John Barry, are responsible for the smooth operation of Molonglo Observatory Synthesis Telescope (MOST) and the day-to-day observing programme of SUMSS. The SUMSS survey is dedicated to Michael Large whose expertise and vision made the project possible. The MOST is operated by the School of Physics with the support of the Australian Research Council and the Science Foundation for Physics within the University of Sydney. 
This scientific work makes use of the Murchison Radio-astronomy Observatory, operated by CSIRO. We acknowledge the Wajarri Yamatji people as the traditional owners of the Observatory site. Support for the MWA comes from the US National Science Foundation (grants AST-0457585, PHY-0835713, CAREER-0847753, and AST-0908884), the Australian Research Council (LIEF grants LE0775621 and LE0882938), the US Air Force Office of Scientific Research (grant FA9550-0510247), and the Centre for All-sky Astrophysics (an Australian Research Council Centre of Excellence funded by grant CE110001020). Support is also provided by the Smithsonian Astrophysical Observatory, the MIT School of Science, the Raman Research Institute, the Australian National University, and the Victoria University of Wellington (via grant MED-E1799 from the New Zealand Ministry of Economic Development and an IBM Shared University Research Grant). The Australian Federal government provides additional support via the Commonwealth Scientific and Industrial Research Organisation (CSIRO), National Collaborative Research Infrastructure Strategy, Education Investment Fund, and the Australia India Strategic Research Fund, and Astronomy Australia Limited, under contract to Curtin University. This work was supported by resources provided by the Pawsey Supercomputing Centre with funding from the Australian Government and the Government of Western Australia. We acknowledge the iVEC Petabyte Data Store, the Initiative in Innovative Computing, and the CUDA Center for Excellence sponsored by NVIDIA at Harvard University, and the International Centre for Radio Astronomy Research (ICRAR), a Joint Venture of Curtin University, and The University of Western Australia, funded by the Western Australian State government.

This publication makes use of data products from the Wide-field Infrared Survey Explorer, which is a joint project of the University of California, Los Angeles, and the Jet Propulsion Laboratory/California Institute of Technology, funded by the National Aeronautics and Space Administration.

This project used public archival data from the Dark Energy Survey (DES). Funding for the DES Projects has been provided by the U.S. Department of Energy, the U.S. National Science Foundation, the Ministry of Science and Education of Spain, the Science and Technology FacilitiesCouncil of the United Kingdom, the Higher Education Funding Council for England, the National Center for Supercomputing Applications at the University of Illinois at Urbana-Champaign, the Kavli Institute of Cosmological Physics at the University of Chicago, the Center for Cosmology and Astro-Particle Physics at the Ohio State University, the Mitchell Institute for Fundamental Physics and Astronomy at Texas A\&M University, Financiadora de Estudos e Projetos, Funda{\c c}{\~a}o Carlos Chagas Filho de Amparo {\`a} Pesquisa do Estado do Rio de Janeiro, Conselho Nacional de Desenvolvimento Cient{\'i}fico e Tecnol{\'o}gico and the Minist{\'e}rio da Ci{\^e}ncia, Tecnologia e Inova{\c c}{\~a}o, the Deutsche Forschungsgemeinschaft, and the Collaborating Institutions in the Dark Energy Survey.
The Collaborating Institutions are Argonne National Laboratory, the University of California at Santa Cruz, the University of Cambridge, Centro de Investigaciones Energ{\'e}ticas, Medioambientales y Tecnol{\'o}gicas-Madrid, the University of Chicago, University College London, the DES-Brazil Consortium, the University of Edinburgh, the Eidgen{\"o}ssische Technische Hochschule (ETH) Z{\"u}rich,  Fermi National Accelerator Laboratory, the University of Illinois at Urbana-Champaign, the Institut de Ci{\`e}ncies de l'Espai (IEEC/CSIC), the Institut de F{\'i}sica d'Altes Energies, Lawrence Berkeley National Laboratory, the Ludwig-Maximilians Universit{\"a}t M{\"u}nchen and the associated Excellence Cluster Universe, the University of Michigan, the National Optical Astronomy Observatory, the University of Nottingham, The Ohio State University, the OzDES Membership Consortium, the University of Pennsylvania, the University of Portsmouth, SLAC National Accelerator Laboratory, Stanford University, the University of Sussex, and Texas A\&M University.
Based in part on observations at Cerro Tololo Inter-American Observatory, National Optical Astronomy Observatory, which is operated by the Association of Universities for Research in Astronomy (AURA) under a cooperative agreement with the National Science Foundation.
The Pan-STARRS1 Surveys (PS1) have been made possible through contributions of the Institute for Astronomy, the University of Hawaii, the Pan-STARRS Project Office, the Max-Planck Society and its participating institutes, the Max Planck Institute for Astronomy, Heidelberg and the Max Planck Institute for Extraterrestrial Physics, Garching, The Johns Hopkins University, Durham University, the University of Edinburgh, Queen's University Belfast, the Harvard-Smithsonian Center for Astrophysics, the Las Cumbres Observatory Global Telescope Network Incorporated, the National Central University of Taiwan, the Space Telescope Science Institute, the National Aeronautics and Space Administration under Grant No. NNX08AR22G issued through the Planetary Science Division of the NASA Science Mission Directorate, the National Science Foundation under Grant No. AST-1238877, the University of Maryland, and Eotvos Lorand University (ELTE).
Based on photographic data obtained using The UK Schmidt Telescope. The UK Schmidt Telescope was operated by the Royal Observatory Edinburgh, with funding from the UK Science and Engineering Research Council, until 1988 June, and thereafter by the Anglo-Australian Observatory. Original plate material is copyright (c) of the Royal Observatory Edinburgh and the Anglo-Australian Observatory. The plates were processed into the present compressed digital form with their permission. The Digitized Sky Survey was produced at the Space Telescope Science Institute under US Government grant NAG W-2166.
We acknowledge the efforts of the staff of the Anglo-Australian Observatory, who have undertaken the observations and developed the 6dF instrument.

SAOImageDS9 development has been made possible by funding from the \chn\ X-ray Science Center (CXC), the High Energy Astrophysics Science Archive Center (HEASARC) and the JWST Mission office at Space Telescope Science Institute \citep{joye03}.
This research has made use of data obtained from the high-energy Astrophysics Science Archive Research Center (HEASARC) provided by NASA’s Goddard Space Flight Center. We acknowledge the use of NASA's SkyView facility (http://skyview.gsfc.nasa.gov) located at NASA Goddard Space Flight Center.
This dataset or service is made available by the Infrared Science Archive (IRSA) at IPAC, which is operated by the California Institute of Technology under contract with the National Aeronautics and Space Administration.
TOPCAT and STILTS astronomical software \citep{taylor05} were used for the preparation and manipulation of the tabular data and the images.
The analysis is partially based on the OCCAM computing facility hosted by C3S\footnote{http://c3s.unito.it/} at UniTO \citep{aldinucci17}.

\end{acknowledgments}

\newpage
\appendix

\begin{figure*}[!th]
\begin{center}
\includegraphics[height=3.8cm,width=8.8cm,angle=0]{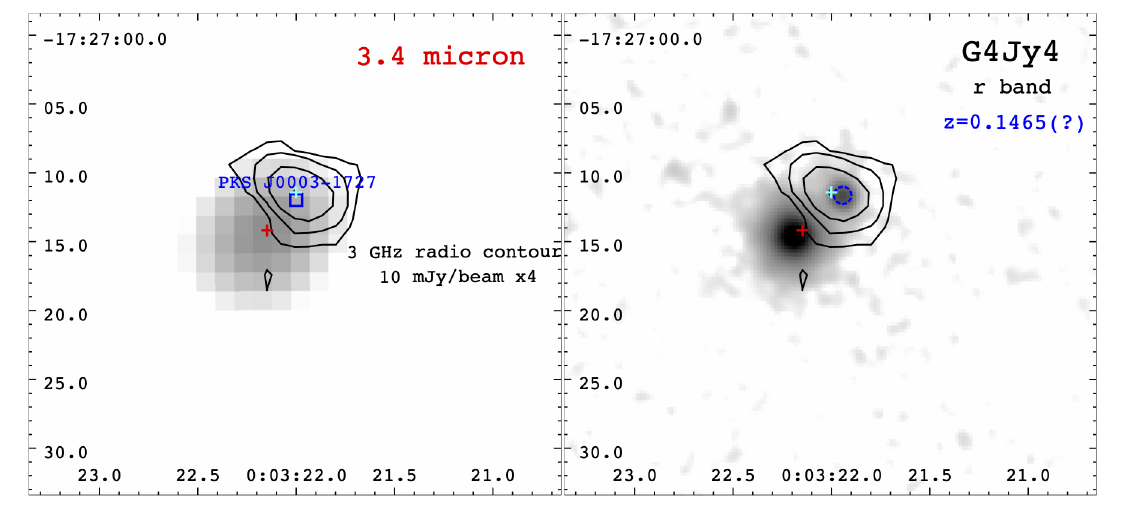}
\includegraphics[height=3.8cm,width=8.8cm,angle=0]{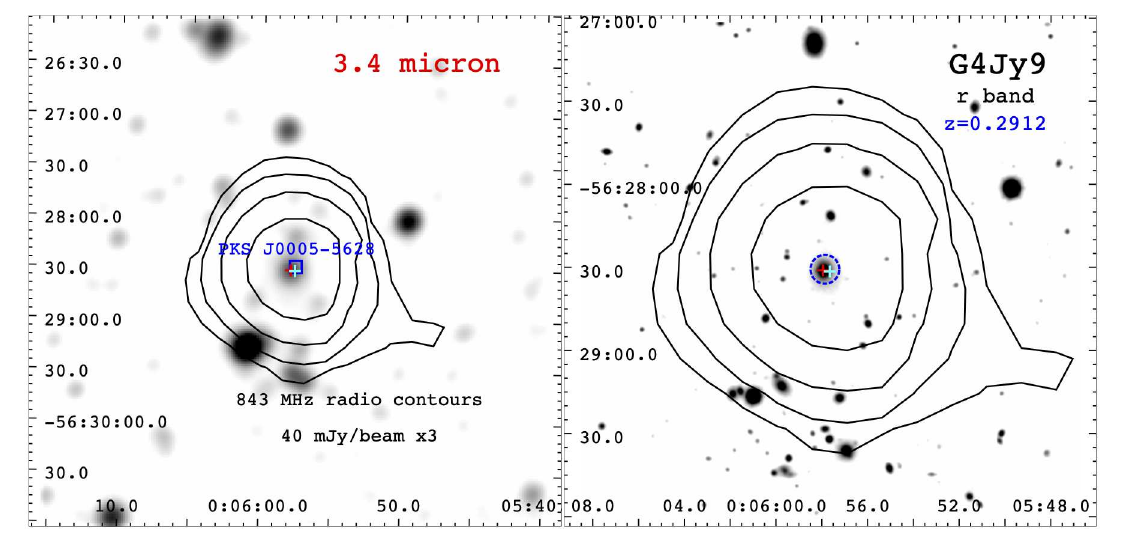}
\includegraphics[height=3.8cm,width=8.6cm,angle=0]{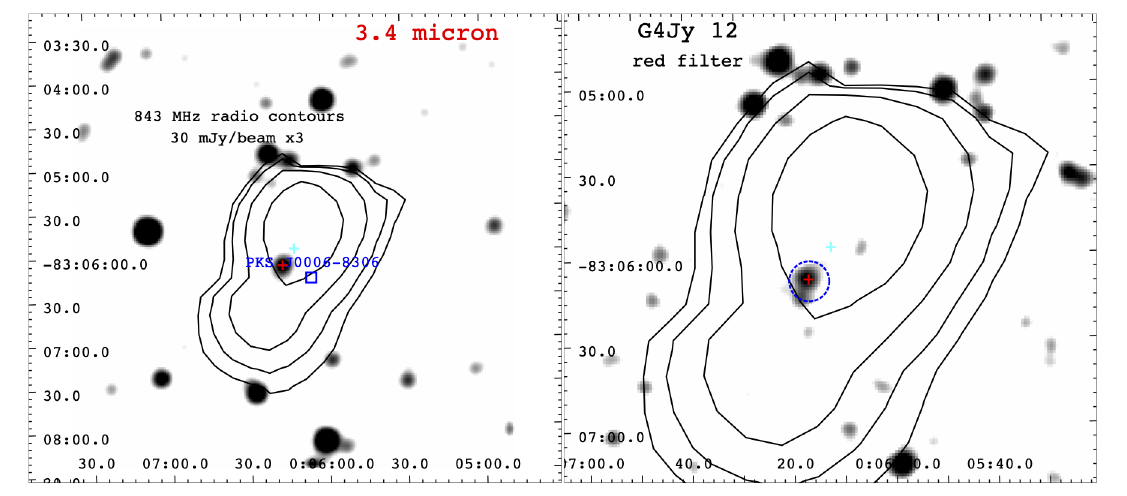}
\includegraphics[height=3.8cm,width=8.8cm,angle=0]{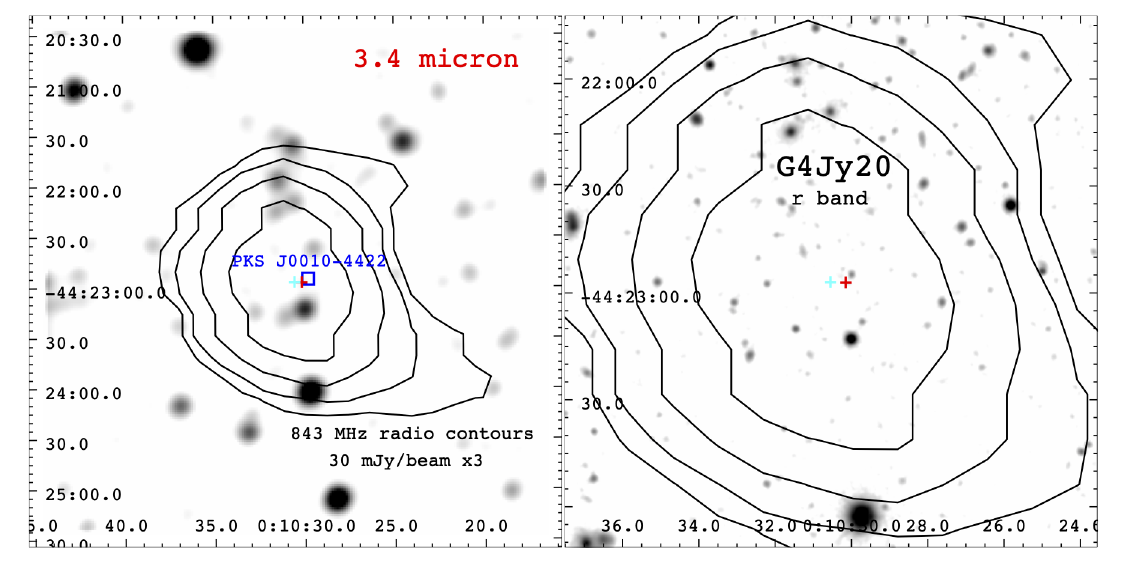}
\includegraphics[height=3.8cm,width=8.8cm,angle=0]{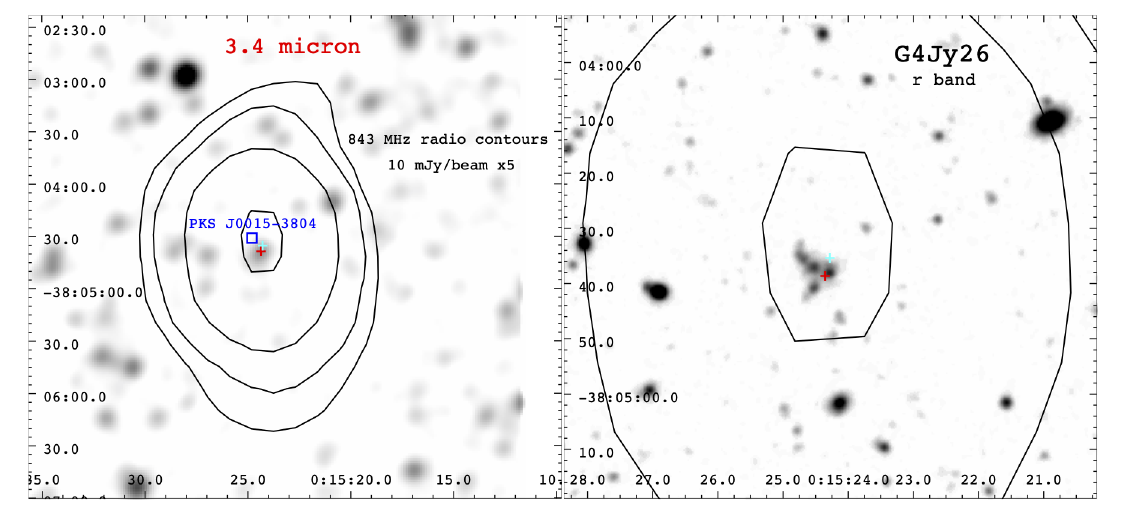}
\includegraphics[height=3.8cm,width=8.8cm,angle=0]{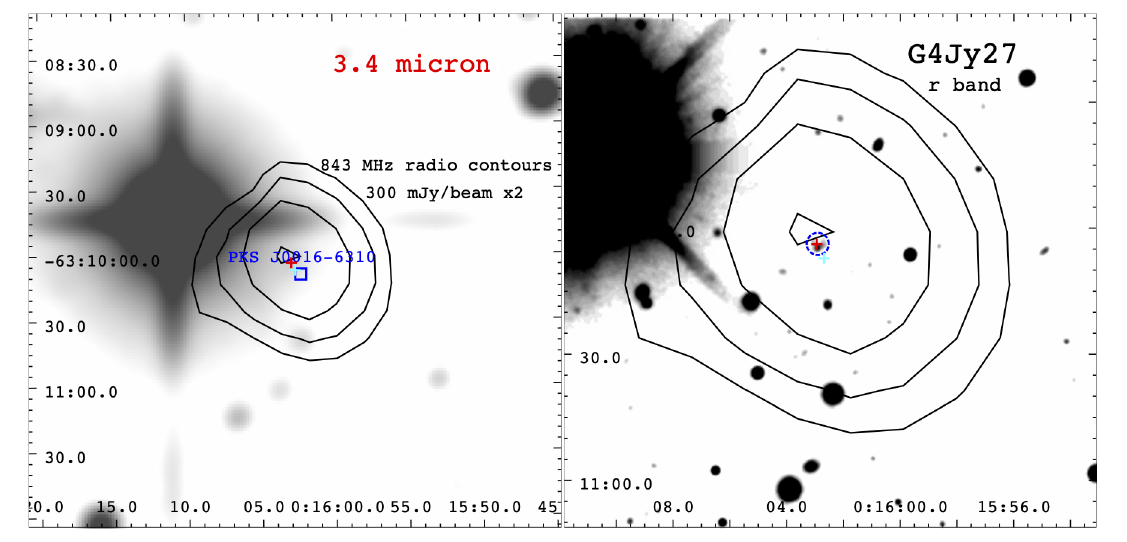}
\includegraphics[height=3.8cm,width=8.8cm,angle=0]{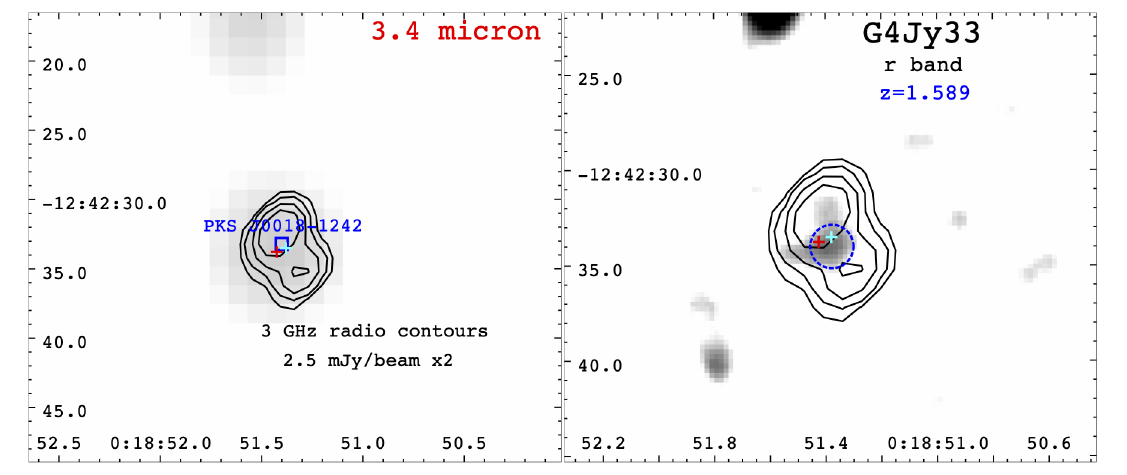}
\includegraphics[height=3.8cm,width=8.8cm,angle=0]{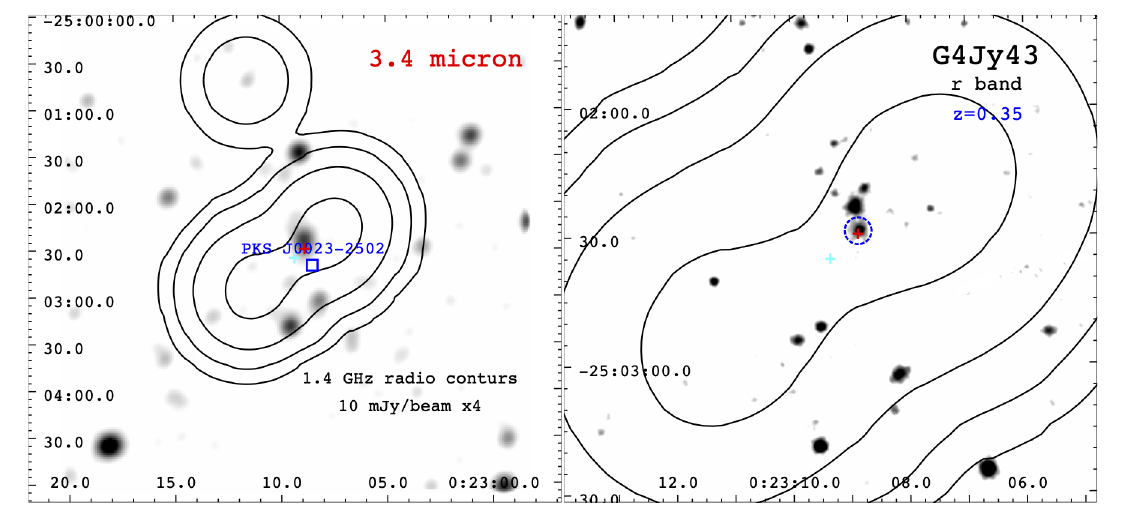}
\includegraphics[height=3.8cm,width=8.8cm,angle=0]{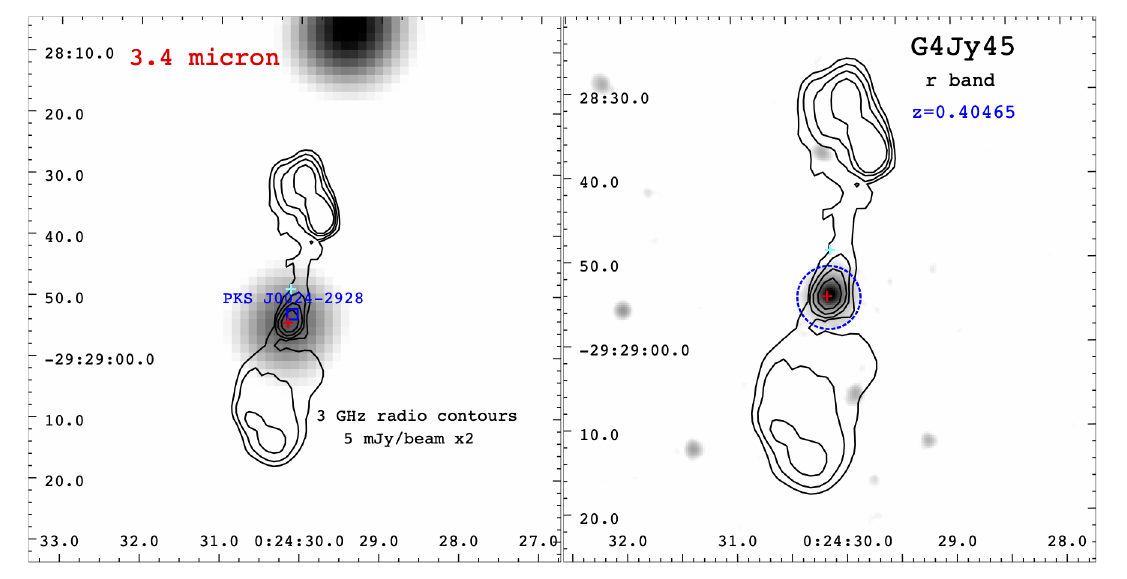}
\includegraphics[height=3.8cm,width=8.8cm,angle=0]{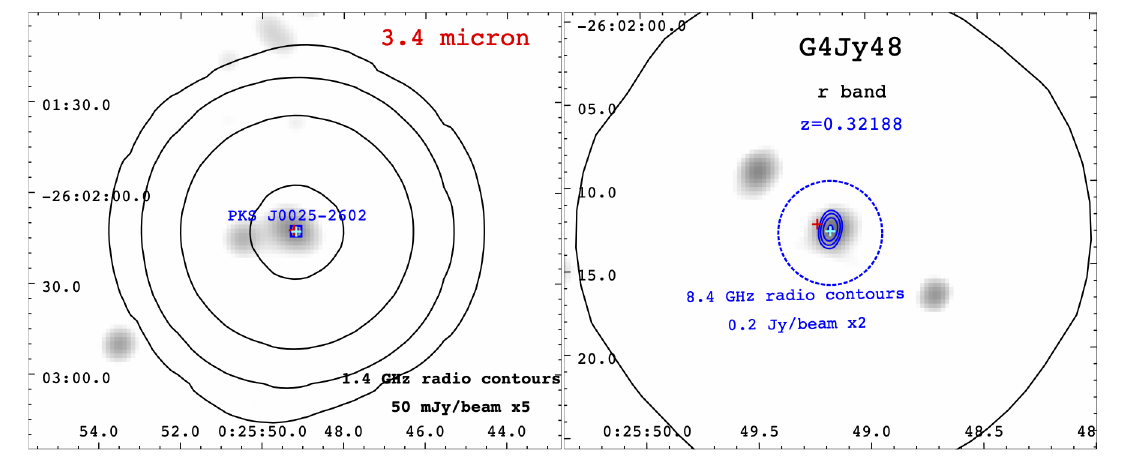}
\includegraphics[height=3.8cm,width=8.8cm,angle=0]{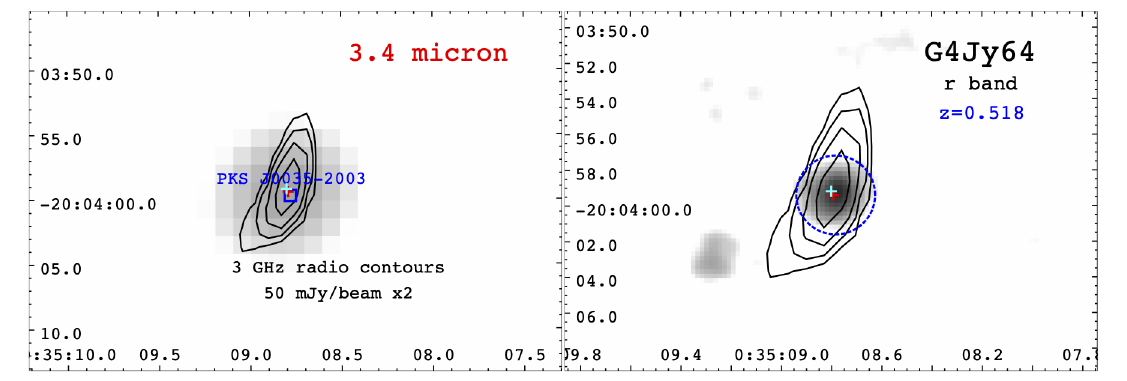}
\includegraphics[height=3.8cm,width=8.8cm,angle=0]{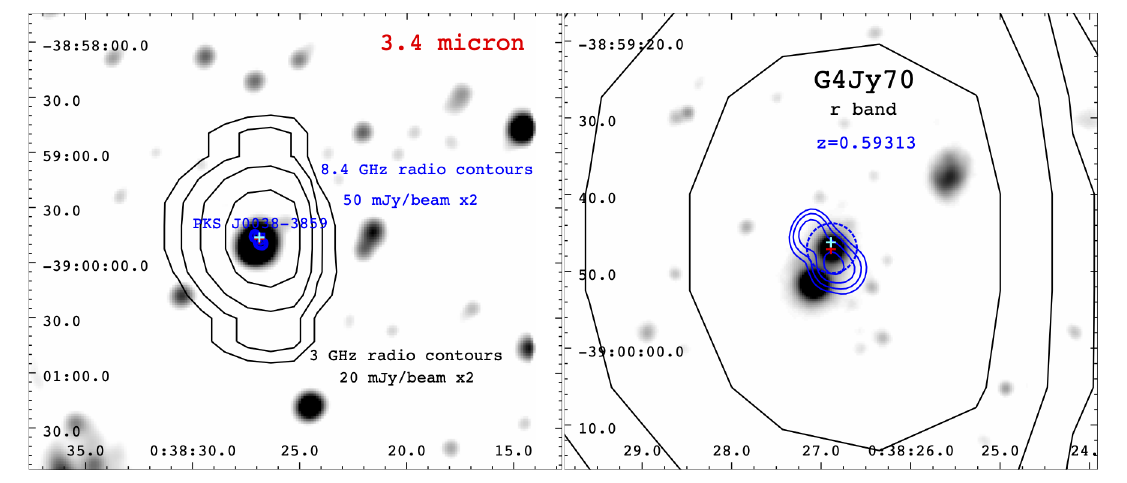}
\caption{Same as Figure~\ref{fig:example1} for the following \cs\ radio sources: \\ 
G4Jy\,4, G4Jy\,9, G4Jy\,12, G4Jy\,20, G4Jy\,26, G4Jy\,27, G4Jy\,33, G4Jy\,43, G4Jy\,45, G4Jy\,48, G4Jy\,64, G4Jy\,70.}
\end{center}
\end{figure*}

\begin{figure*}[!th]
\begin{center}
\includegraphics[height=3.8cm,width=8.8cm,angle=0]{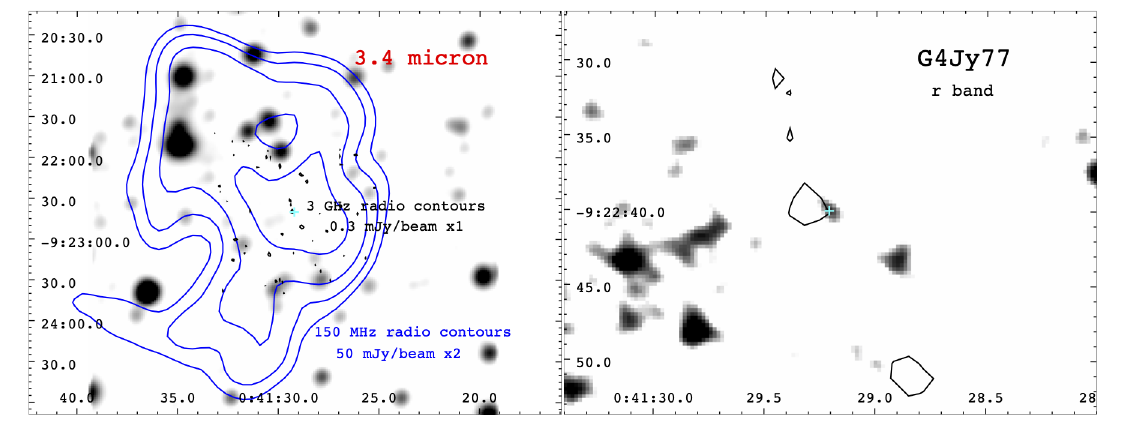}
\includegraphics[height=3.8cm,width=8.8cm,angle=0]{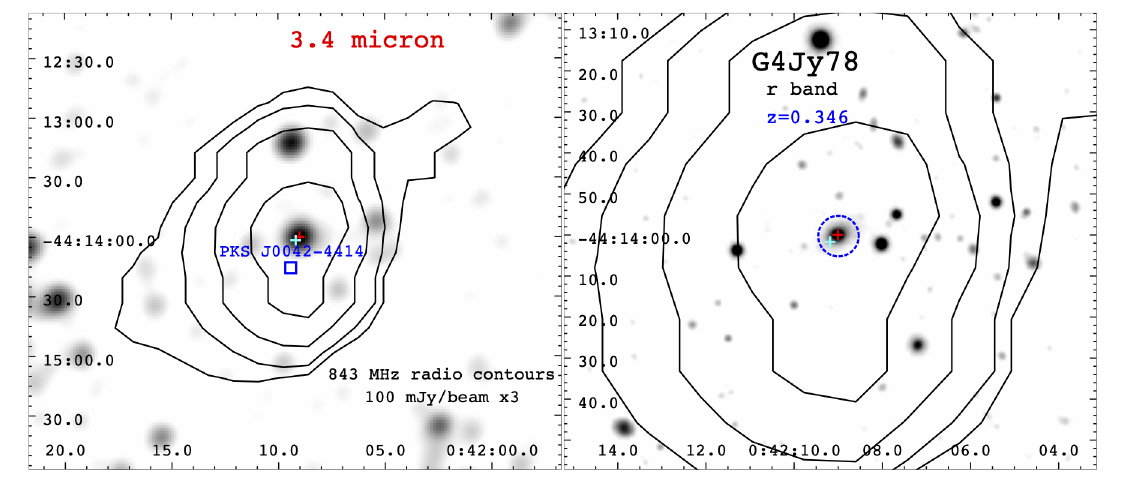}
\includegraphics[height=3.8cm,width=8.8cm,angle=0]{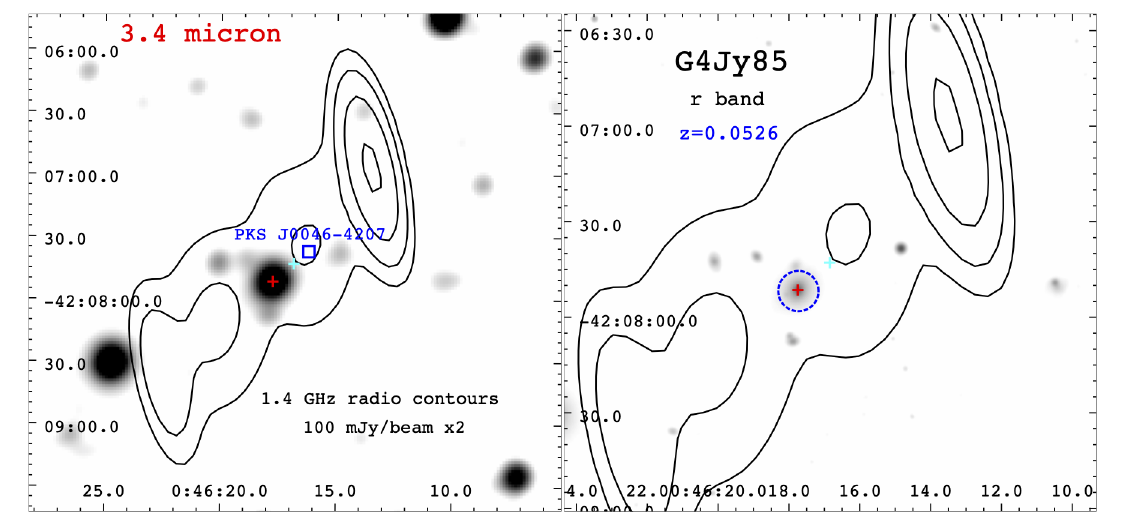}
\includegraphics[height=3.8cm,width=8.8cm,angle=0]{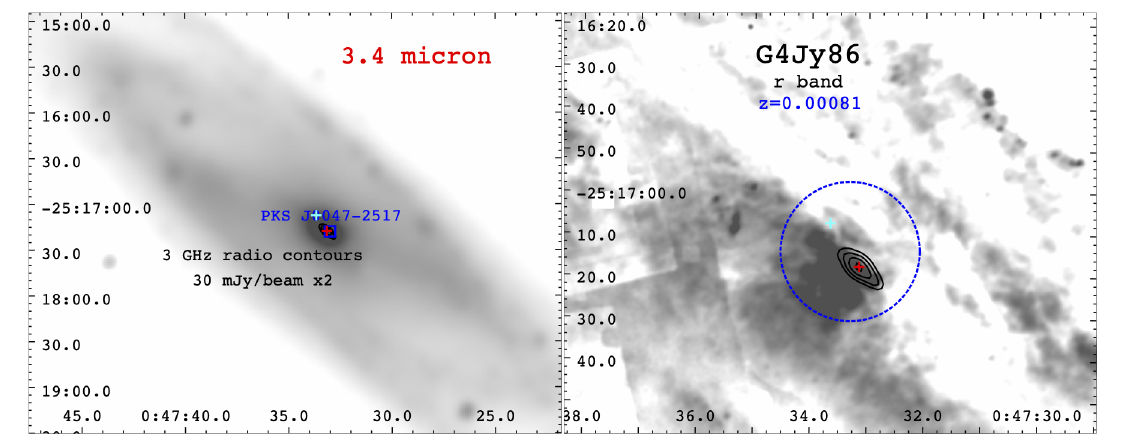}
\includegraphics[height=3.8cm,width=8.8cm,angle=0]{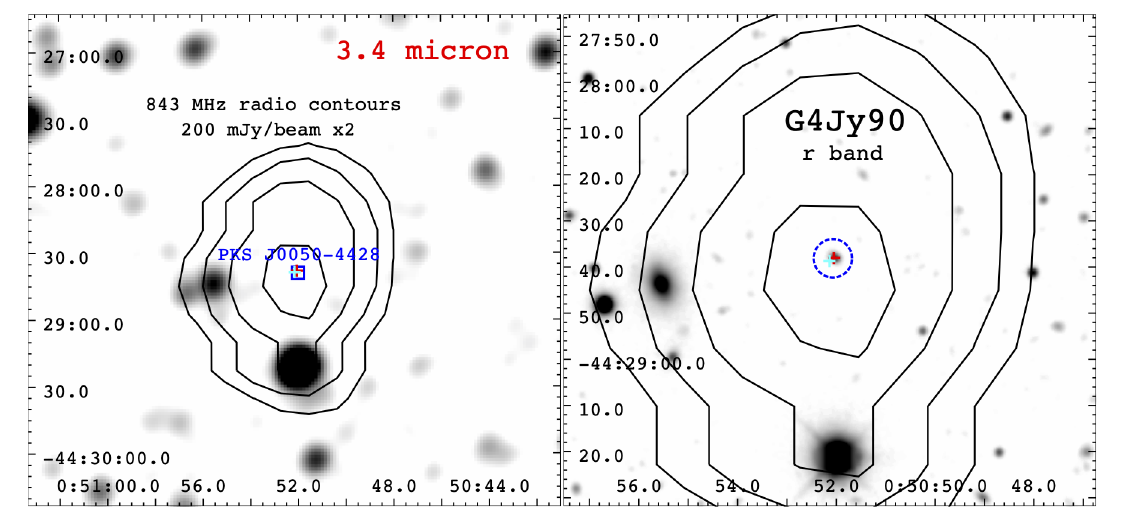}
\includegraphics[height=3.8cm,width=8.8cm,angle=0]{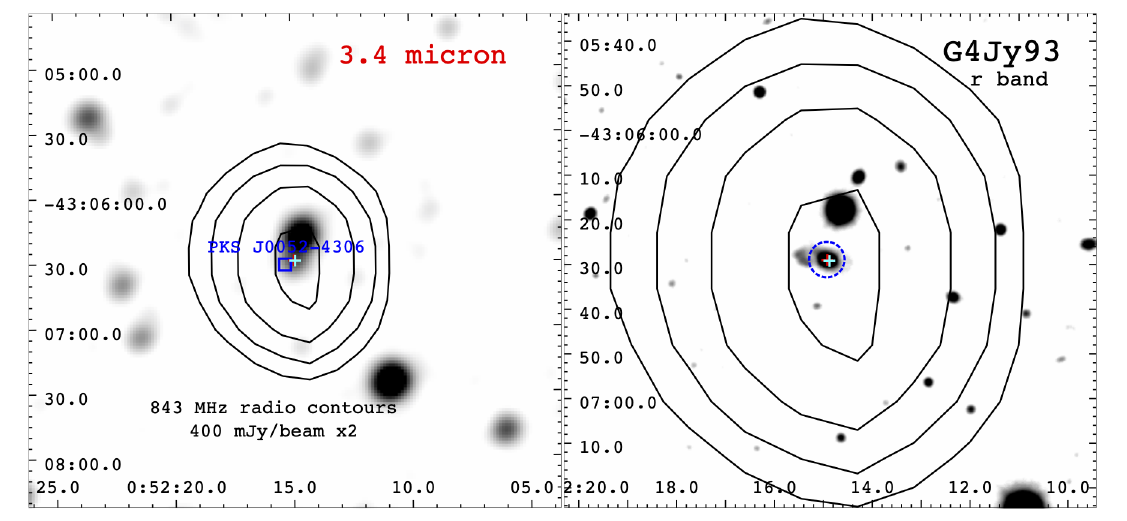}
\includegraphics[height=3.8cm,width=8.8cm,angle=0]{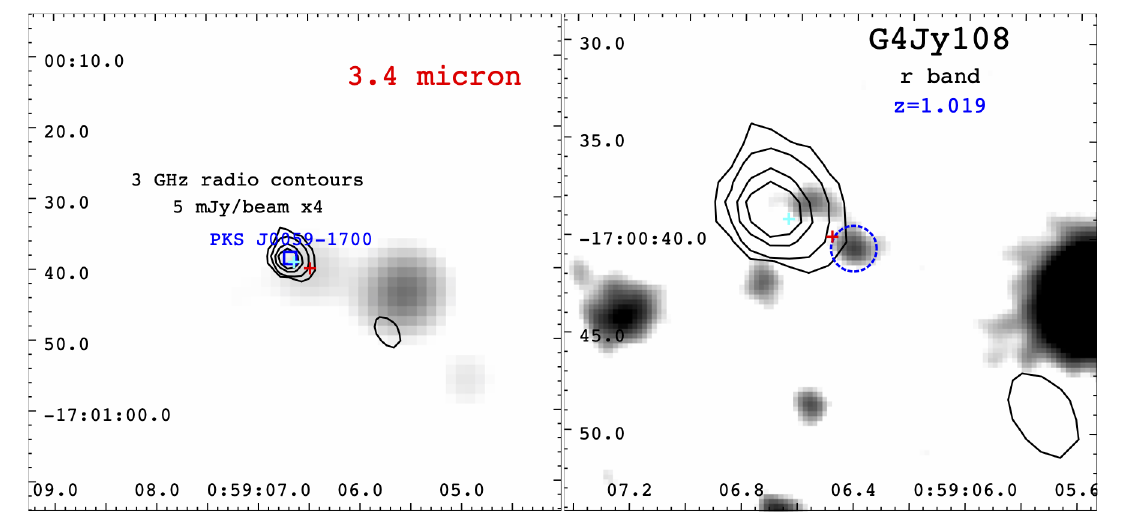}
\includegraphics[height=3.8cm,width=8.8cm,angle=0]{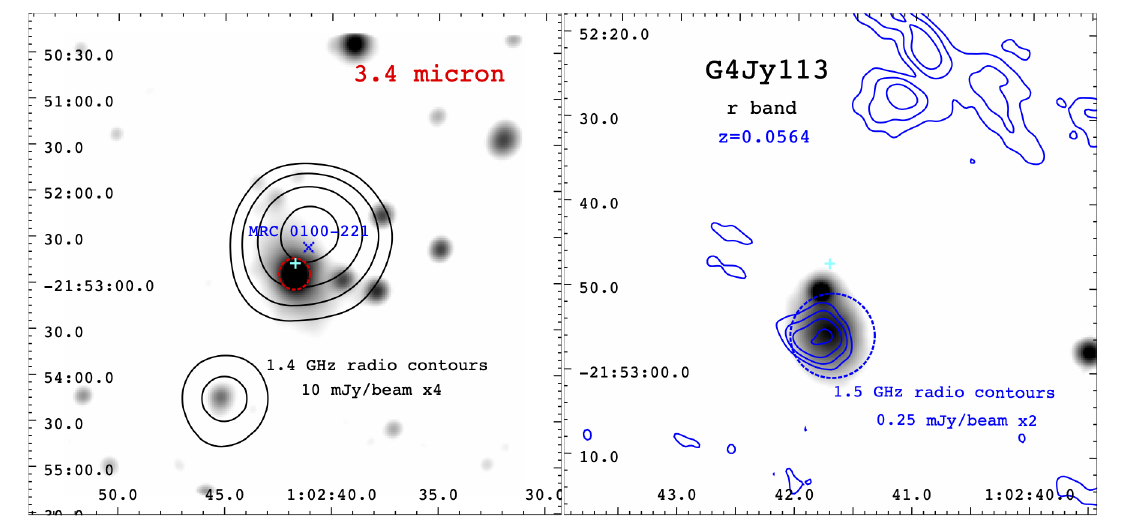}
\includegraphics[height=3.8cm,width=8.8cm,angle=0]{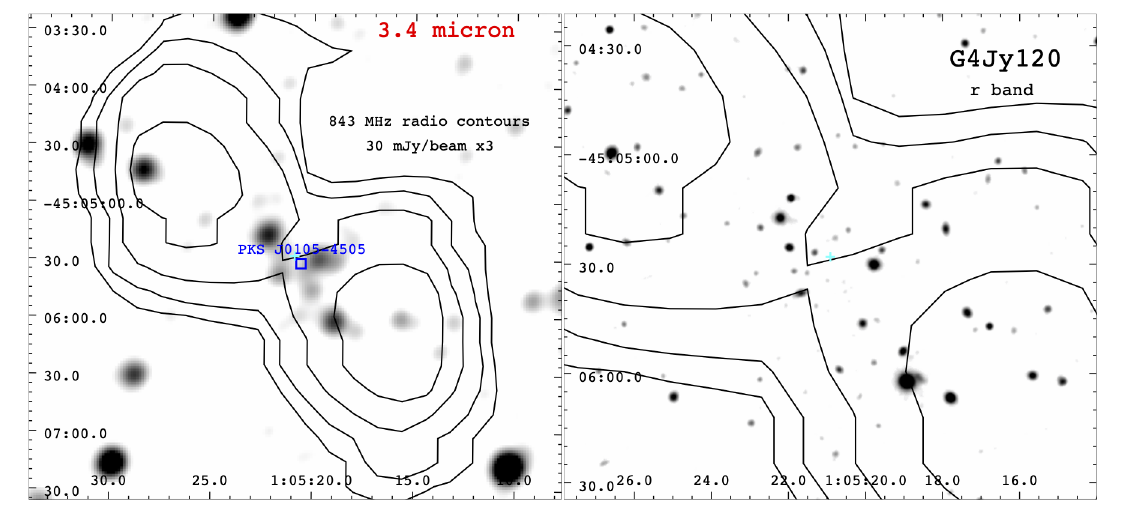}
\includegraphics[height=3.8cm,width=8.8cm,angle=0]{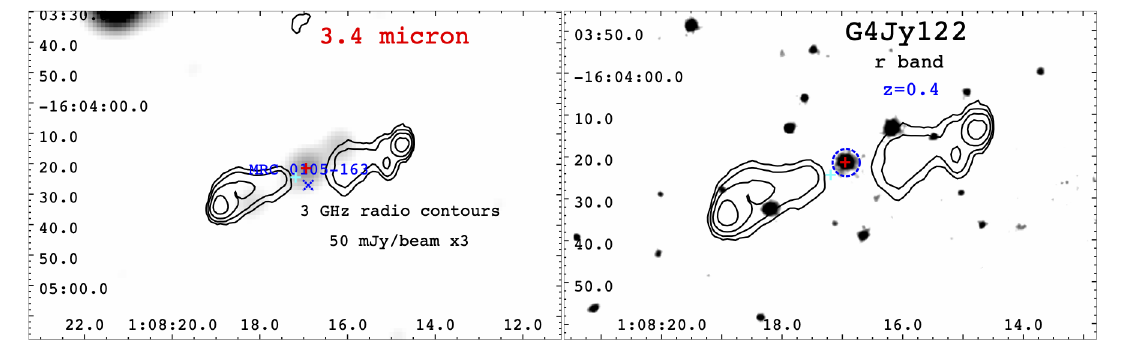}
\includegraphics[height=3.8cm,width=8.8cm,angle=0]{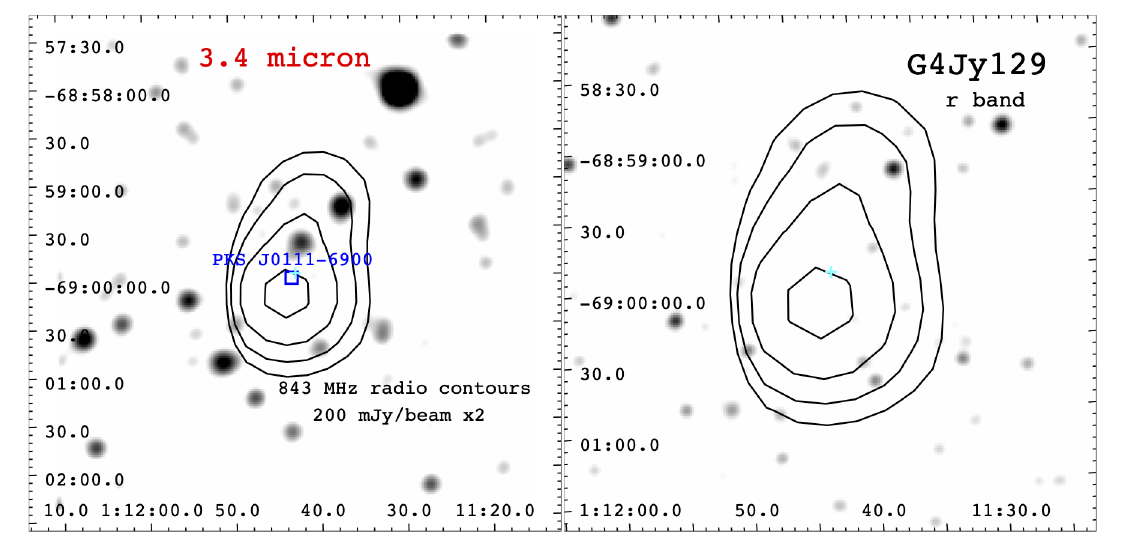}
\includegraphics[height=3.8cm,width=8.8cm,angle=0]{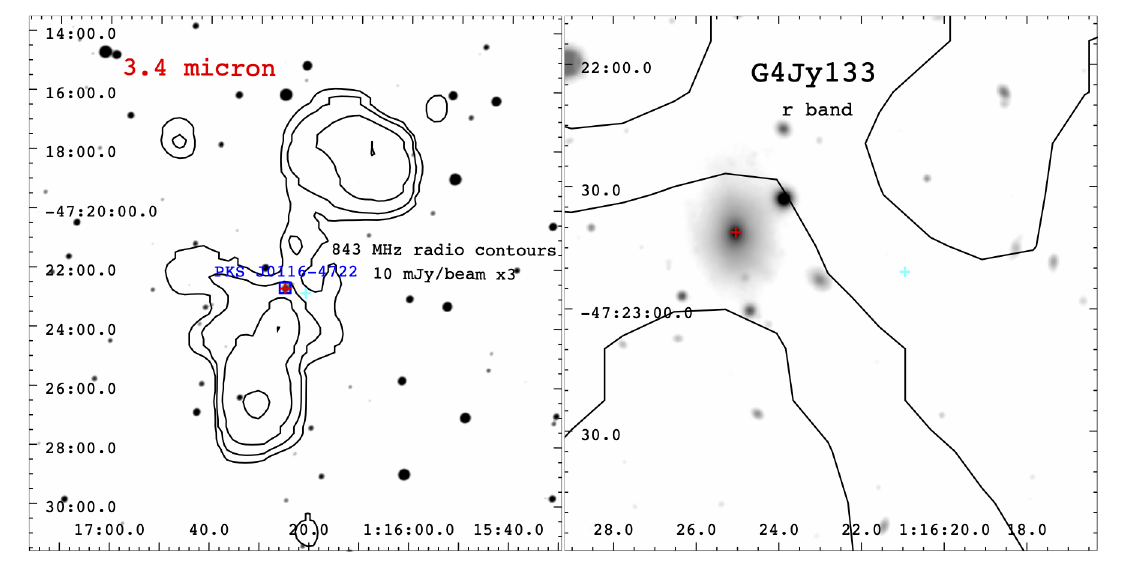}
\caption{Same as Figure~\ref{fig:example1} for the following \cs\ radio sources: \\ 
G4Jy\,77, G4Jy\,78, G4Jy\,85, G4Jy\,86, G4Jy\,90, G4Jy\,93, G4Jy\,108, G4Jy\,113, G4Jy\,120, G4Jy\,122, G4Jy\,129, G4Jy\,133.}
\end{center}
\end{figure*}

\begin{figure*}[!th]
\begin{center}
\includegraphics[height=3.8cm,width=8.8cm,angle=0]{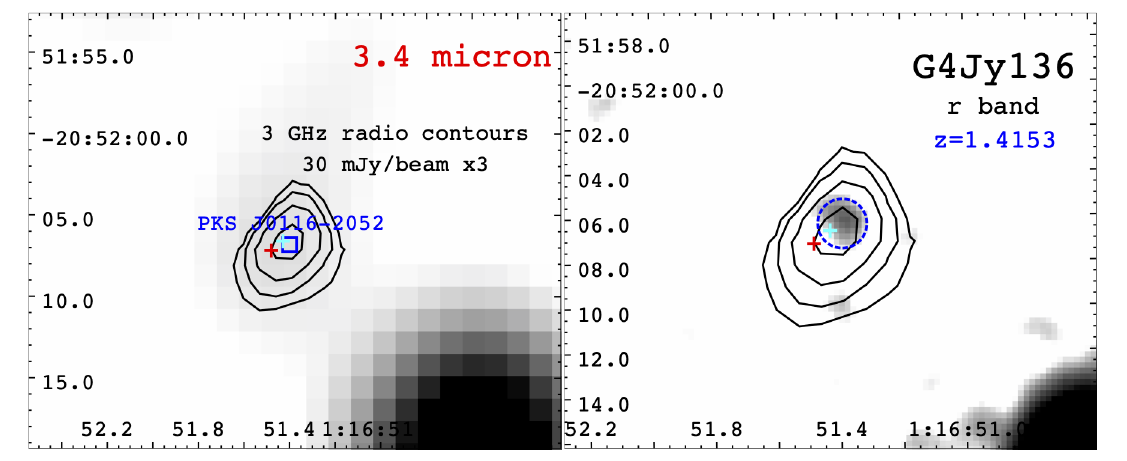}
\includegraphics[height=3.8cm,width=8.8cm,angle=0]{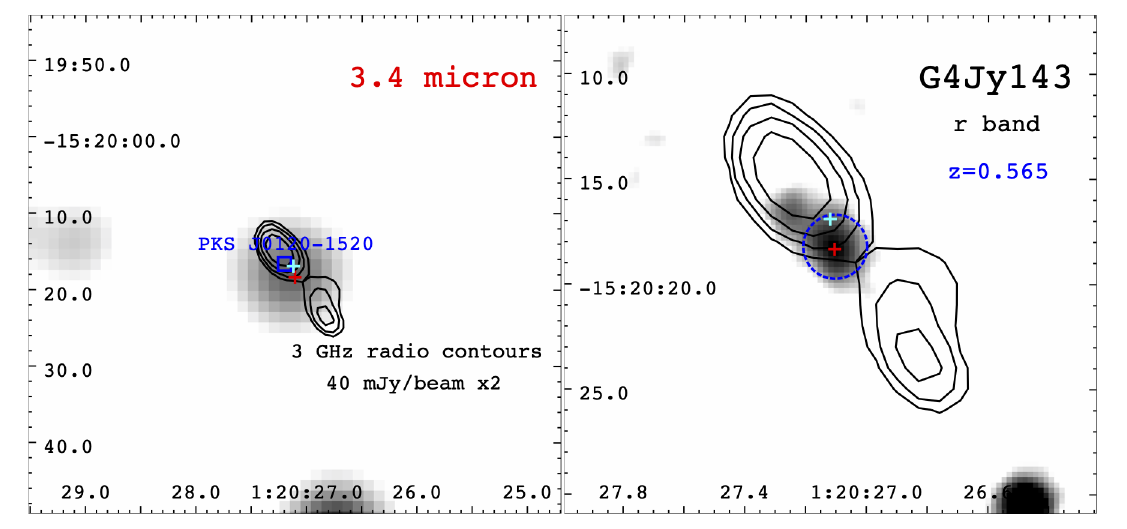}
\includegraphics[height=3.8cm,width=8.8cm,angle=0]{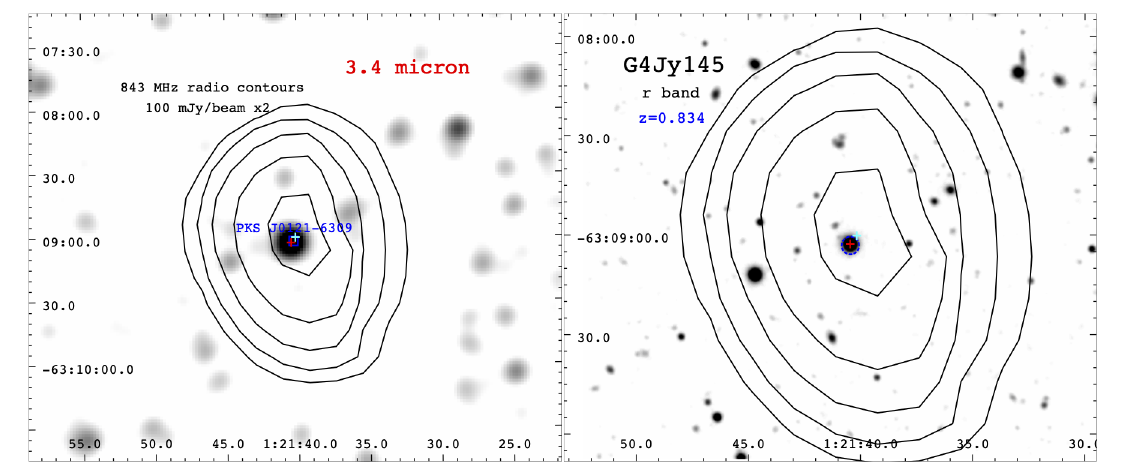}
\includegraphics[height=3.8cm,width=8.8cm,angle=0]{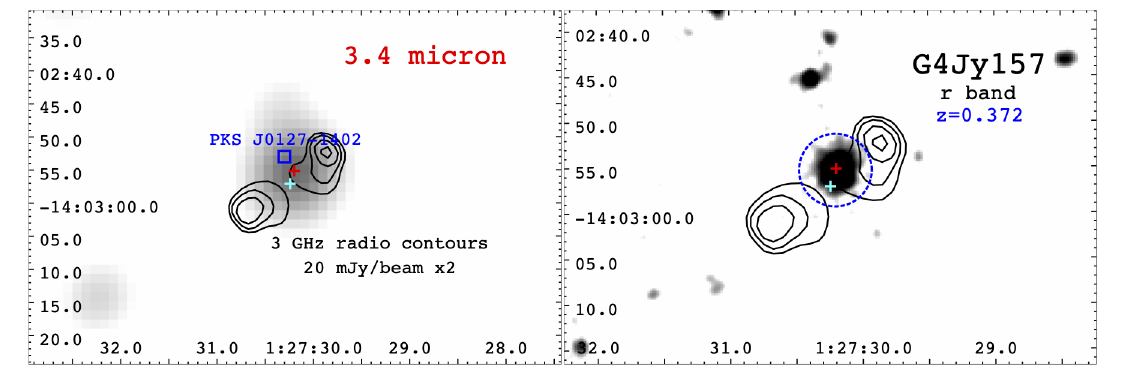}
\includegraphics[height=3.8cm,width=8.8cm,angle=0]{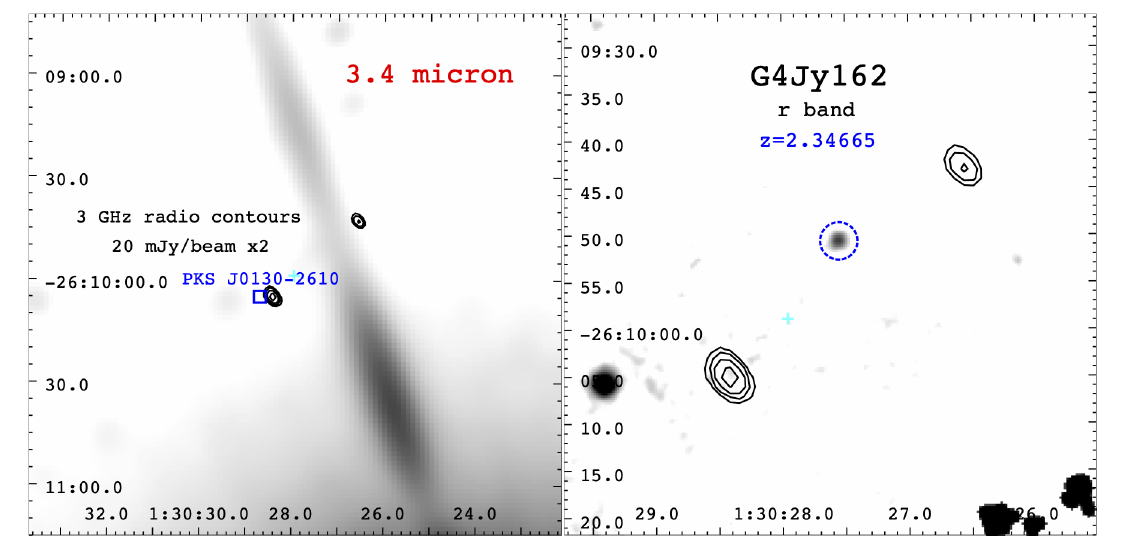}
\includegraphics[height=3.8cm,width=8.8cm,angle=0]{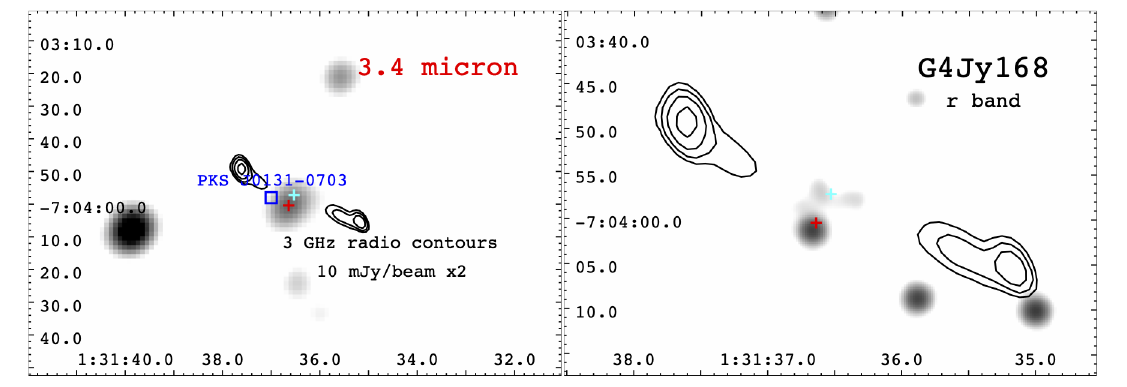}
\includegraphics[height=3.8cm,width=8.8cm,angle=0]{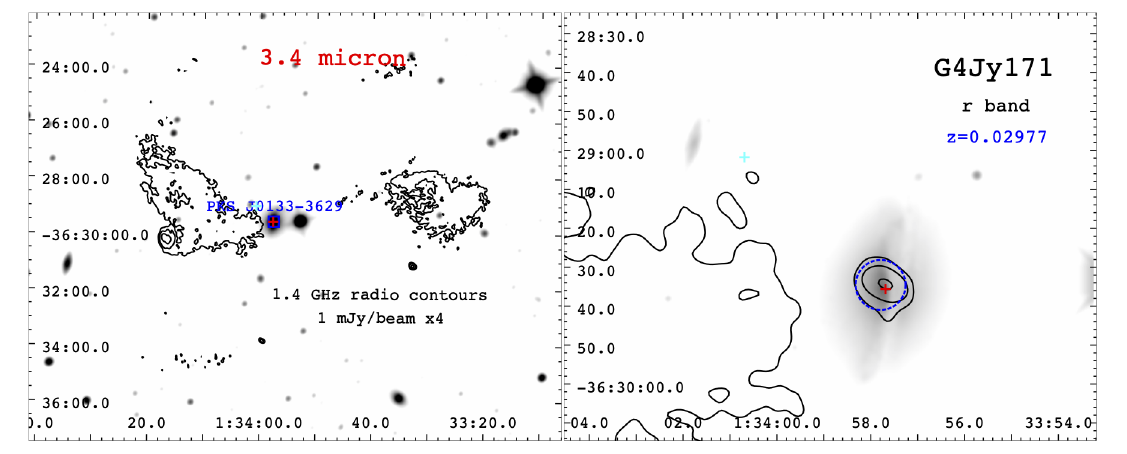}
\includegraphics[height=3.8cm,width=8.8cm,angle=0]{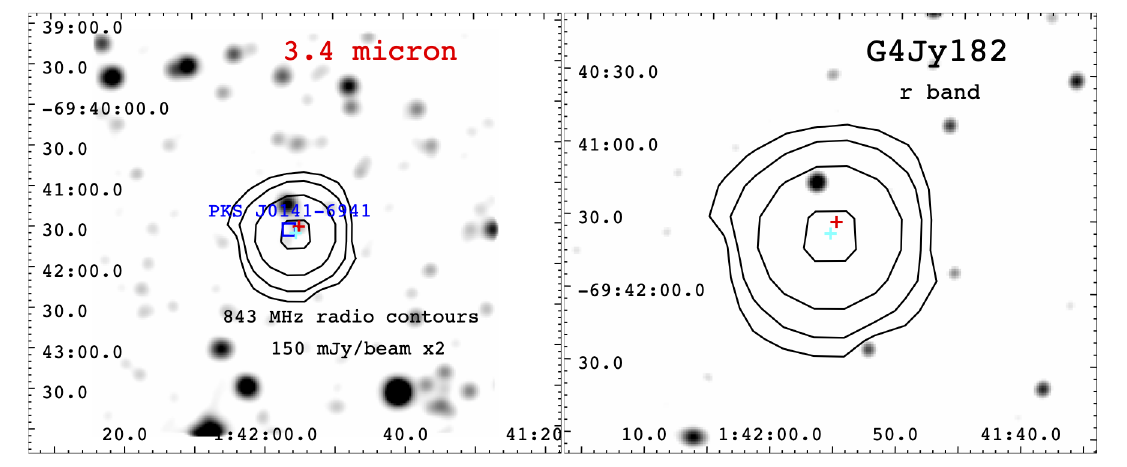}
\includegraphics[height=3.8cm,width=8.8cm,angle=0]{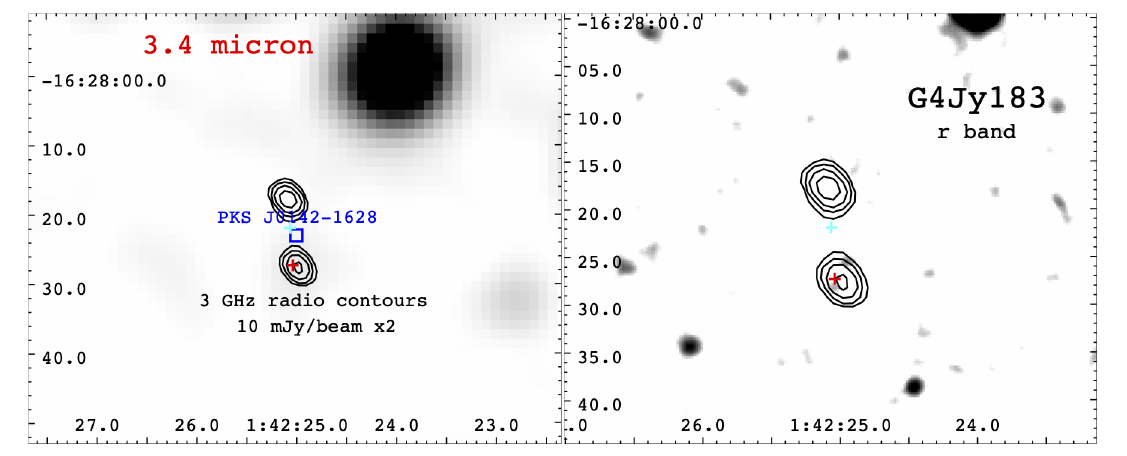}
\includegraphics[height=3.8cm,width=8.8cm,angle=0]{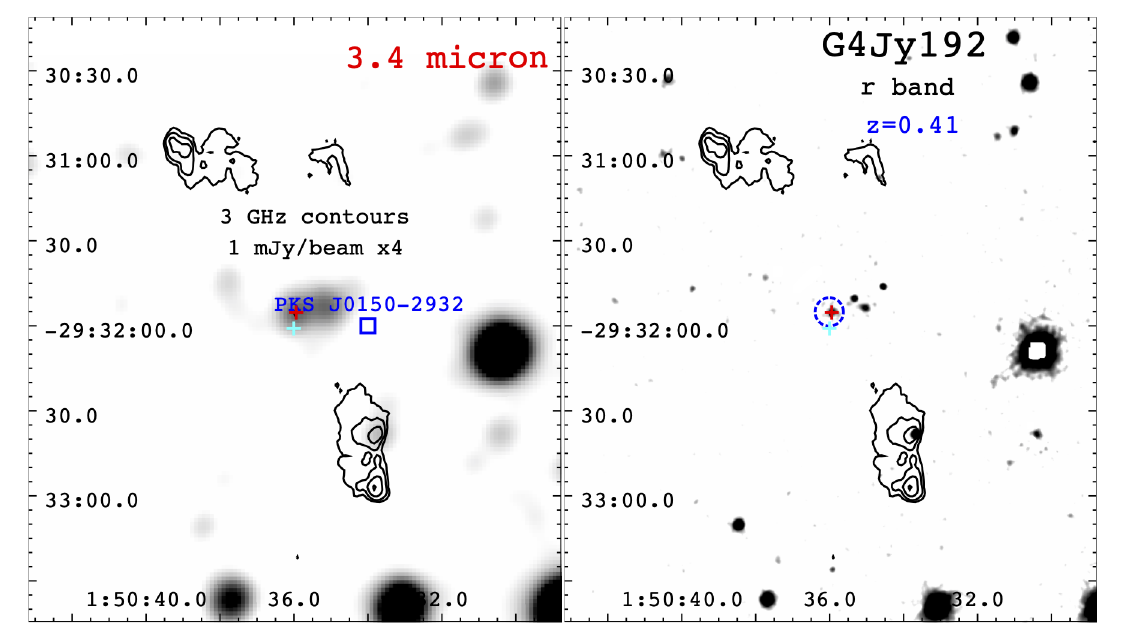}
\includegraphics[height=3.8cm,width=8.8cm,angle=0]{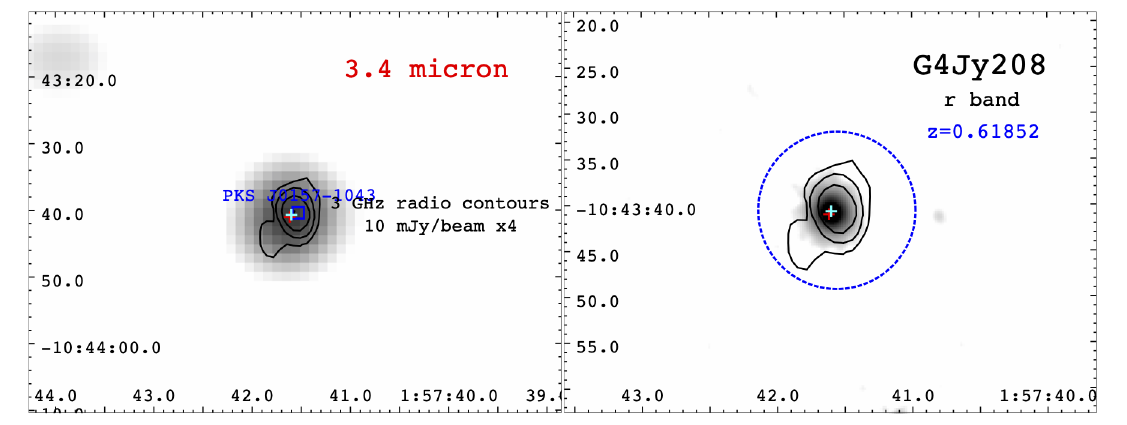}
\includegraphics[height=3.8cm,width=8.8cm,angle=0]{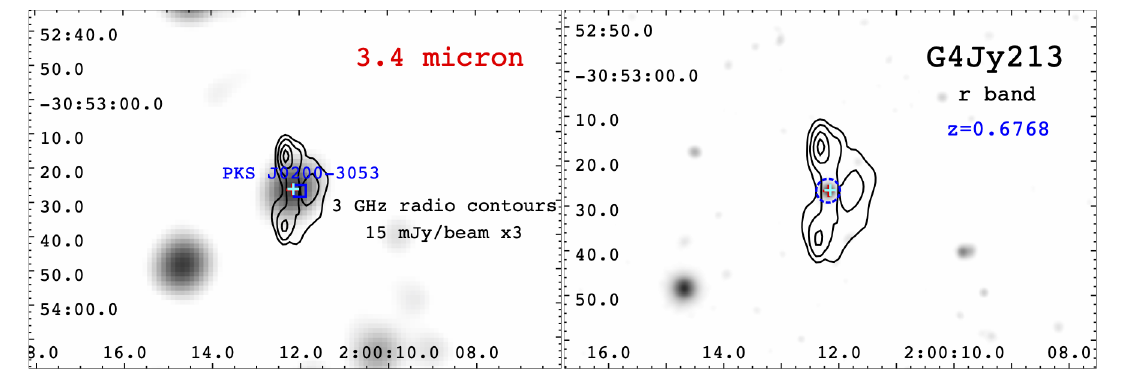}
\caption{Same as Figure~\ref{fig:example1} for the following \cs\ radio sources: \\ 
G4Jy\,136, G4Jy\,143, G4Jy\,145, G4Jy\,157, G4Jy\,162, G4Jy\,168, G4Jy\,171, G4Jy\,182, G4Jy\,183, G4Jy\,192, G4Jy\,208, G4Jy\,213.}
\end{center}
\end{figure*}

\begin{figure*}[!th]
\begin{center}
\includegraphics[height=3.8cm,width=8.8cm,angle=0]{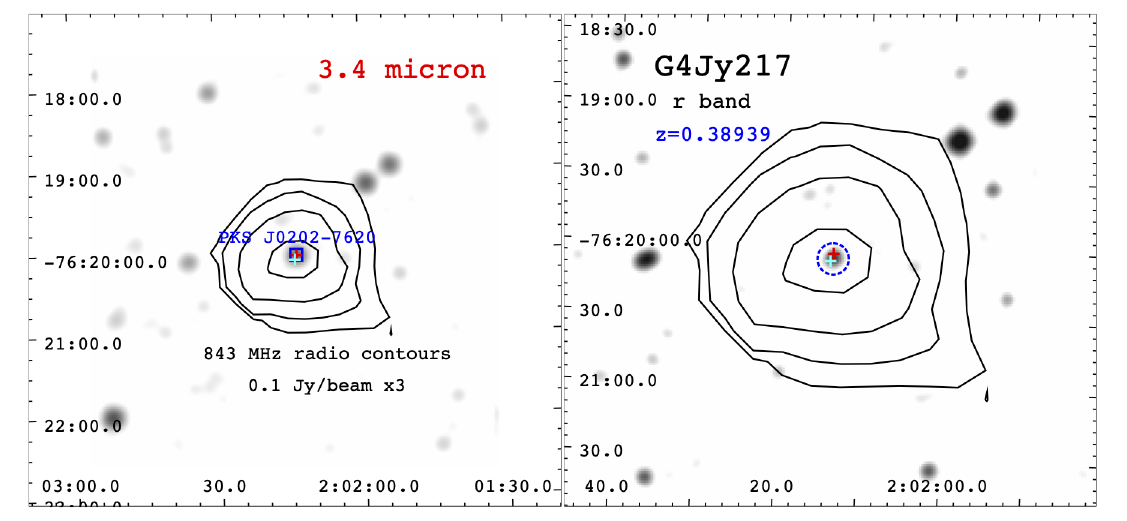}
\includegraphics[height=3.8cm,width=8.8cm,angle=0]{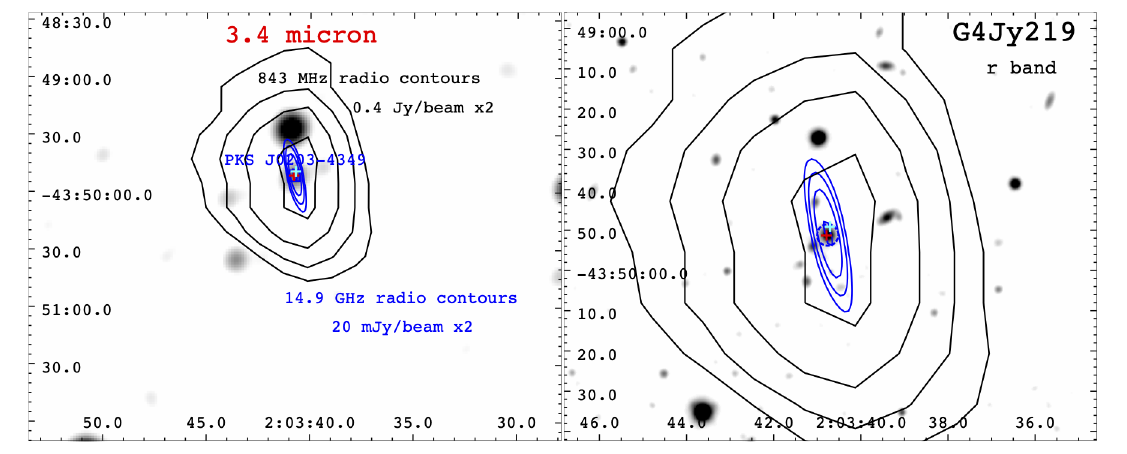}
\includegraphics[height=3.8cm,width=8.8cm,angle=0]{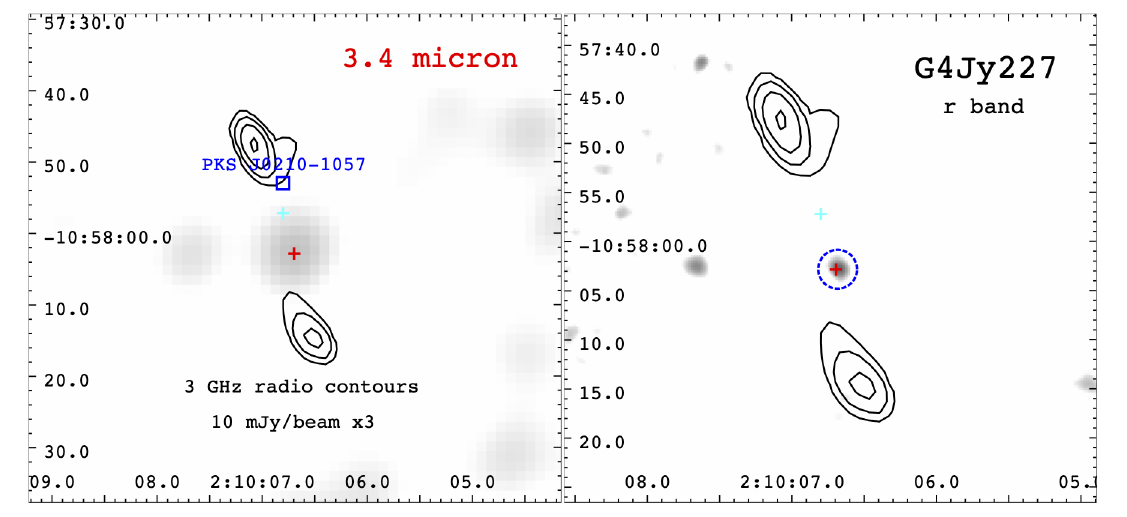}
\includegraphics[height=3.8cm,width=8.8cm,angle=0]{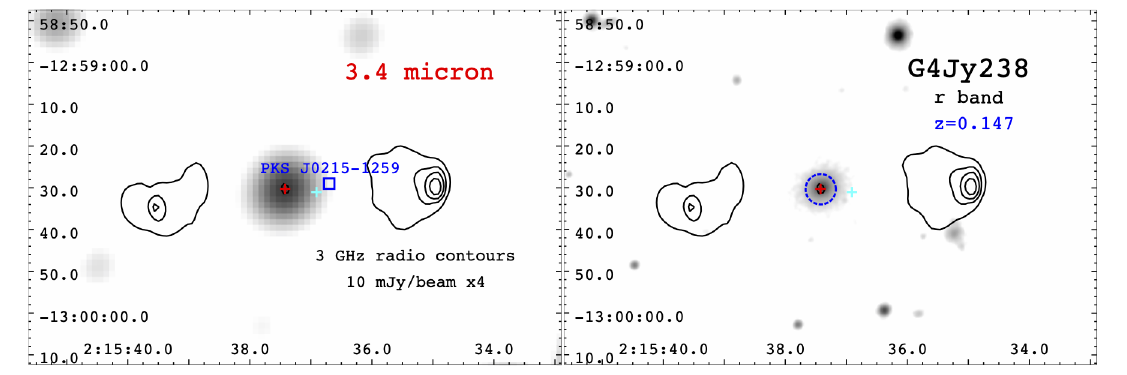}
\includegraphics[height=3.8cm,width=8.8cm,angle=0]{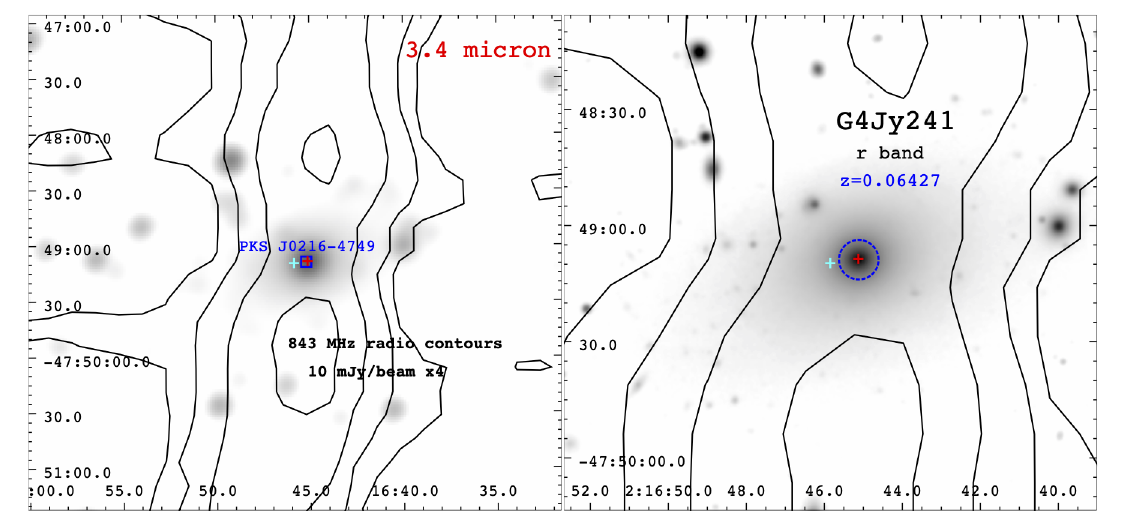}
\includegraphics[height=3.8cm,width=8.8cm,angle=0]{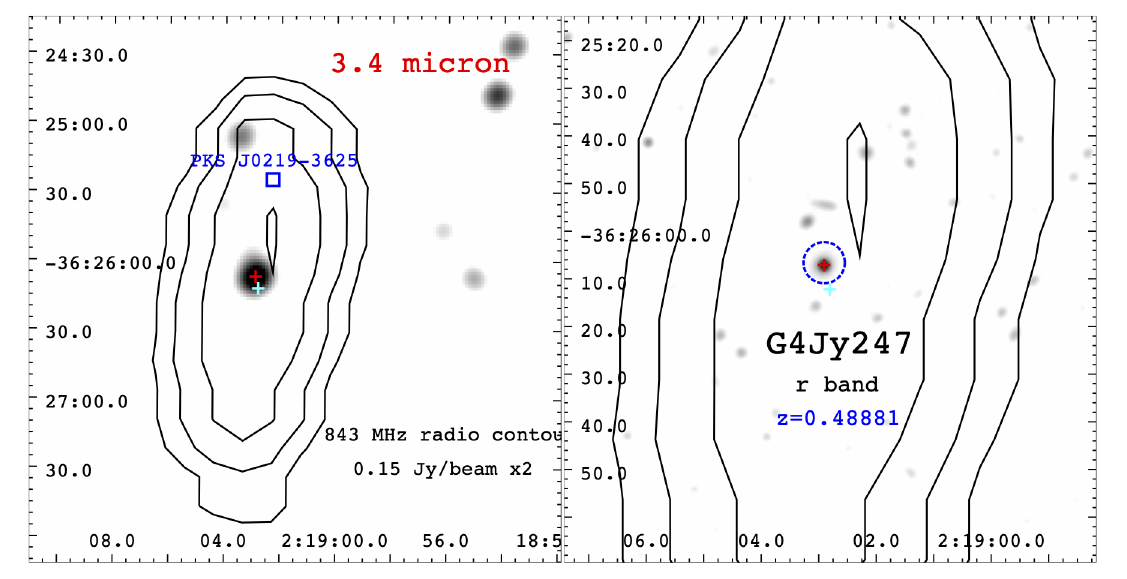}
\includegraphics[height=3.8cm,width=8.8cm,angle=0]{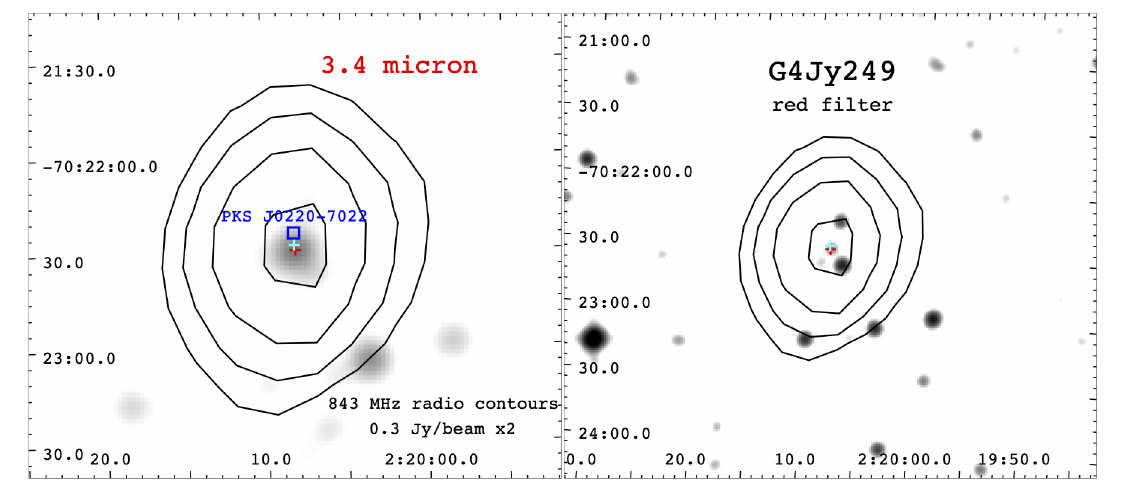}
\includegraphics[height=3.8cm,width=8.8cm,angle=0]{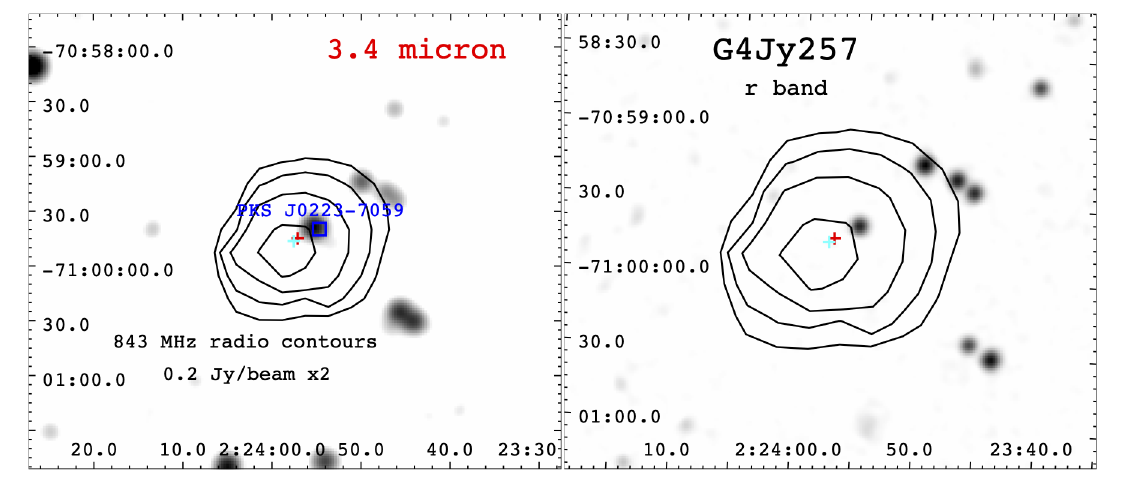}
\includegraphics[height=3.8cm,width=8.8cm,angle=0]{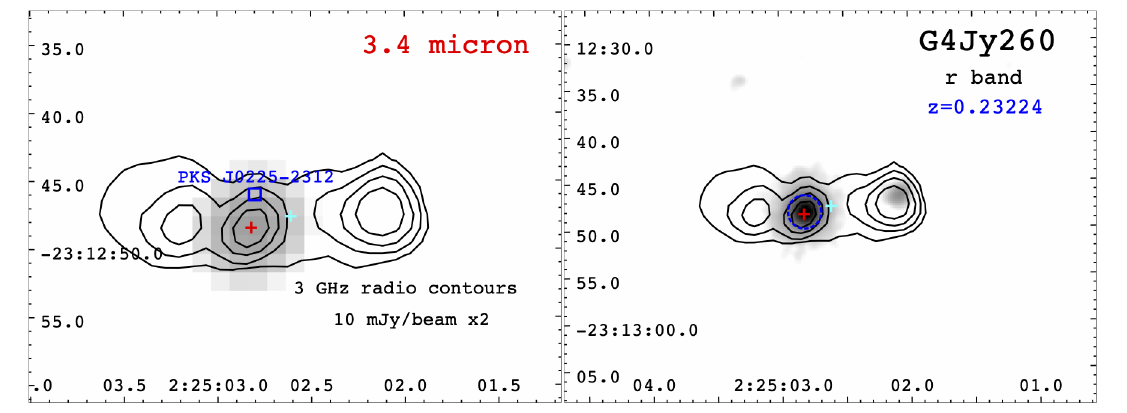}
\includegraphics[height=3.8cm,width=8.8cm,angle=0]{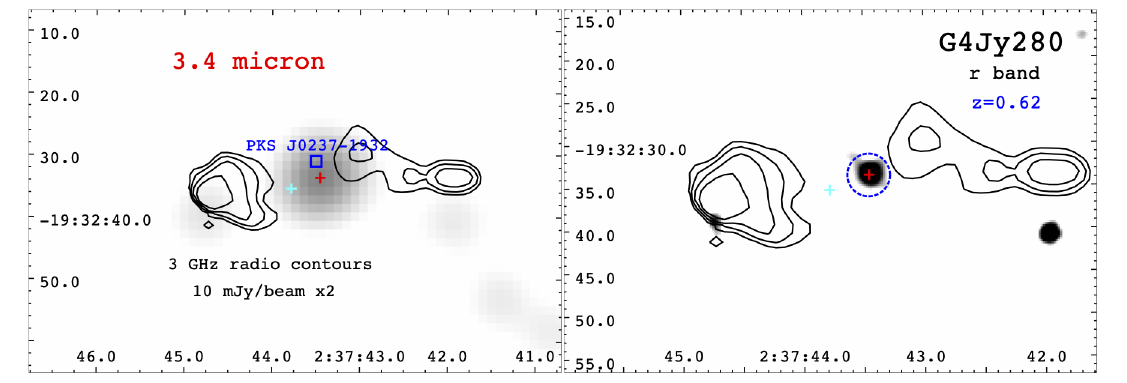}
\includegraphics[height=3.8cm,width=8.8cm,angle=0]{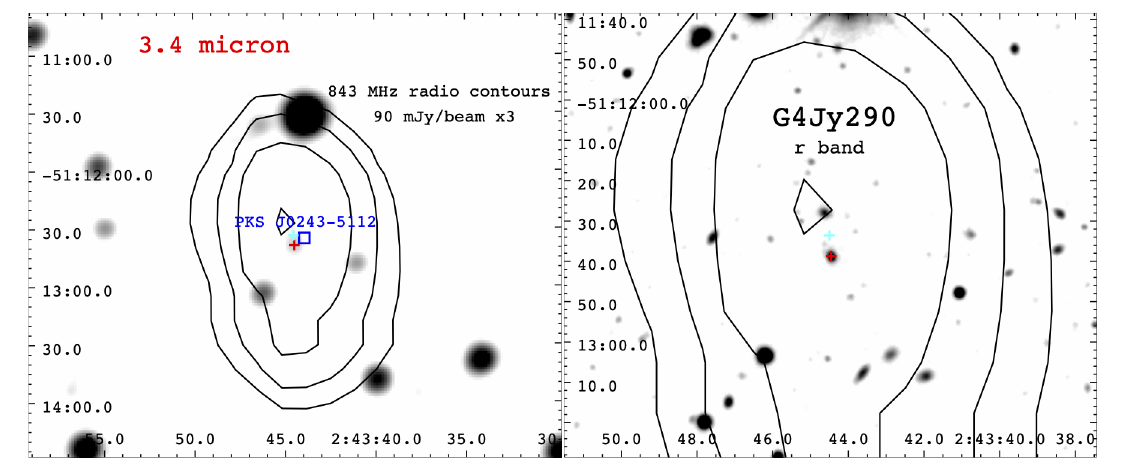}
\includegraphics[height=3.8cm,width=8.8cm,angle=0]{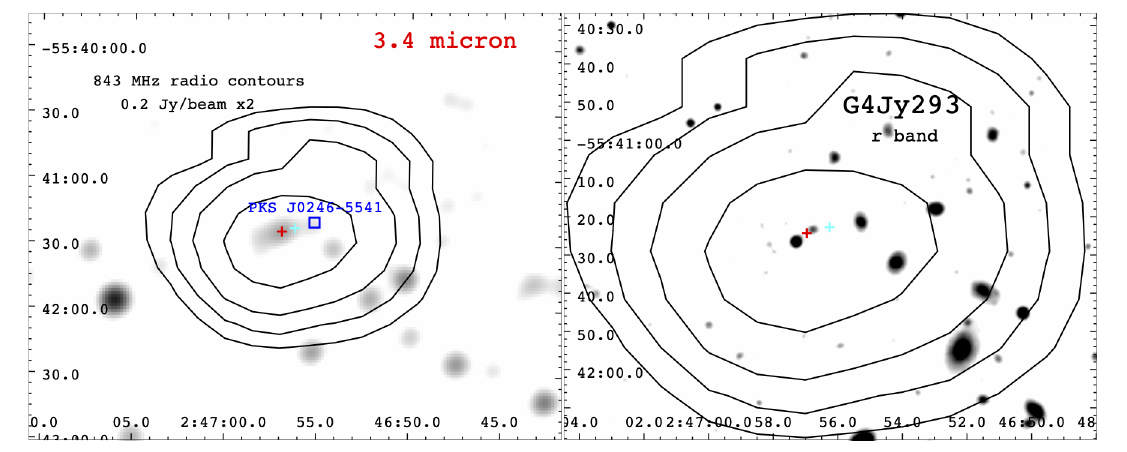}
\caption{Same as Figure~\ref{fig:example1} for the following \cs\ radio sources: \\ 
G4Jy\,217, G4Jy\,219, G4Jy\,227, G4Jy\,238, G4Jy\,241, G4Jy\,247, G4Jy\,249, G4Jy\,257, G4Jy\,260, G4Jy\,280, G4Jy\,290, G4Jy\,293.}
\end{center}
\end{figure*}

\begin{figure*}[!th]
\begin{center}
\includegraphics[height=3.8cm,width=8.8cm,angle=0]{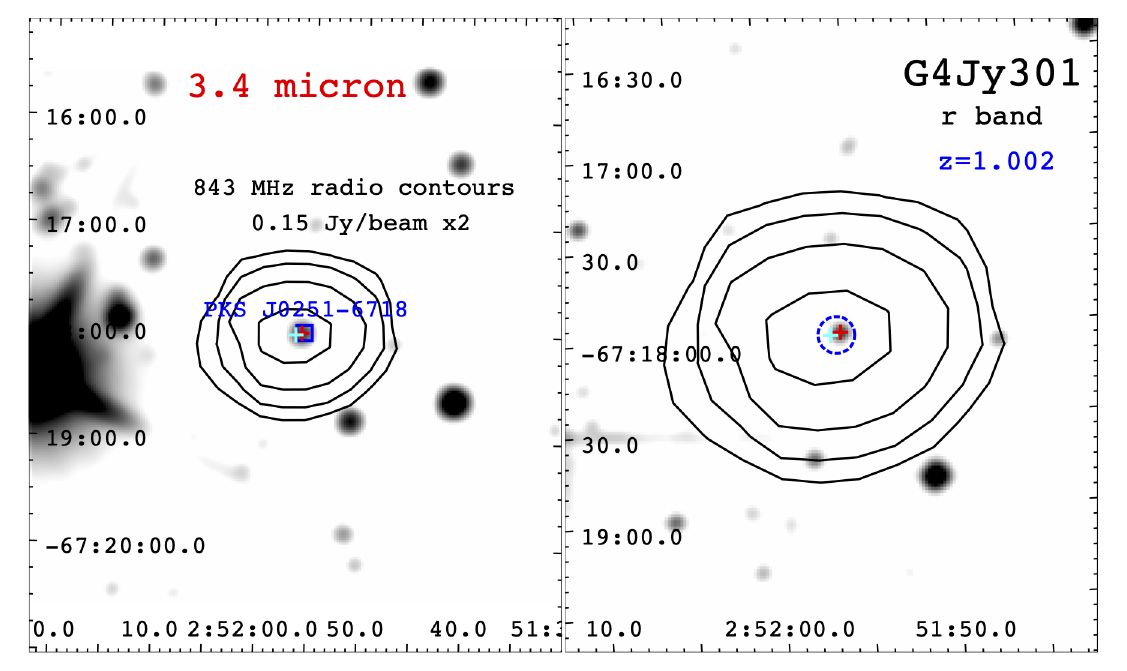}
\includegraphics[height=3.8cm,width=8.8cm,angle=0]{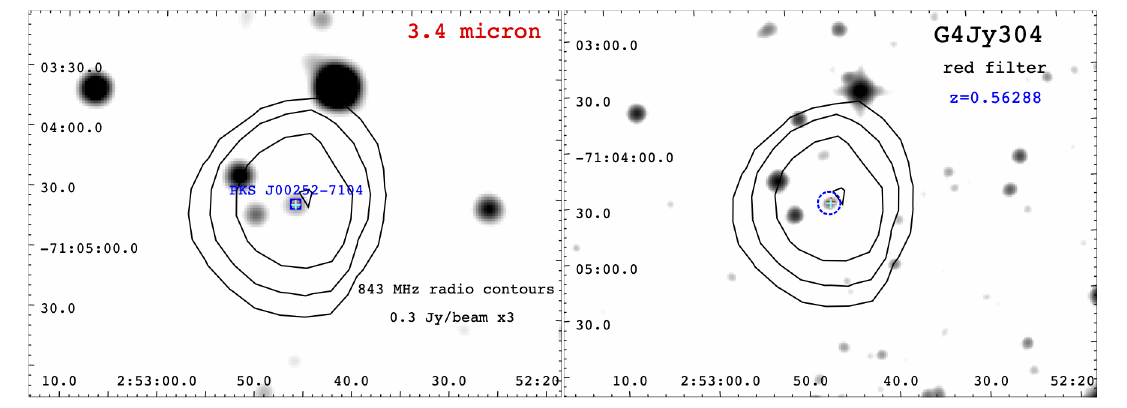}
\includegraphics[height=3.8cm,width=8.8cm,angle=0]{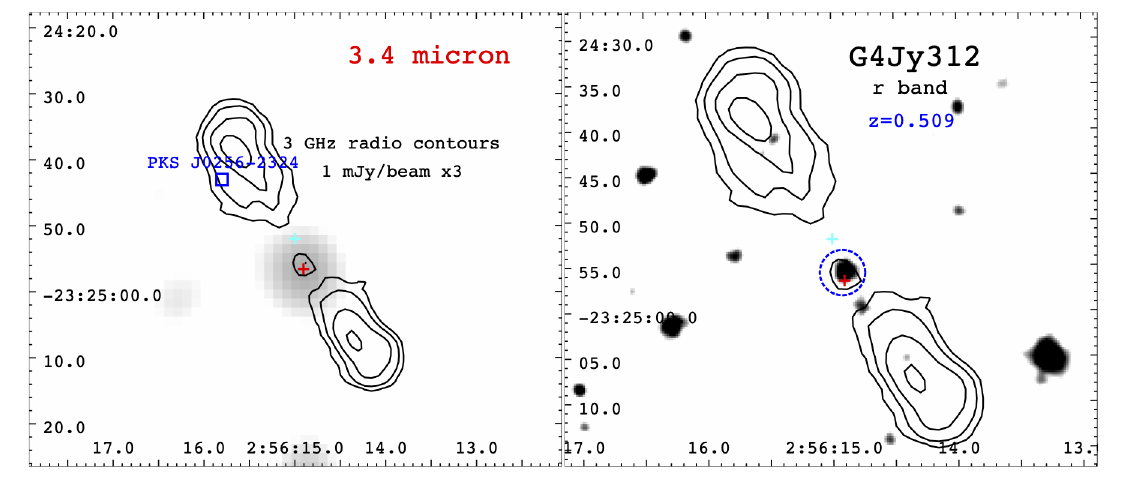}
\includegraphics[height=3.8cm,width=8.8cm,angle=0]{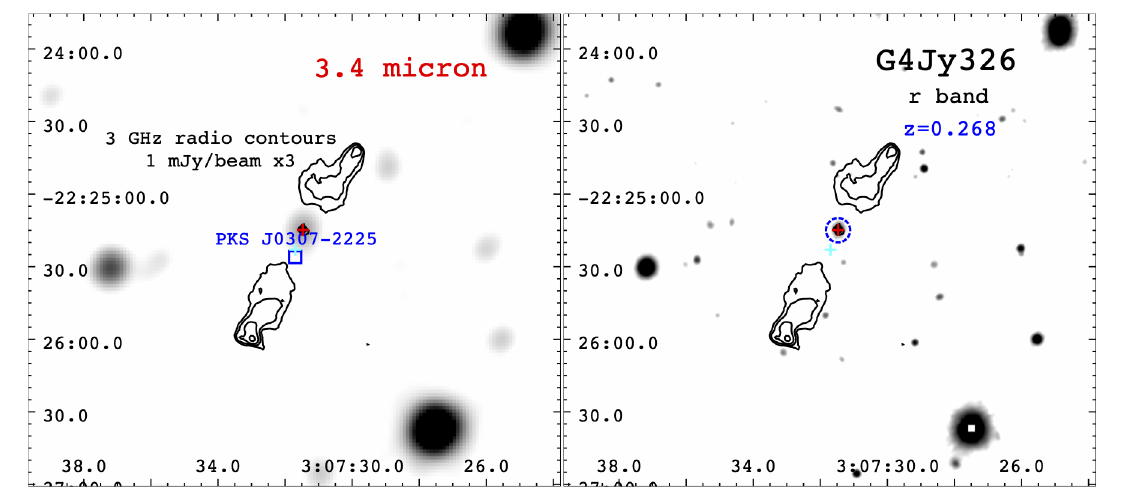}
\includegraphics[height=3.8cm,width=8.8cm,angle=0]{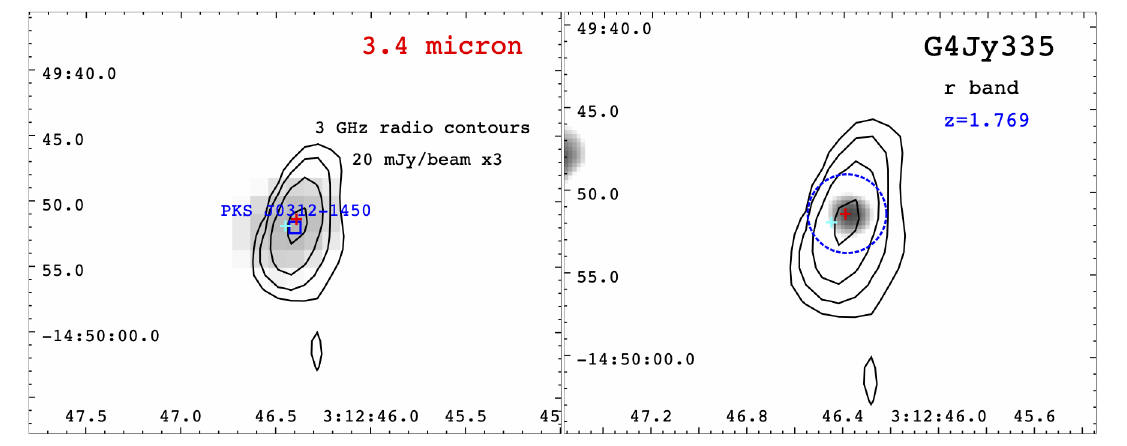}
\includegraphics[height=3.8cm,width=8.8cm,angle=0]{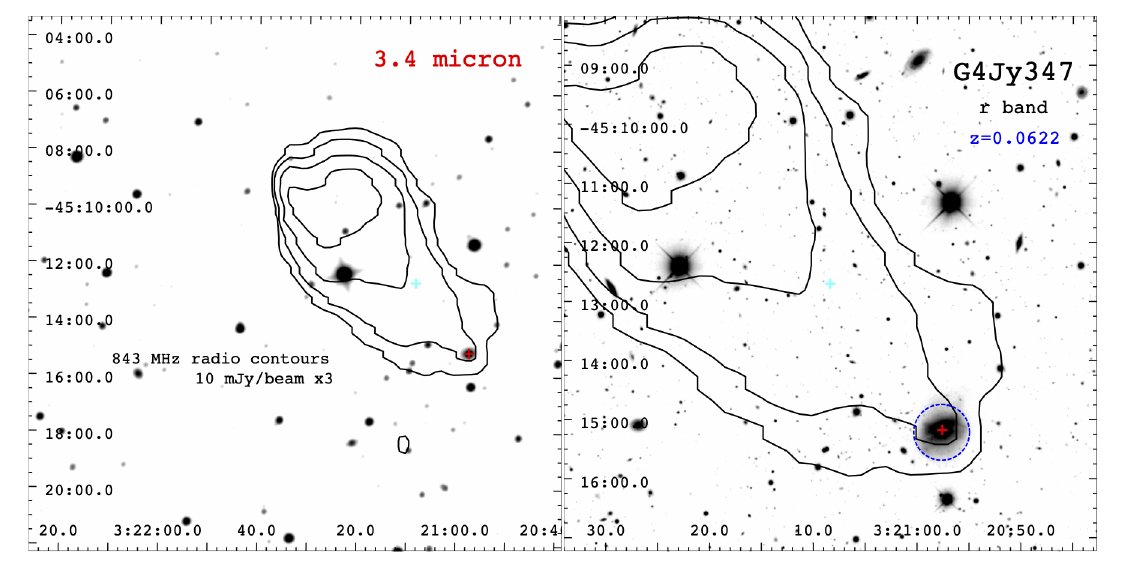}
\includegraphics[height=3.8cm,width=8.8cm,angle=0]{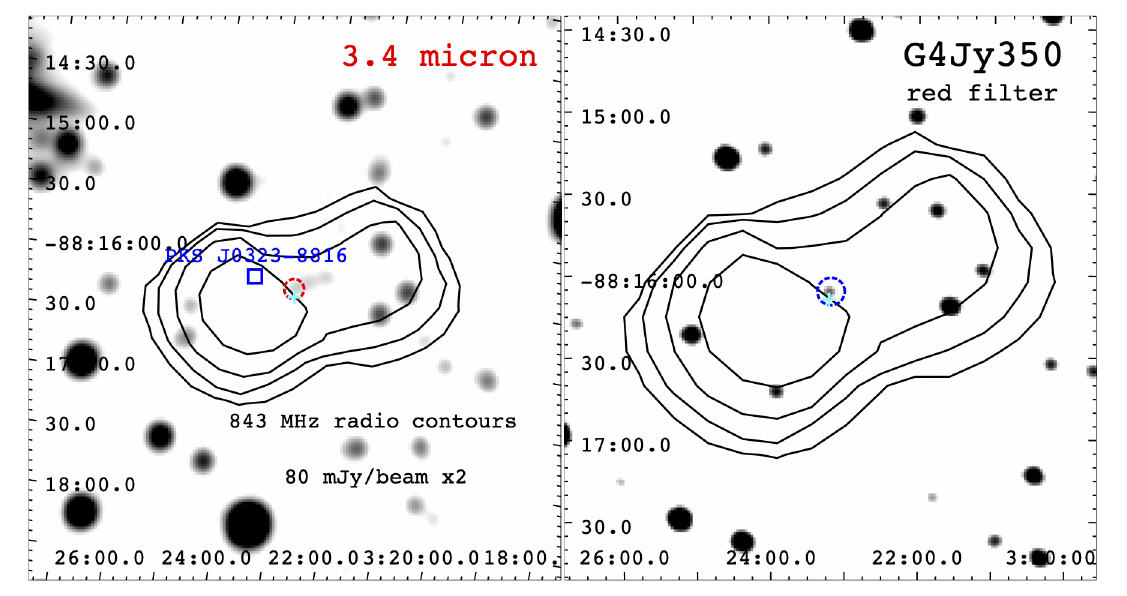}
\includegraphics[height=3.8cm,width=8.8cm,angle=0]{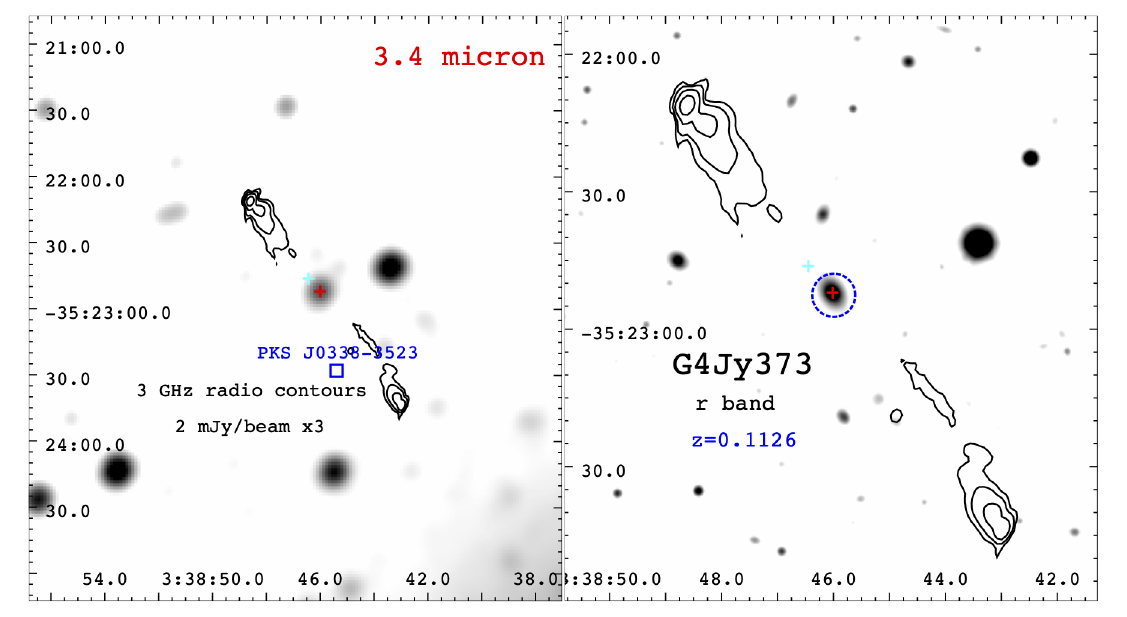}
\includegraphics[height=3.8cm,width=8.8cm,angle=0]{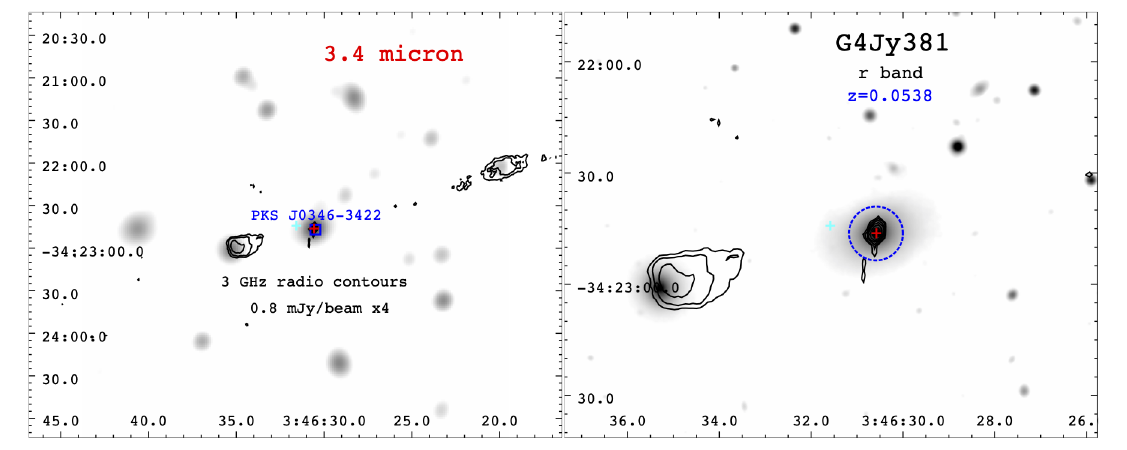}
\includegraphics[height=3.8cm,width=8.8cm,angle=0]{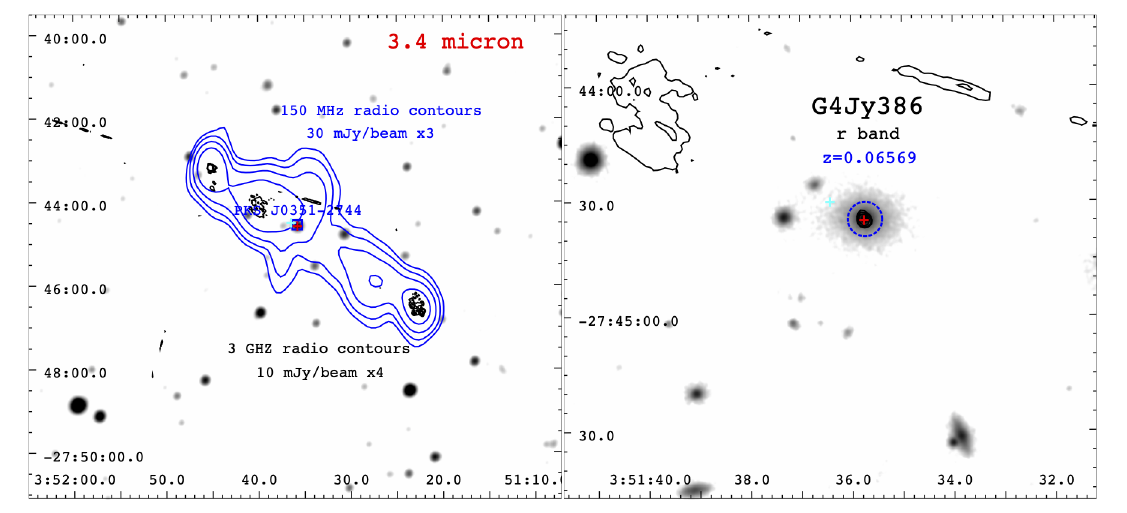}
\includegraphics[height=3.8cm,width=8.8cm,angle=0]{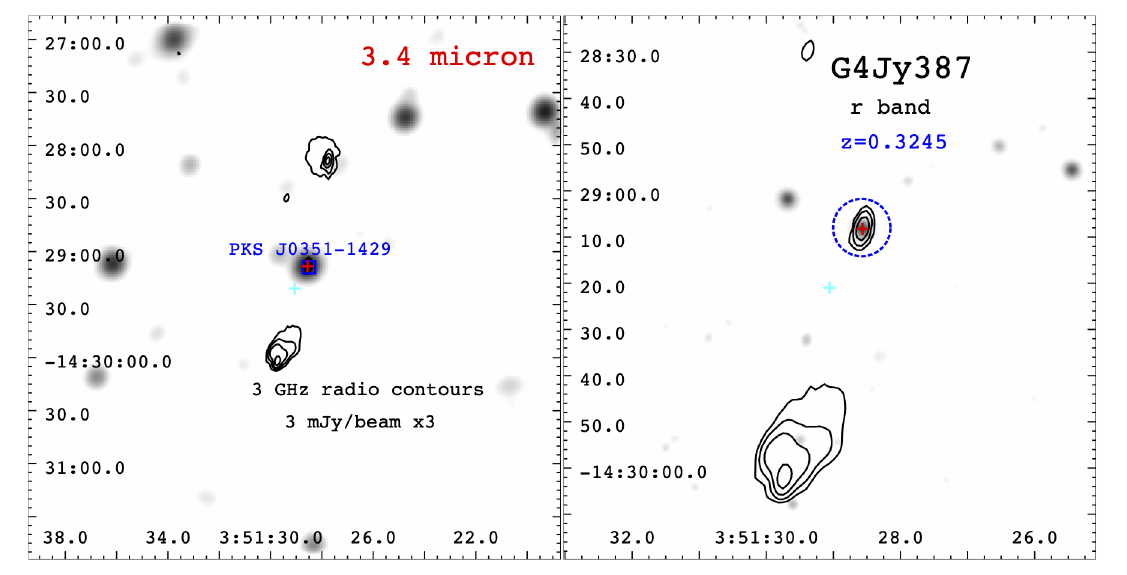}
\includegraphics[height=3.8cm,width=8.8cm,angle=0]{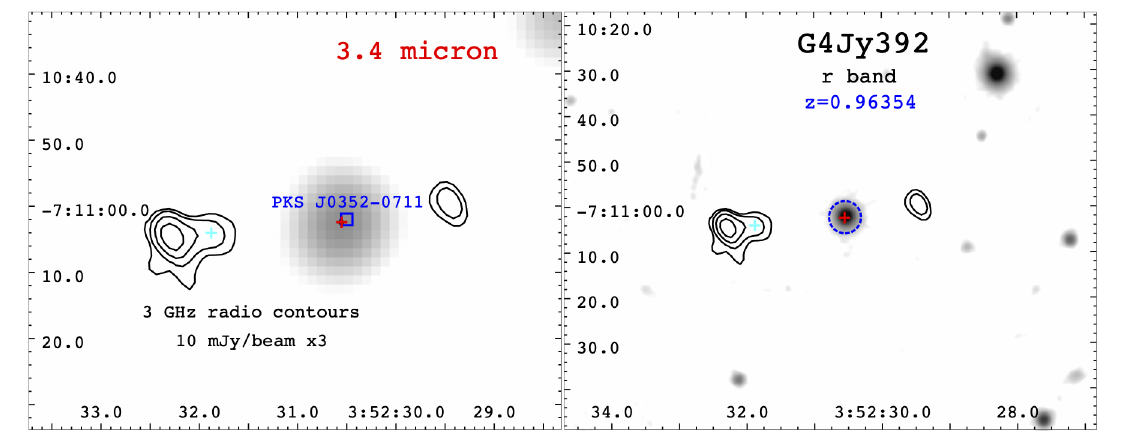}
\caption{Same as Figure~\ref{fig:example1} for the following \cs\ radio sources: \\ 
G4Jy\,301, G4Jy\,304, G4Jy\,312, G4Jy\,326, G4Jy\,335, G4Jy\,347, G4Jy\,350, G4Jy\,373, G4Jy\,381, G4Jy\,386, G4Jy\,387, G4Jy\,392.}
\end{center}
\end{figure*}

\begin{figure*}[!th]
\begin{center}
\includegraphics[height=3.8cm,width=8.8cm,angle=0]{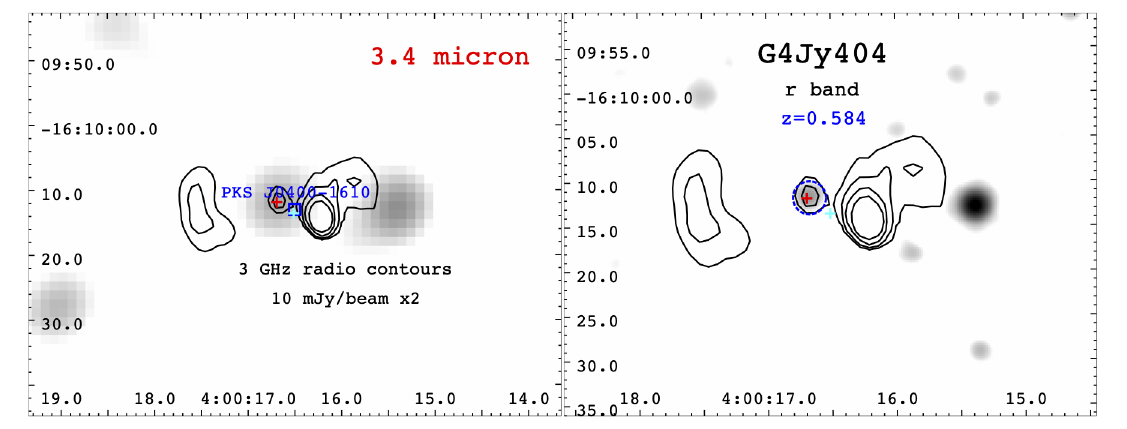}
\includegraphics[height=3.8cm,width=8.8cm,angle=0]{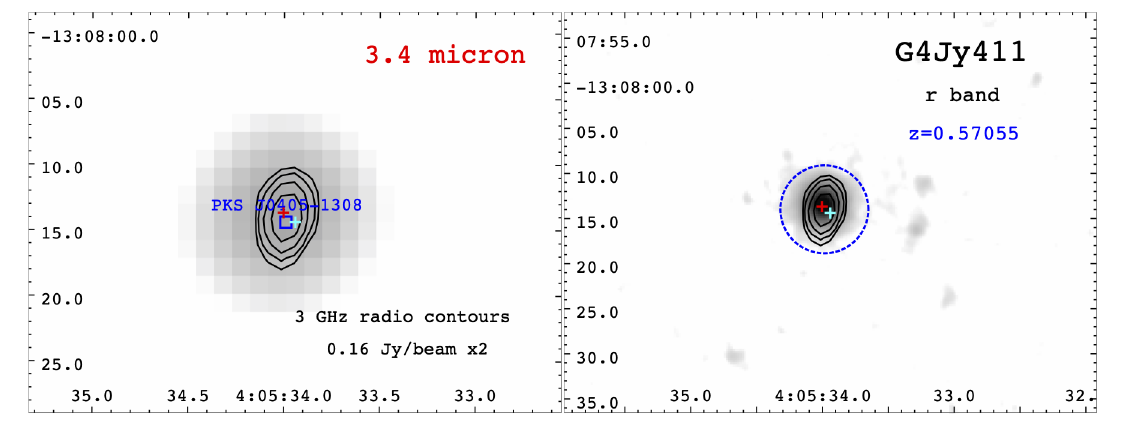}
\includegraphics[height=3.8cm,width=8.8cm,angle=0]{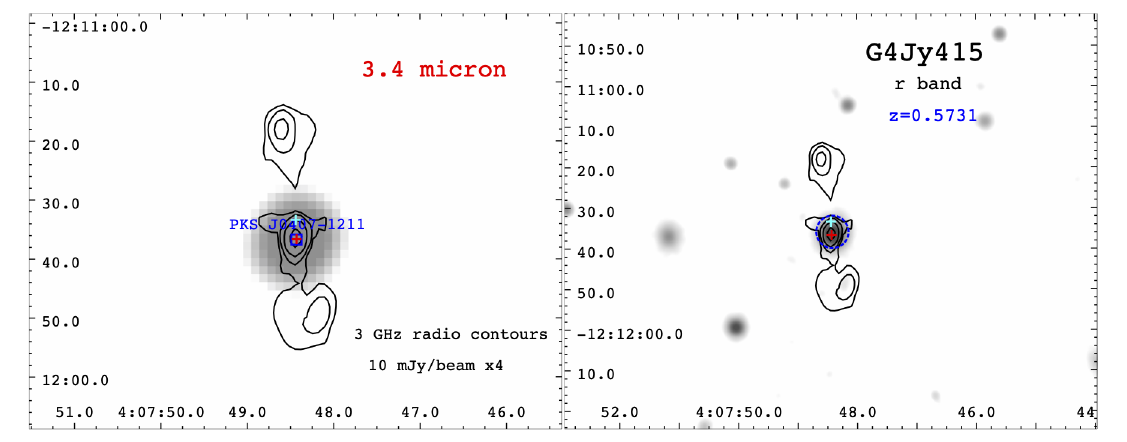}
\includegraphics[height=3.8cm,width=8.8cm,angle=0]{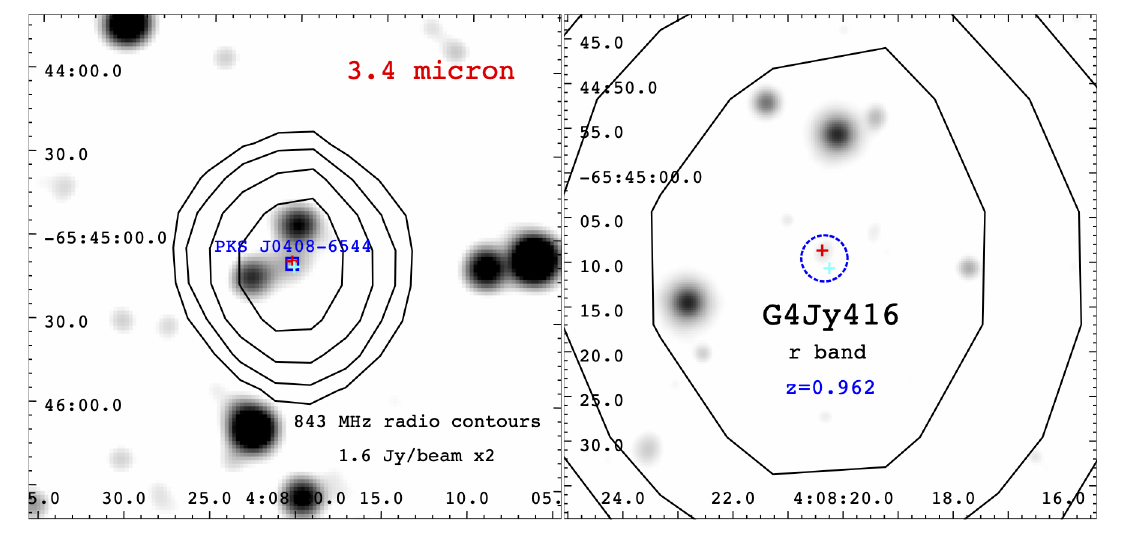}
\includegraphics[height=3.8cm,width=8.8cm,angle=0]{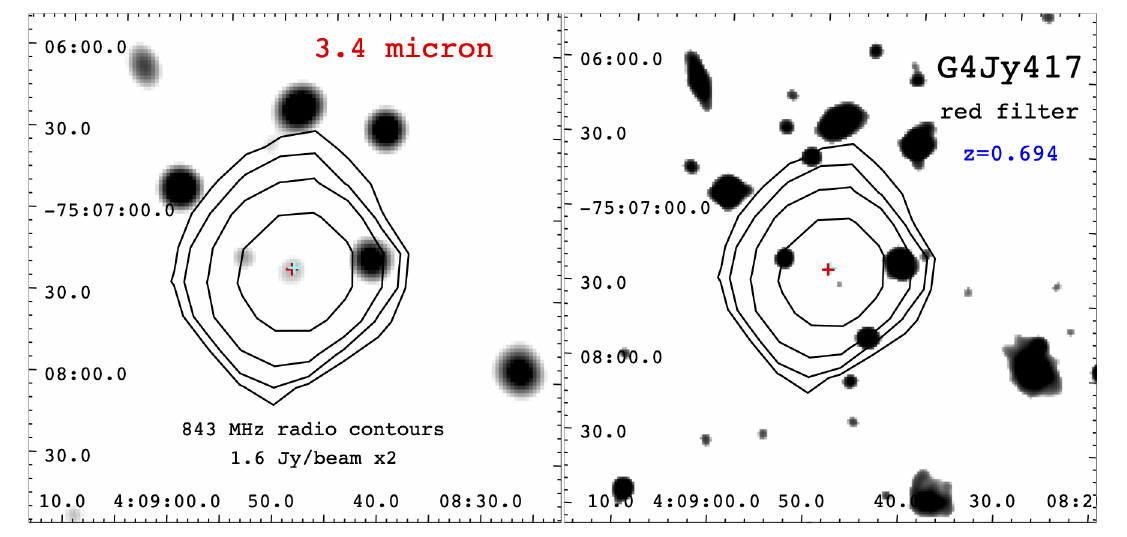}
\includegraphics[height=3.8cm,width=8.8cm,angle=0]{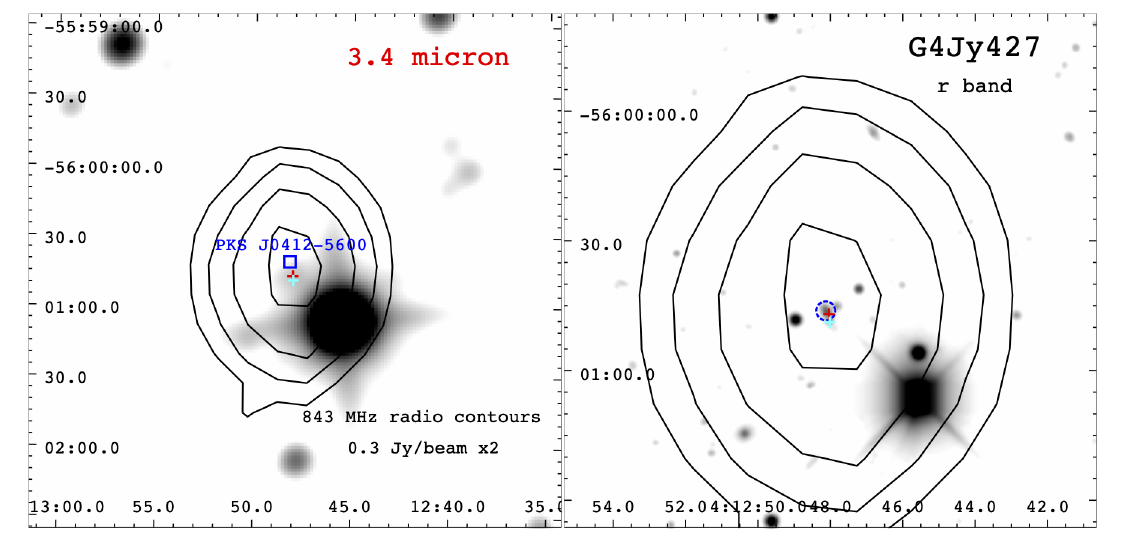}
\includegraphics[height=3.8cm,width=8.8cm,angle=0]{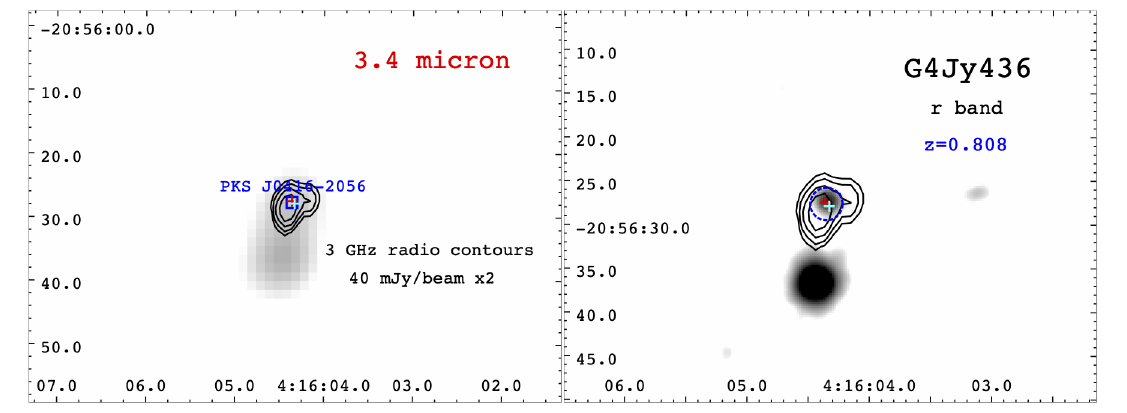}
\includegraphics[height=3.8cm,width=8.8cm,angle=0]{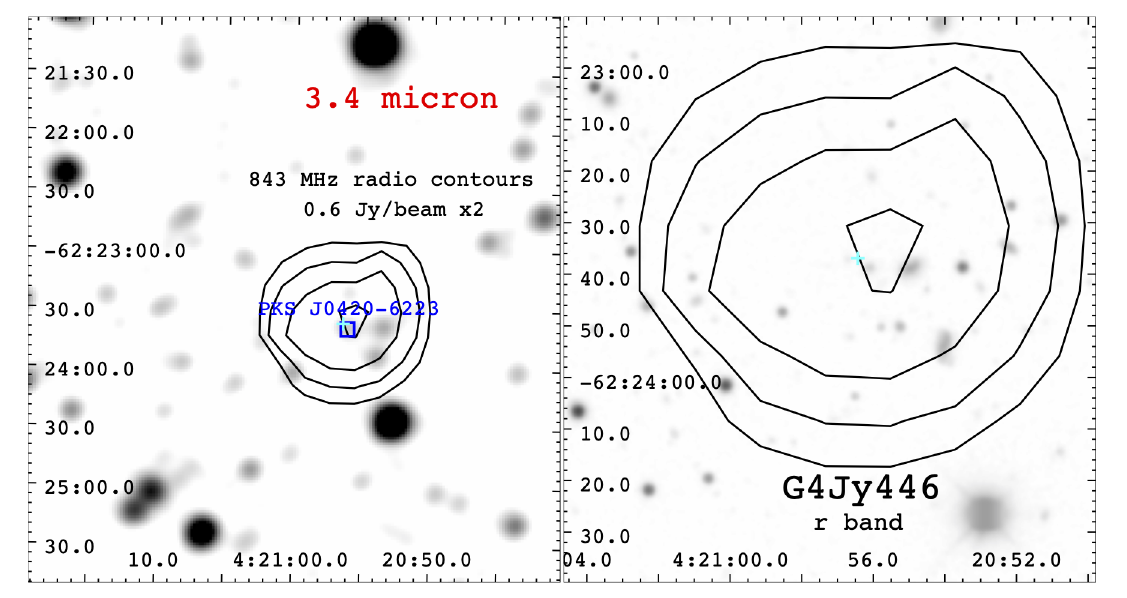}
\includegraphics[height=3.8cm,width=8.8cm,angle=0]{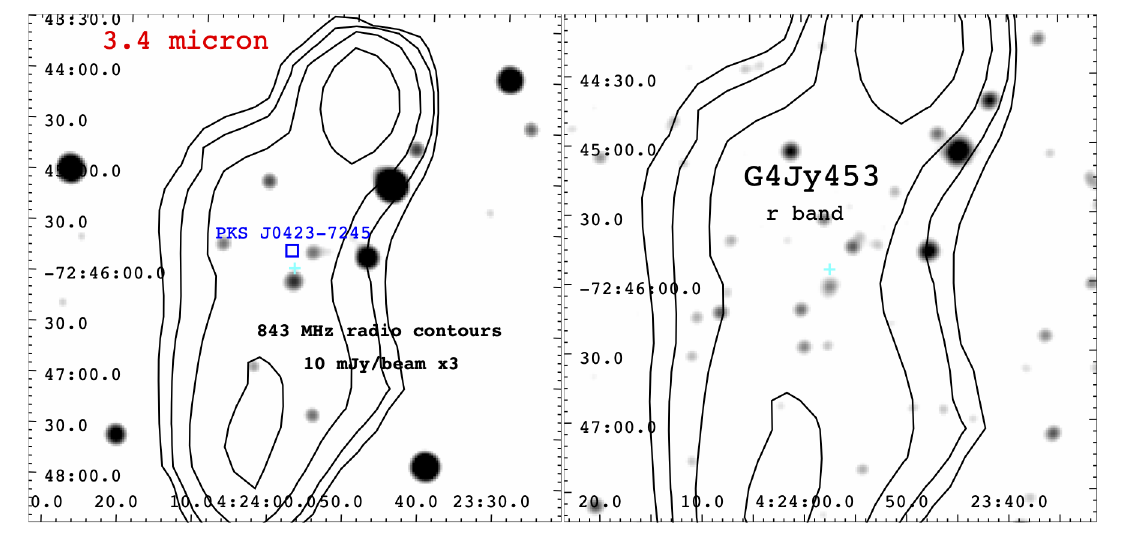}
\includegraphics[height=3.8cm,width=8.8cm,angle=0]{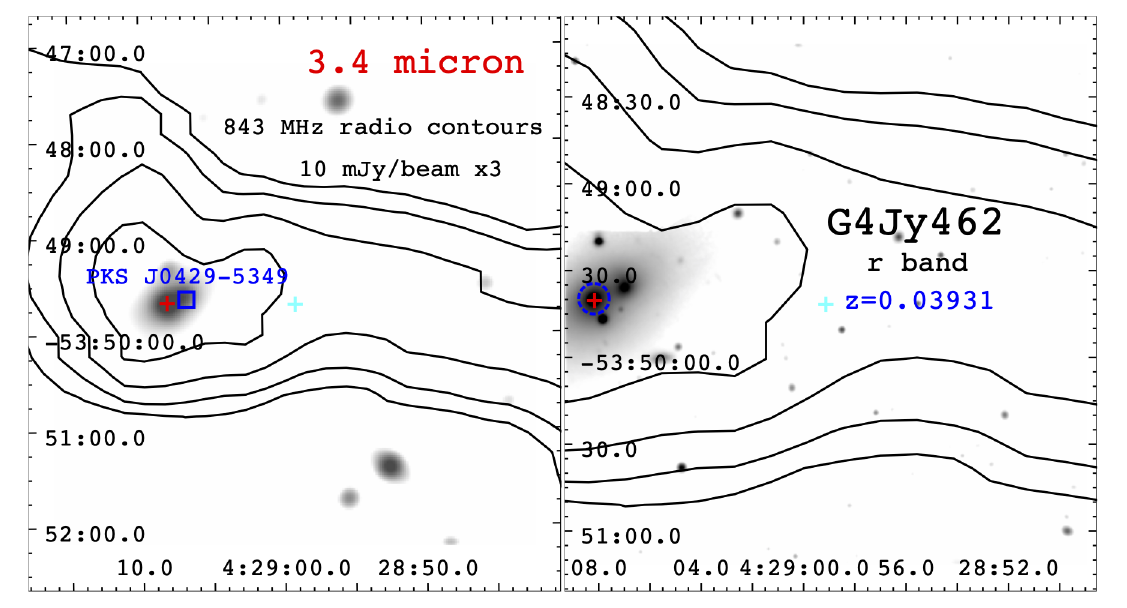}
\includegraphics[height=3.8cm,width=8.8cm,angle=0]{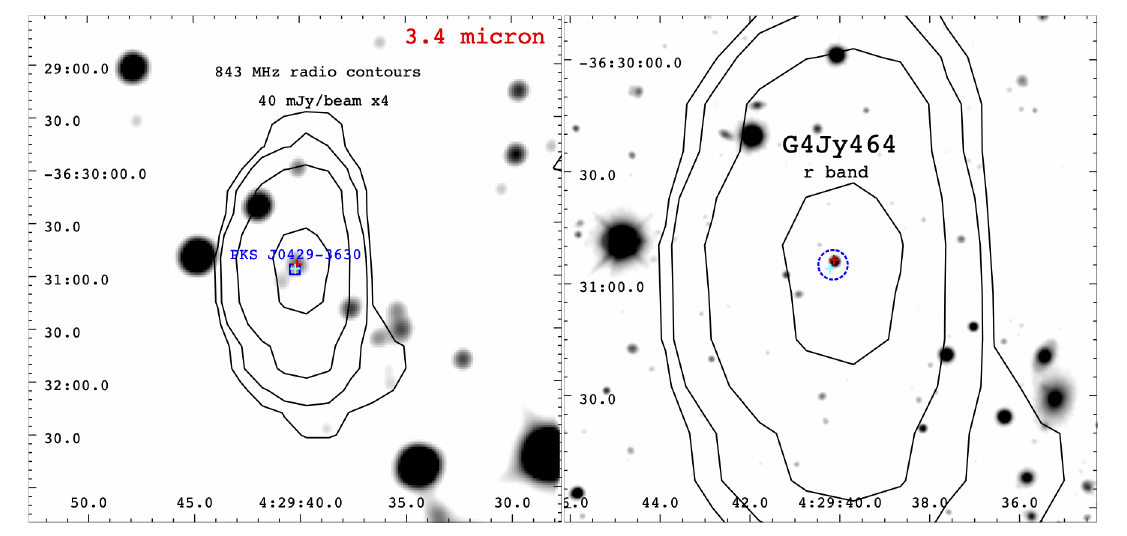}
\includegraphics[height=3.8cm,width=8.8cm,angle=0]{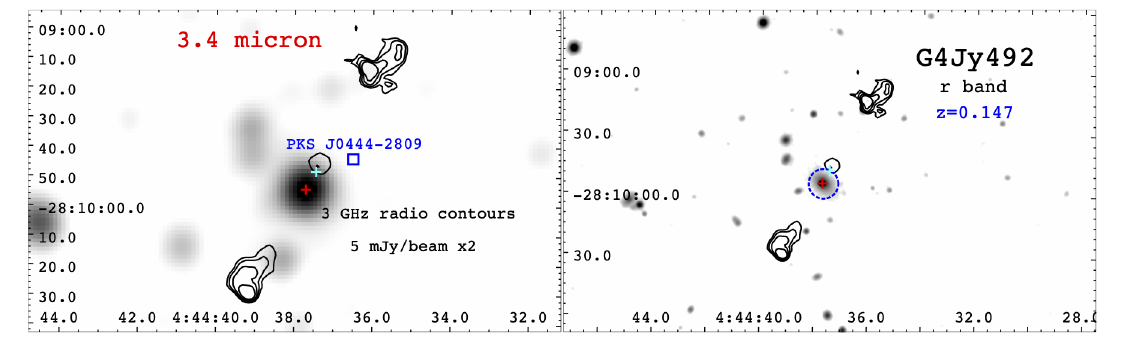}
\caption{Same as Figure~\ref{fig:example1} for the following \cs\ radio sources: \\ 
G4Jy\,404, G4Jy\,411, G4Jy\,415, G4Jy\,416, G4Jy\,417, G4Jy\,427, G4Jy\,436, G4Jy\,446, G4Jy\,453, G4Jy\,462, G4Jy\,464, G4Jy\,492.}
\end{center}
\end{figure*}

\begin{figure*}[!th]
\begin{center}
\includegraphics[height=3.8cm,width=8.8cm,angle=0]{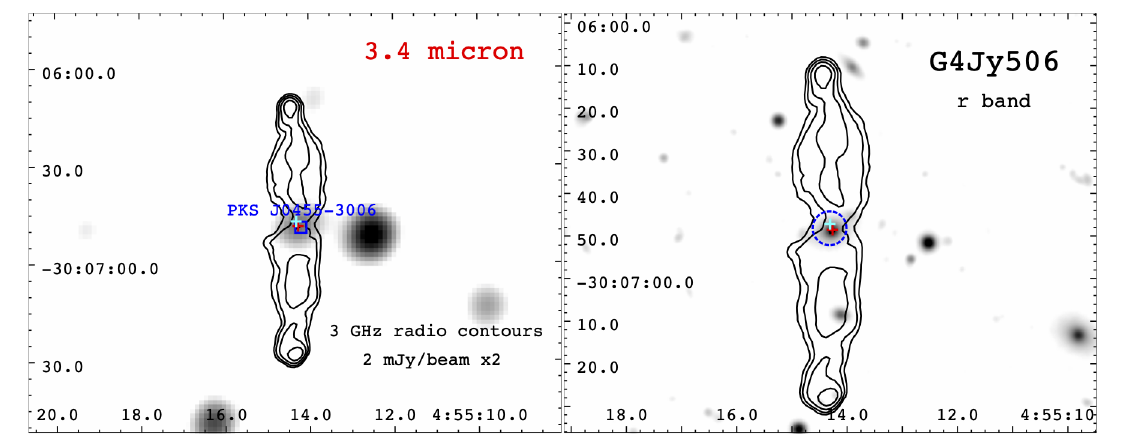}
\includegraphics[height=3.8cm,width=8.8cm,angle=0]{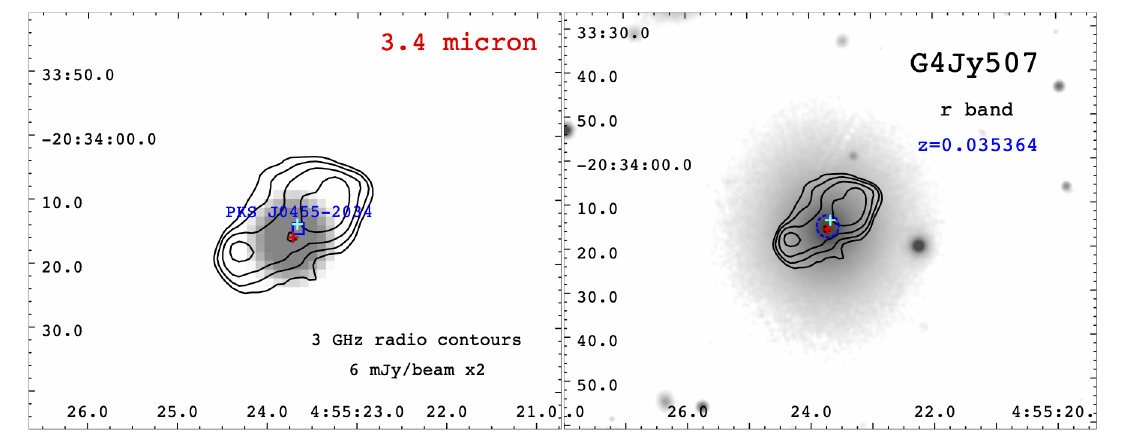}
\includegraphics[height=3.8cm,width=8.8cm,angle=0]{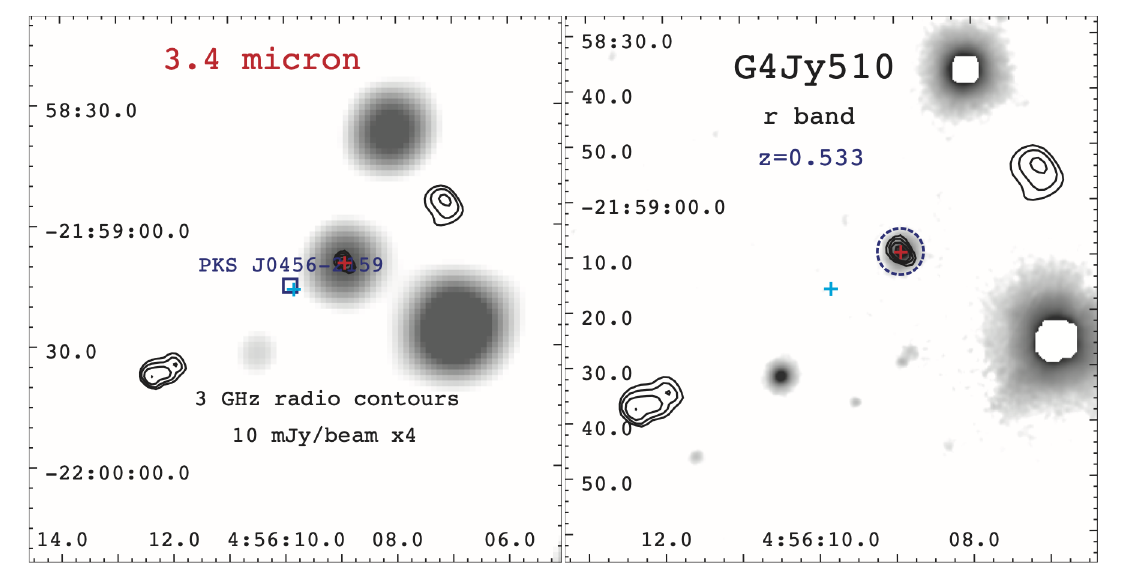}
\includegraphics[height=3.8cm,width=8.8cm,angle=0]{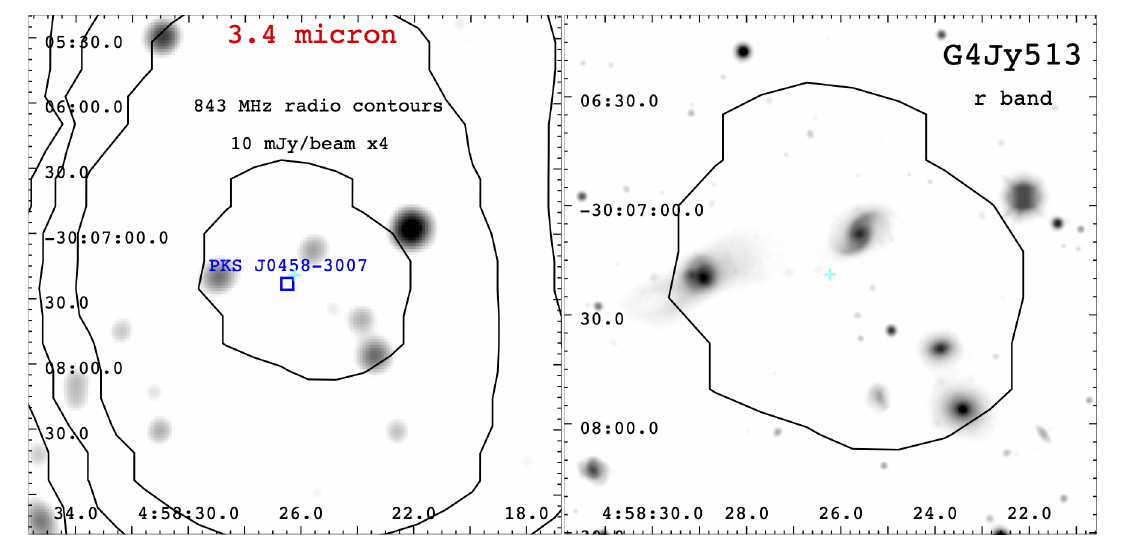}
\includegraphics[height=3.8cm,width=8.8cm,angle=0]{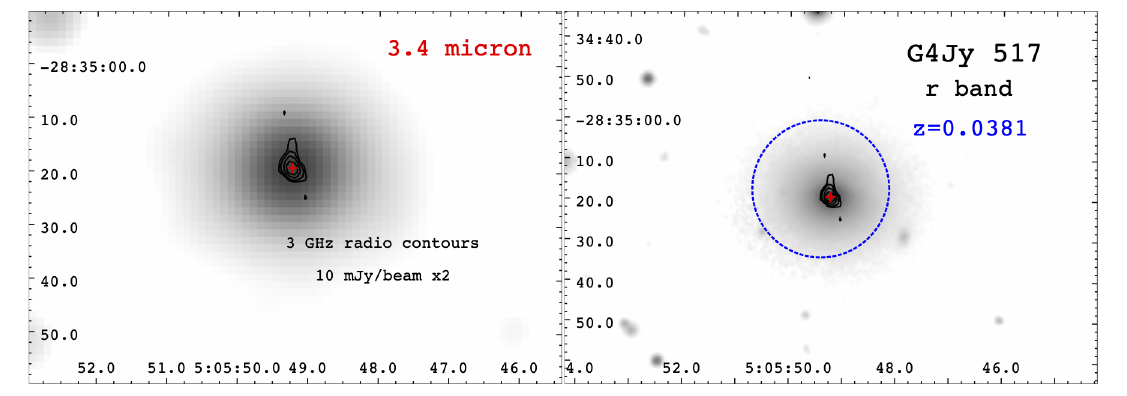}
\includegraphics[height=3.8cm,width=8.8cm,angle=0]{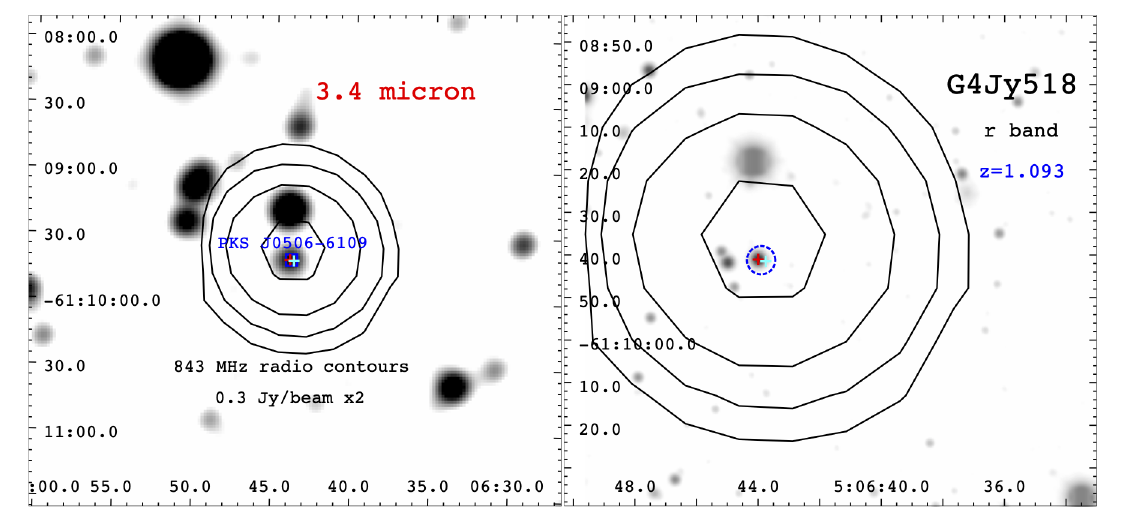}
\includegraphics[height=3.8cm,width=8.8cm,angle=0]{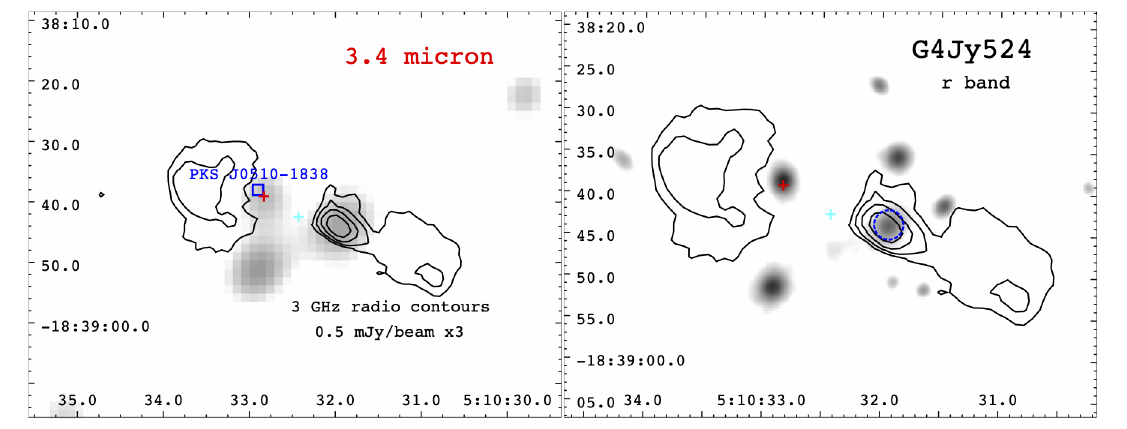}
\includegraphics[height=3.8cm,width=8.8cm,angle=0]{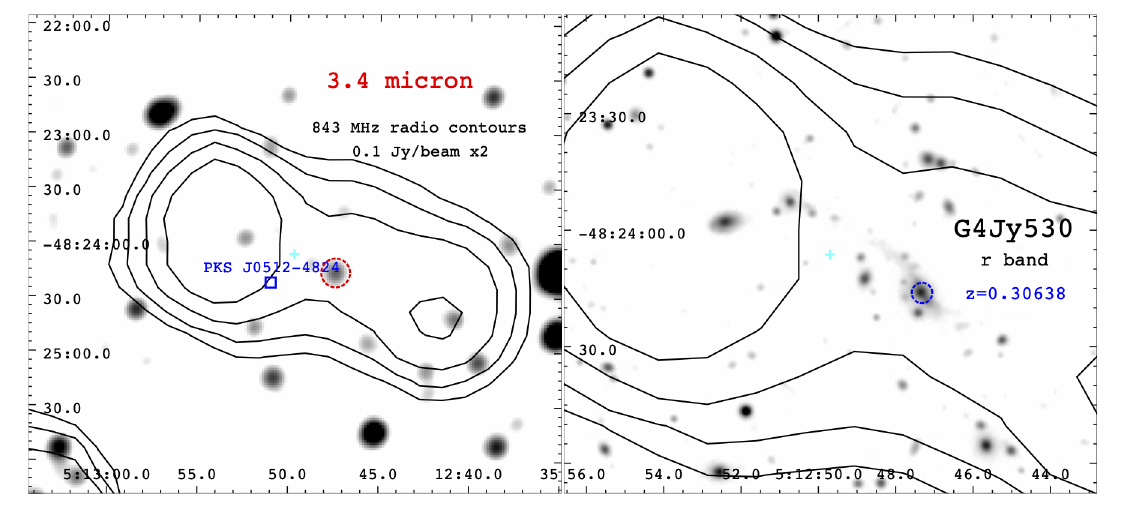}
\includegraphics[height=3.8cm,width=8.8cm,angle=0]{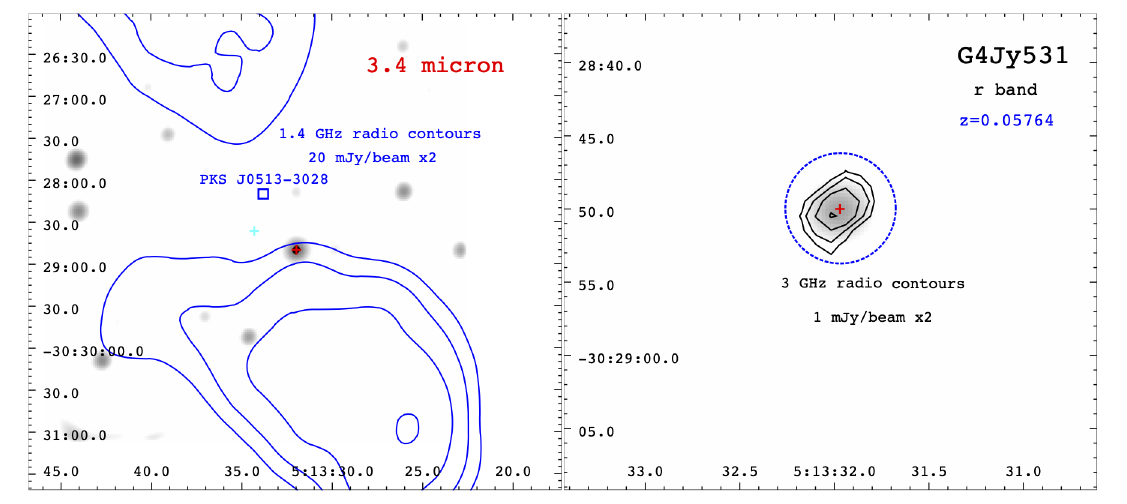}
\includegraphics[height=3.8cm,width=8.8cm,angle=0]{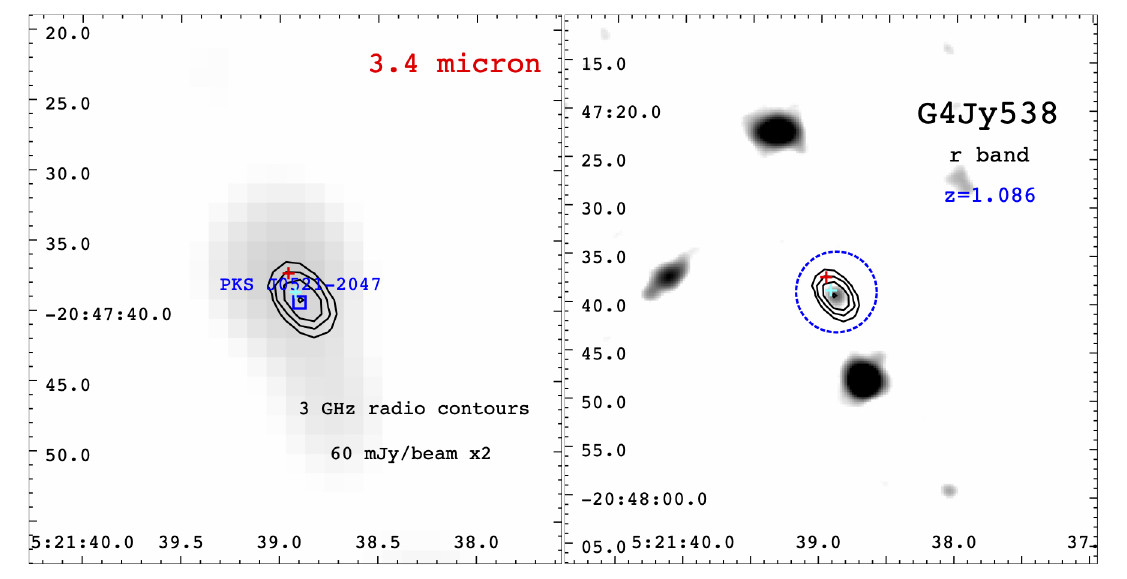}
\includegraphics[height=3.8cm,width=8.8cm,angle=0]{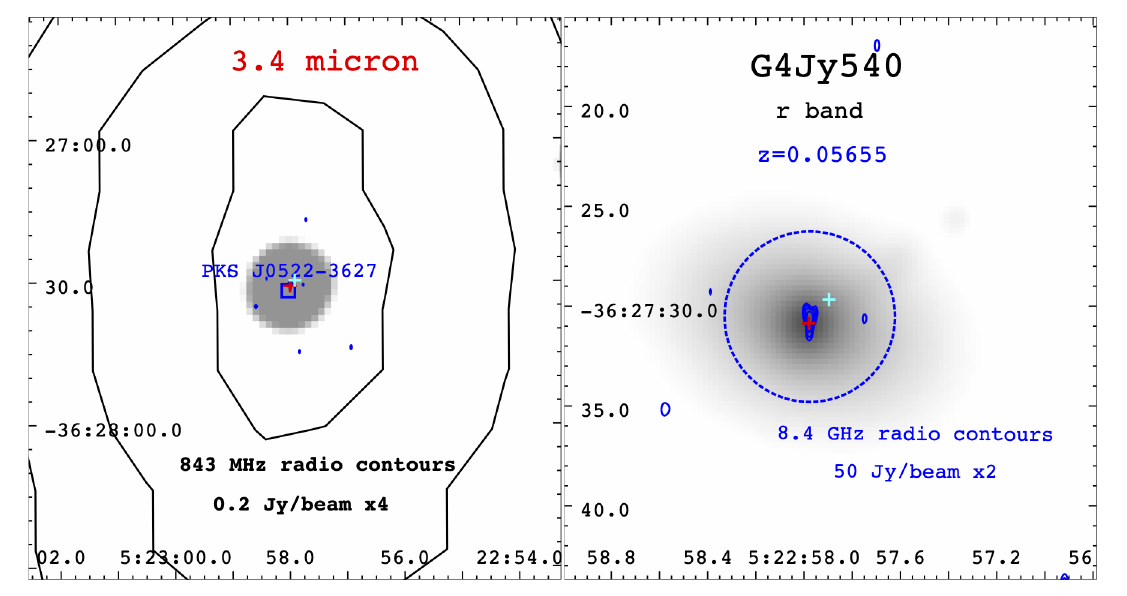}
\includegraphics[height=3.8cm,width=8.8cm,angle=0]{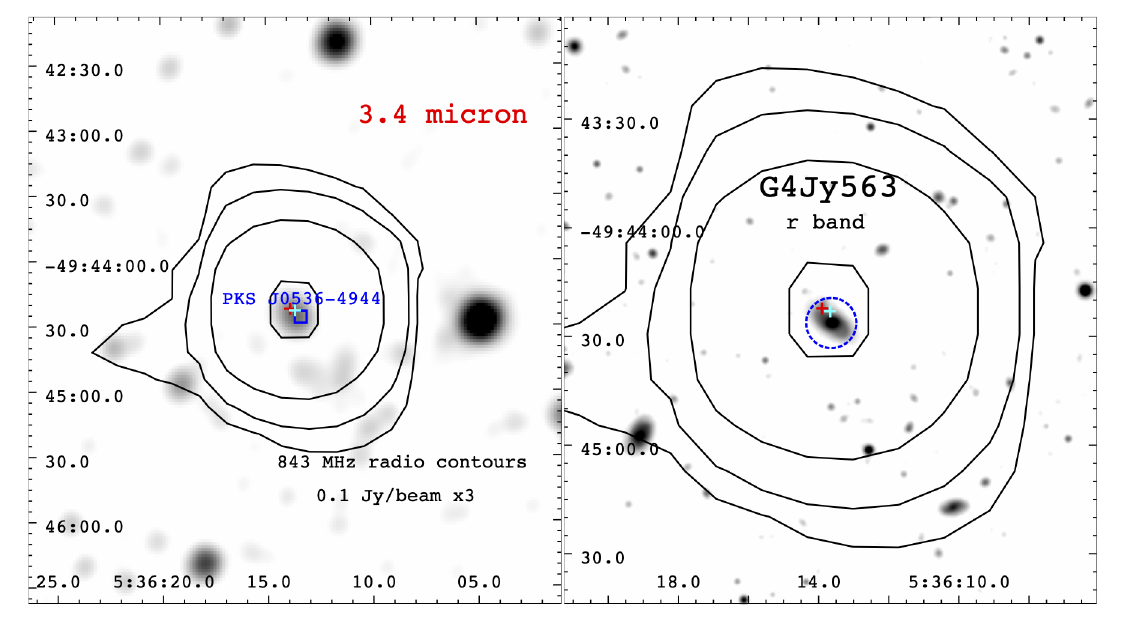}
\caption{Same as Figure~\ref{fig:example1} for the following \cs\ radio sources: \\ 
G4Jy\,506, G4Jy\,507, G4Jy\,510, G4Jy\,513, G4Jy\,517, G4Jy\,518, G4Jy\,524, G4Jy\,530, G4Jy\,531, G4Jy\,538, G4Jy\,540, G4Jy\,563.}
\end{center}
\end{figure*}

\begin{figure*}[!th]
\begin{center}
\includegraphics[height=3.8cm,width=8.8cm,angle=0]{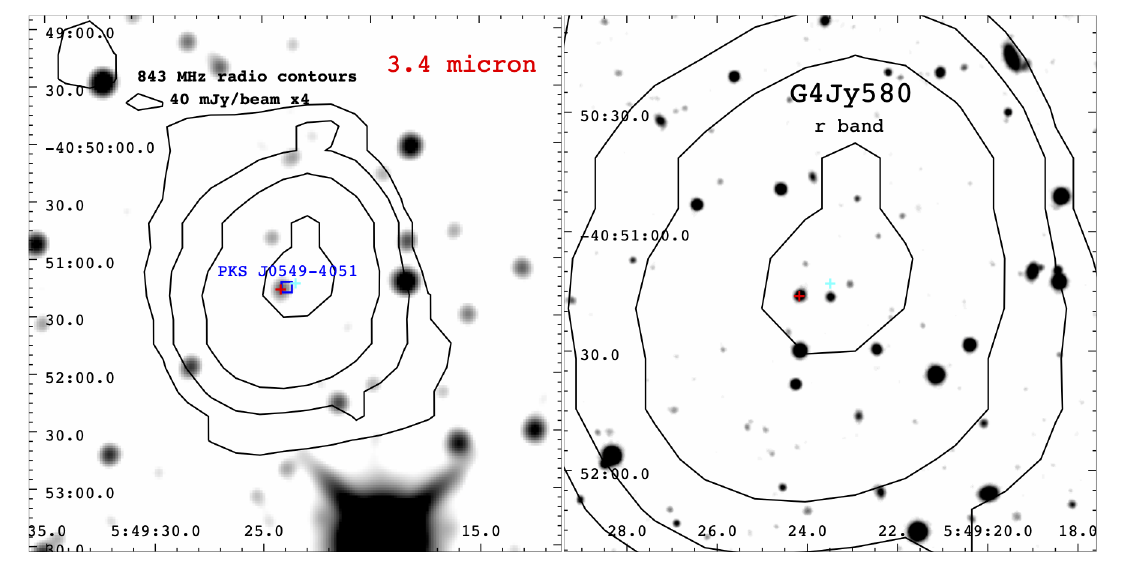}
\includegraphics[height=3.8cm,width=8.8cm,angle=0]{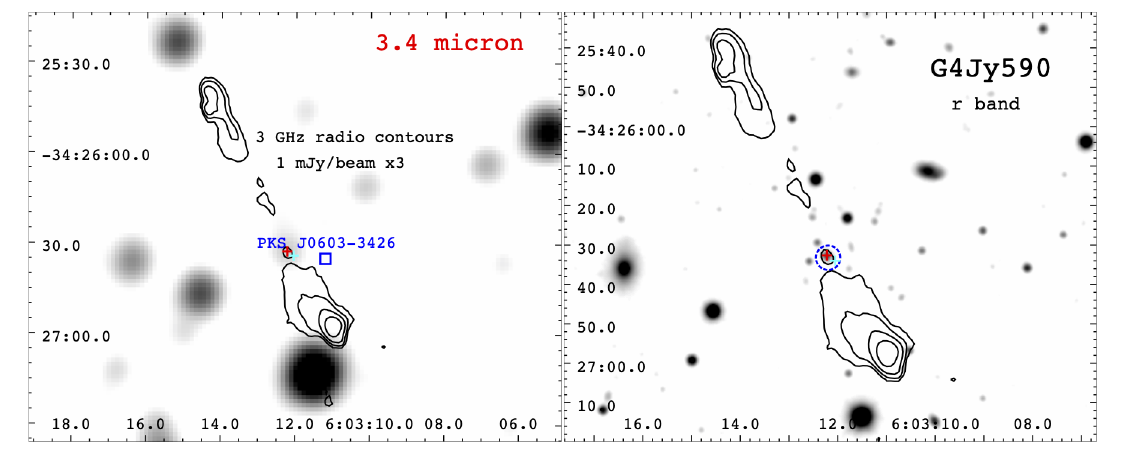}
\includegraphics[height=3.8cm,width=8.8cm,angle=0]{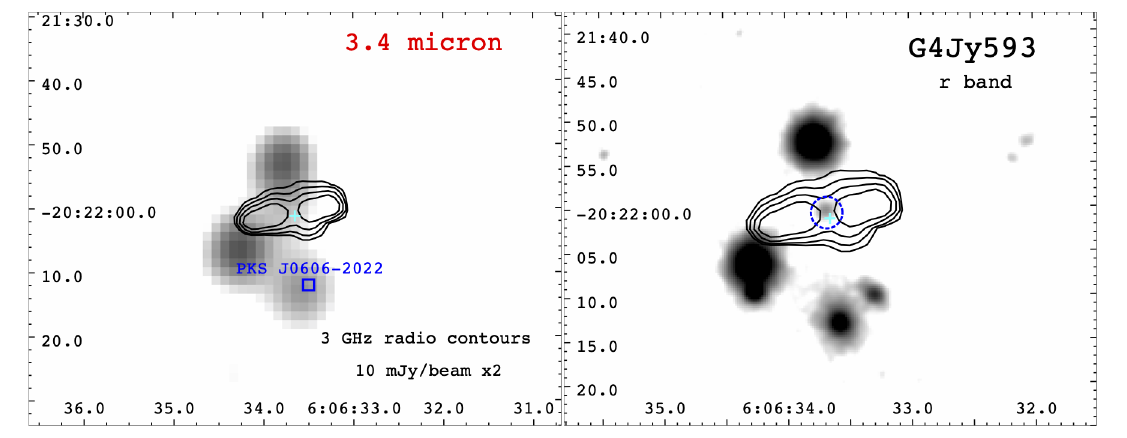}
\includegraphics[height=3.8cm,width=8.8cm,angle=0]{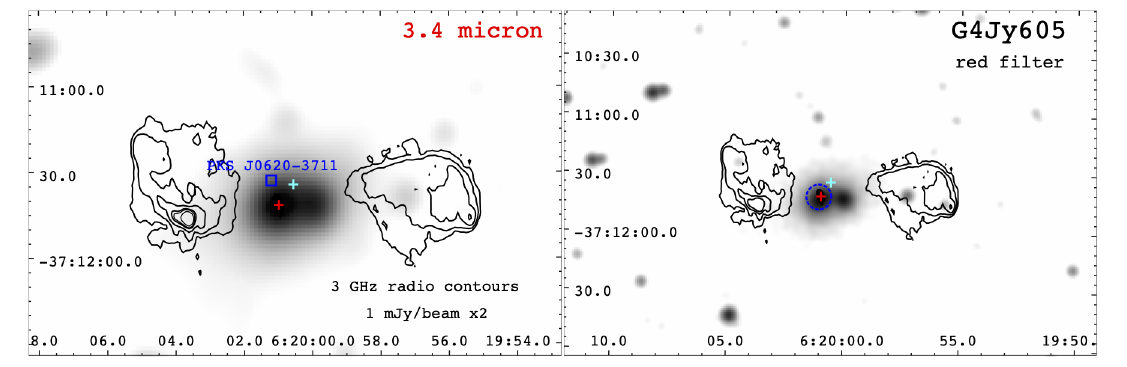}
\includegraphics[height=3.8cm,width=8.8cm,angle=0]{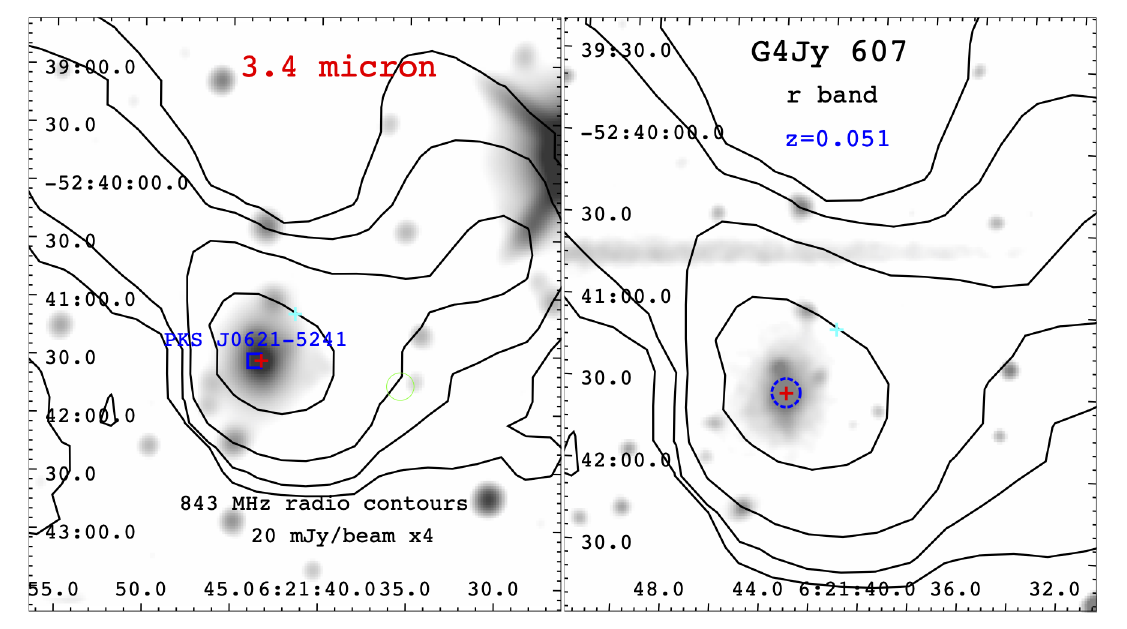}
\includegraphics[height=3.8cm,width=8.8cm,angle=0]{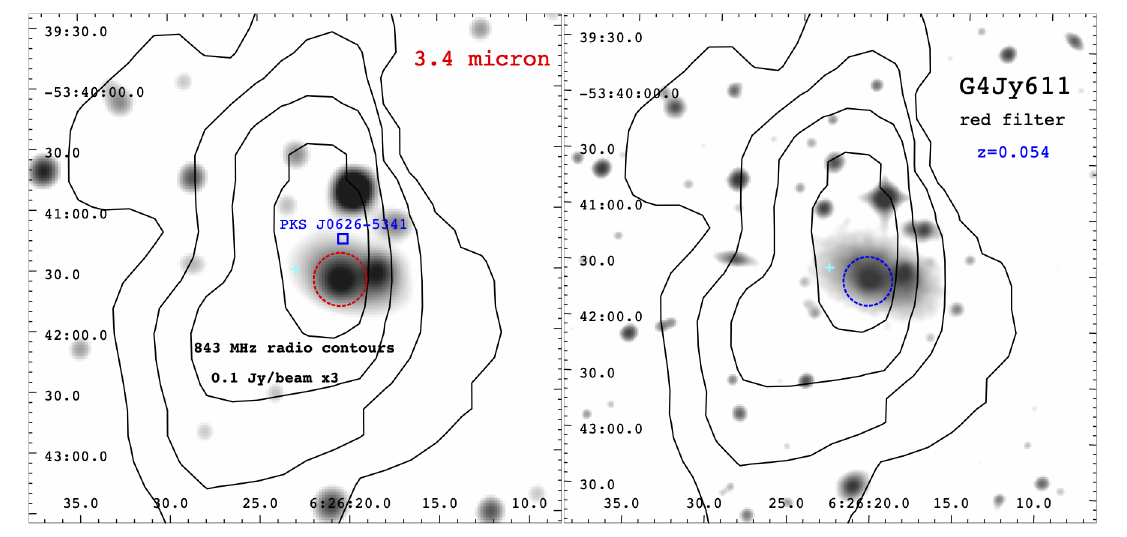}
\includegraphics[height=3.8cm,width=8.8cm,angle=0]{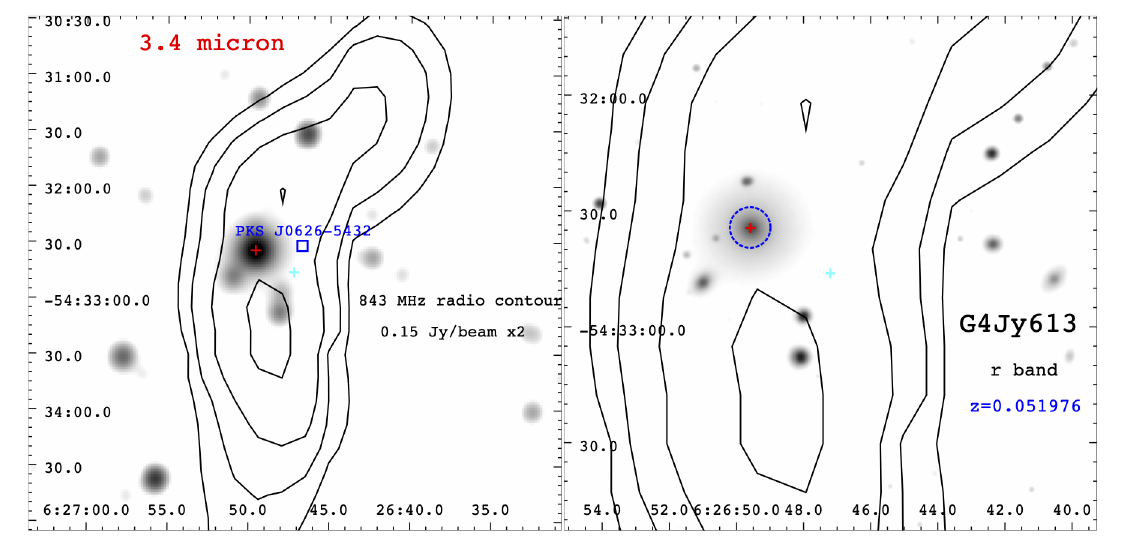}
\includegraphics[height=3.8cm,width=8.8cm,angle=0]{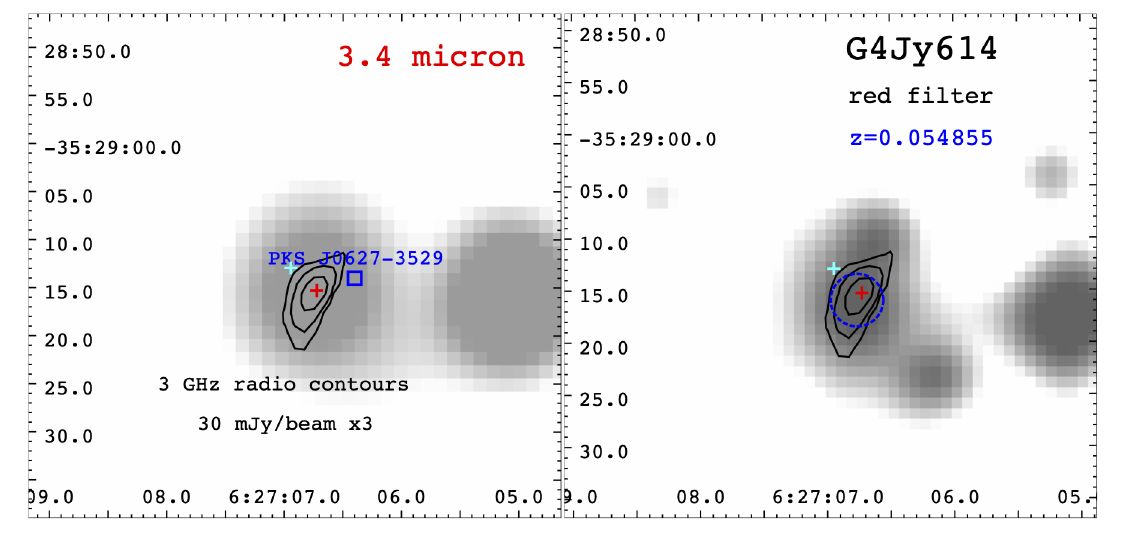}
\includegraphics[height=3.8cm,width=8.8cm,angle=0]{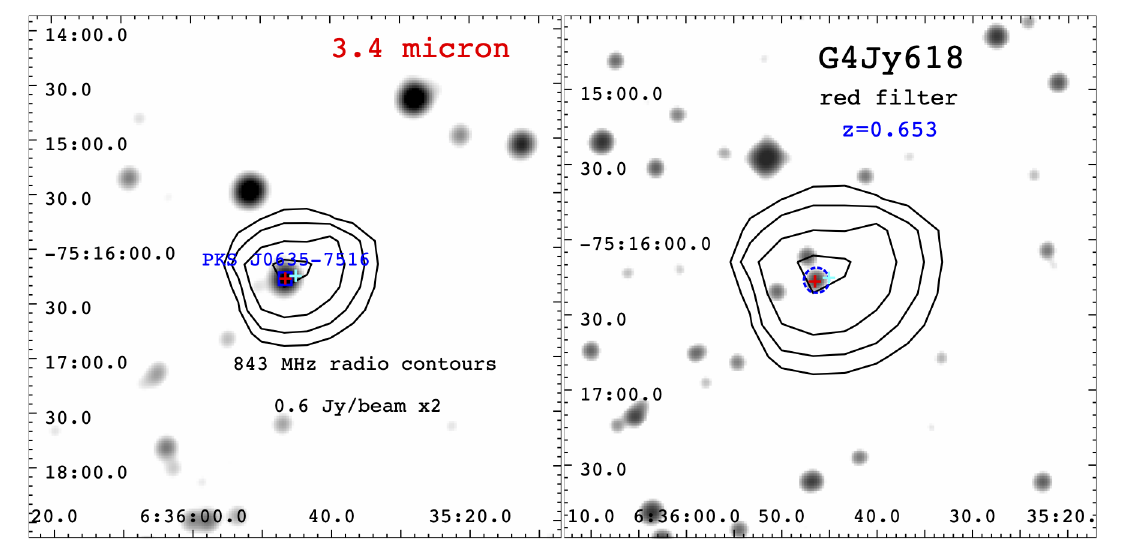}
\includegraphics[height=3.8cm,width=8.8cm,angle=0]{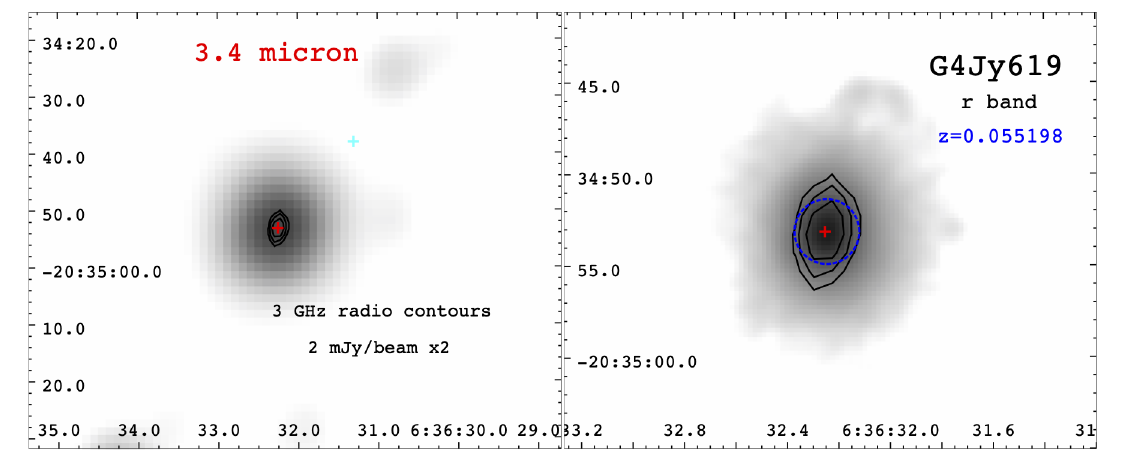}
\includegraphics[height=3.8cm,width=8.8cm,angle=0]{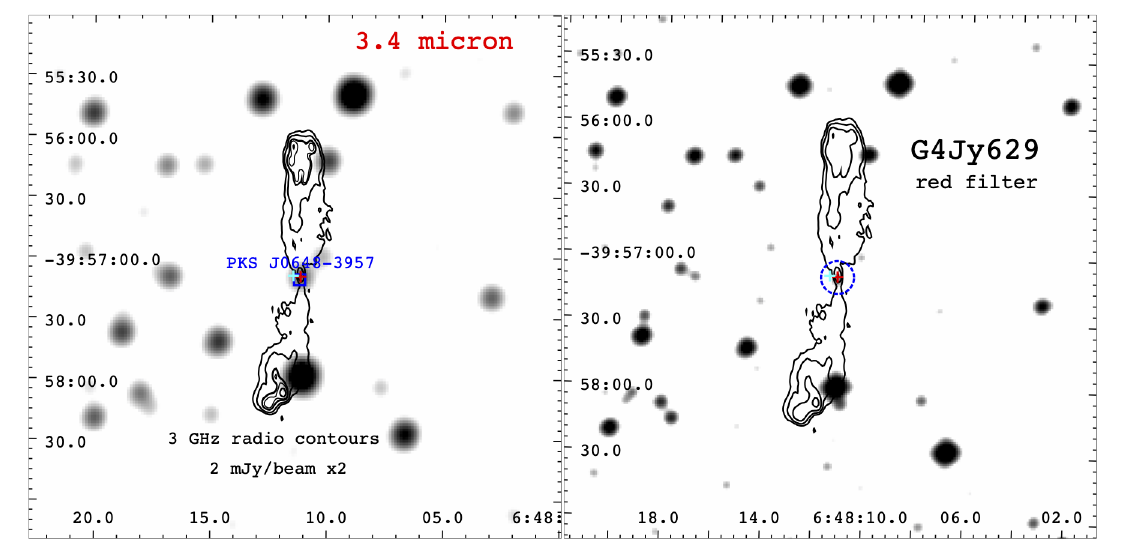}
\includegraphics[height=3.8cm,width=8.8cm,angle=0]{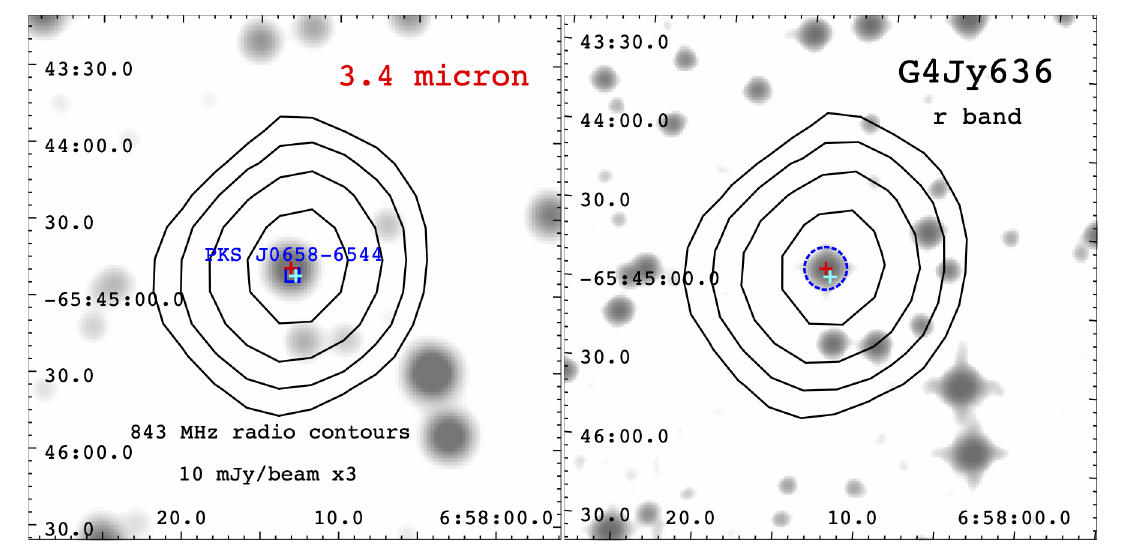}
\caption{Same as Figure~\ref{fig:example1} for the following \cs\ radio sources: \\ 
G4Jy\,580, G4Jy\,590, G4Jy\,593, G4Jy\,605, G4Jy\,607, G4Jy\,611, G4Jy\,613, G4Jy\,614, G4Jy\,618, G4Jy\,619, G4Jy\,629, G4Jy\,636.}
\end{center}
\end{figure*}

\begin{figure*}[!th]
\begin{center}
\includegraphics[height=3.8cm,width=8.8cm,angle=0]{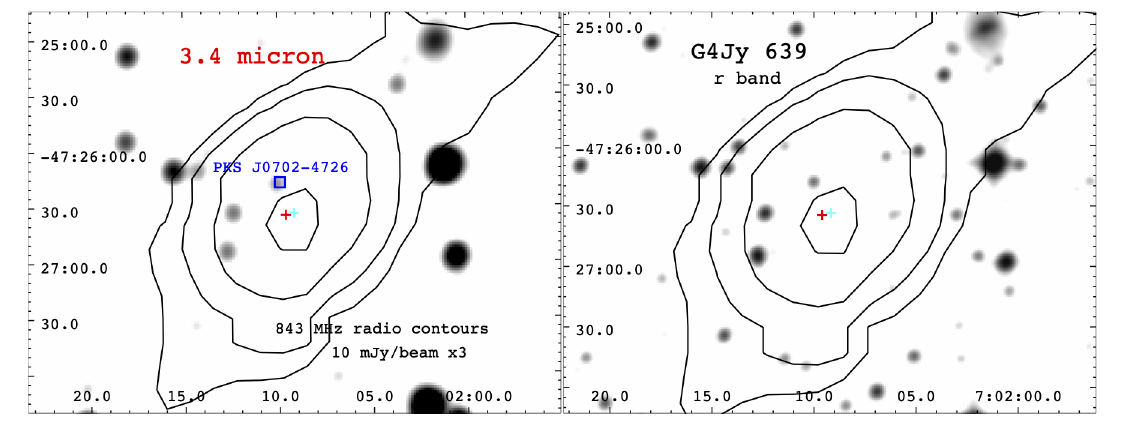}
\includegraphics[height=3.8cm,width=8.8cm,angle=0]{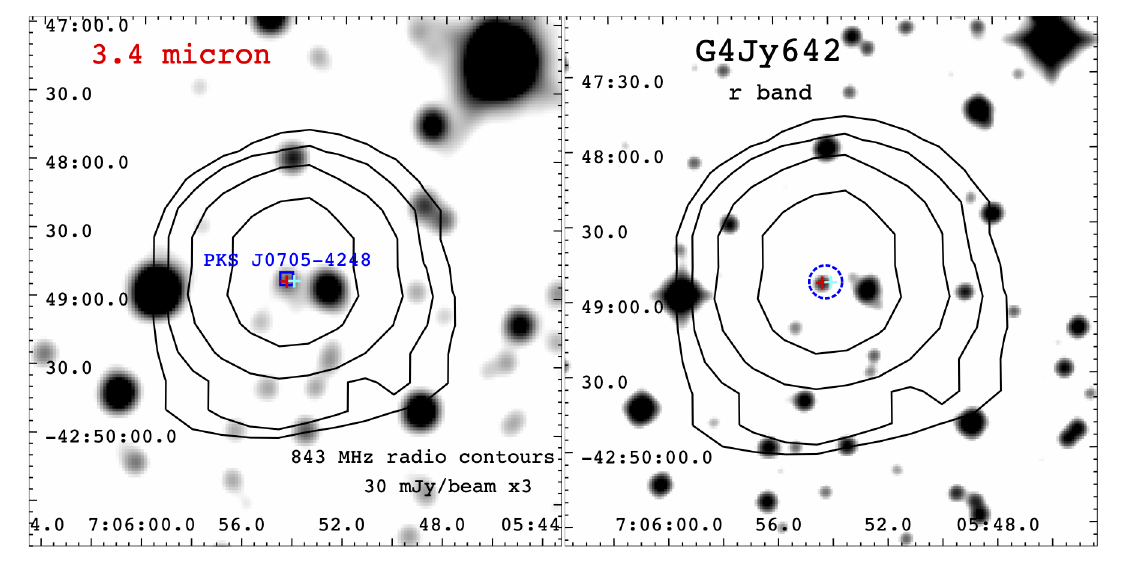}
\includegraphics[height=3.8cm,width=8.8cm,angle=0]{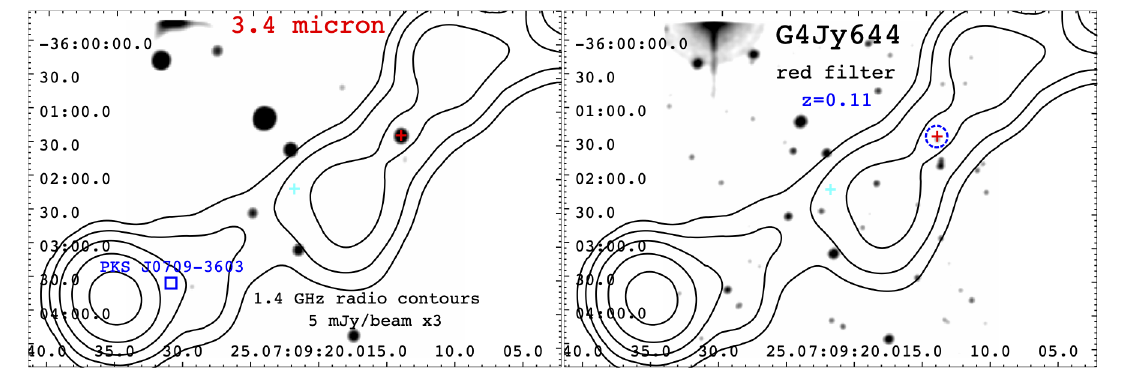}
\includegraphics[height=3.8cm,width=8.8cm,angle=0]{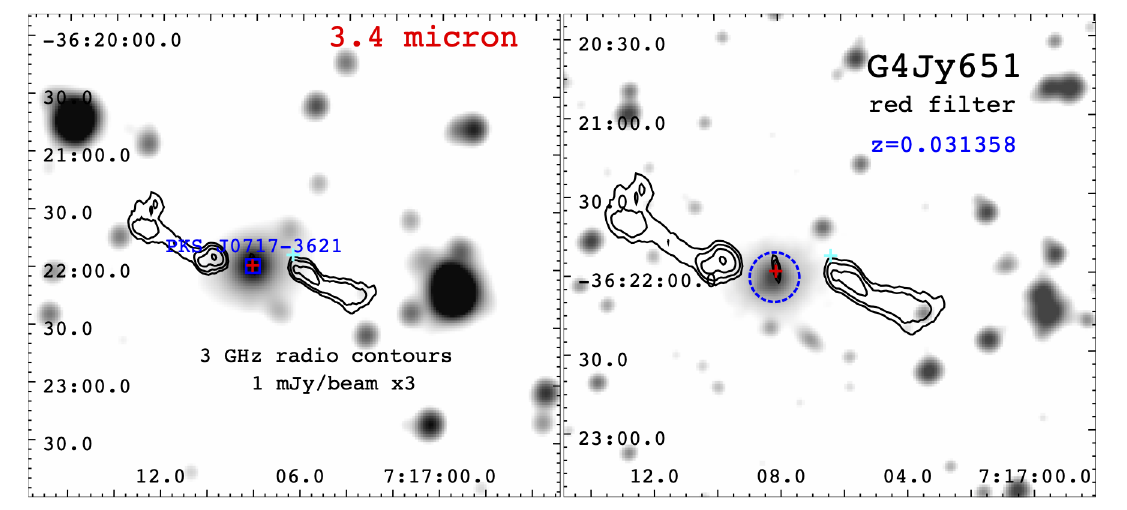}
\includegraphics[height=3.8cm,width=8.8cm,angle=0]{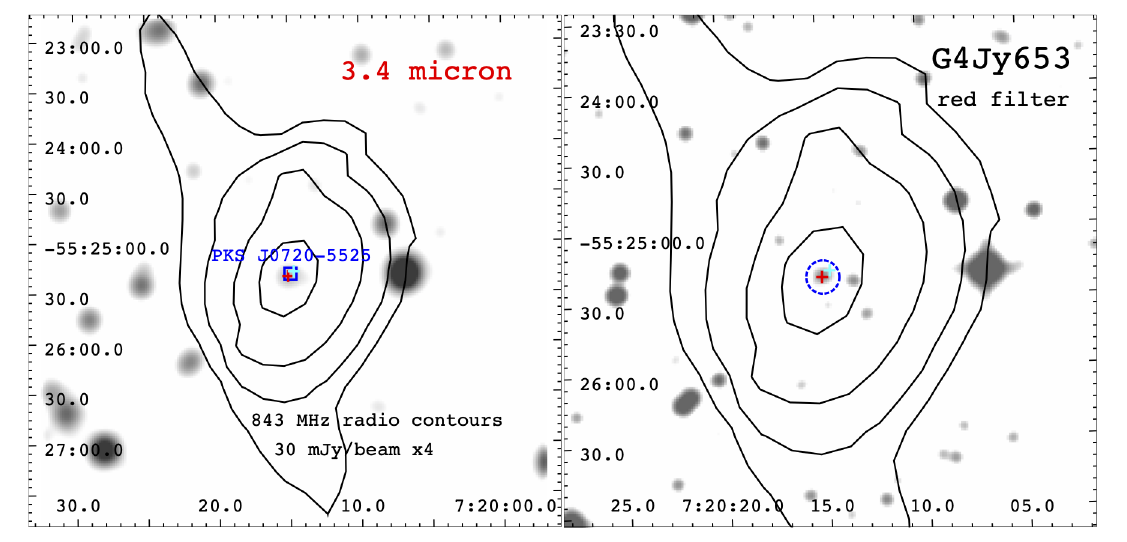}
\includegraphics[height=3.8cm,width=8.8cm,angle=0]{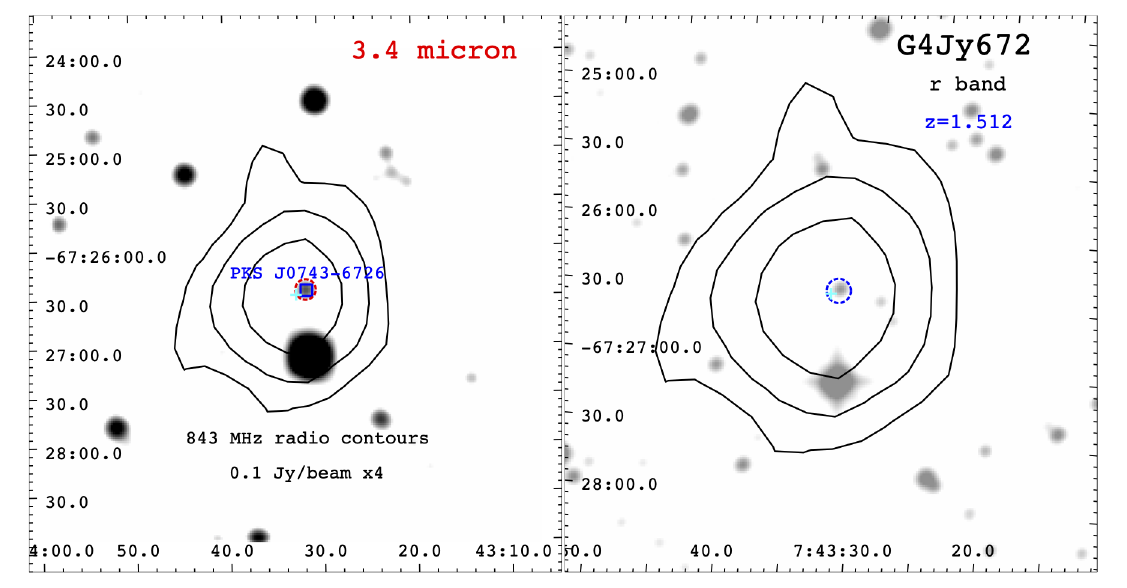}
\includegraphics[height=3.8cm,width=8.8cm,angle=0]{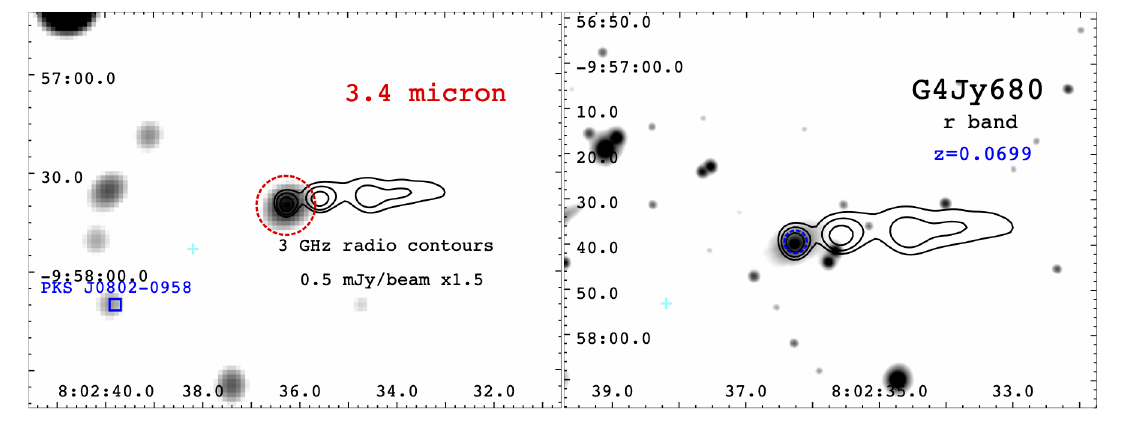}
\includegraphics[height=3.8cm,width=8.8cm,angle=0]{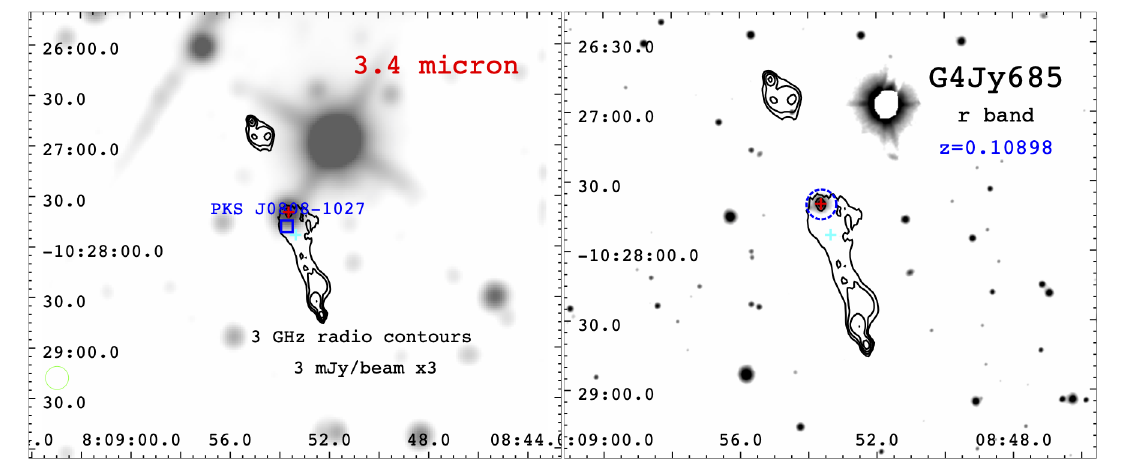}
\includegraphics[height=3.8cm,width=8.8cm,angle=0]{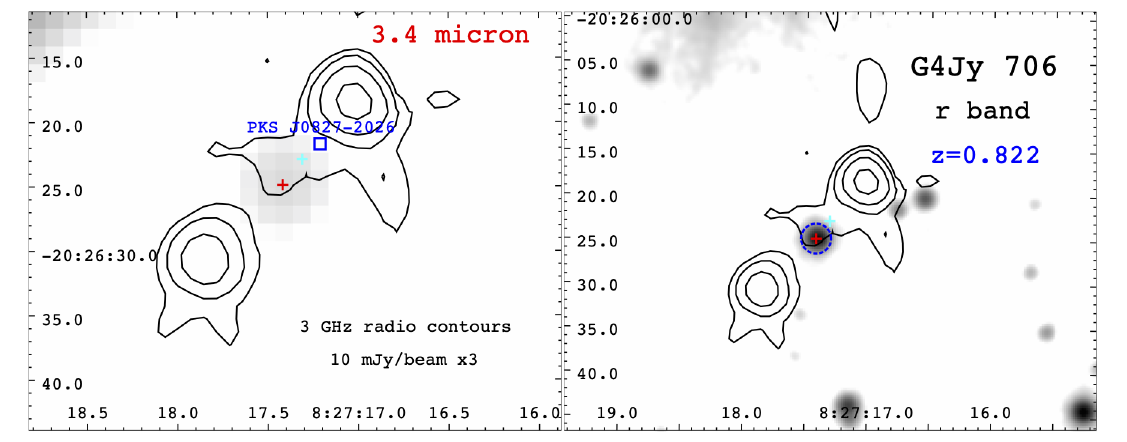}
\includegraphics[height=3.8cm,width=8.8cm,angle=0]{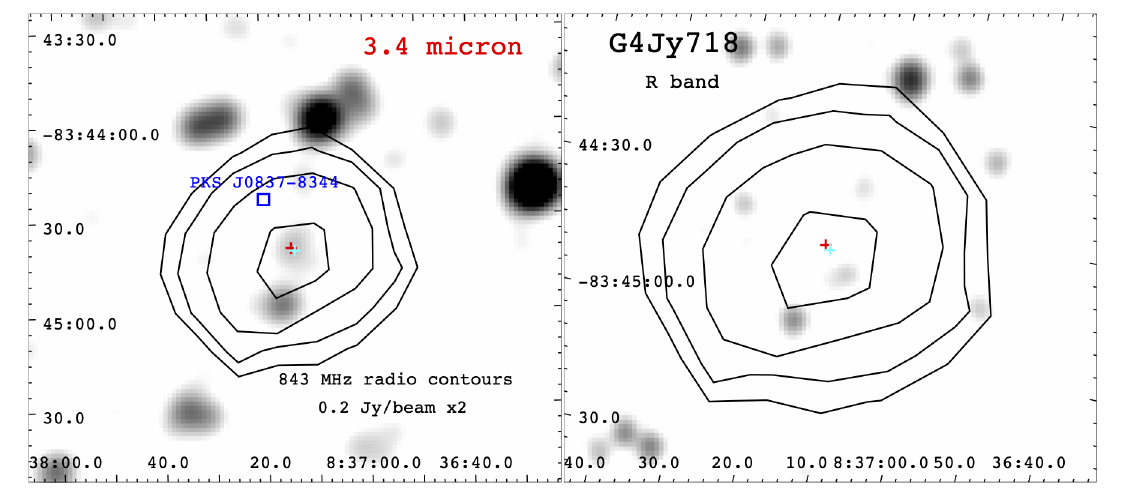}
\includegraphics[height=3.8cm,width=8.8cm,angle=0]{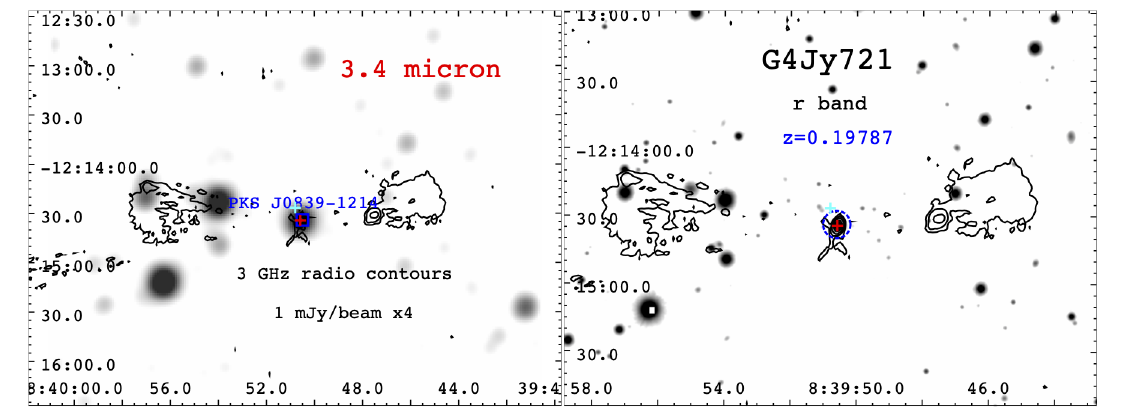}
\includegraphics[height=3.8cm,width=8.8cm,angle=0]{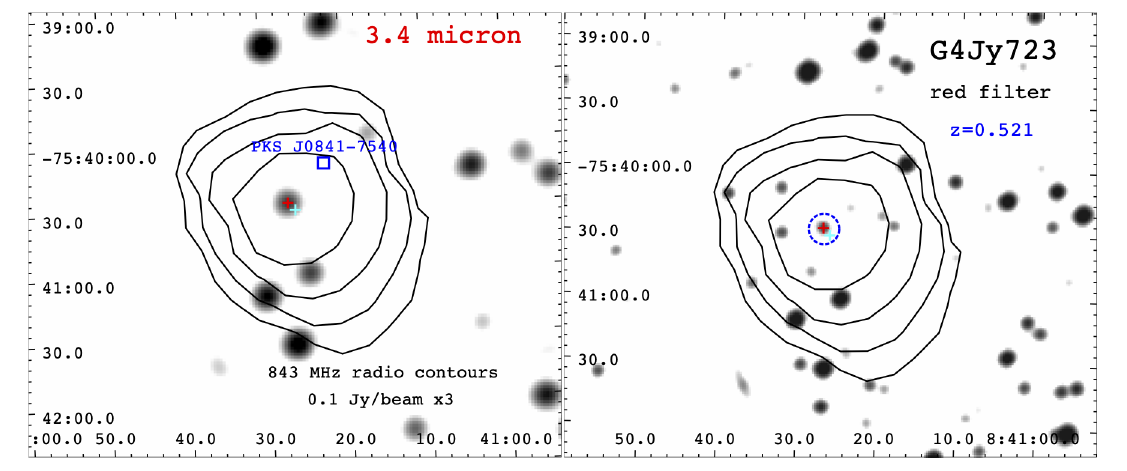}
\caption{Same as Figure~\ref{fig:example1} for the following \cs\ radio sources: \\ 
G4Jy\,639, G4Jy\,642, G4Jy\,644, G4Jy\,651, G4Jy\,653, G4Jy\,672, G4Jy\,680, G4Jy\,685, G4Jy\,706, G4Jy\,718, G4Jy\,721, G4Jy\,723.}
\end{center}
\end{figure*}

\begin{figure*}[!th]
\begin{center}
\includegraphics[height=3.8cm,width=8.8cm,angle=0]{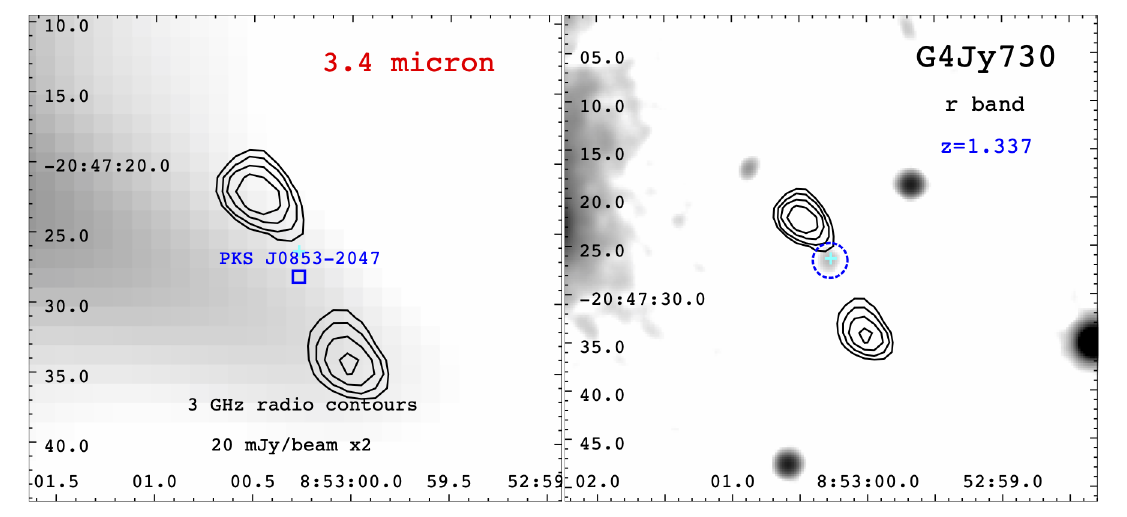}
\includegraphics[height=3.8cm,width=8.8cm,angle=0]{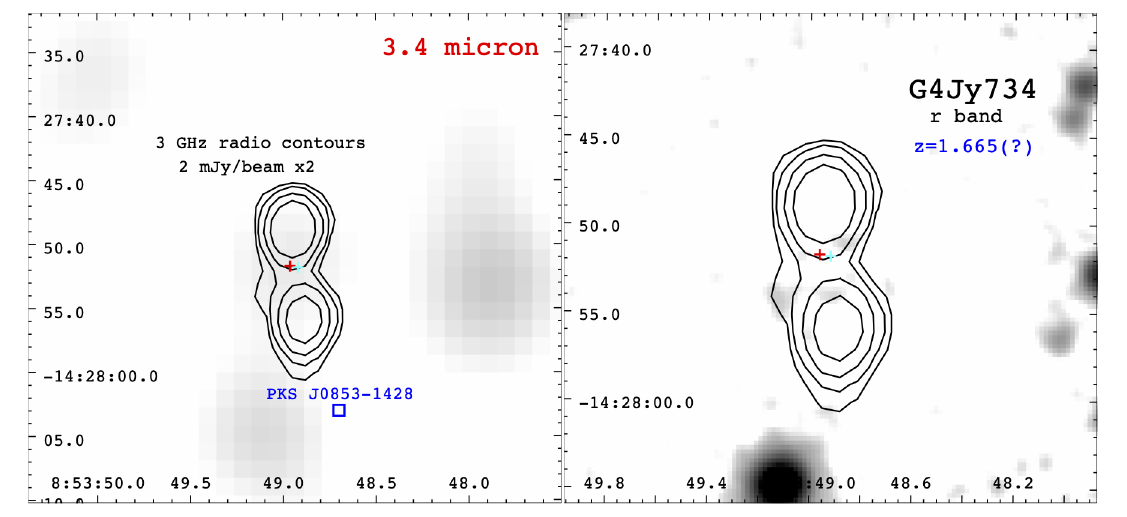}
\includegraphics[height=3.8cm,width=8.8cm,angle=0]{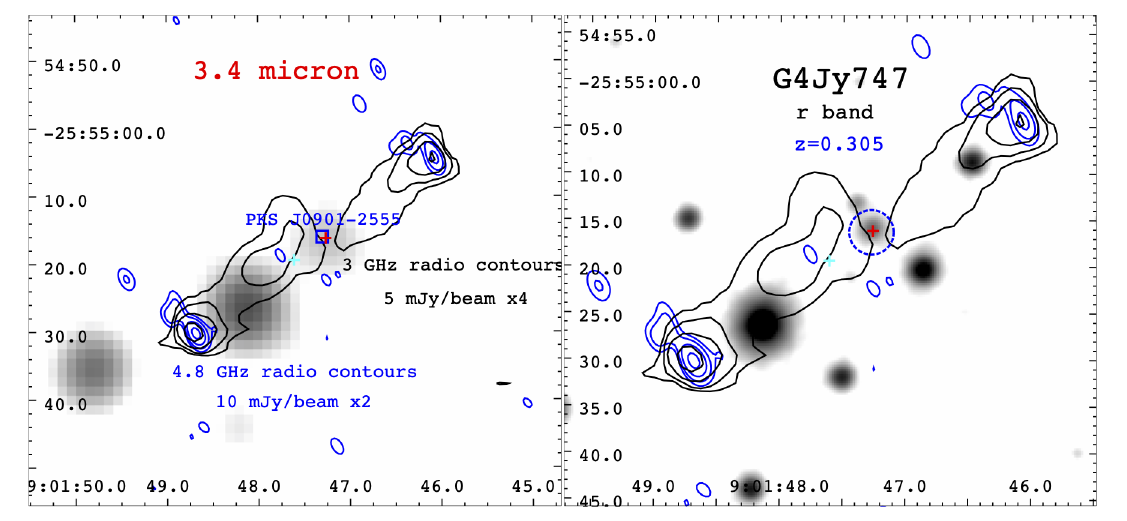}
\includegraphics[height=3.8cm,width=8.8cm,angle=0]{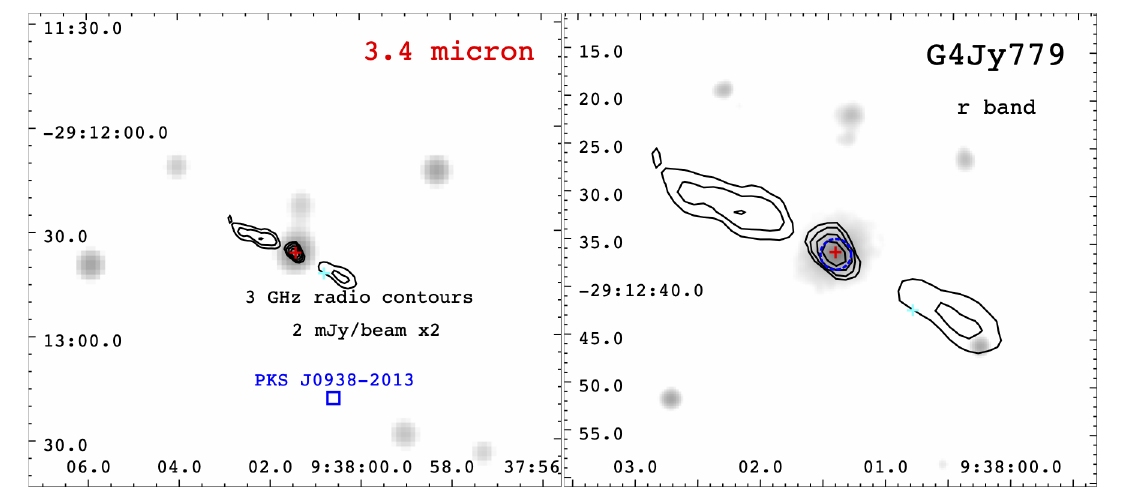}
\includegraphics[height=3.8cm,width=8.8cm,angle=0]{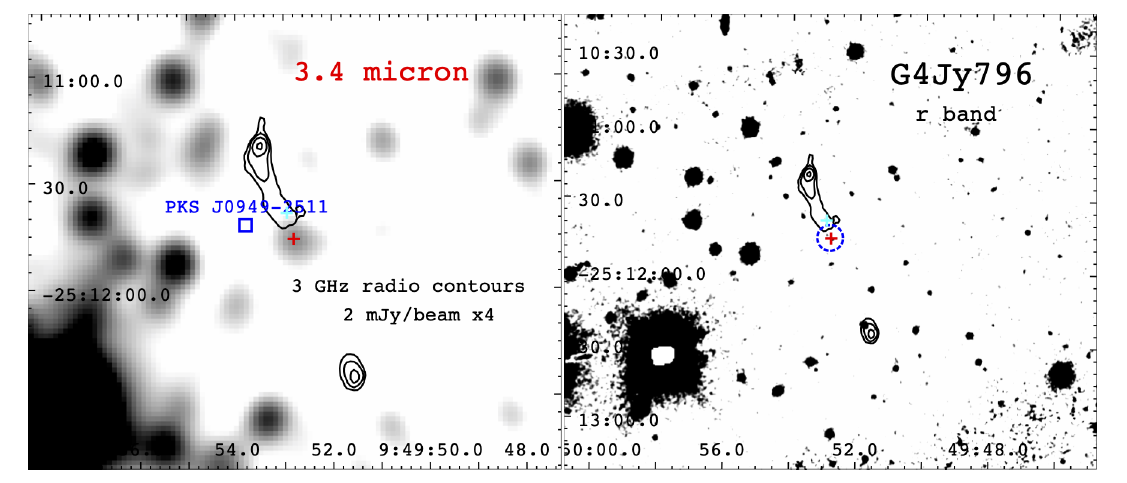}
\includegraphics[height=3.8cm,width=8.8cm,angle=0]{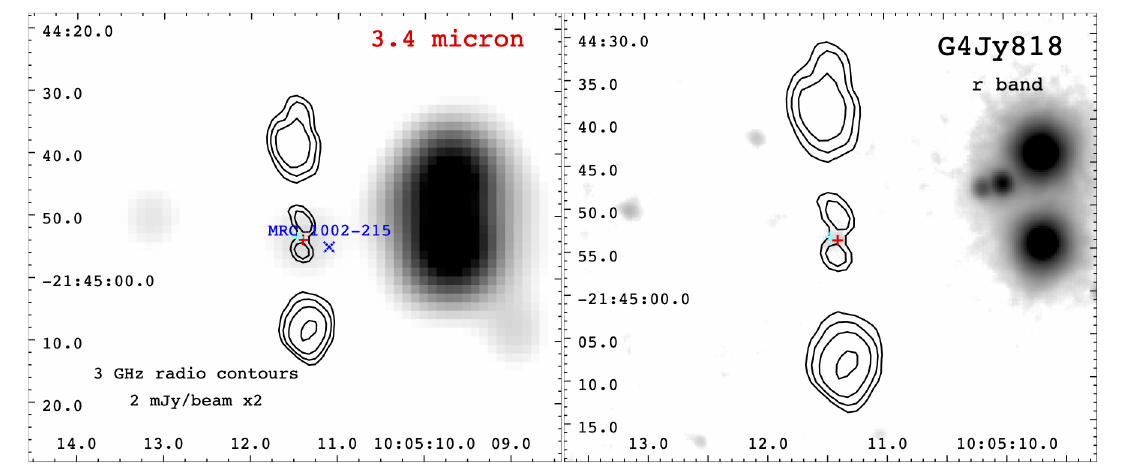}
\includegraphics[height=3.8cm,width=8.8cm,angle=0]{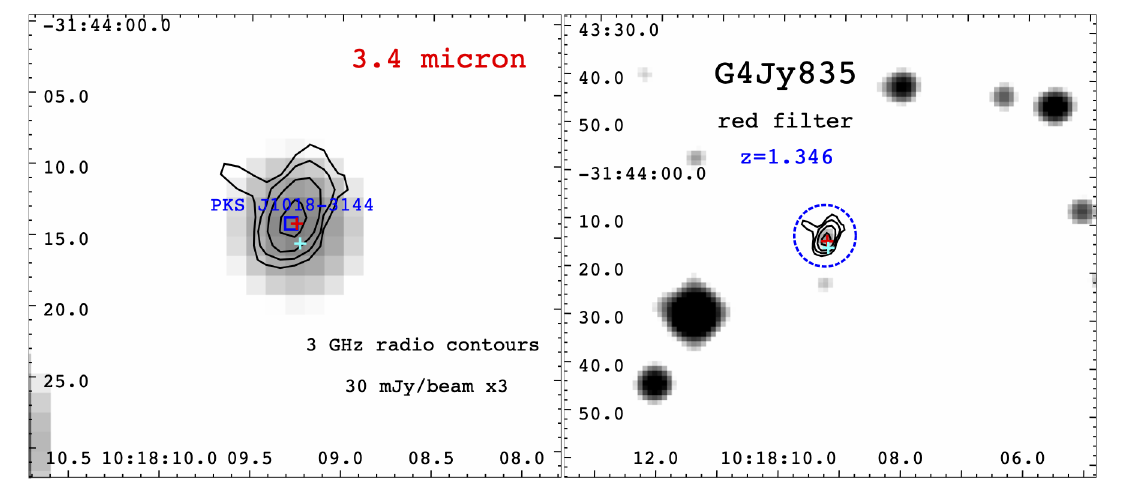}
\includegraphics[height=3.8cm,width=8.8cm,angle=0]{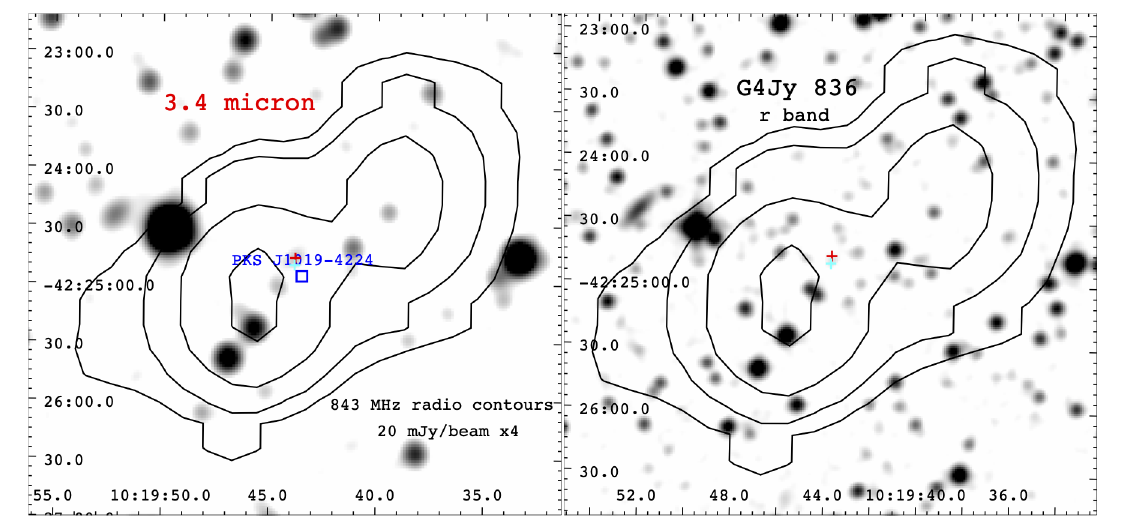}
\includegraphics[height=3.8cm,width=8.8cm,angle=0]{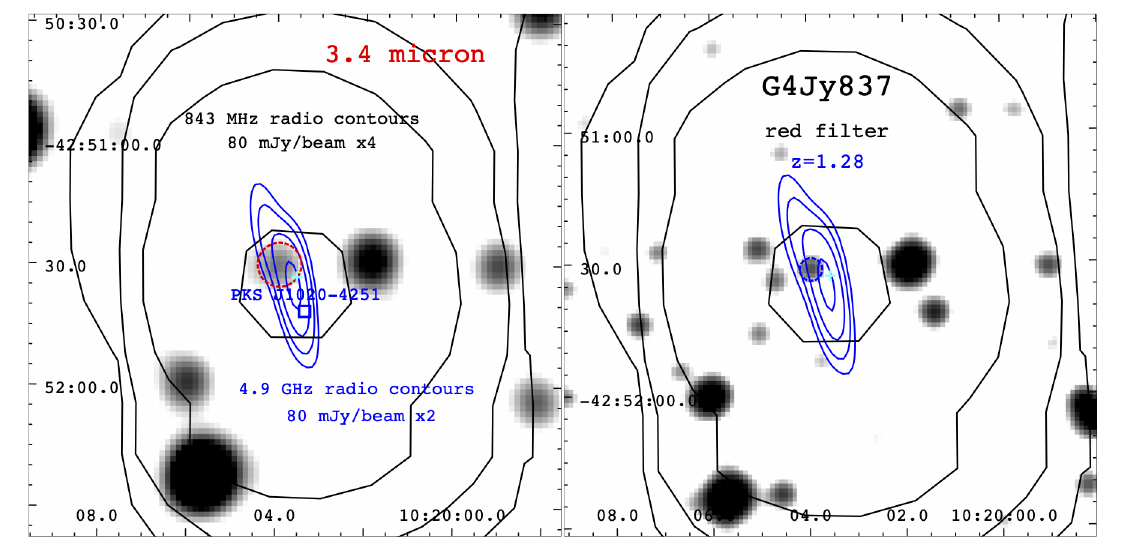}
\includegraphics[height=3.8cm,width=8.8cm,angle=0]{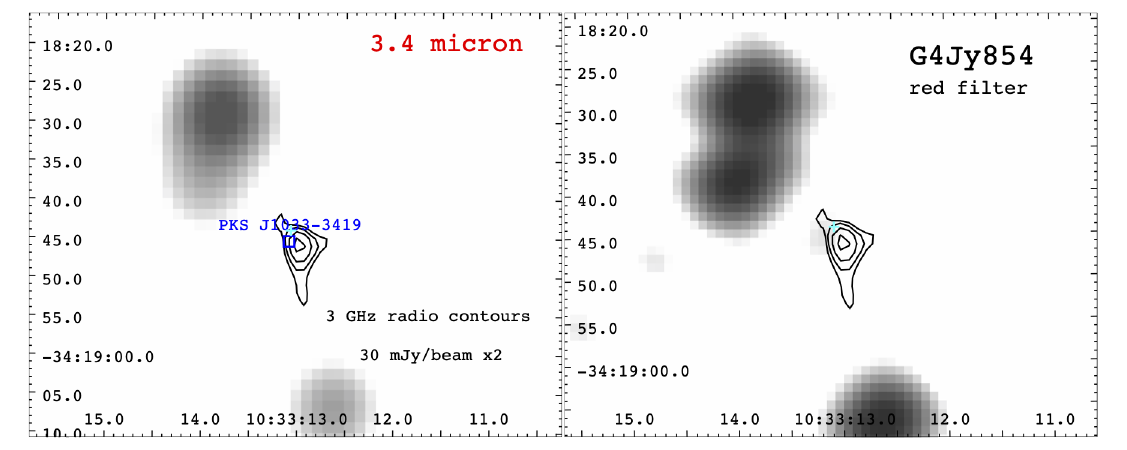}
\includegraphics[height=3.8cm,width=8.8cm,angle=0]{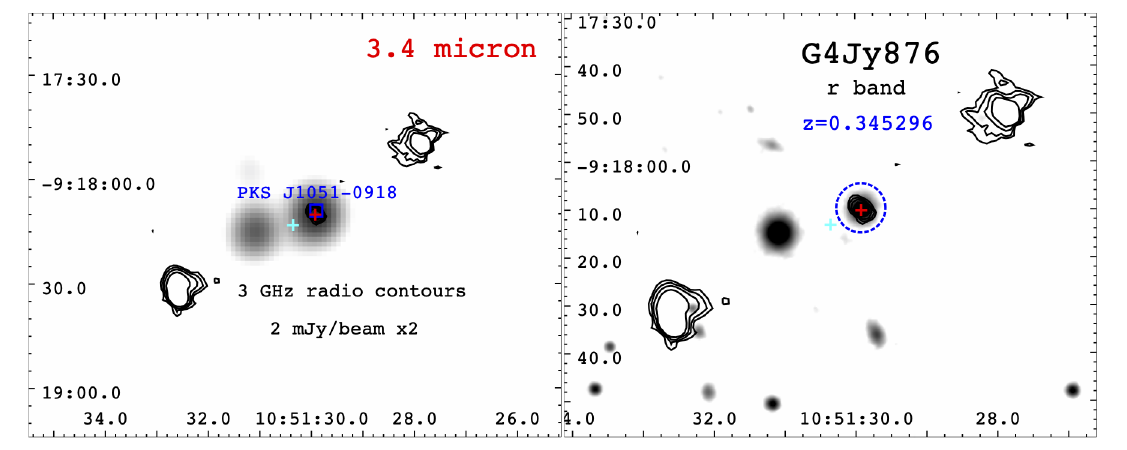}
\includegraphics[height=3.8cm,width=8.8cm,angle=0]{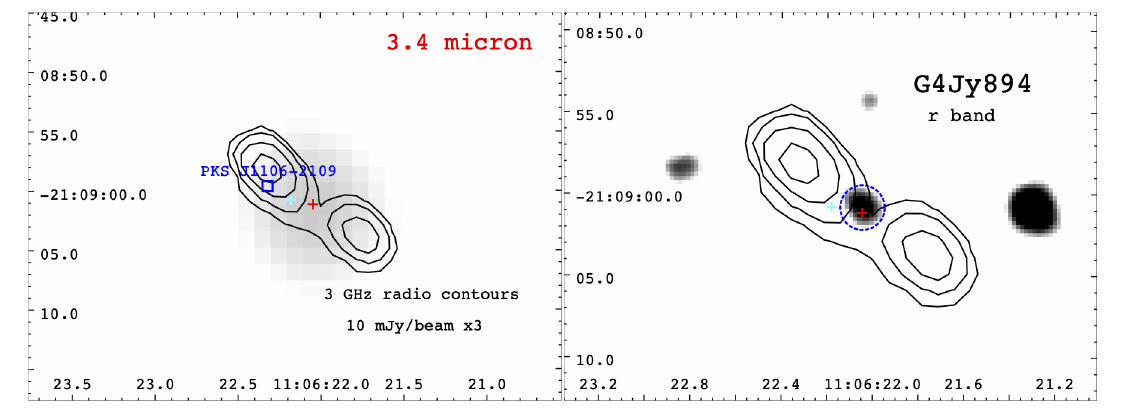}
\end{center}
\caption{Same as Figure~\ref{fig:example1} for the following \cs\ radio sources: \\ 
G4Jy\,730, G4Jy\,734, G4Jy\,747, G4Jy\,779, G4Jy\,796, G4Jy\,818, G4Jy\,835, G4Jy\,836, G4Jy\,837, G4Jy\,854, G4Jy\,876, G4Jy\,894.}
\label{fig:}
\end{figure*}

\begin{figure*}[!th]
\begin{center}
\includegraphics[height=3.8cm,width=8.8cm,angle=0]{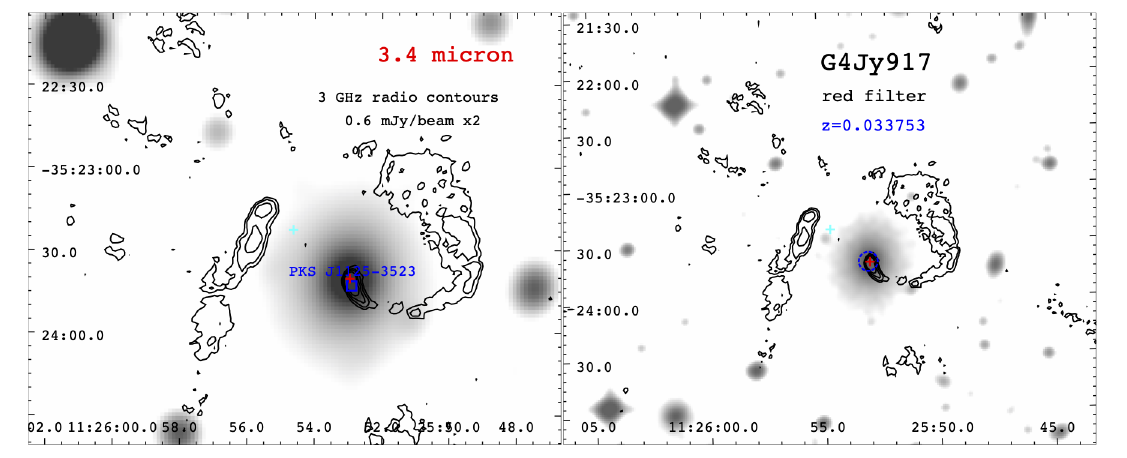}
\includegraphics[height=3.8cm,width=8.8cm,angle=0]{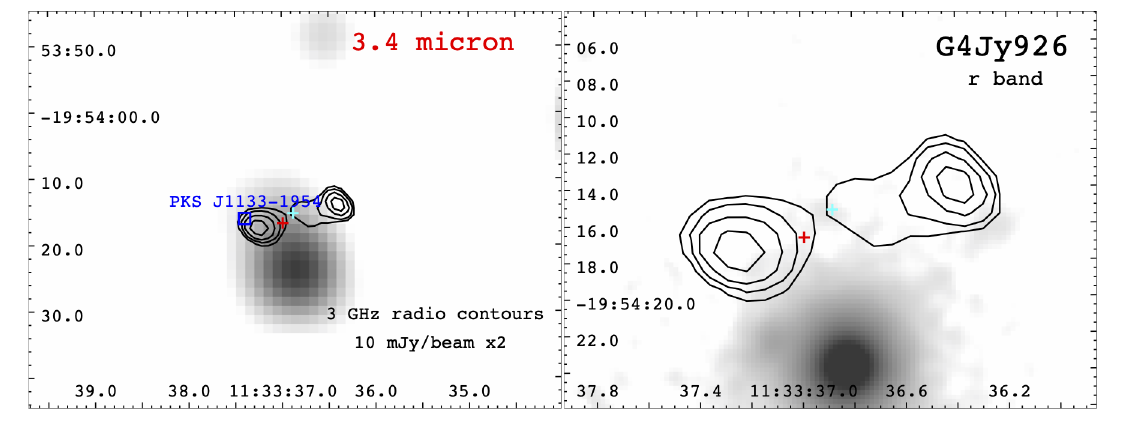}
\includegraphics[height=3.8cm,width=8.8cm,angle=0]{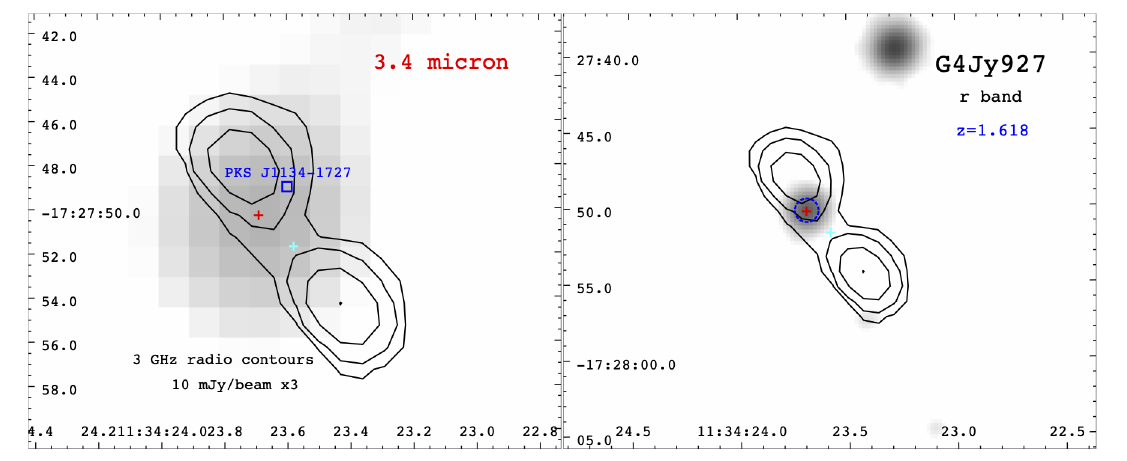}
\includegraphics[height=3.8cm,width=8.8cm,angle=0]{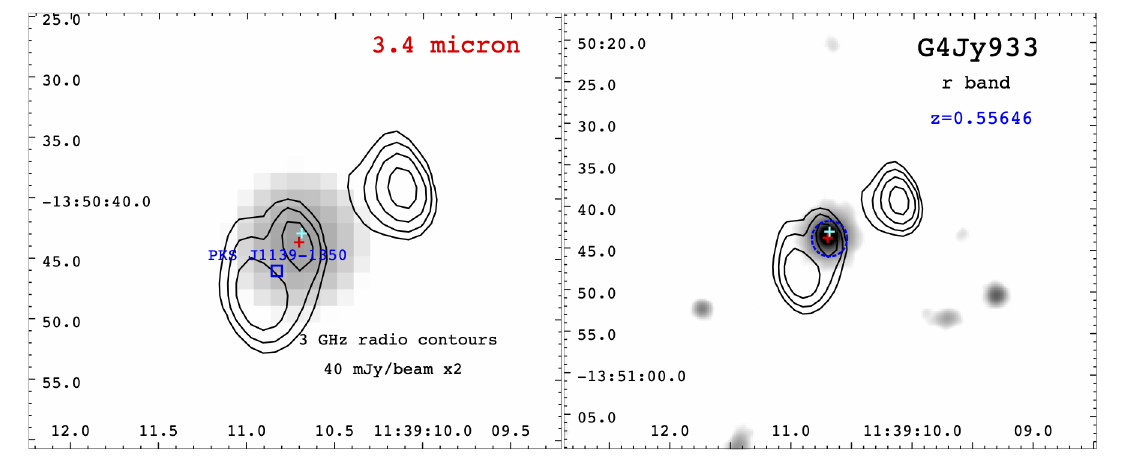}
\includegraphics[height=3.8cm,width=8.8cm,angle=0]{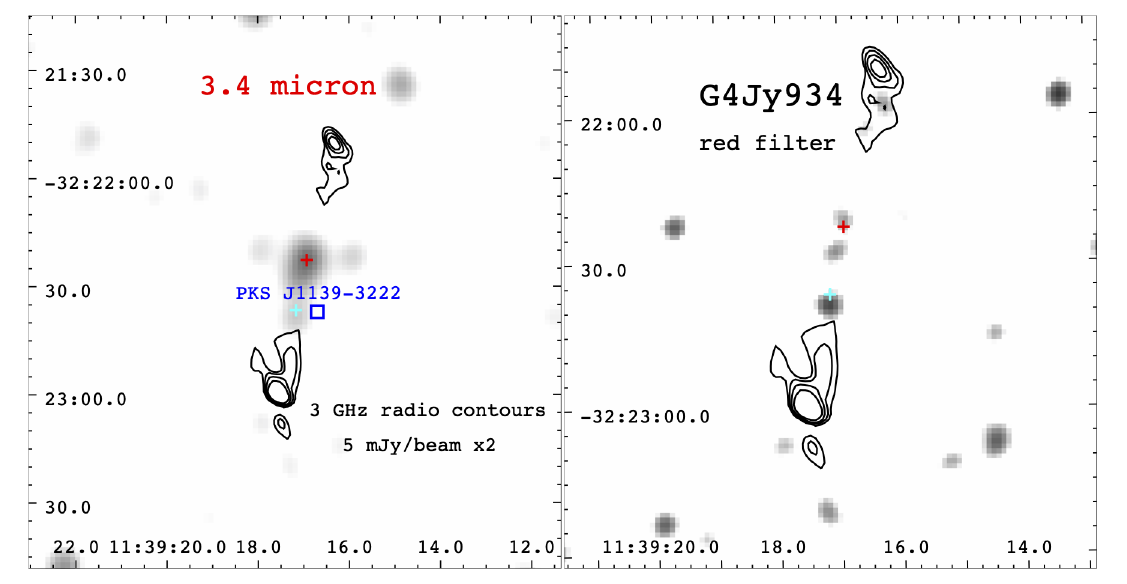}
\includegraphics[height=3.8cm,width=8.8cm,angle=0]{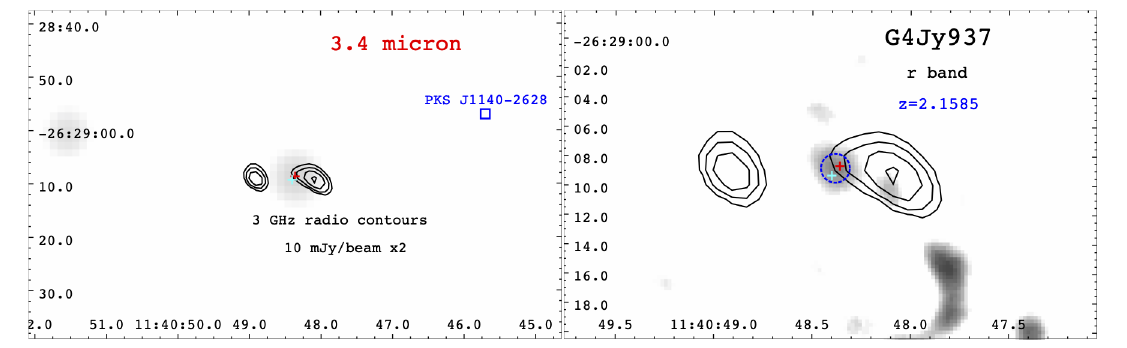}
\includegraphics[height=3.8cm,width=8.8cm,angle=0]{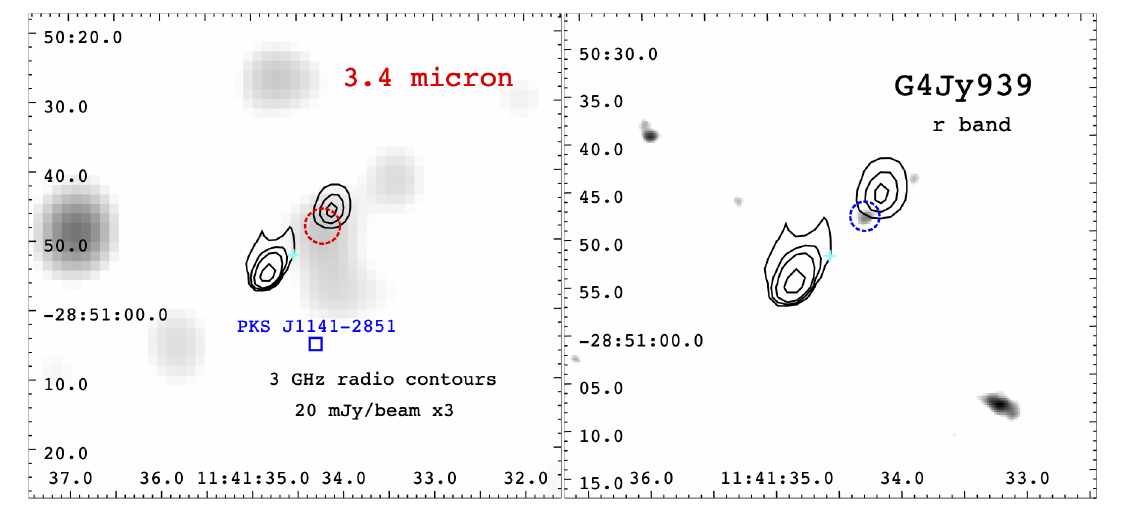}
\includegraphics[height=3.8cm,width=8.8cm,angle=0]{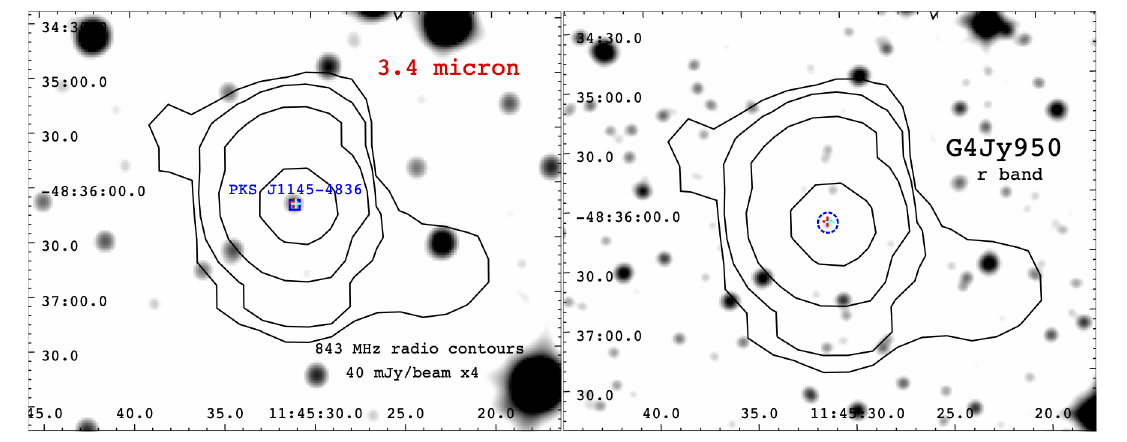}
\includegraphics[height=3.8cm,width=8.8cm,angle=0]{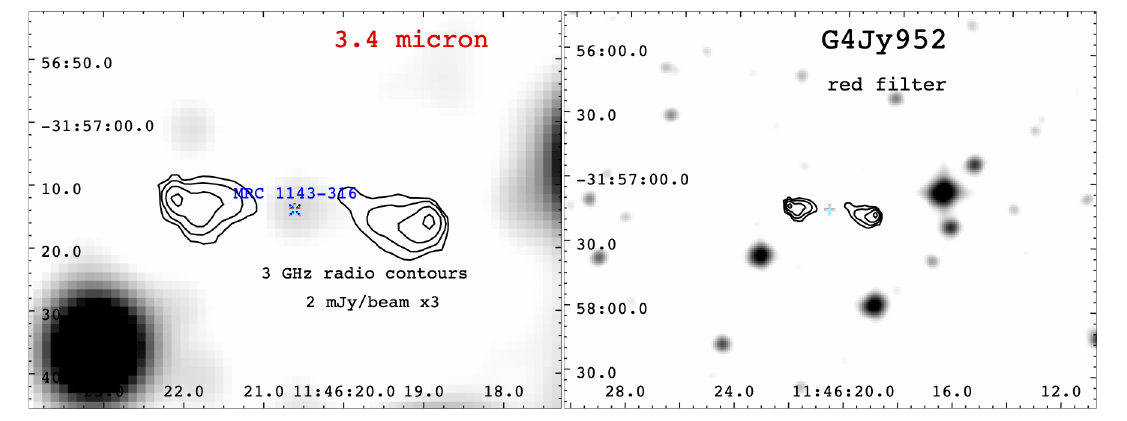}
\includegraphics[height=3.8cm,width=8.8cm,angle=0]{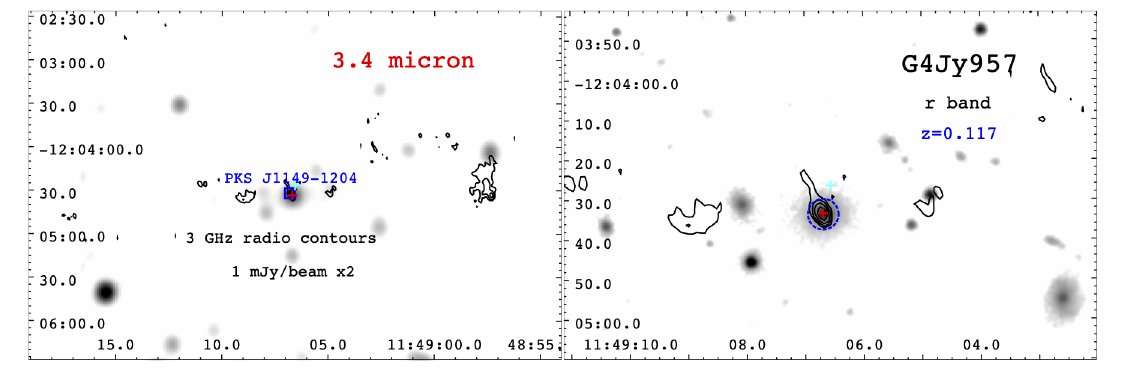}
\includegraphics[height=3.8cm,width=8.8cm,angle=0]{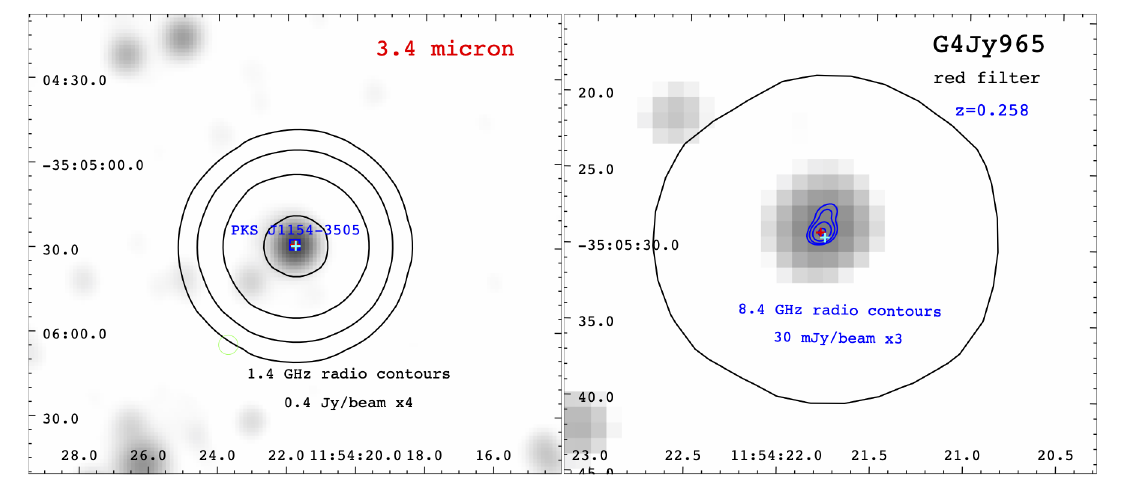}
\includegraphics[height=3.8cm,width=8.8cm,angle=0]{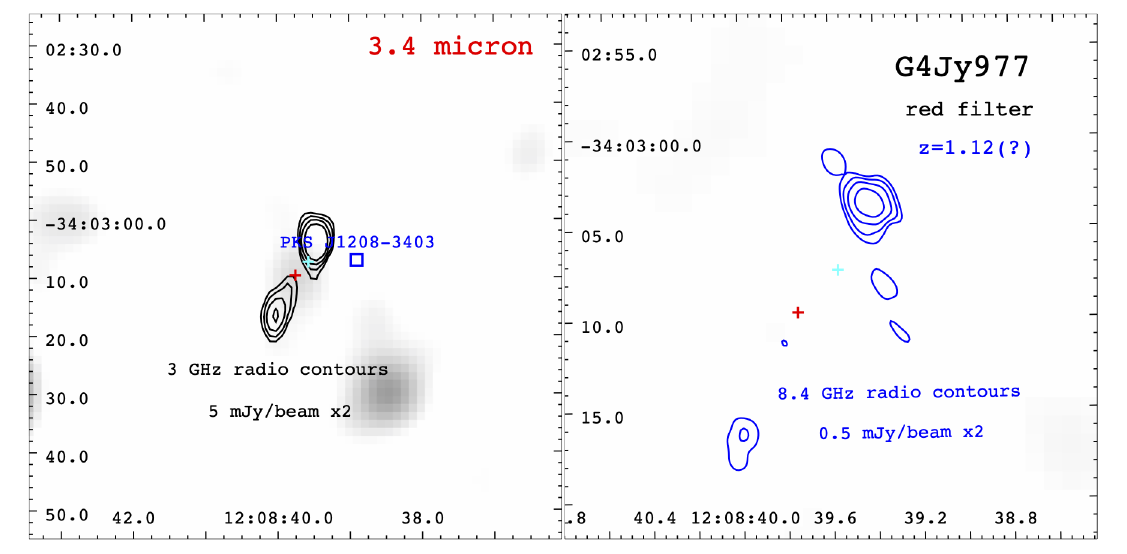}
\caption{Same as Figure~\ref{fig:example1} for the following \cs\ radio sources: \\ 
G4Jy\,917, G4Jy\,926, G4Jy\,927, G4Jy\,933, G4Jy\,934, G4Jy\,937, G4Jy\,939, G4Jy\,950, G4Jy\,952, G4Jy\,957, G4Jy\,965, G4Jy\,977.}
\end{center}
\end{figure*}

\begin{figure*}[!th]
\begin{center}
\includegraphics[height=3.8cm,width=8.8cm,angle=0]{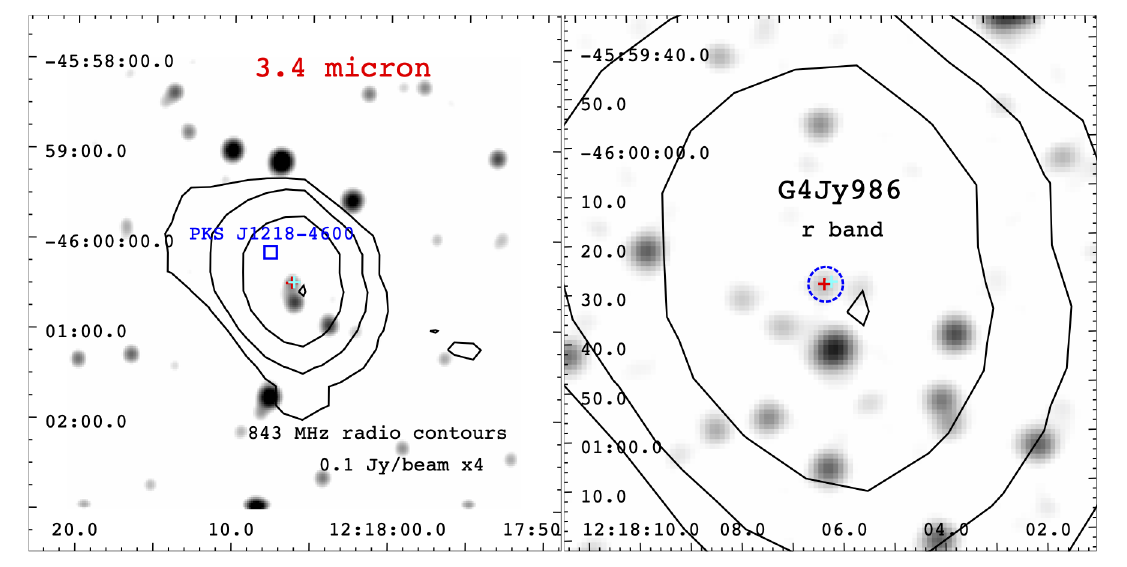}
\includegraphics[height=3.8cm,width=8.8cm,angle=0]{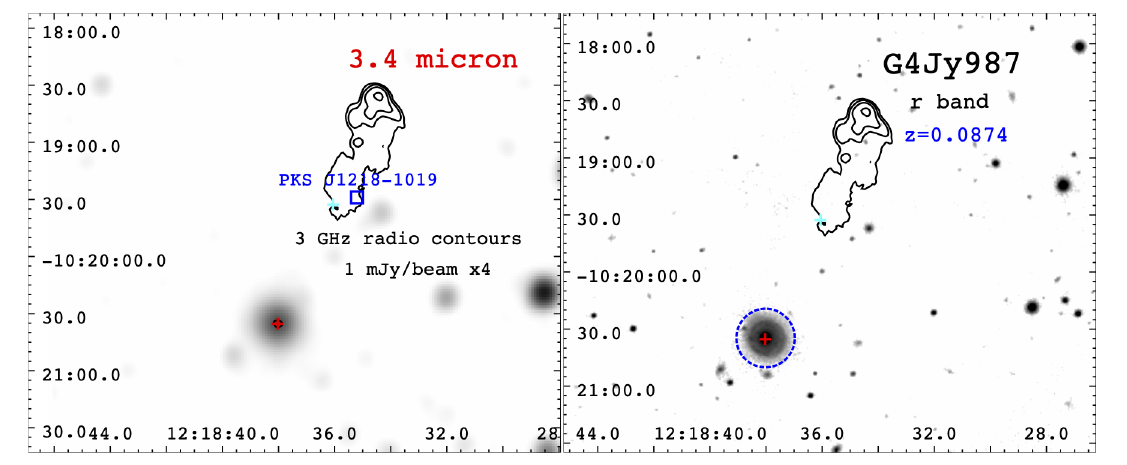}
\includegraphics[height=3.8cm,width=8.8cm,angle=0]{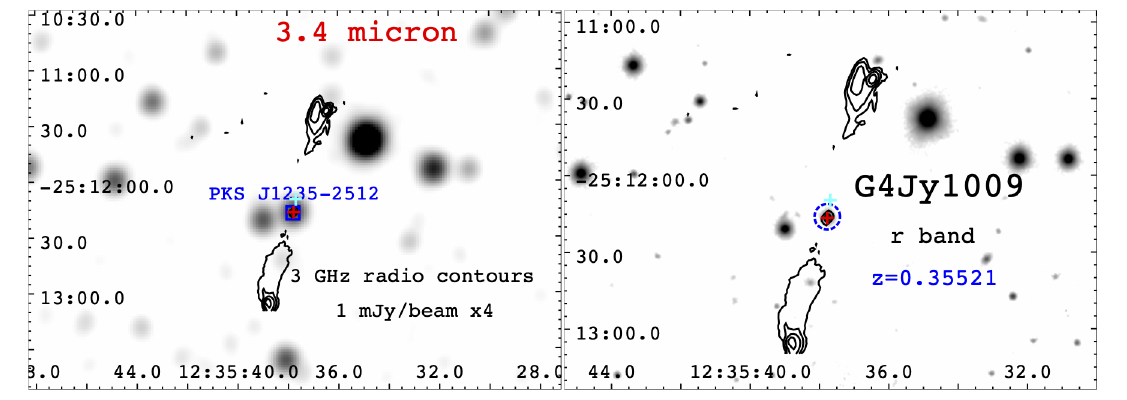}
\includegraphics[height=3.8cm,width=8.8cm,angle=0]{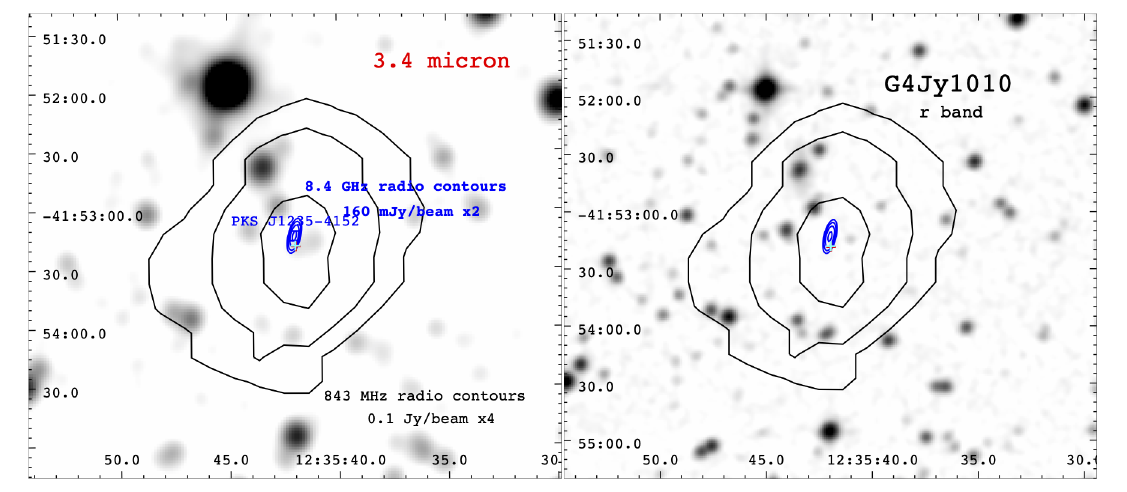}
\includegraphics[height=3.8cm,width=8.8cm,angle=0]{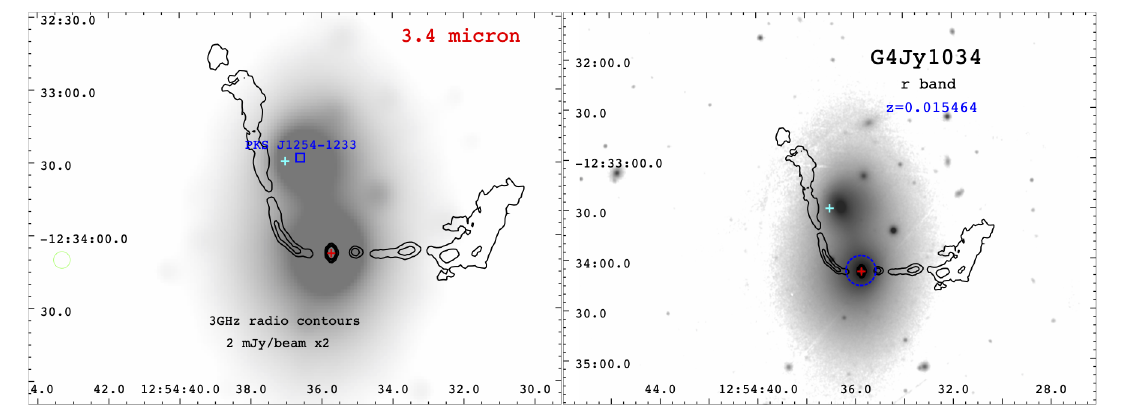}
\includegraphics[height=3.8cm,width=8.8cm,angle=0]{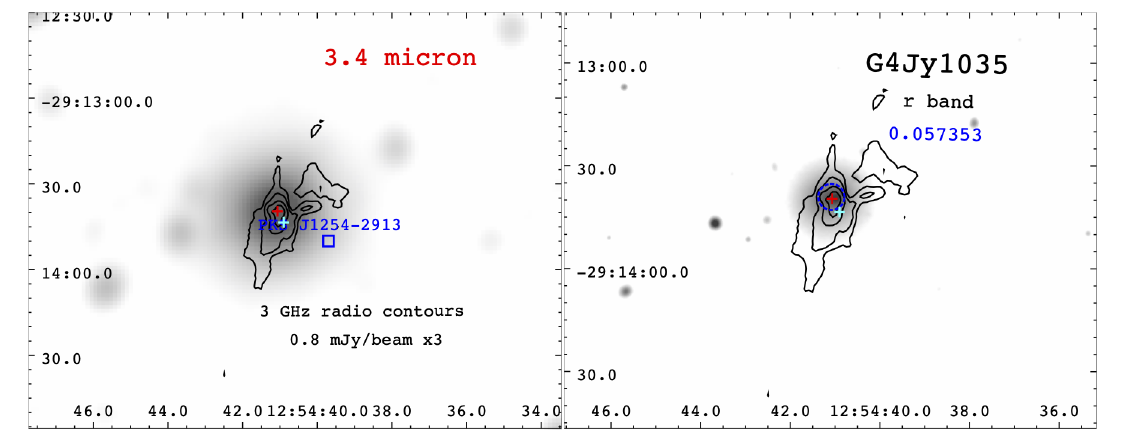}
\includegraphics[height=3.8cm,width=8.8cm,angle=0]{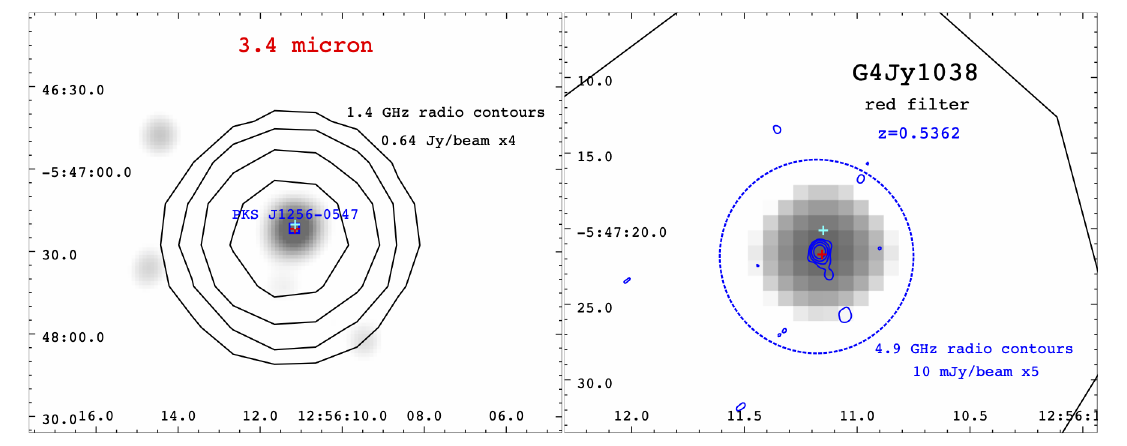}
\includegraphics[height=3.8cm,width=8.8cm,angle=0]{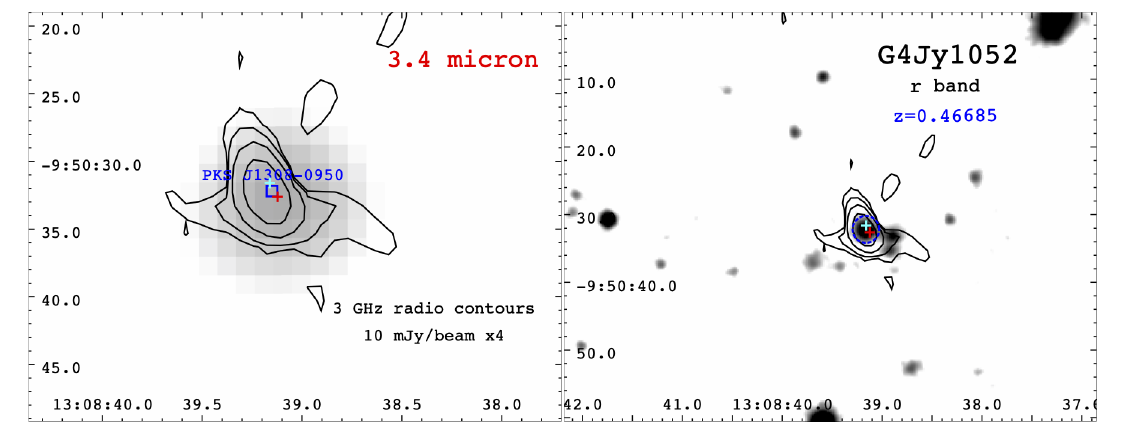}
\includegraphics[height=3.8cm,width=8.8cm,angle=0]{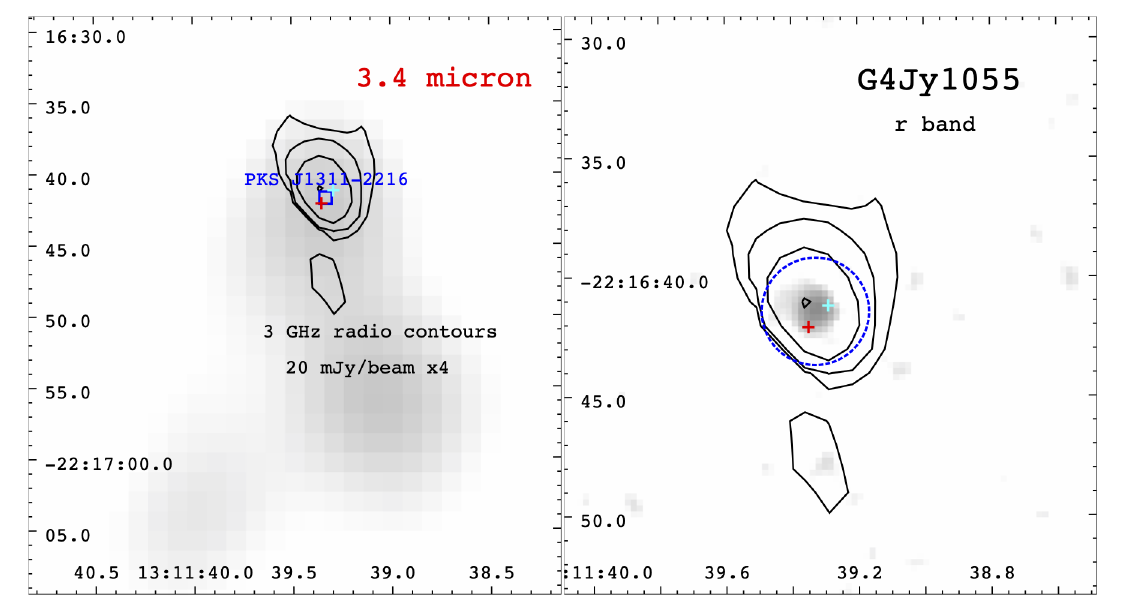}
\includegraphics[height=3.8cm,width=8.8cm,angle=0]{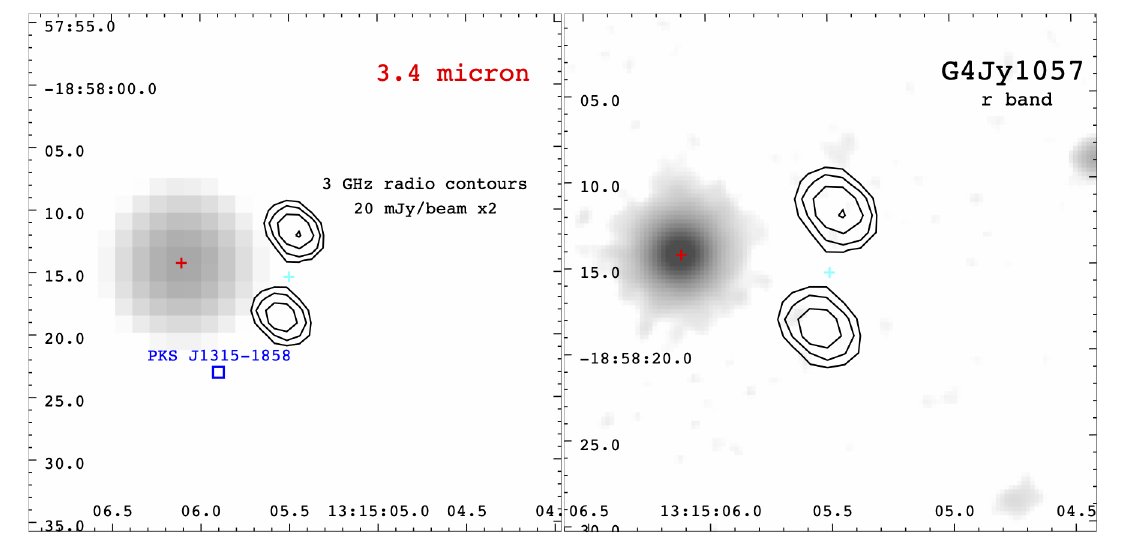}
\includegraphics[height=3.8cm,width=8.8cm,angle=0]{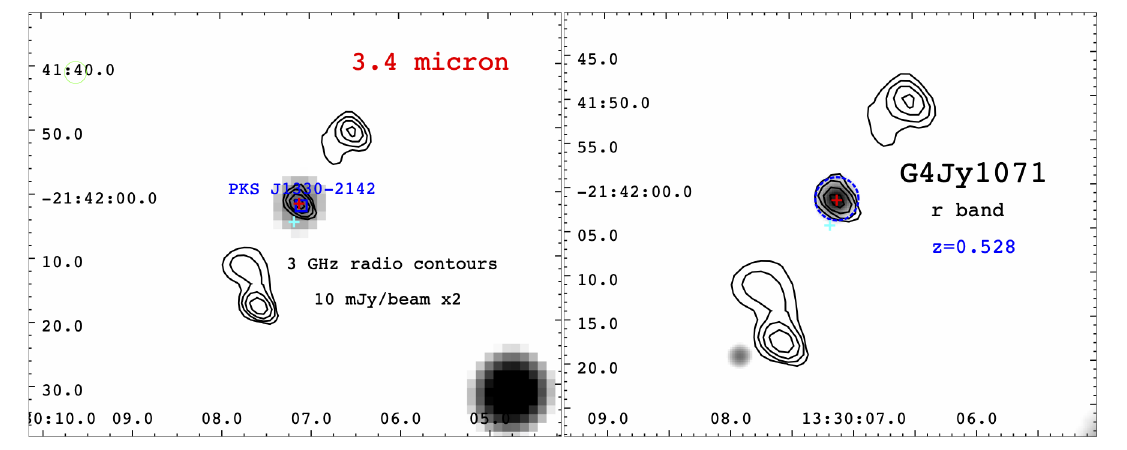}
\includegraphics[height=3.8cm,width=8.8cm,angle=0]{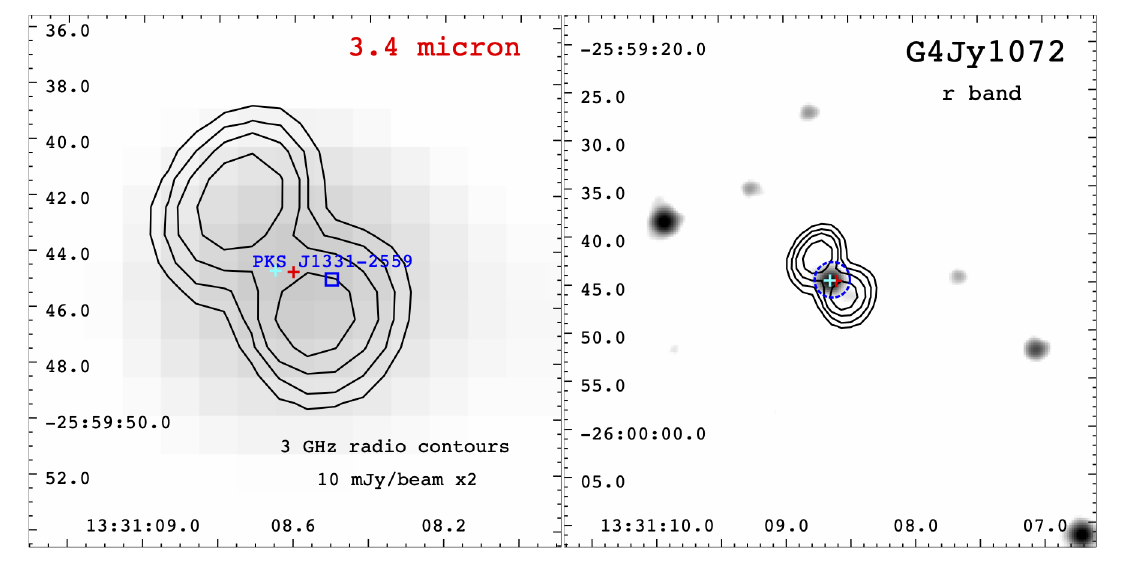}
\caption{Same as Figure~\ref{fig:example1} for the following \cs\ radio sources: \\ 
G4Jy\,986, G4Jy\,987, G4Jy\,1009, G4Jy\,1010, G4Jy\,1034, G4Jy\,1035, G4Jy\,1038, G4Jy\,1052, G4Jy\,1055, G4Jy\,1057, G4Jy\,1071, G4Jy\,1072.}
\end{center}
\end{figure*}

\begin{figure*}[!th]
\begin{center}
\includegraphics[height=3.8cm,width=8.8cm,angle=0]{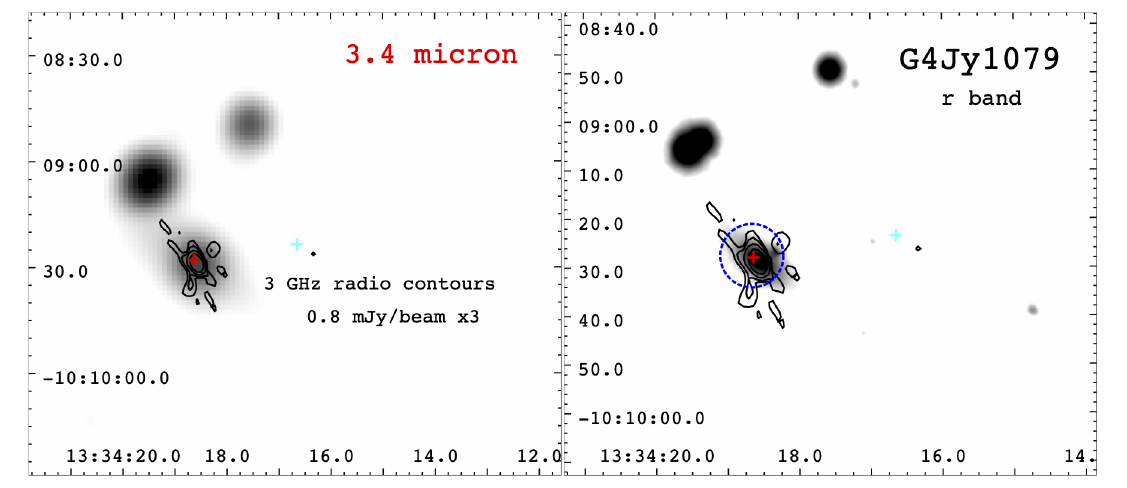}
\includegraphics[height=3.8cm,width=8.8cm,angle=0]{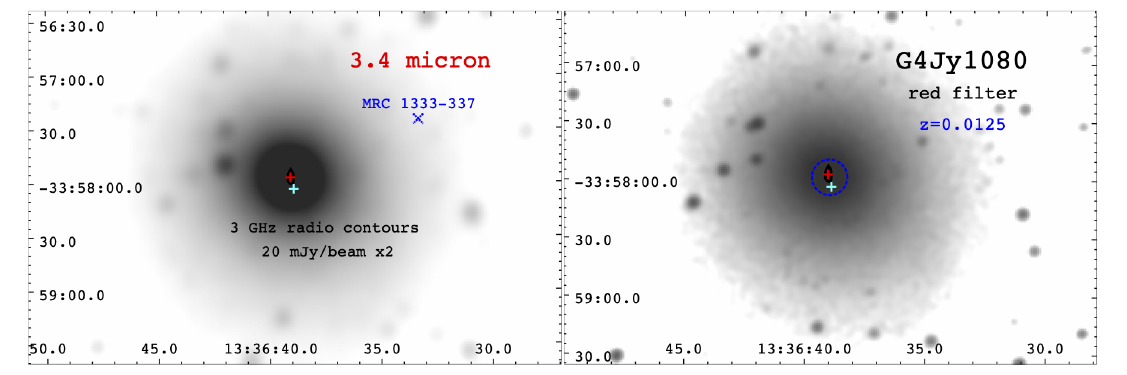}
\includegraphics[height=3.8cm,width=8.8cm,angle=0]{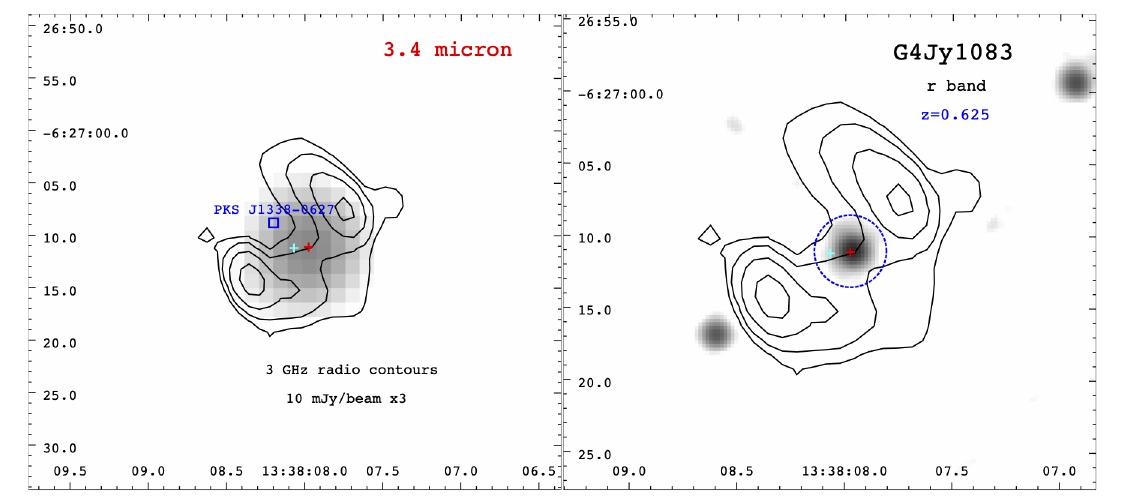}
\includegraphics[height=3.8cm,width=8.8cm,angle=0]{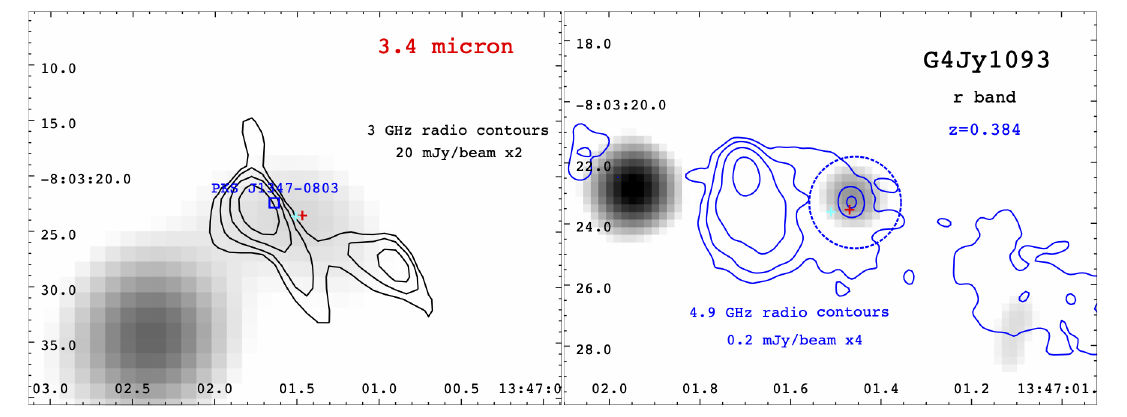}
\includegraphics[height=3.8cm,width=8.8cm,angle=0]{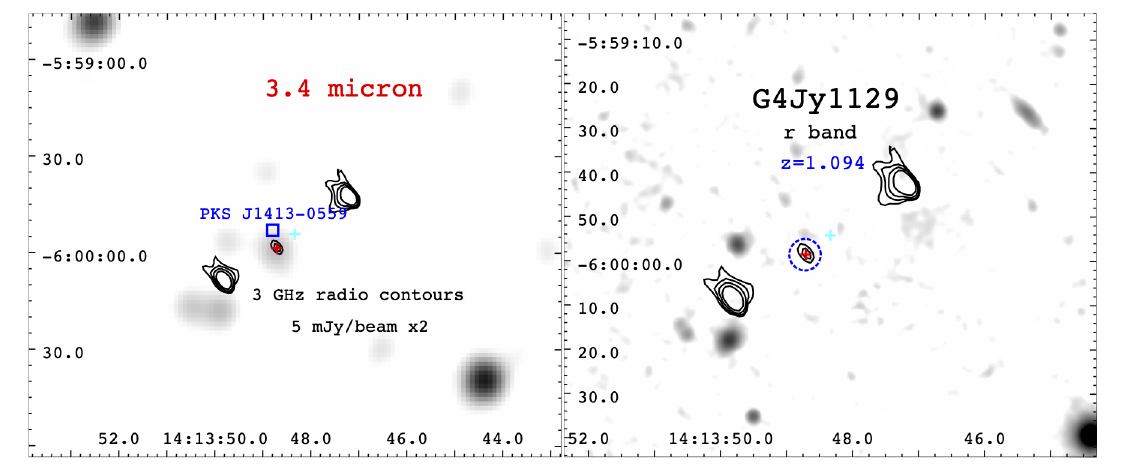}
\includegraphics[height=3.8cm,width=8.8cm,angle=0]{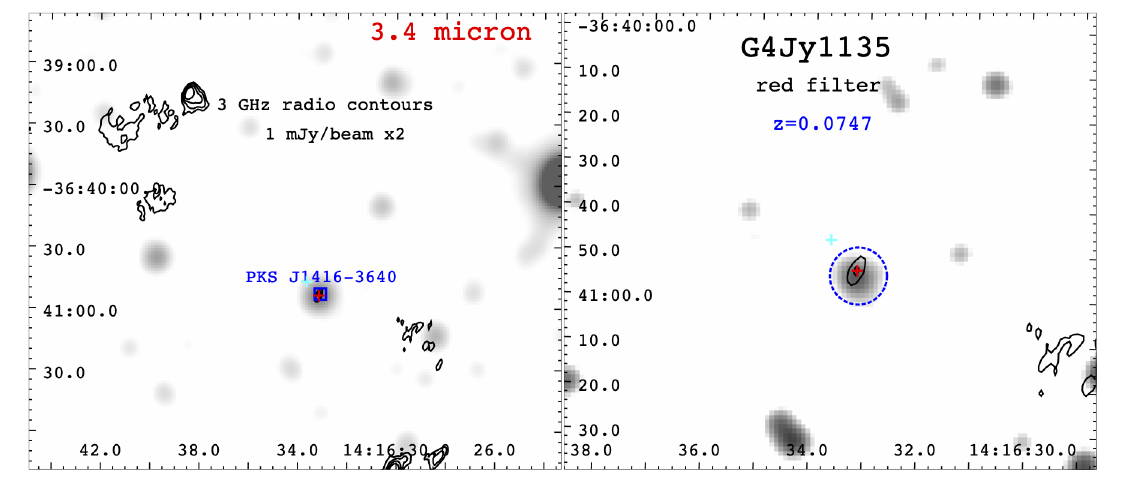}
\includegraphics[height=3.8cm,width=8.8cm,angle=0]{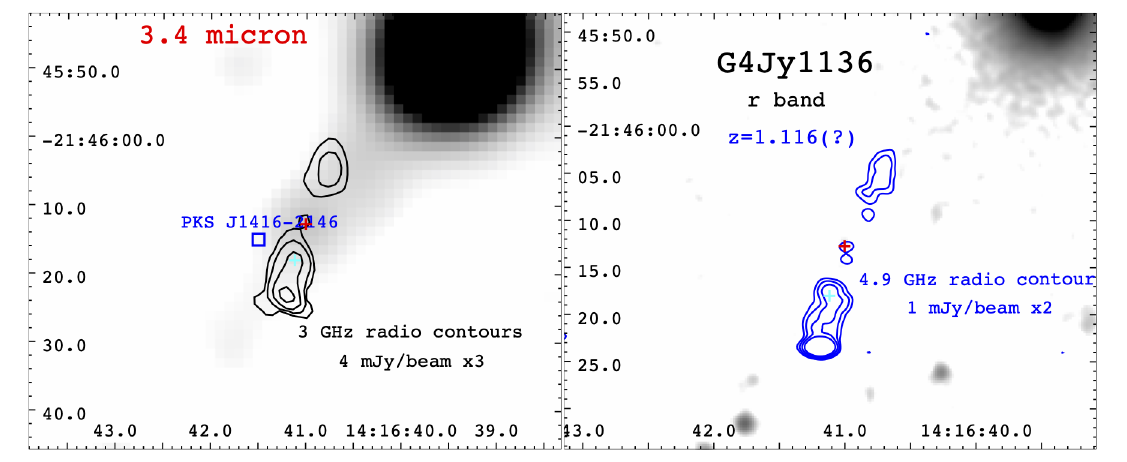}
\includegraphics[height=3.8cm,width=8.8cm,angle=0]{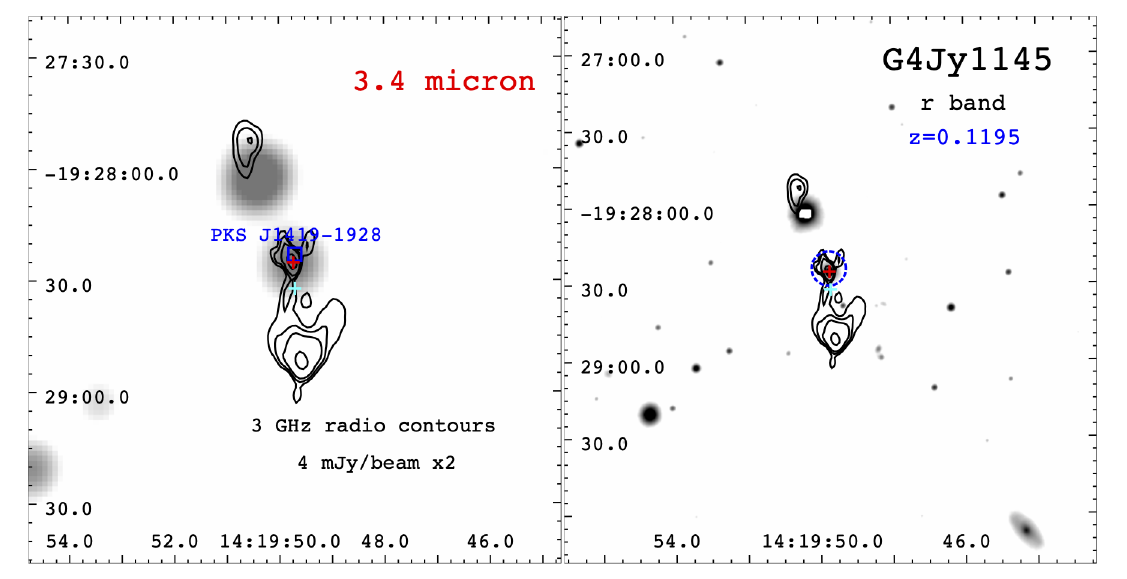}
\includegraphics[height=3.8cm,width=8.8cm,angle=0]{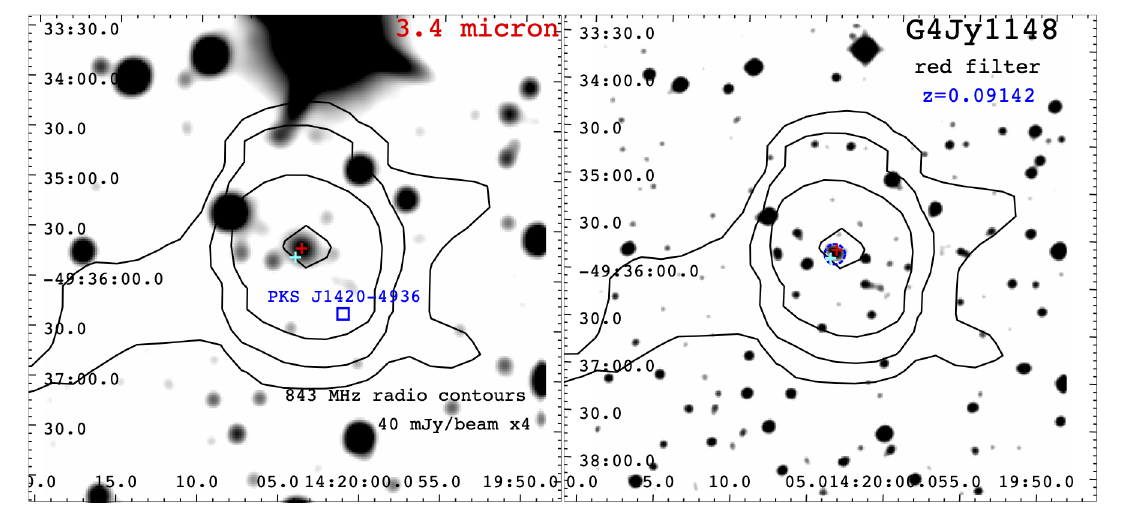}
\includegraphics[height=3.8cm,width=8.8cm,angle=0]{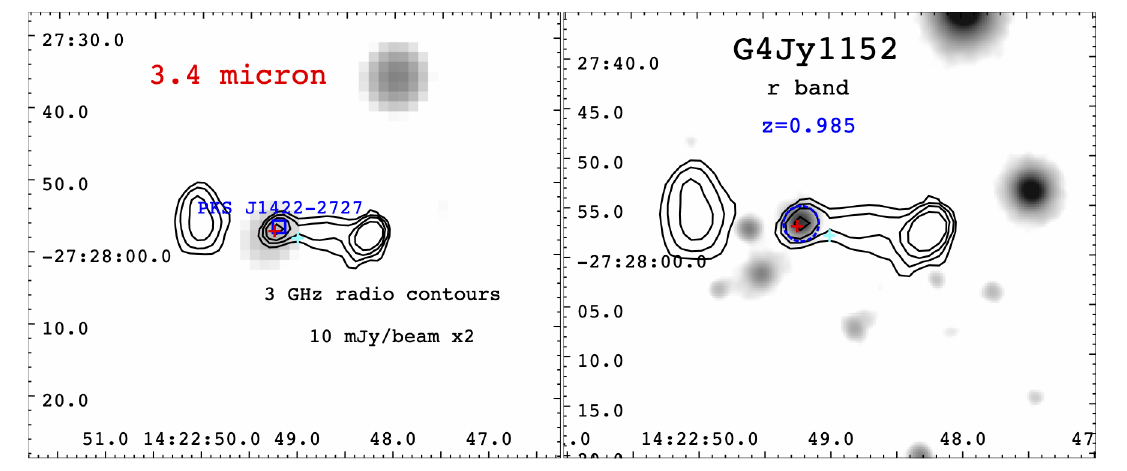}
\includegraphics[height=3.8cm,width=8.8cm,angle=0]{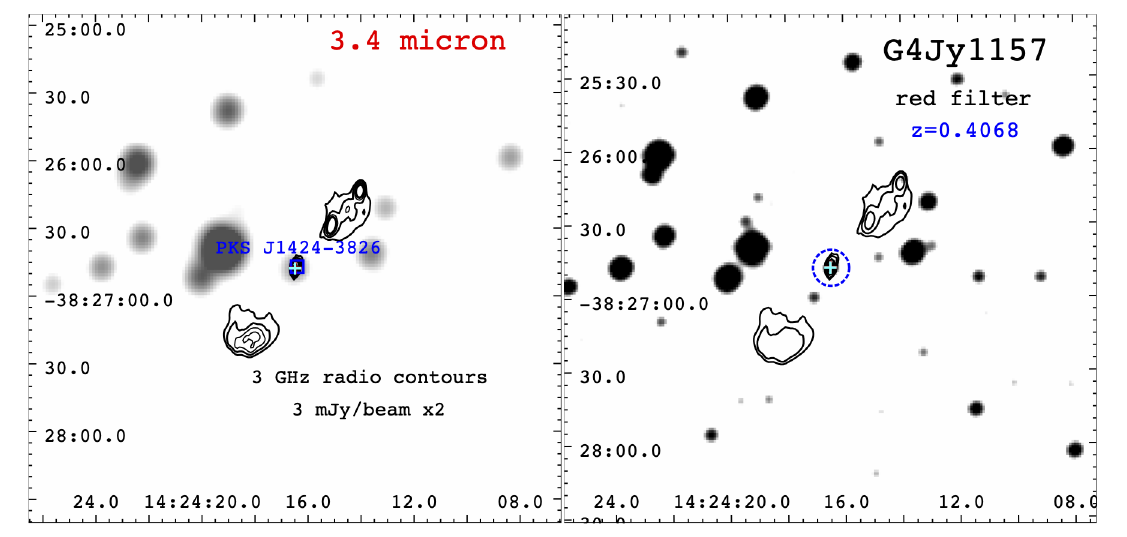}
\includegraphics[height=3.8cm,width=8.8cm,angle=0]{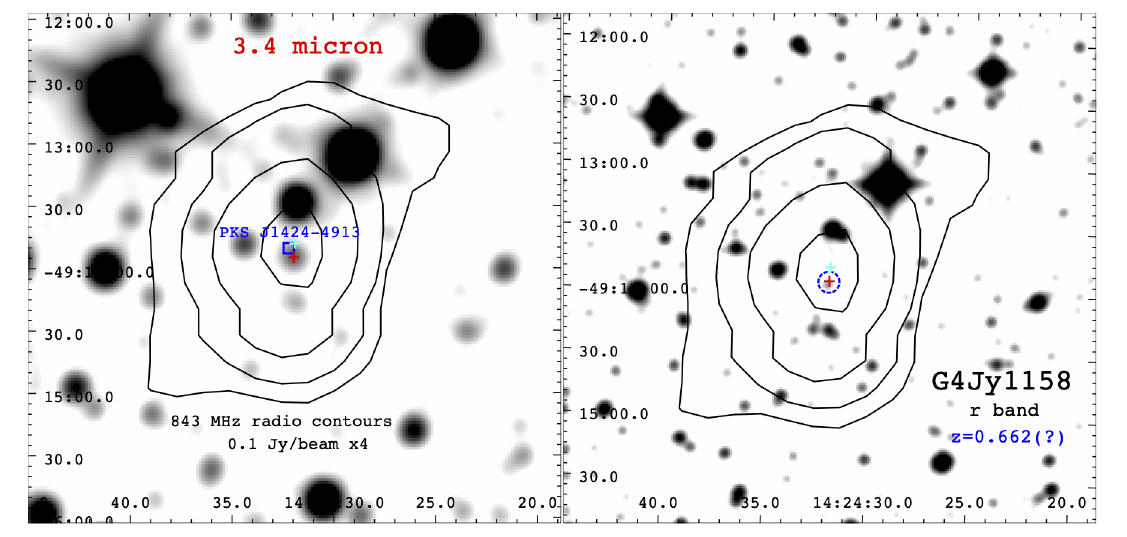}
\caption{Same as Figure~\ref{fig:example1} for the following \cs\ radio sources: \\ 
G4Jy\,1079, G4Jy\,1080, G4Jy\,1083, G4Jy\,1093, G4Jy\,1129, G4Jy\,1135, G4Jy\,1136, G4Jy\,1145, G4Jy\,1148, G4Jy\,1152, G4Jy\,1157, G4Jy\,1158.}
\end{center}
\end{figure*}

\begin{figure*}[!th]
\begin{center}
\includegraphics[height=3.8cm,width=8.8cm,angle=0]{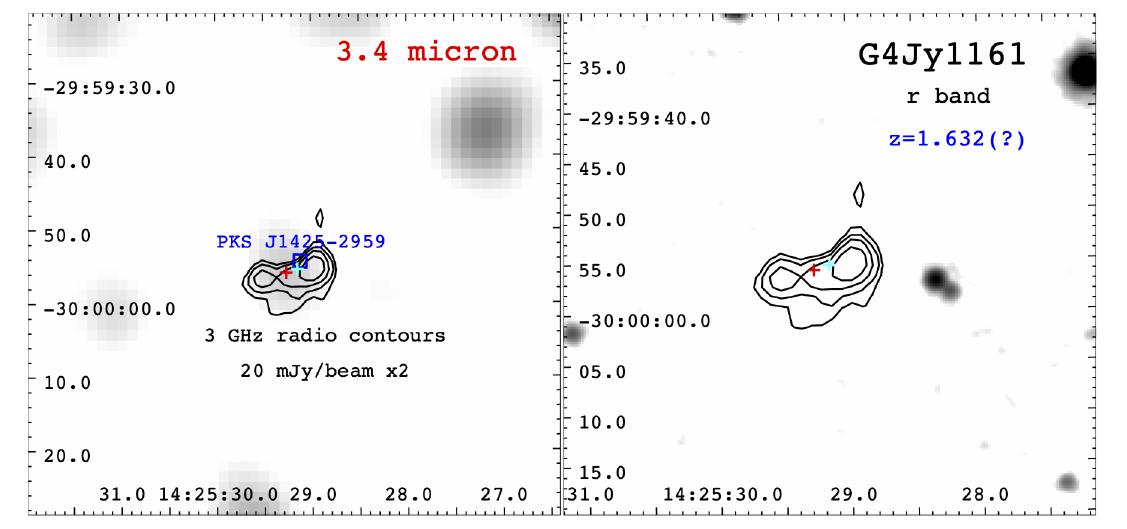}
\includegraphics[height=3.8cm,width=8.8cm,angle=0]{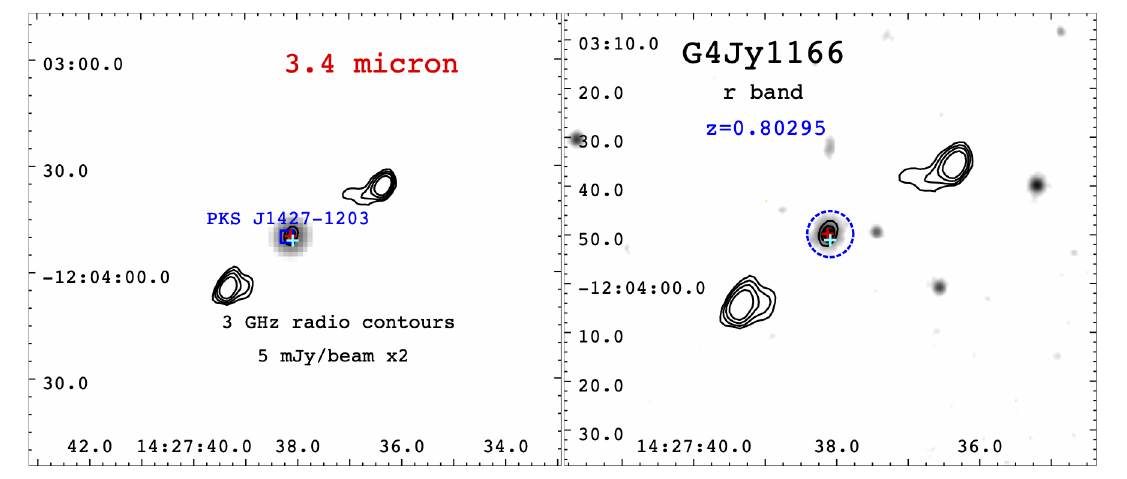}
\includegraphics[height=3.8cm,width=8.8cm,angle=0]{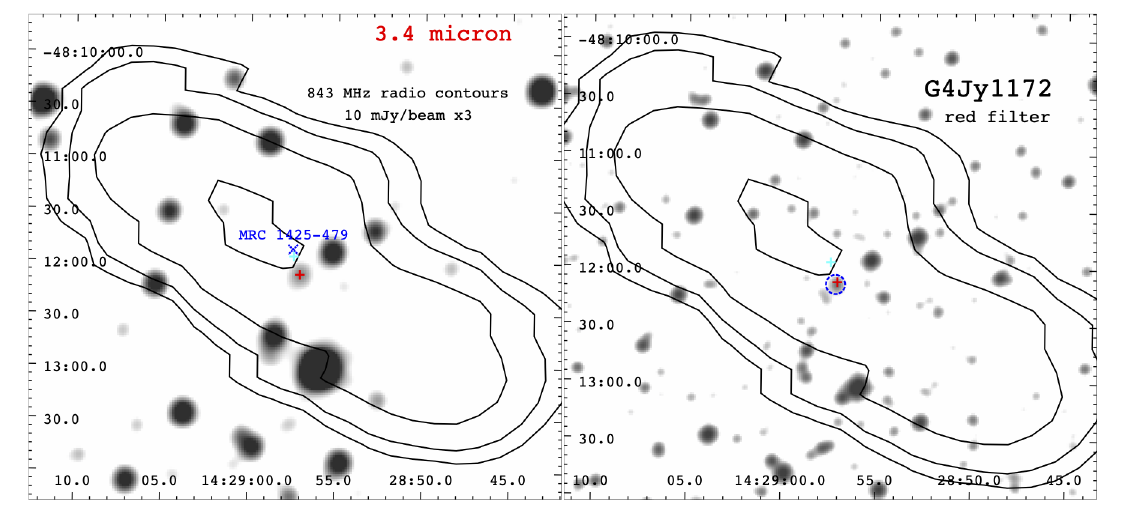}
\includegraphics[height=3.8cm,width=8.8cm,angle=0]{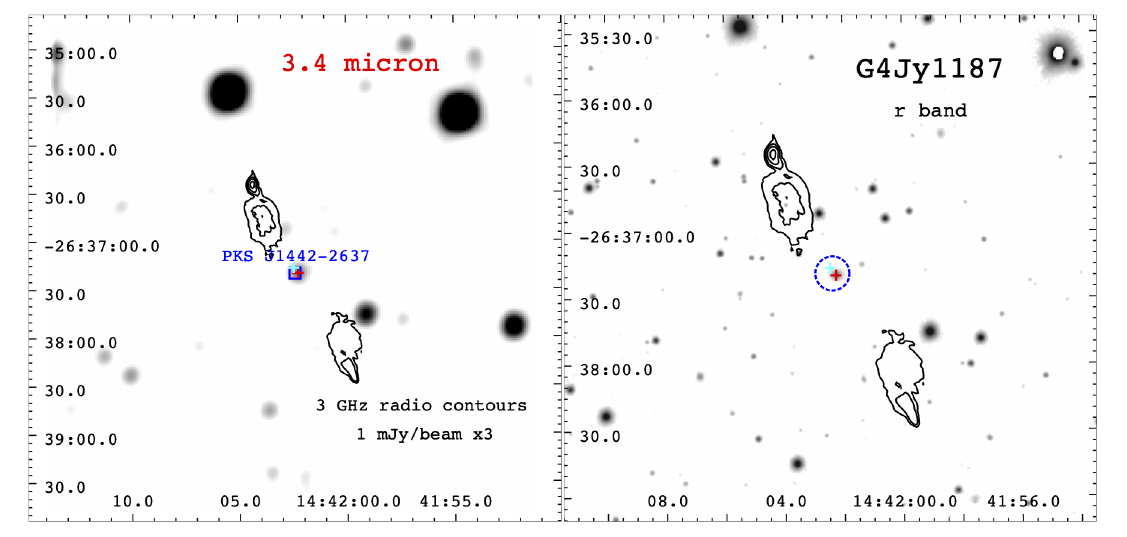}
\includegraphics[height=3.8cm,width=8.8cm,angle=0]{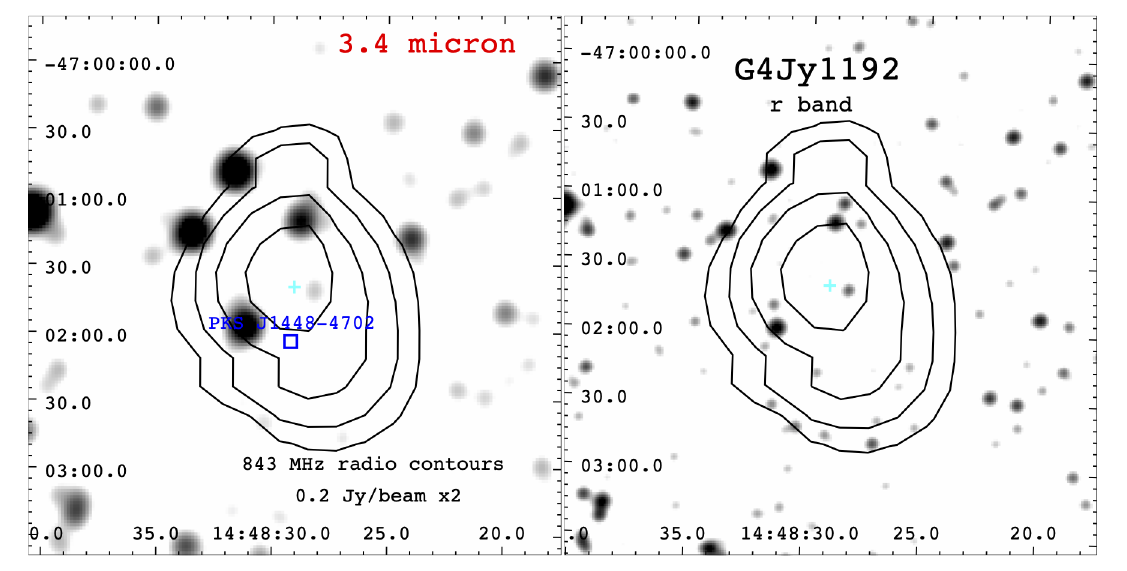}
\includegraphics[height=3.8cm,width=8.8cm,angle=0]{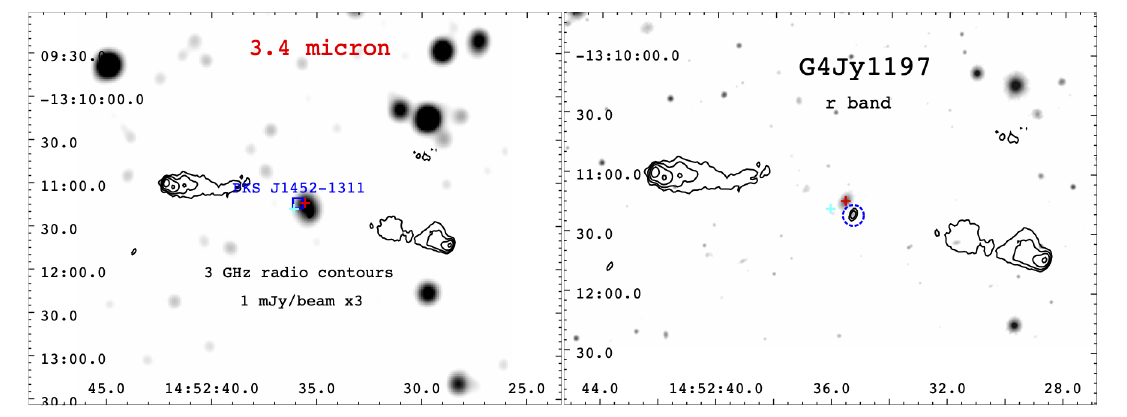}
\includegraphics[height=3.8cm,width=8.8cm,angle=0]{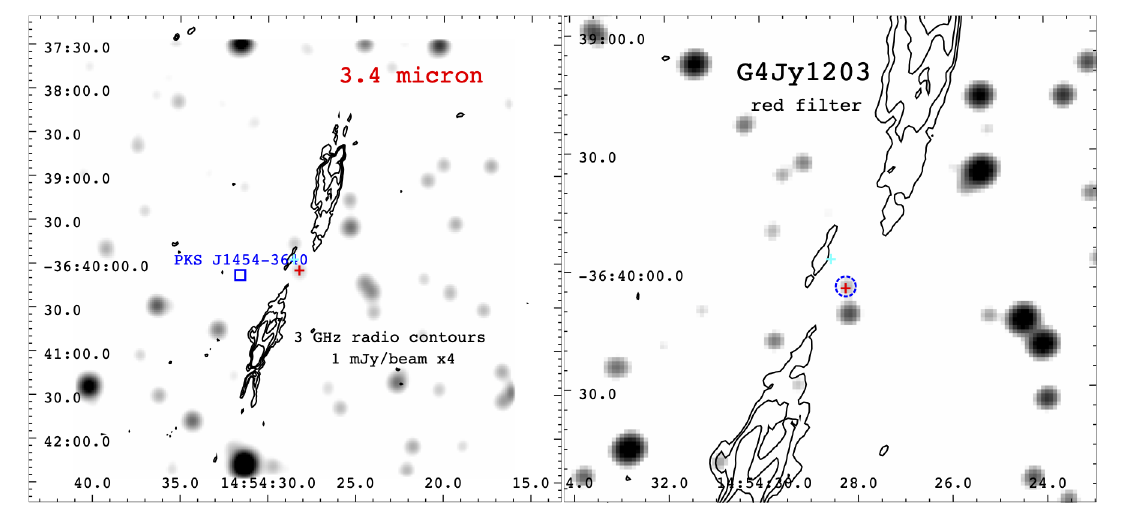}
\includegraphics[height=3.8cm,width=8.8cm,angle=0]{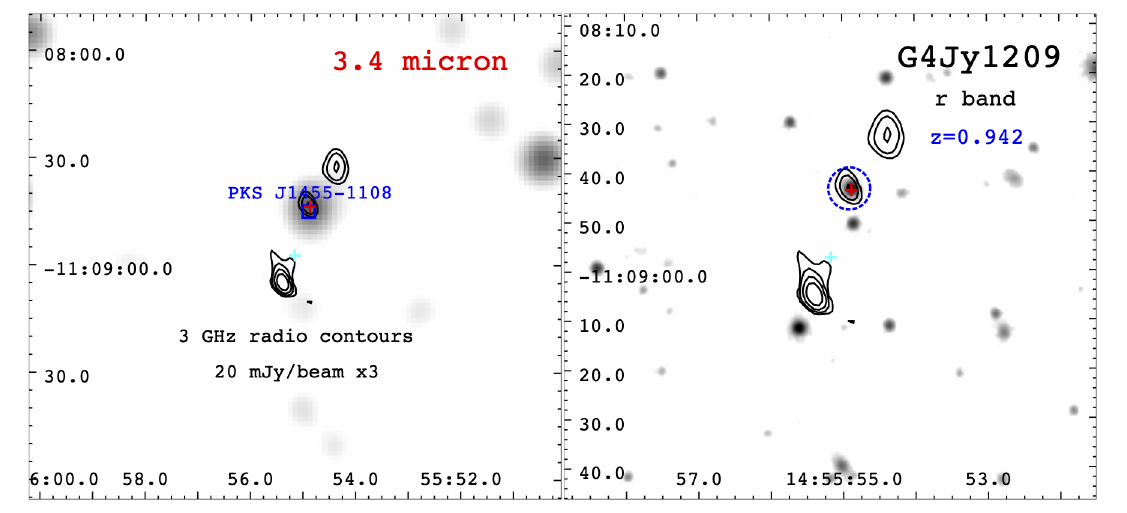}
\includegraphics[height=3.8cm,width=8.8cm,angle=0]{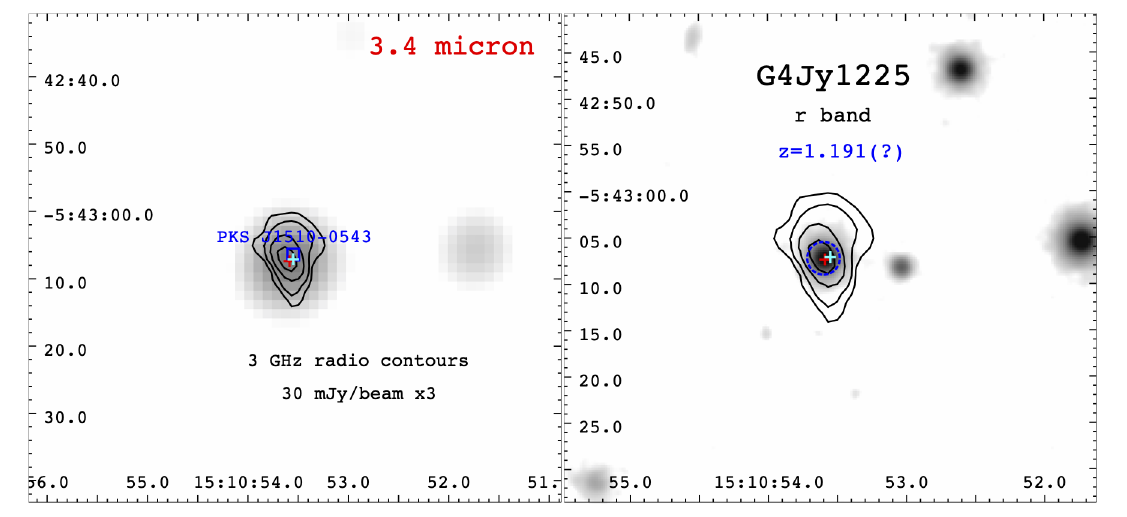}
\includegraphics[height=3.8cm,width=8.8cm,angle=0]{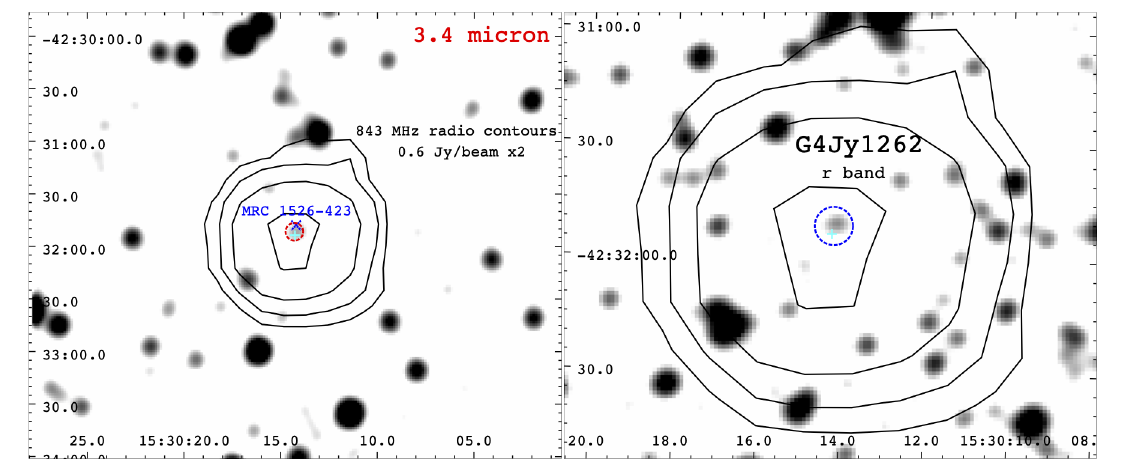}
\includegraphics[height=3.8cm,width=8.8cm,angle=0]{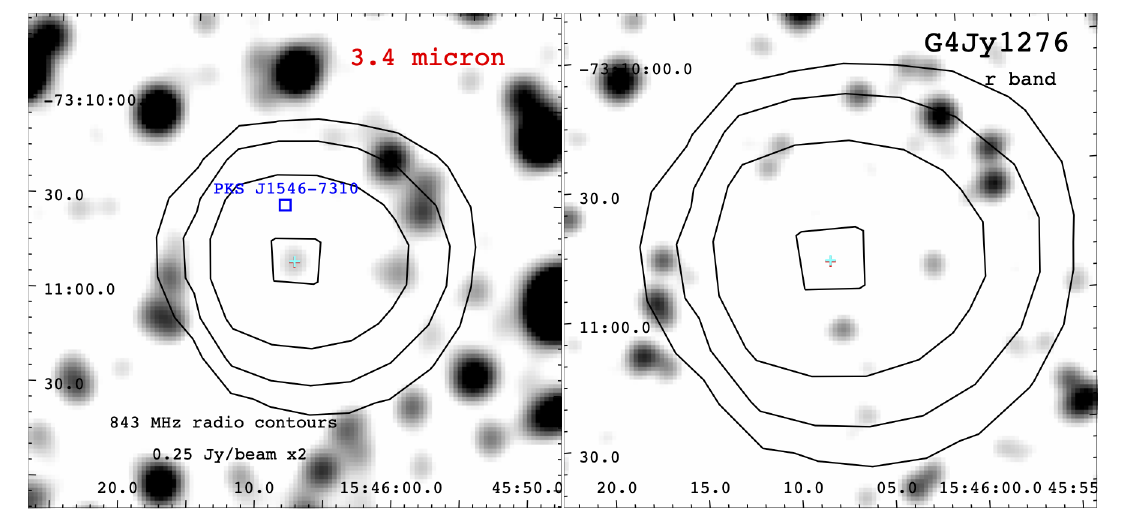}
\includegraphics[height=3.8cm,width=8.8cm,angle=0]{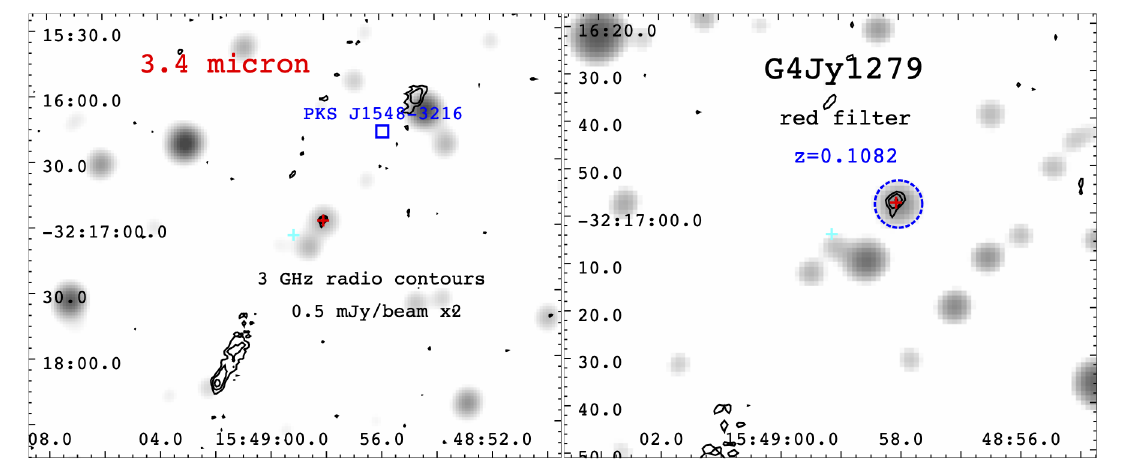}
\caption{Same as Figure~\ref{fig:example1} for the following \cs\ radio sources: \\ 
G4Jy\,1161, G4Jy\,1166, G4Jy\,1172, G4Jy\,1187, G4Jy\,1192, G4Jy\,1197, G4Jy\,1203, G4Jy\,1209, G4Jy\,1225, G4Jy\,1262, G4Jy\,1276, G4Jy\,1279.}
\end{center}
\end{figure*}

\begin{figure*}[!th]
\begin{center}
\includegraphics[height=3.8cm,width=8.8cm,angle=0]{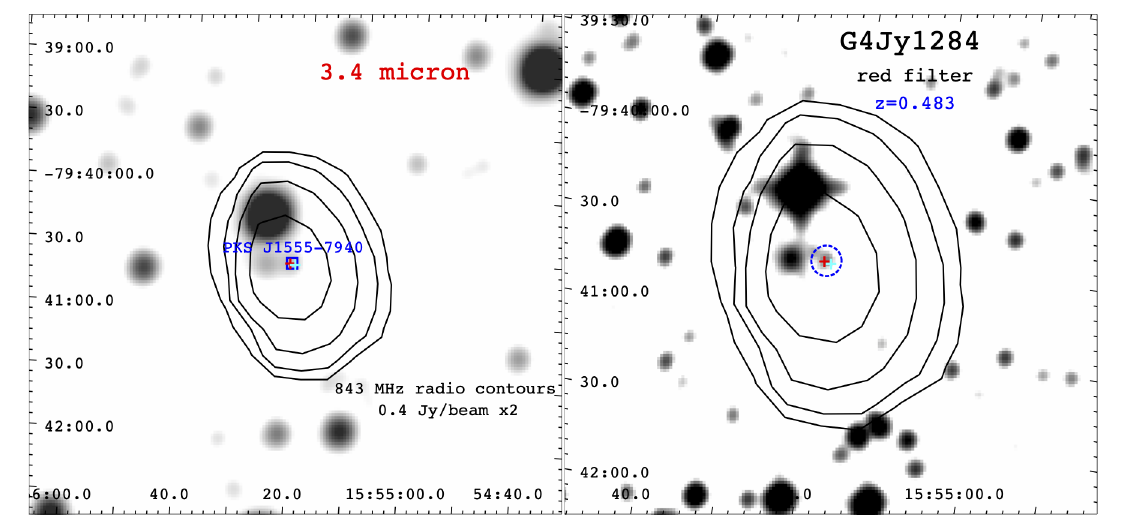}
\includegraphics[height=3.8cm,width=8.8cm,angle=0]{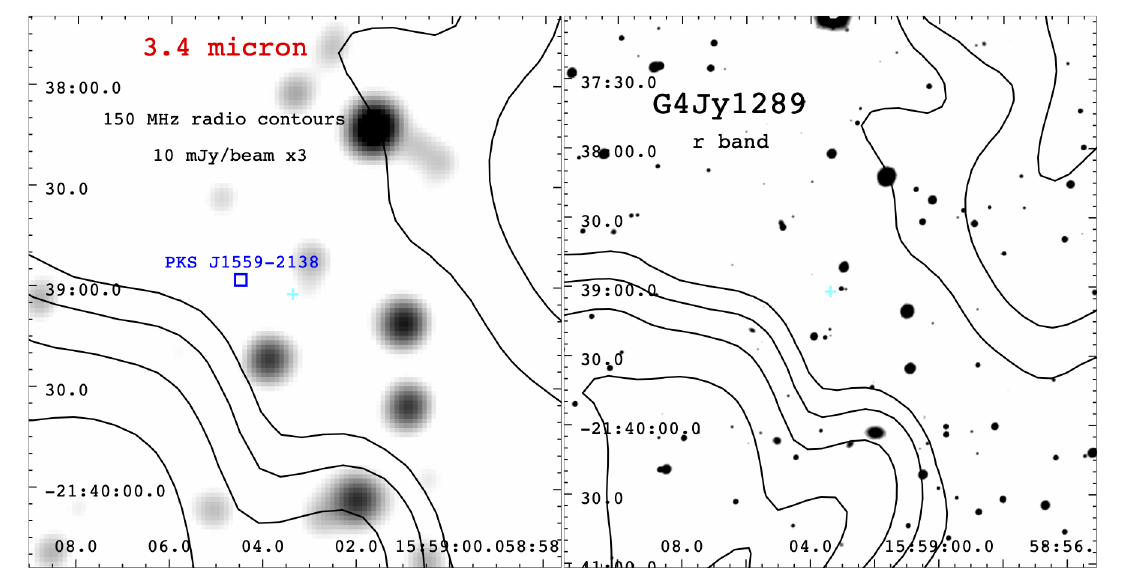}
\includegraphics[height=3.8cm,width=8.8cm,angle=0]{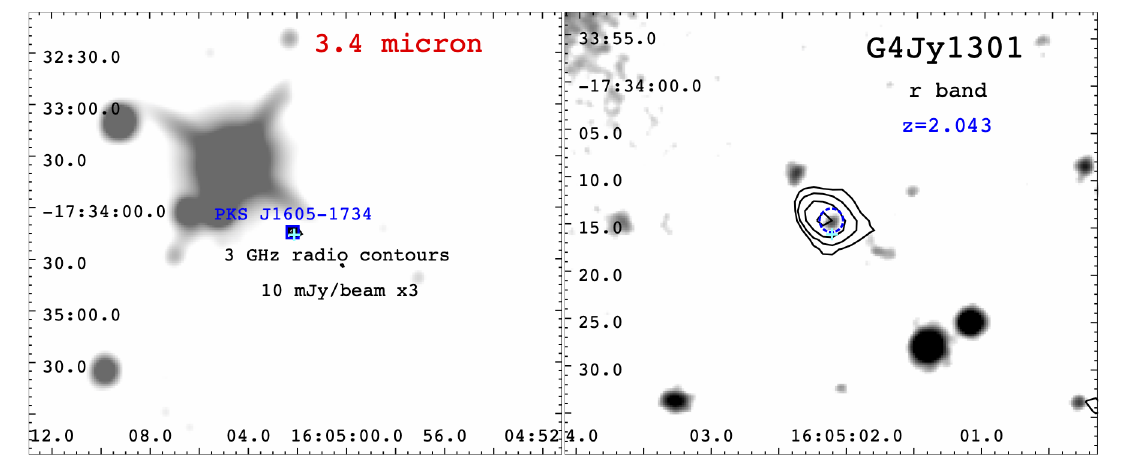}
\includegraphics[height=3.8cm,width=8.8cm,angle=0]{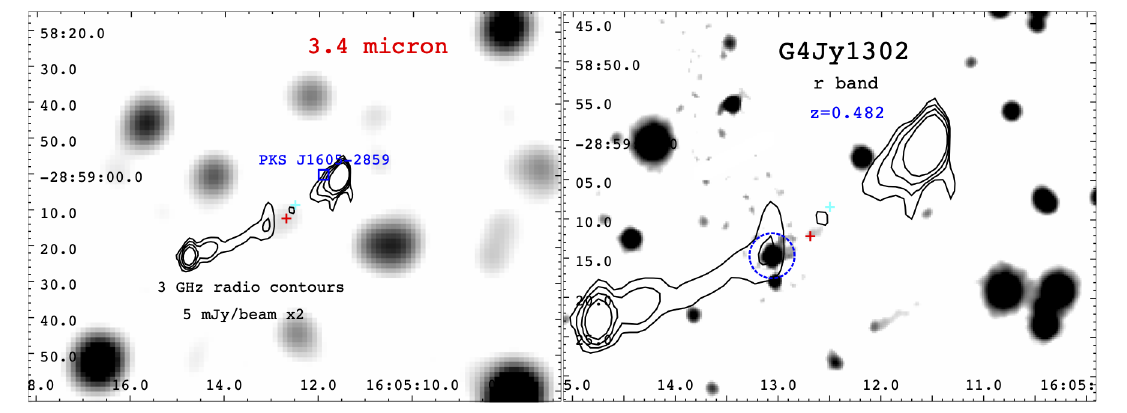}
\includegraphics[height=3.8cm,width=8.8cm,angle=0]{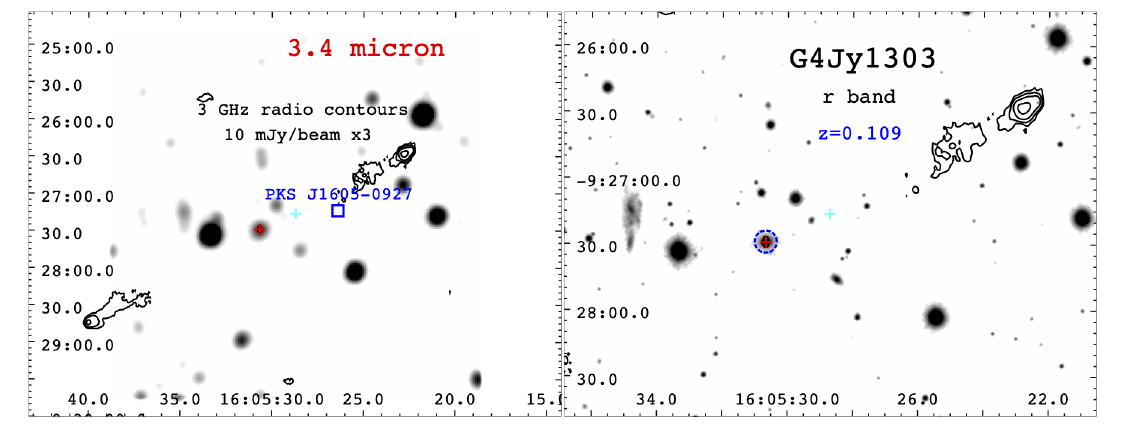}
\includegraphics[height=3.8cm,width=8.8cm,angle=0]{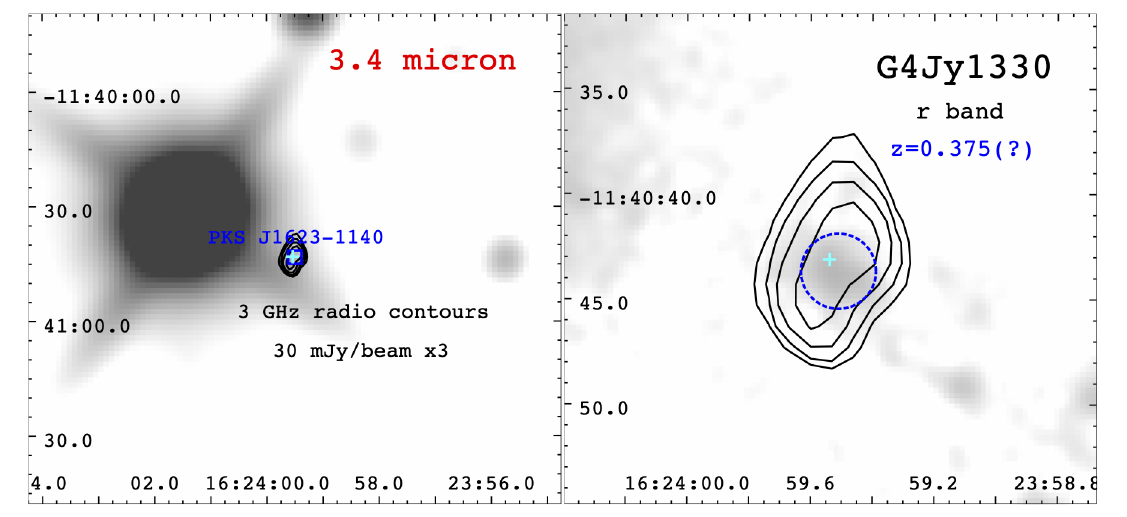}
\includegraphics[height=3.8cm,width=8.8cm,angle=0]{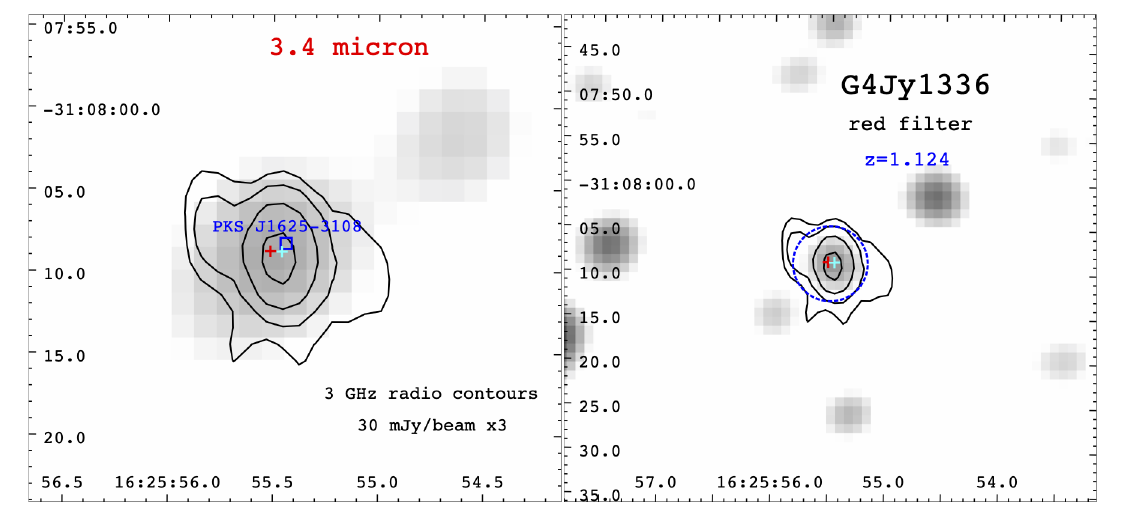}
\includegraphics[height=3.8cm,width=8.8cm,angle=0]{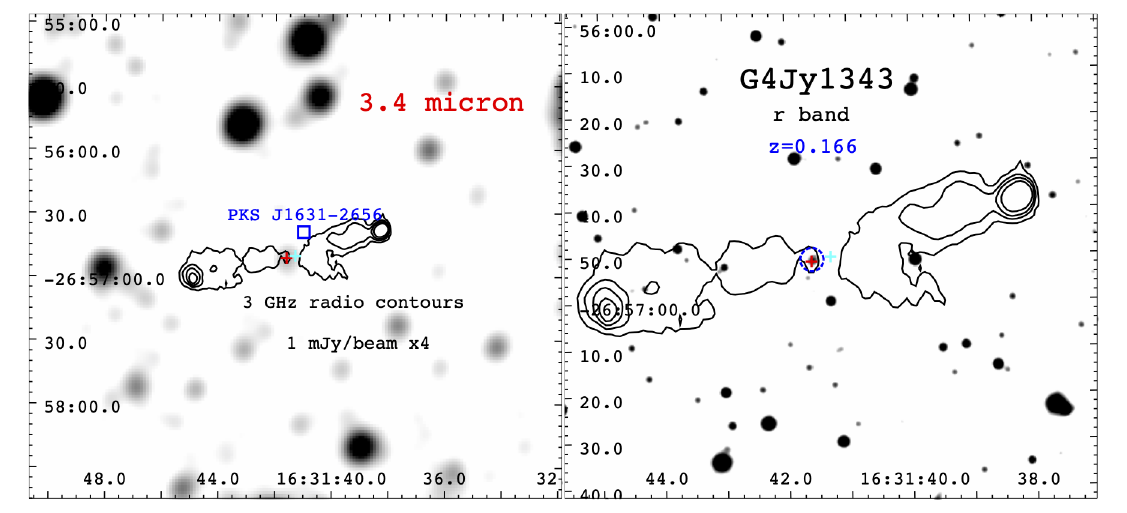}
\includegraphics[height=3.8cm,width=8.8cm,angle=0]{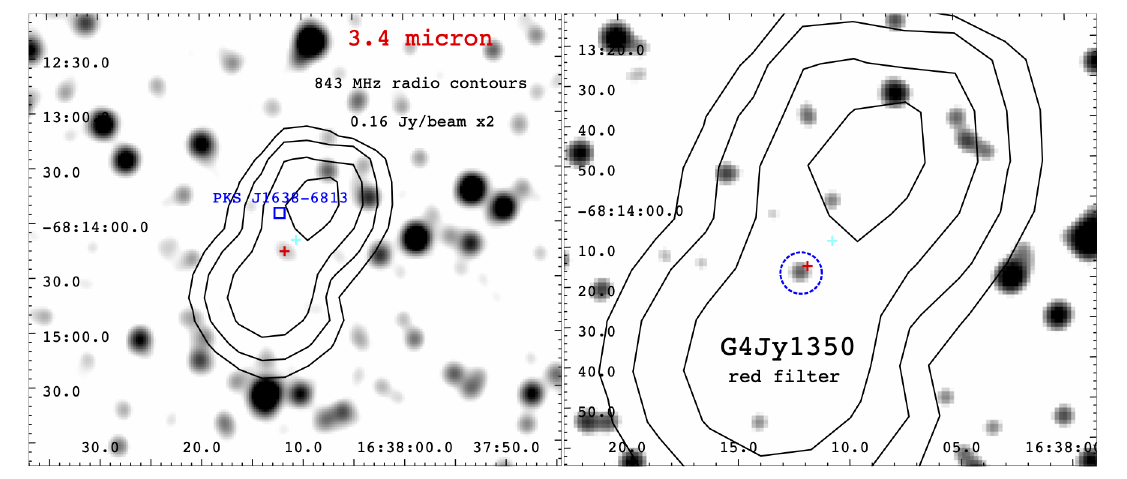}
\includegraphics[height=3.8cm,width=8.8cm,angle=0]{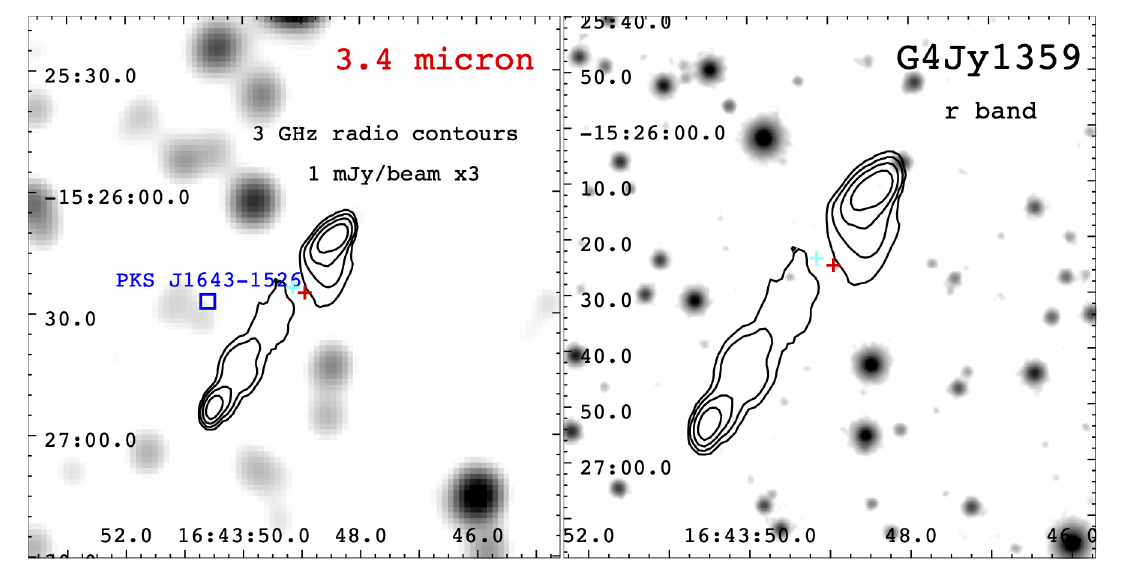}
\includegraphics[height=3.8cm,width=8.8cm,angle=0]{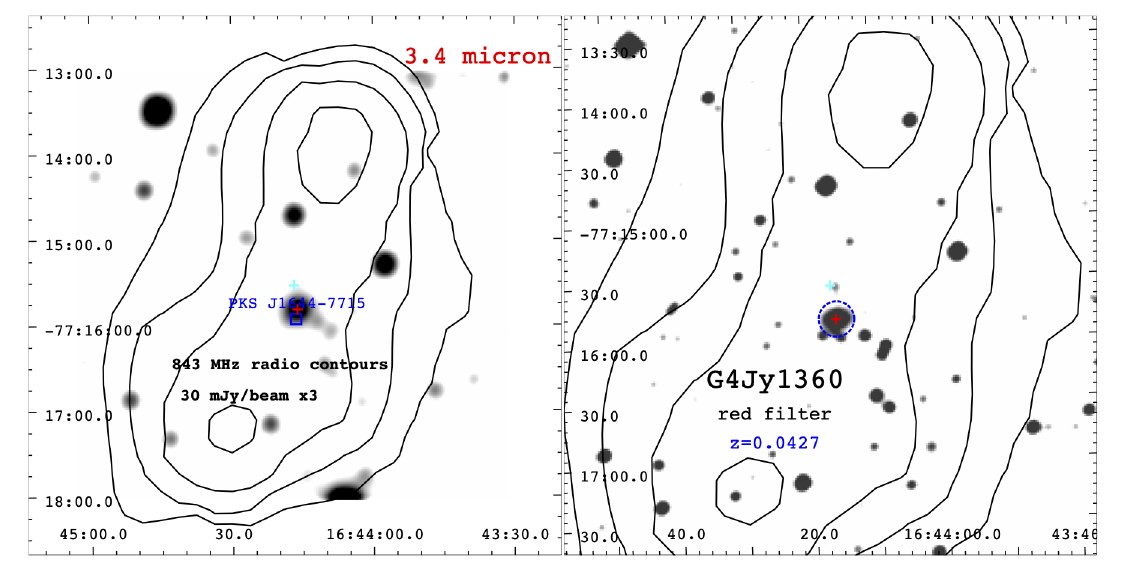}
\includegraphics[height=3.8cm,width=8.8cm,angle=0]{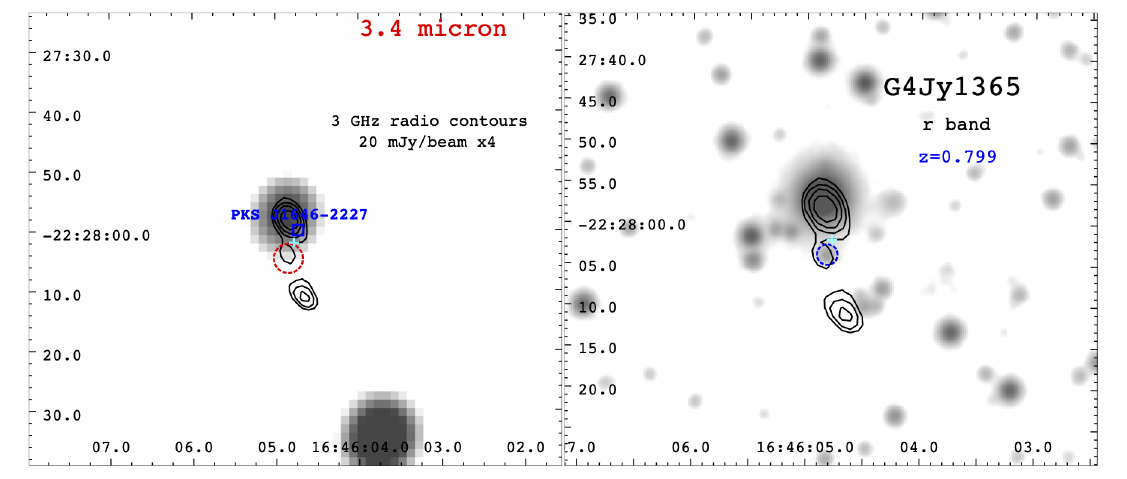}
\caption{Same as Figure~\ref{fig:example1} for the following \cs\ radio sources: \\ 
G4Jy\,1284, G4Jy\,1289, G4Jy\,1301, G4Jy\,1302, G4Jy\,1303, G4Jy\,1330, G4Jy\,1336, G4Jy\,1343, G4Jy\,1350, G4Jy\,1359, G4Jy\,1360, G4Jy\,1365.}
\end{center}
\end{figure*}

\begin{figure*}[!th]
\begin{center}
\includegraphics[height=3.8cm,width=8.8cm,angle=0]{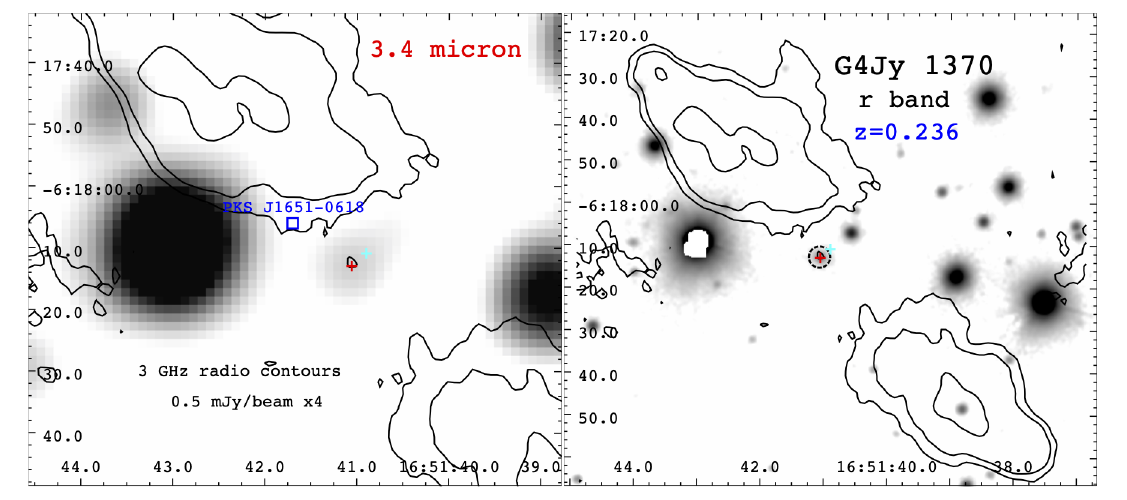}
\includegraphics[height=3.8cm,width=8.8cm,angle=0]{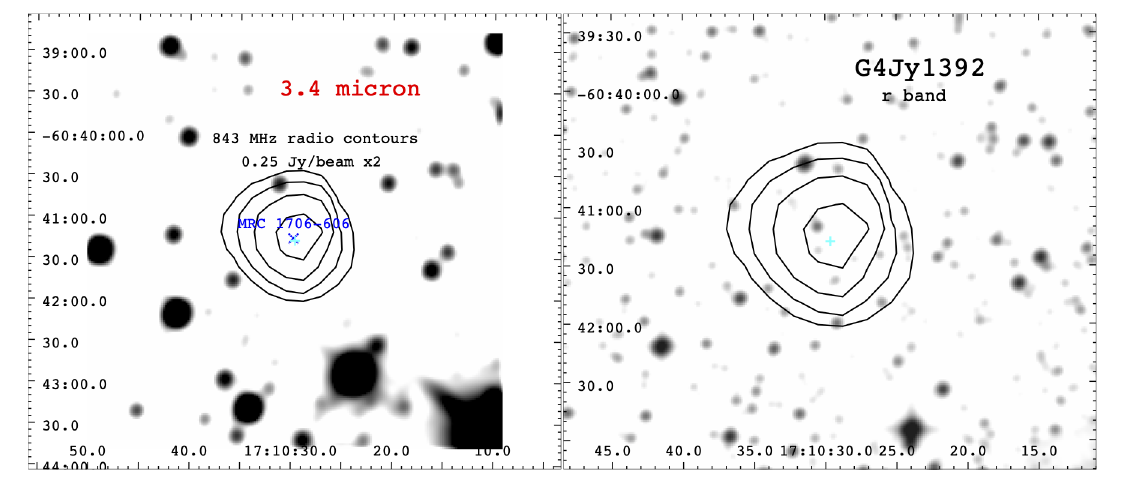}
\includegraphics[height=3.8cm,width=8.8cm,angle=0]{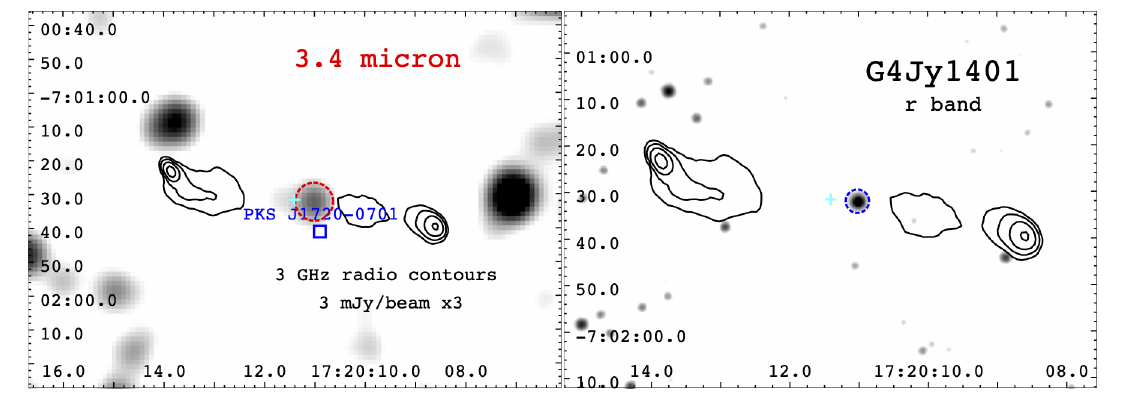}
\includegraphics[height=3.8cm,width=8.8cm,angle=0]{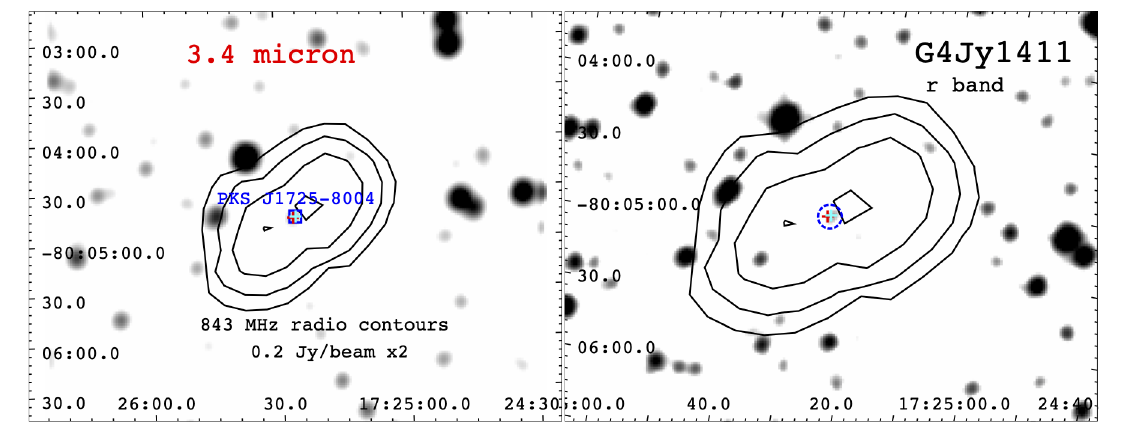}
\includegraphics[height=3.8cm,width=8.8cm,angle=0]{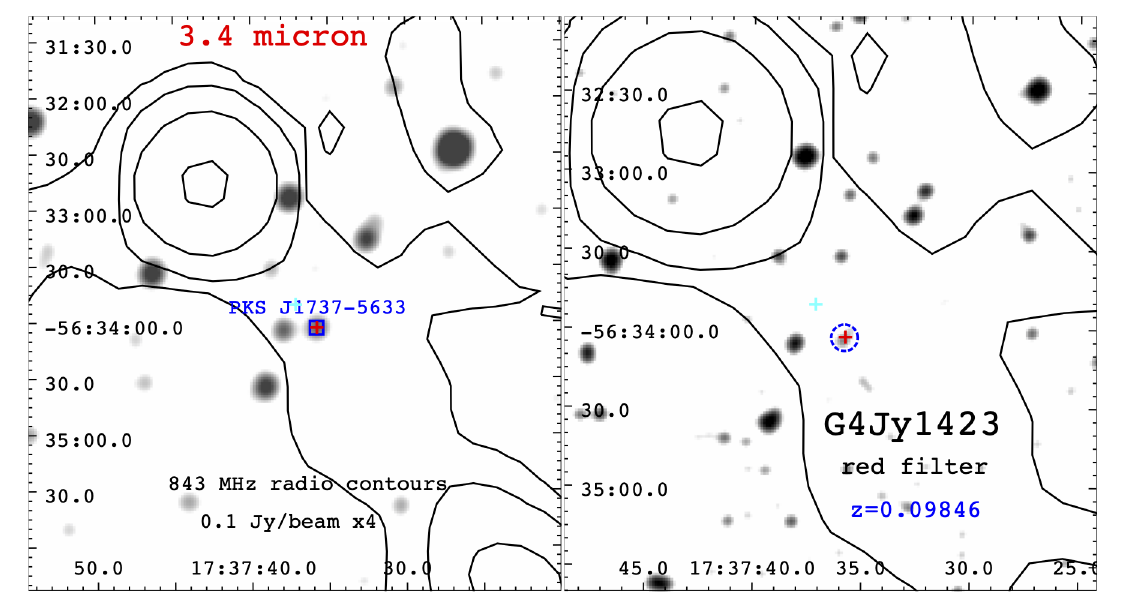}
\includegraphics[height=3.8cm,width=8.8cm,angle=0]{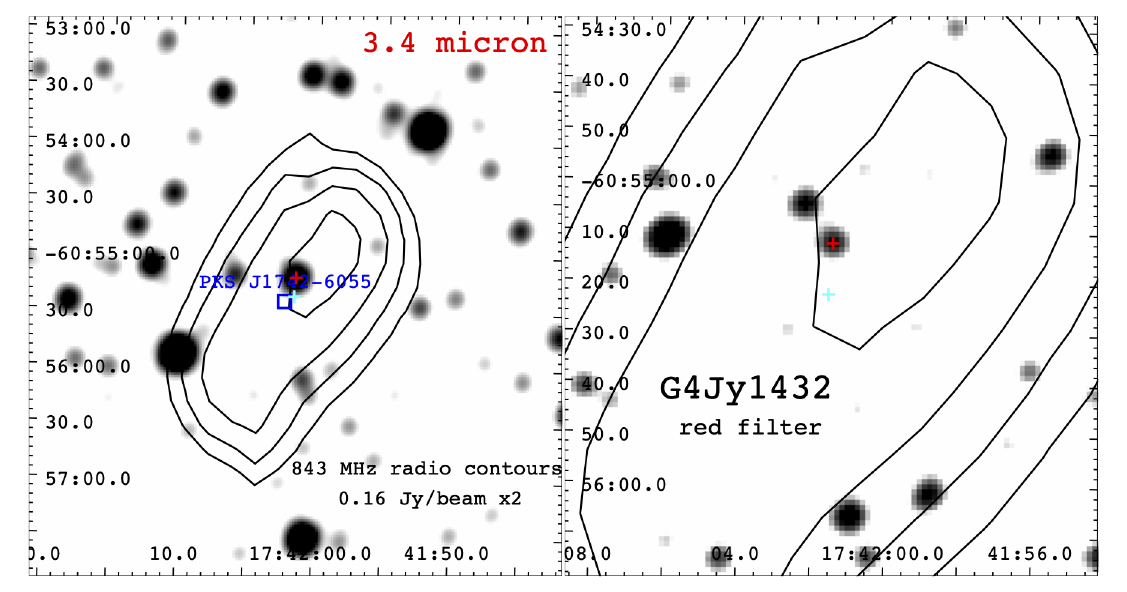}
\includegraphics[height=3.8cm,width=8.8cm,angle=0]{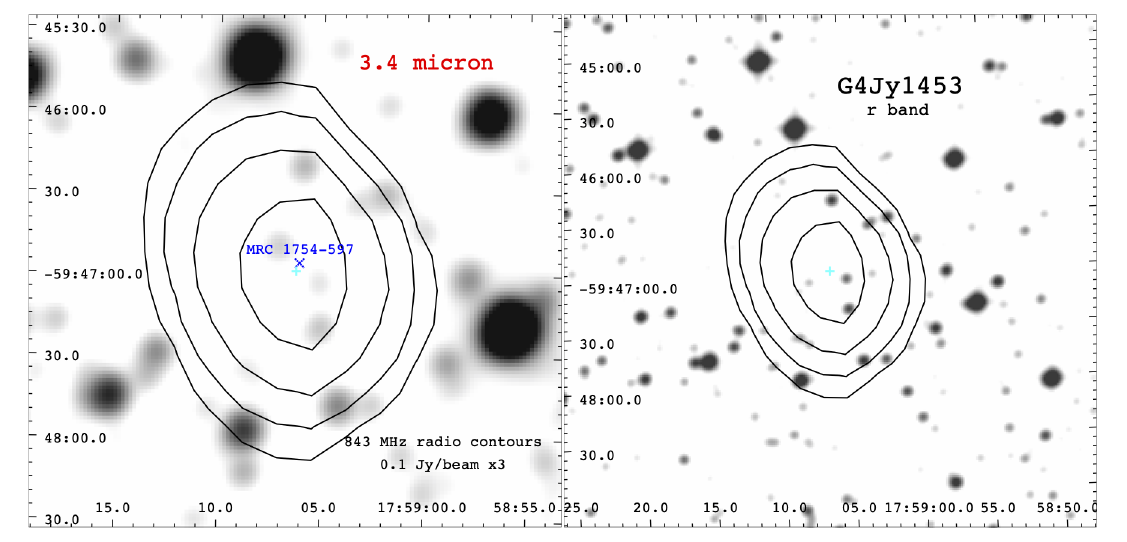}
\includegraphics[height=3.8cm,width=8.8cm,angle=0]{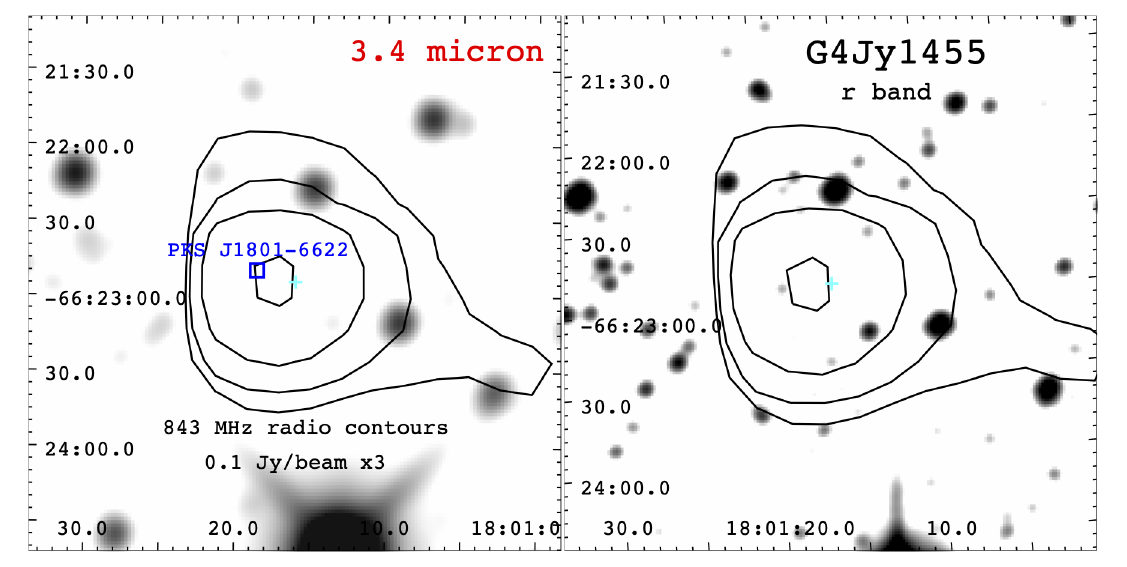}
\includegraphics[height=3.8cm,width=8.8cm,angle=0]{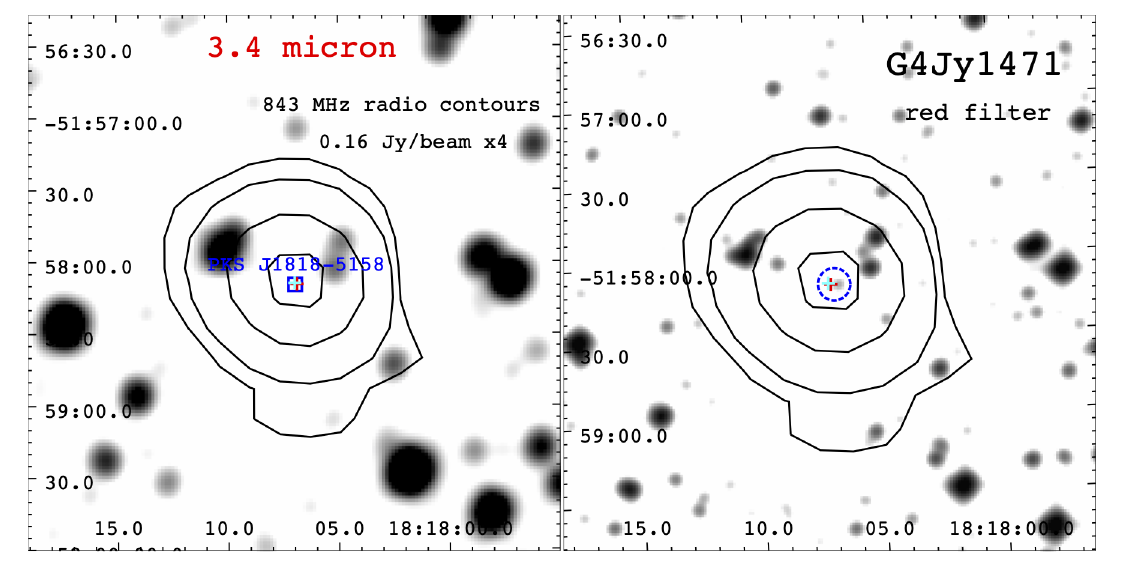}
\includegraphics[height=3.8cm,width=8.8cm,angle=0]{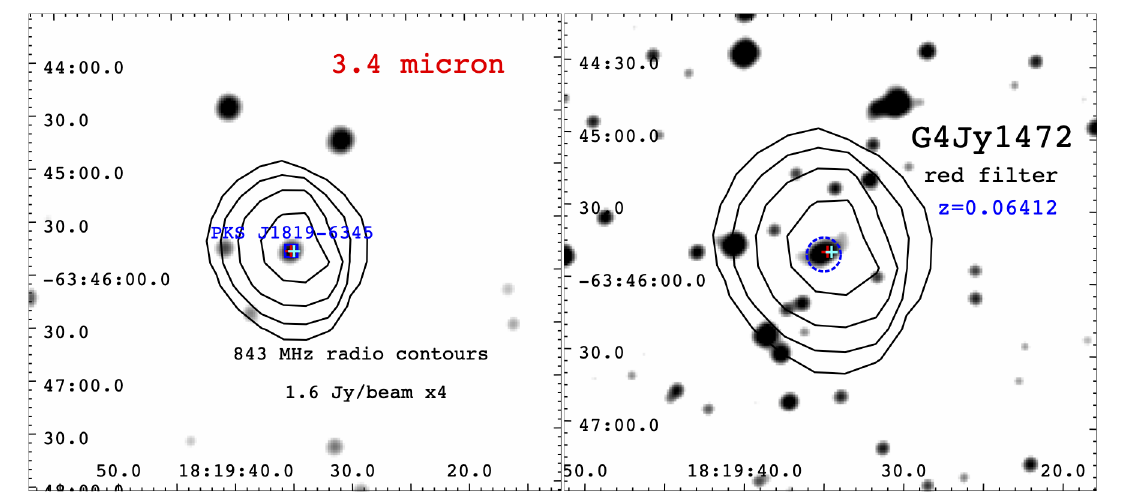}
\includegraphics[height=3.8cm,width=8.8cm,angle=0]{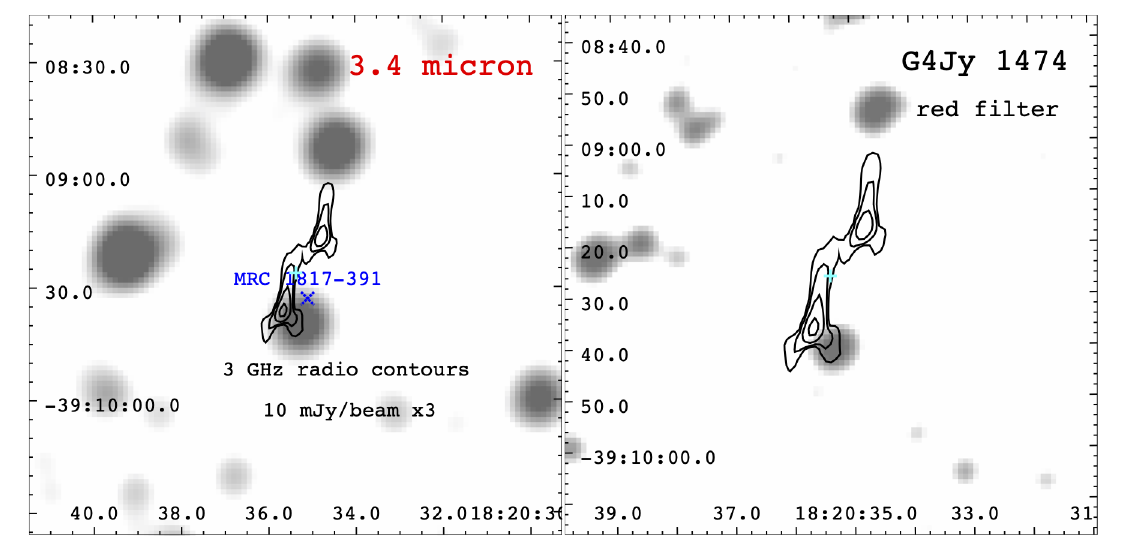}
\includegraphics[height=3.8cm,width=8.8cm,angle=0]{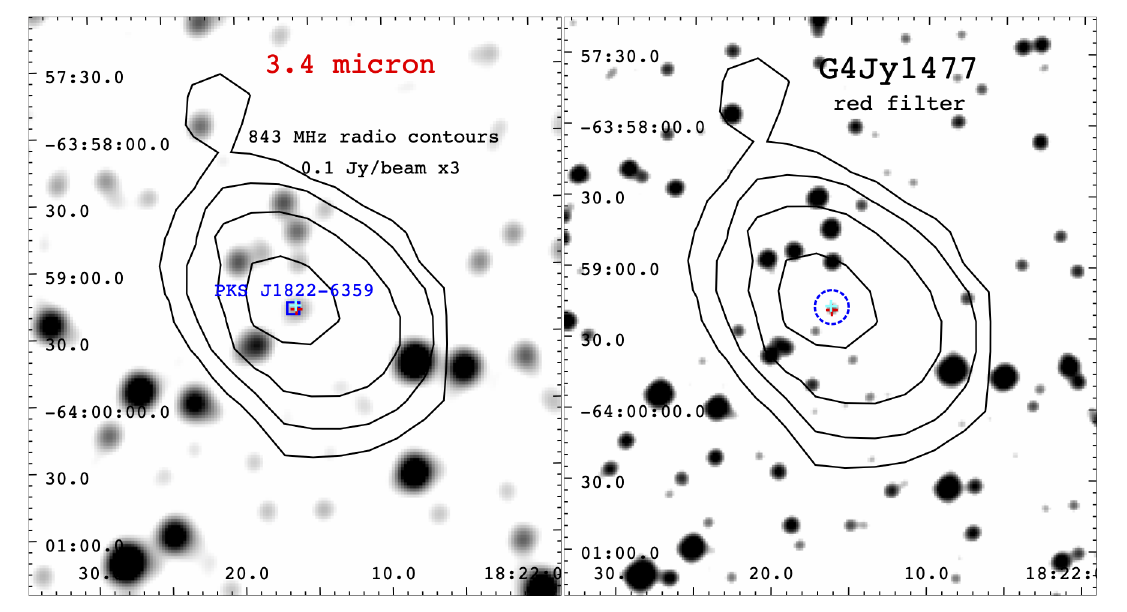}
\caption{Same as Figure~\ref{fig:example1} for the following \cs\ radio sources: \\ 
G4Jy\,1370, G4Jy\,1392, G4Jy\,1401, G4Jy\,1411, G4Jy\,1423, G4Jy\,1432, G4Jy\,1453, G4Jy\,1455, G4Jy\,1471, G4Jy\,1472, G4Jy\,1474, G4Jy\,1477.}
\end{center}
\end{figure*}

\begin{figure*}[!th]
\begin{center}
\includegraphics[height=3.8cm,width=8.8cm,angle=0]{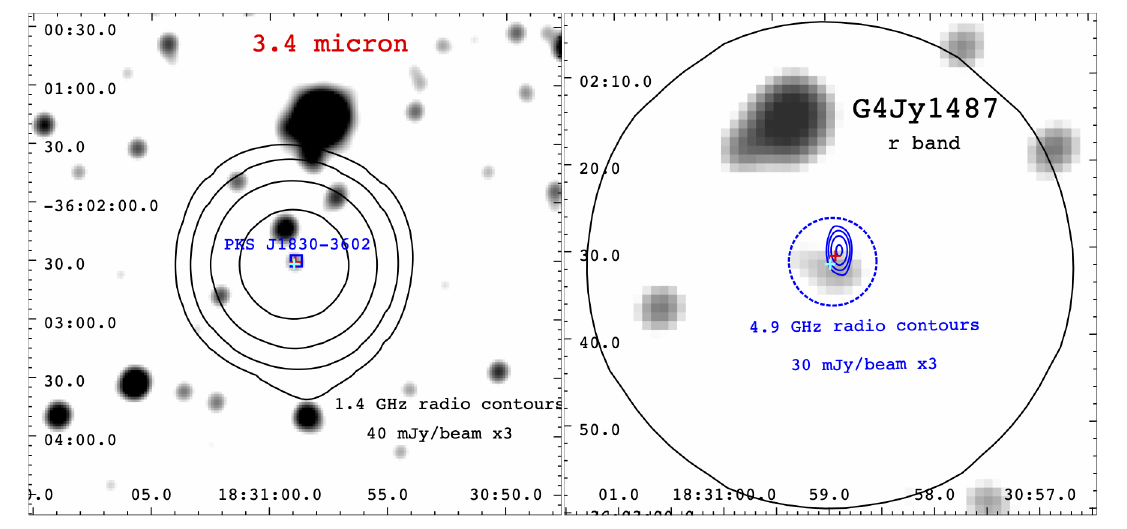}
\includegraphics[height=3.8cm,width=8.8cm,angle=0]{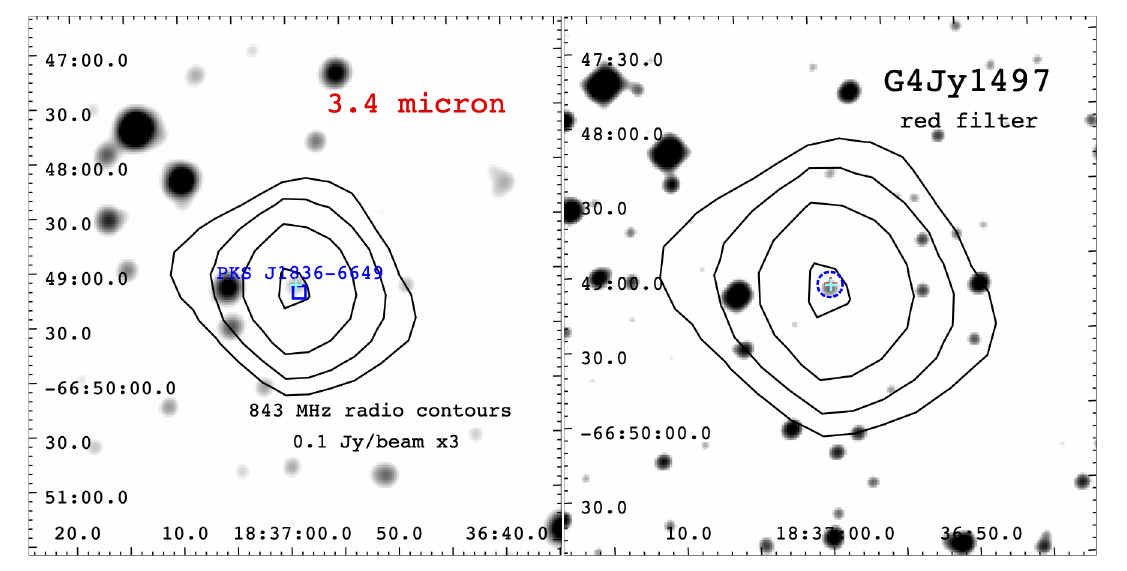}
\includegraphics[height=3.8cm,width=8.8cm,angle=0]{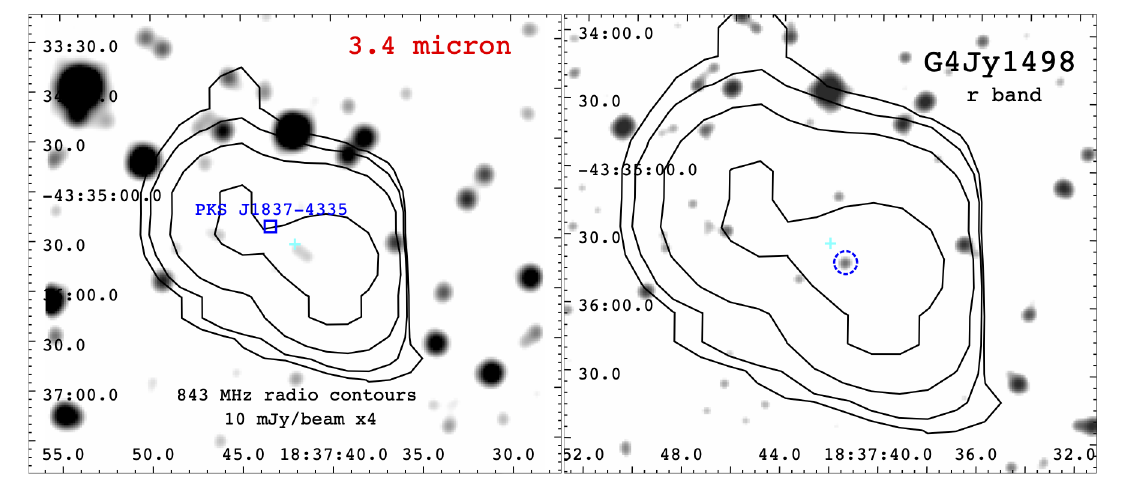}
\includegraphics[height=3.8cm,width=8.8cm,angle=0]{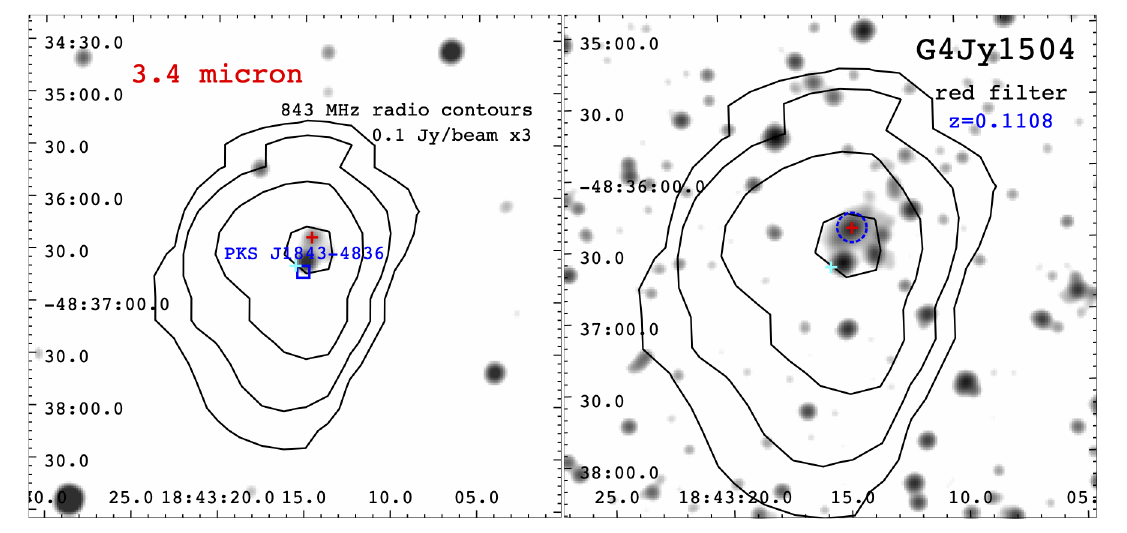}
\includegraphics[height=3.8cm,width=8.8cm,angle=0]{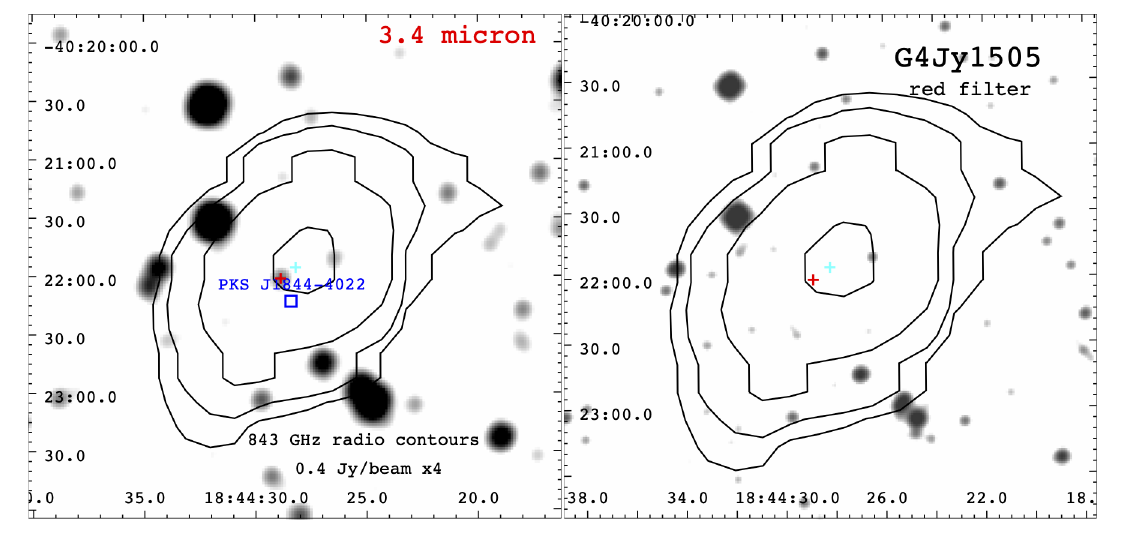}
\includegraphics[height=3.8cm,width=8.8cm,angle=0]{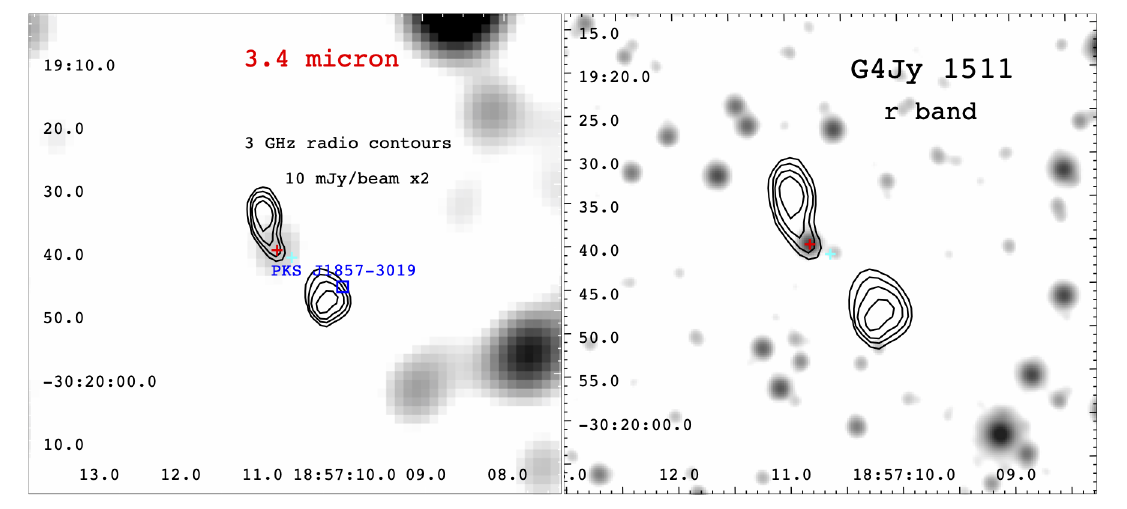}
\includegraphics[height=3.8cm,width=8.8cm,angle=0]{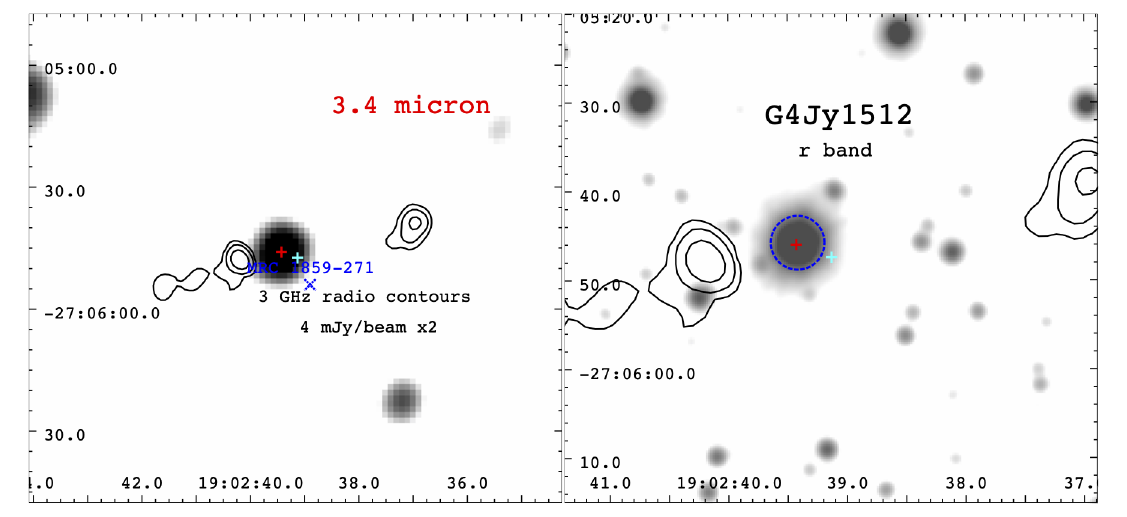}
\includegraphics[height=3.8cm,width=8.8cm,angle=0]{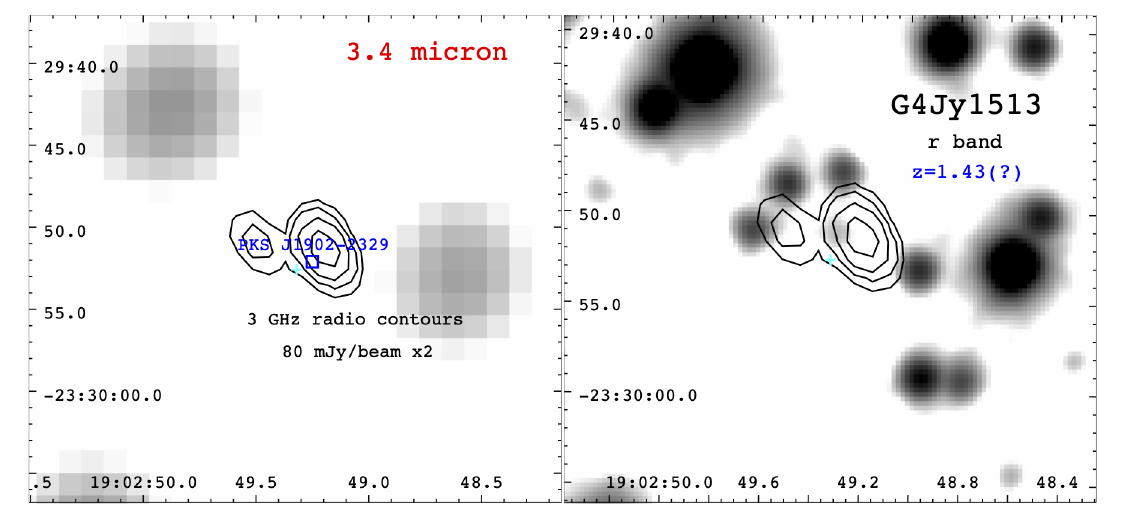}
\includegraphics[height=3.8cm,width=8.8cm,angle=0]{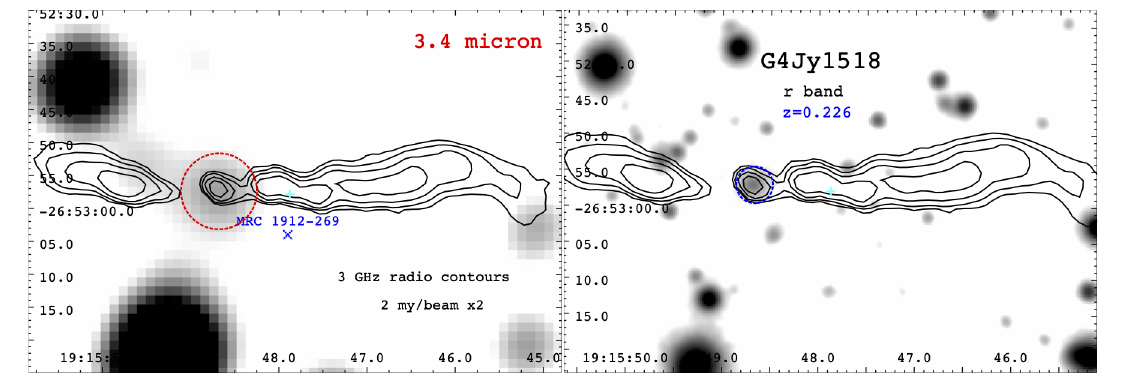}
\includegraphics[height=3.8cm,width=8.8cm,angle=0]{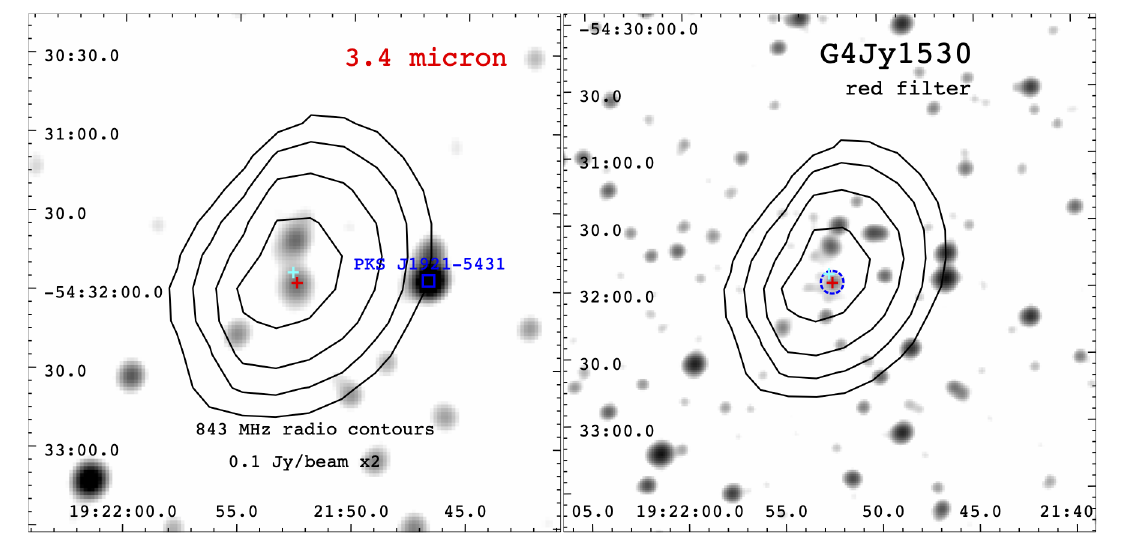}
\includegraphics[height=3.8cm,width=8.8cm,angle=0]{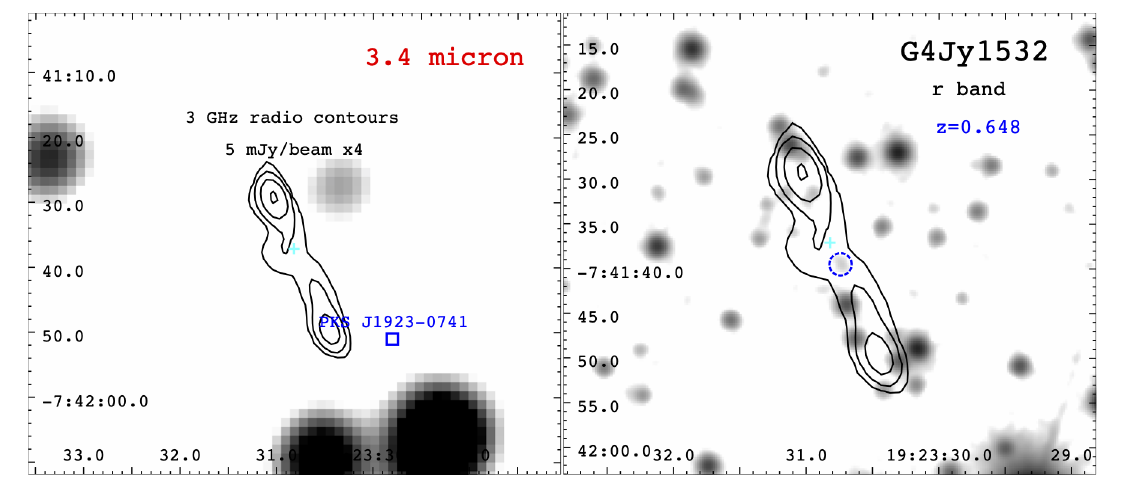}
\includegraphics[height=3.8cm,width=8.8cm,angle=0]{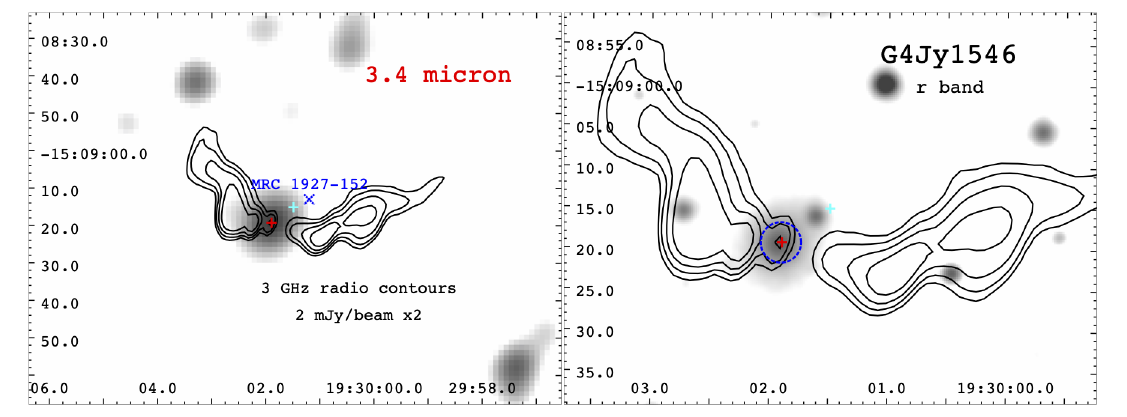}
\caption{Same as Figure~\ref{fig:example1} for the following \cs\ radio sources: \\ 
G4Jy\,1487, G4Jy\,1497, G4Jy\,1498, G4Jy\,1504, G4Jy\,1505, G4Jy\,1511, G4Jy\,1512, G4Jy\,1513, G4Jy\,1518, G4Jy\,1530, G4Jy\,1532, G4Jy\,1546.}
\end{center}
\end{figure*}

\begin{figure*}[!th]
\begin{center}
\includegraphics[height=3.8cm,width=8.8cm,angle=0]{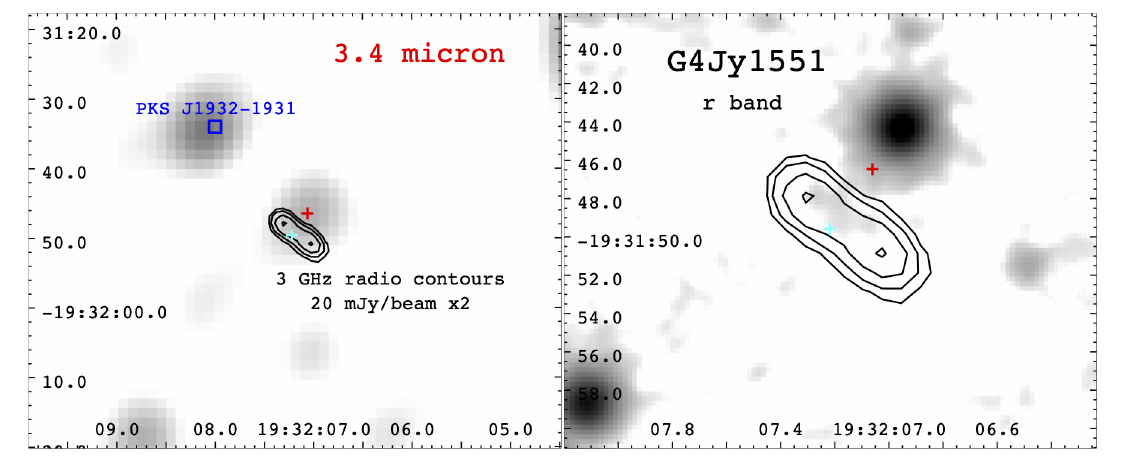}
\includegraphics[height=3.8cm,width=8.8cm,angle=0]{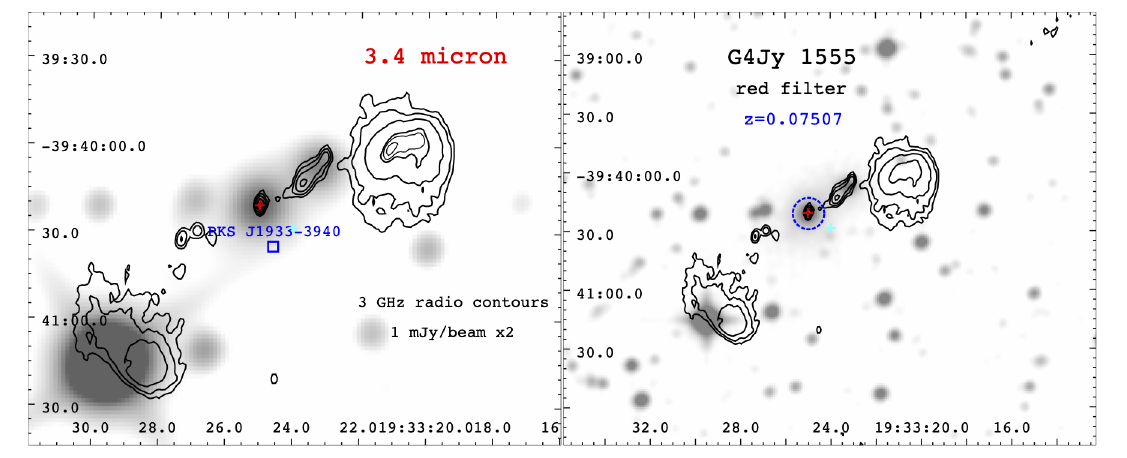}
\includegraphics[height=3.8cm,width=8.8cm,angle=0]{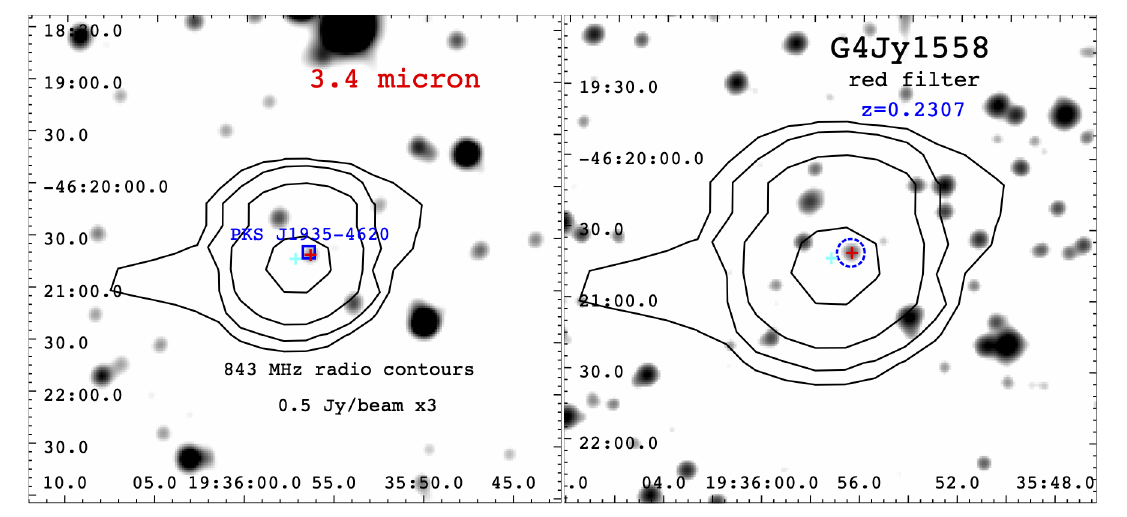}
\includegraphics[height=3.8cm,width=8.8cm,angle=0]{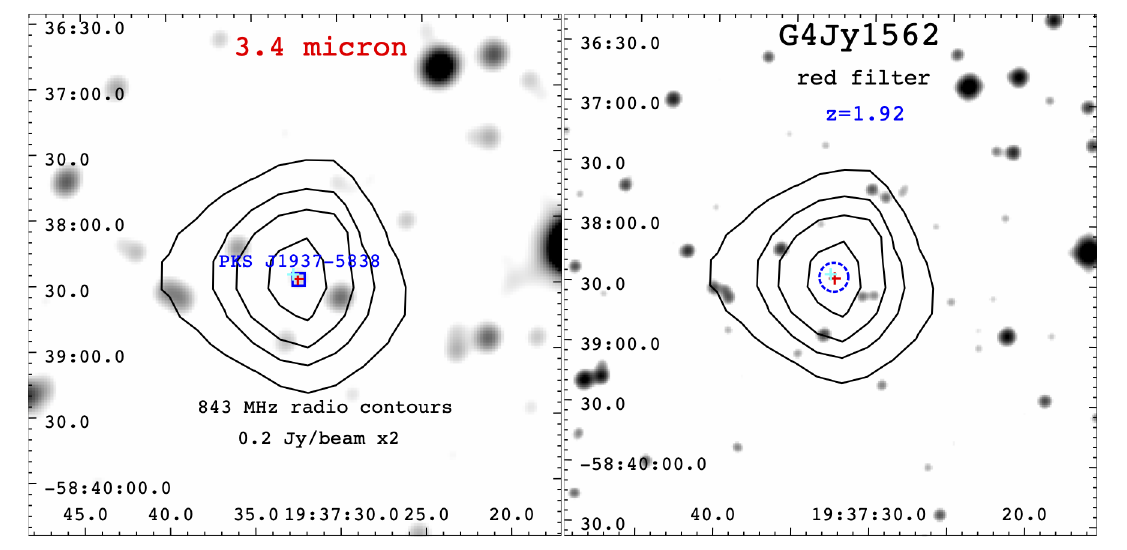}
\includegraphics[height=3.8cm,width=8.8cm,angle=0]{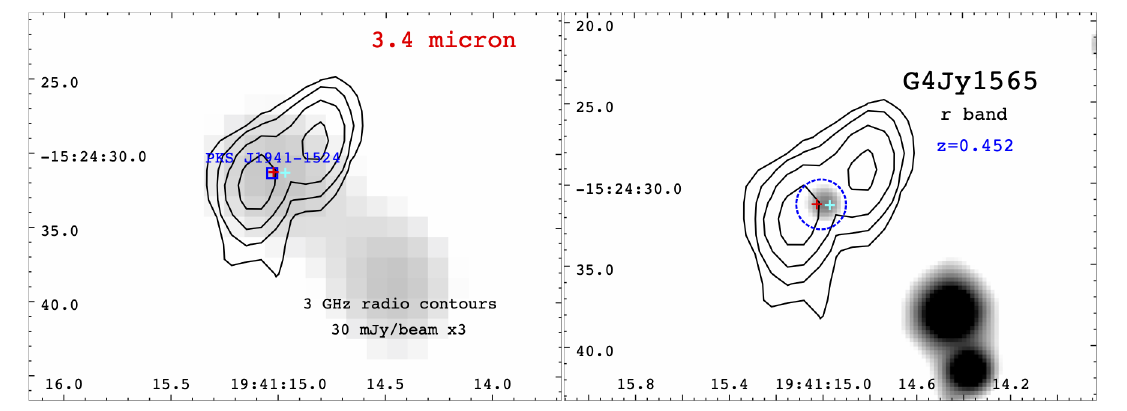}
\includegraphics[height=3.8cm,width=8.8cm,angle=0]{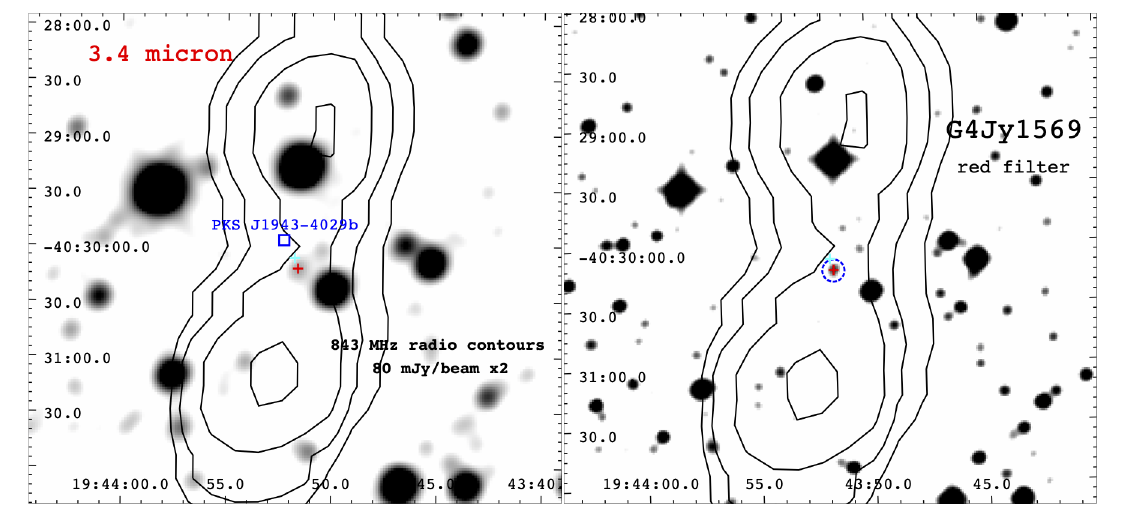}
\includegraphics[height=3.8cm,width=8.8cm,angle=0]{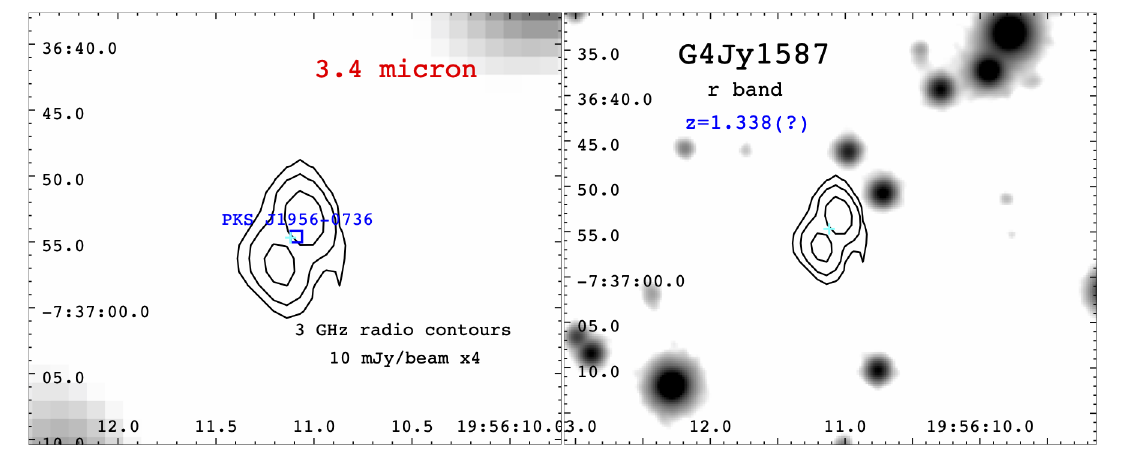}
\includegraphics[height=3.8cm,width=8.8cm,angle=0]{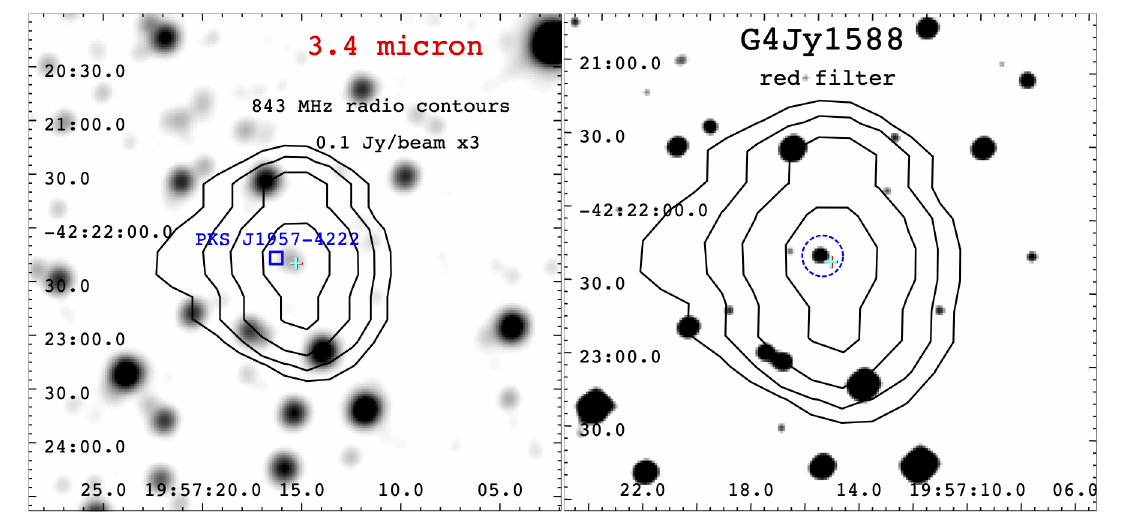}
\includegraphics[height=3.8cm,width=8.8cm,angle=0]{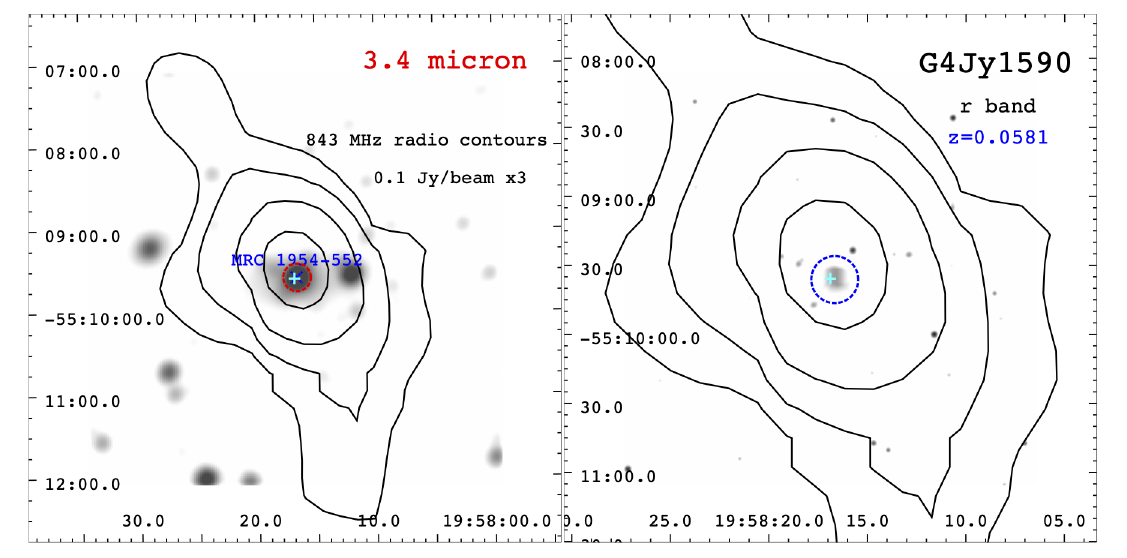}
\includegraphics[height=3.8cm,width=8.8cm,angle=0]{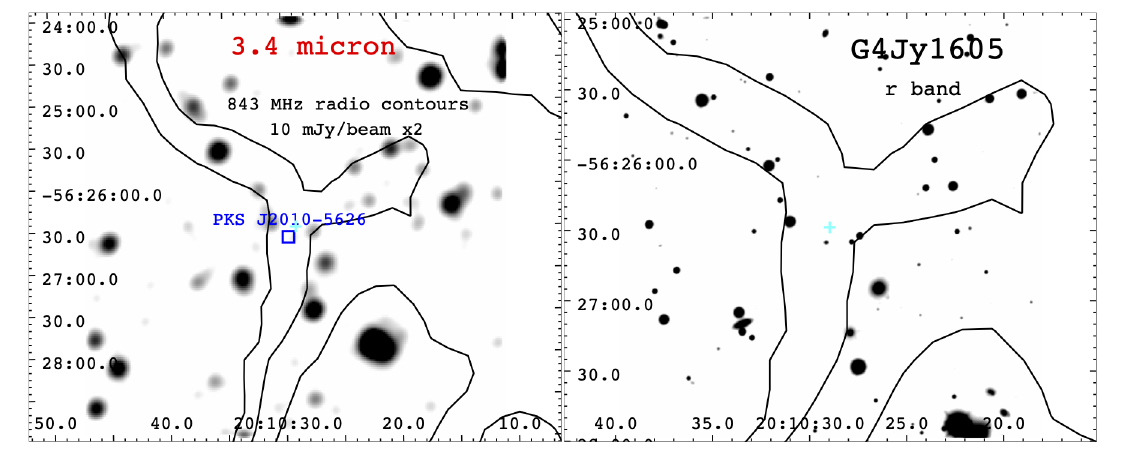}
\includegraphics[height=3.8cm,width=8.8cm,angle=0]{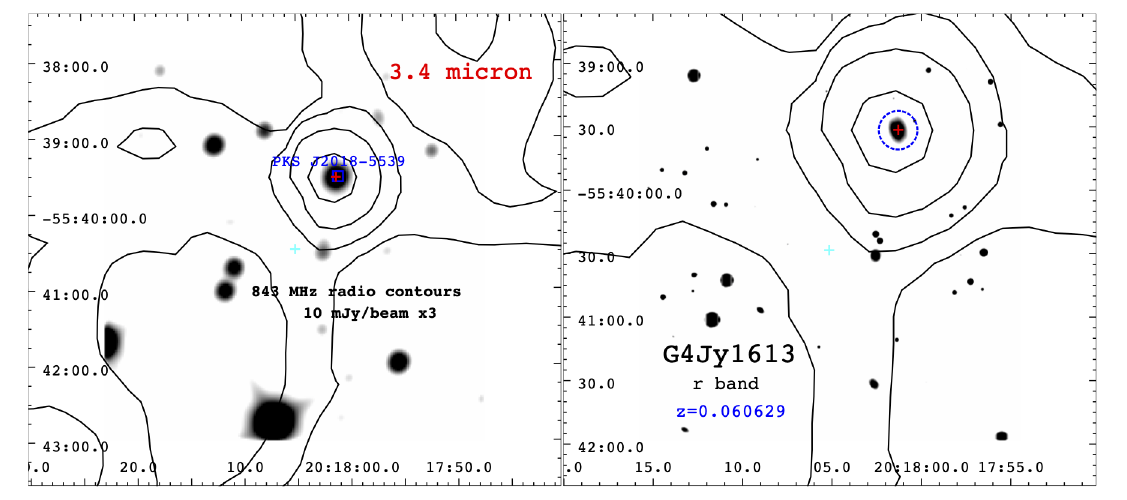}
\includegraphics[height=3.8cm,width=8.8cm,angle=0]{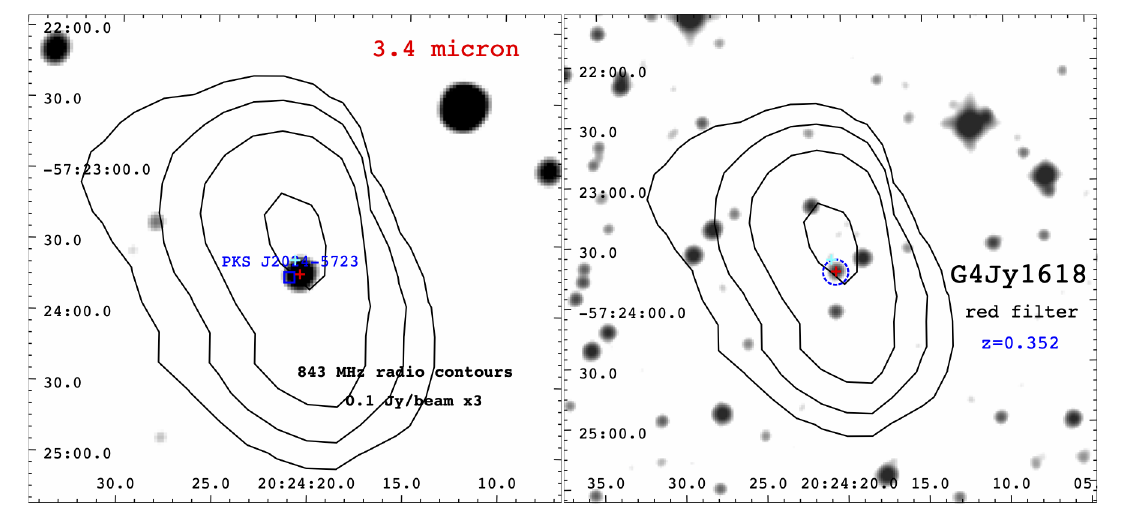}
\caption{Same as Figure~\ref{fig:example1} for the following \cs\ radio sources: \\ 
G4Jy\,1551, G4Jy\,1555, G4Jy\,1558, G4Jy\,1562, G4Jy\,1565, G4Jy\,1569, G4Jy\,1587, G4Jy\,1588, G4Jy\,1590, G4Jy\,1605, G4Jy\,1613, G4Jy\,1618.}
\end{center}
\end{figure*}

\begin{figure*}[!th]
\begin{center}
\includegraphics[height=3.8cm,width=8.8cm,angle=0]{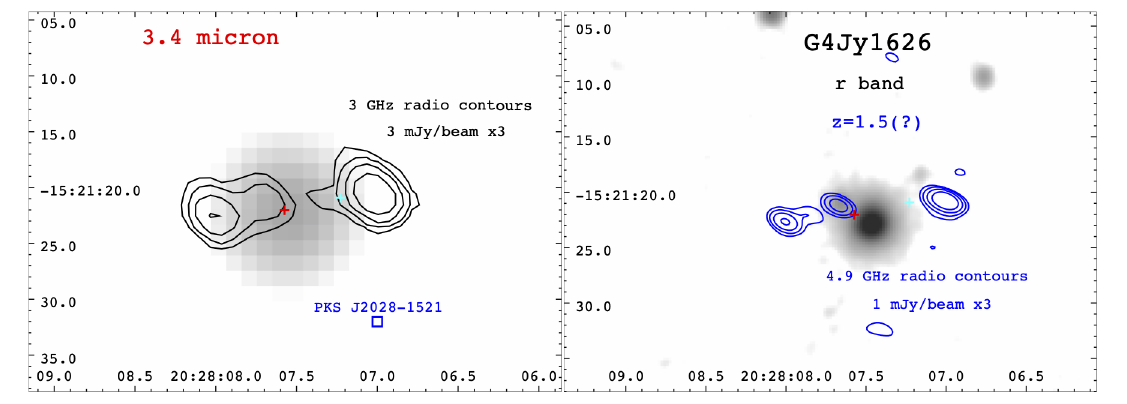}
\includegraphics[height=3.8cm,width=8.8cm,angle=0]{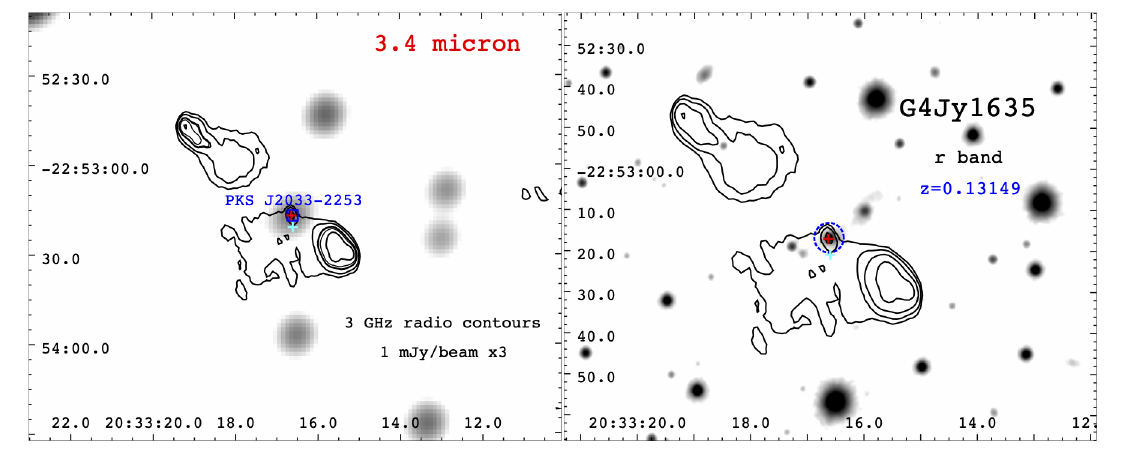}
\includegraphics[height=3.8cm,width=8.8cm,angle=0]{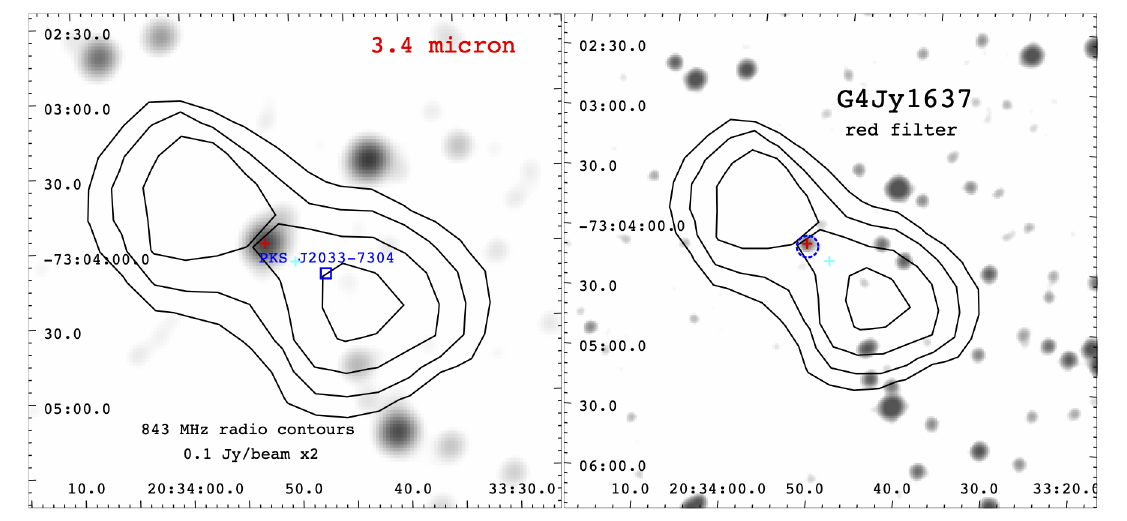}
\includegraphics[height=3.8cm,width=8.8cm,angle=0]{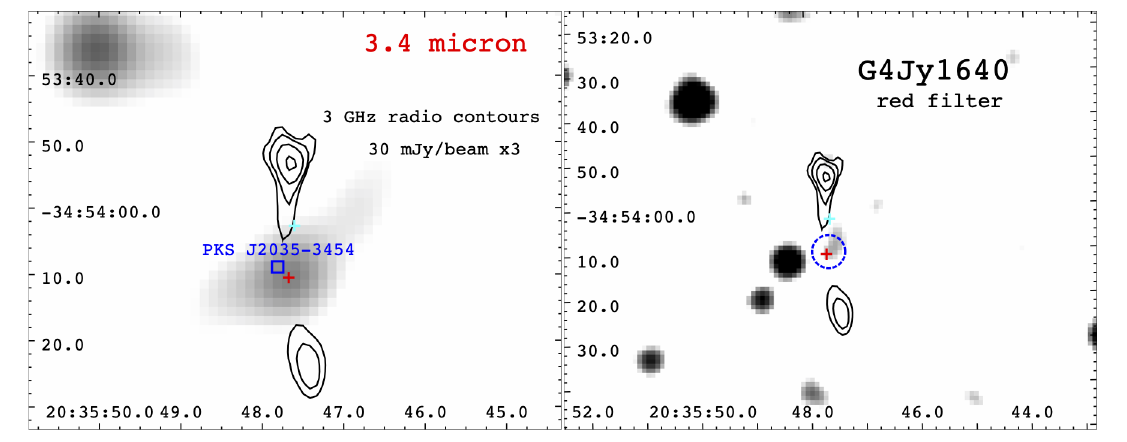}
\includegraphics[height=3.8cm,width=8.8cm,angle=0]{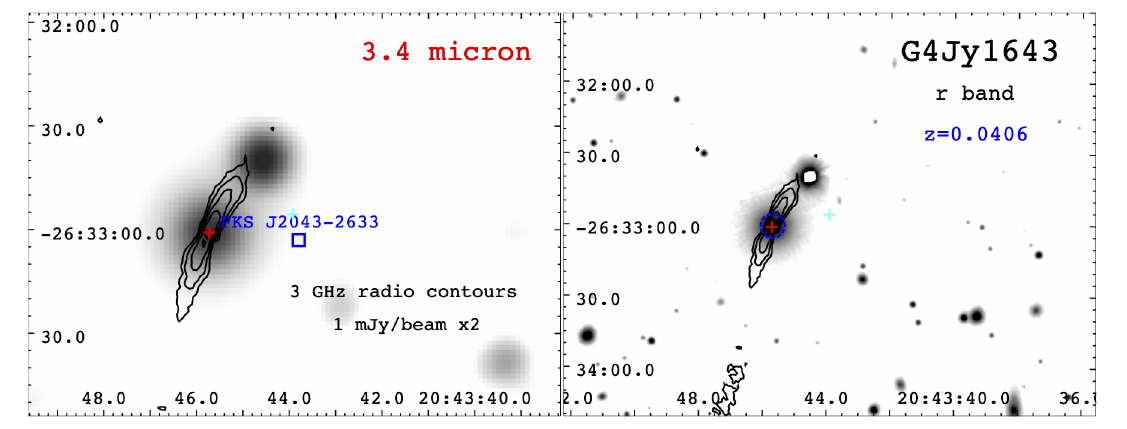}
\includegraphics[height=3.8cm,width=8.8cm,angle=0]{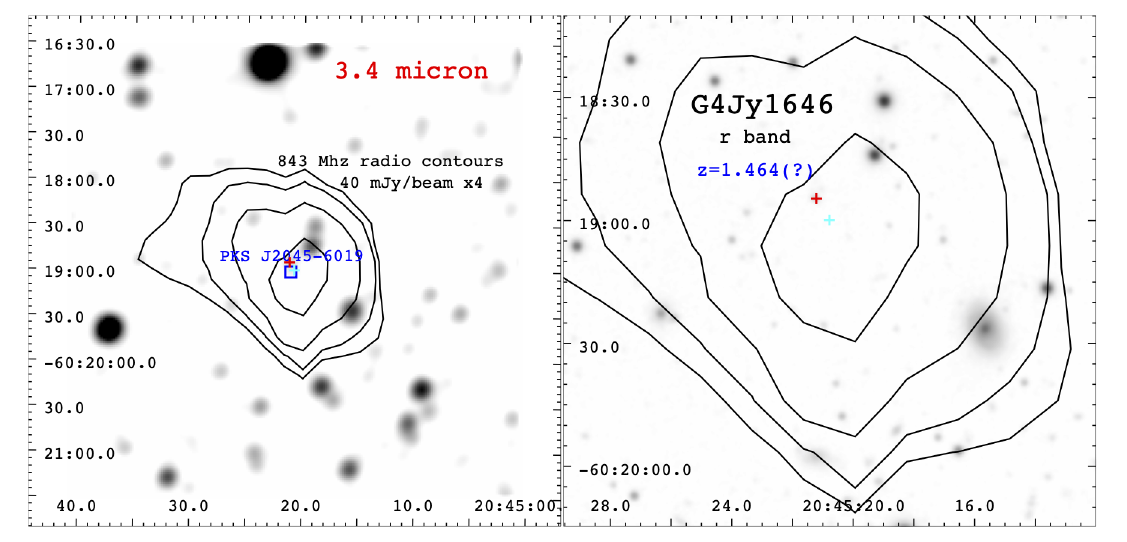}
\includegraphics[height=3.8cm,width=8.8cm,angle=0]{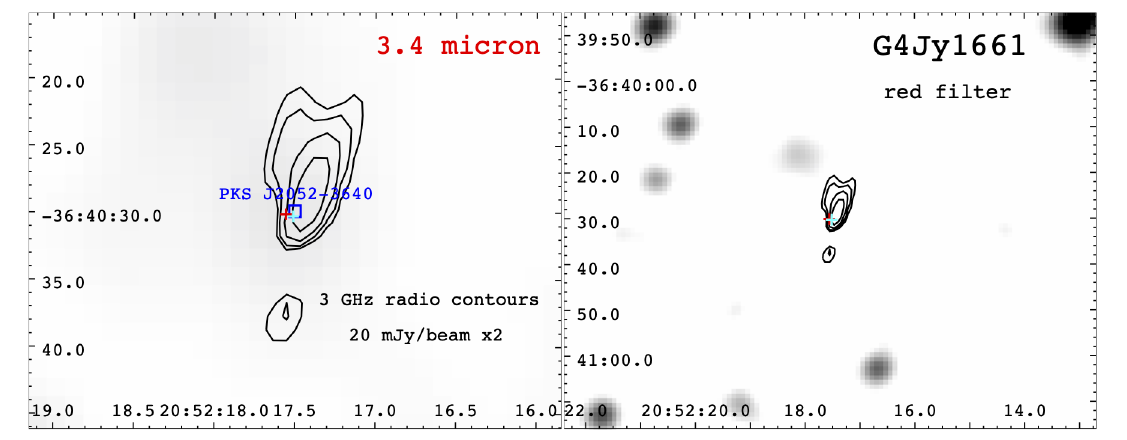}
\includegraphics[height=3.8cm,width=8.8cm,angle=0]{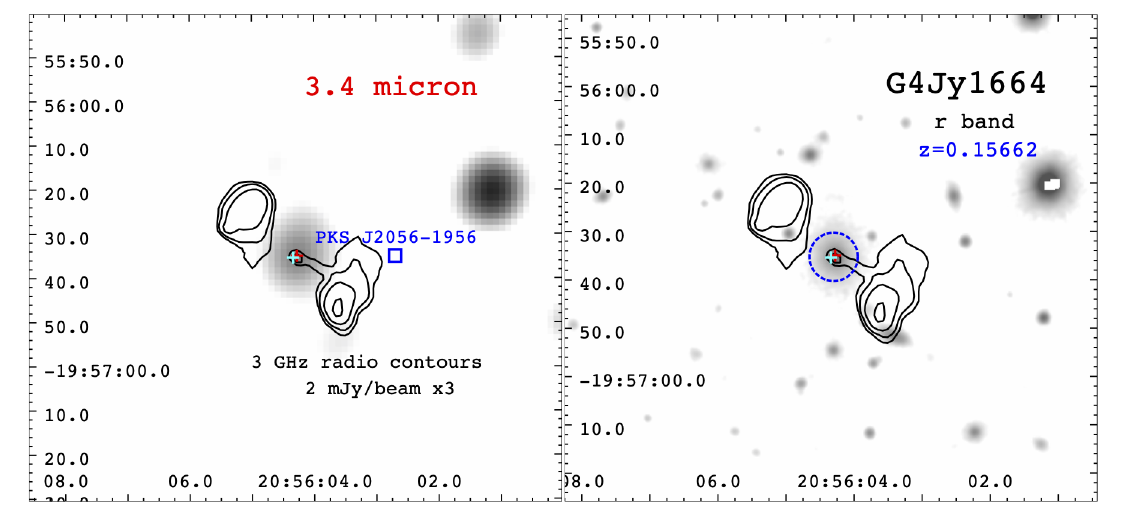}
\includegraphics[height=3.8cm,width=8.8cm,angle=0]{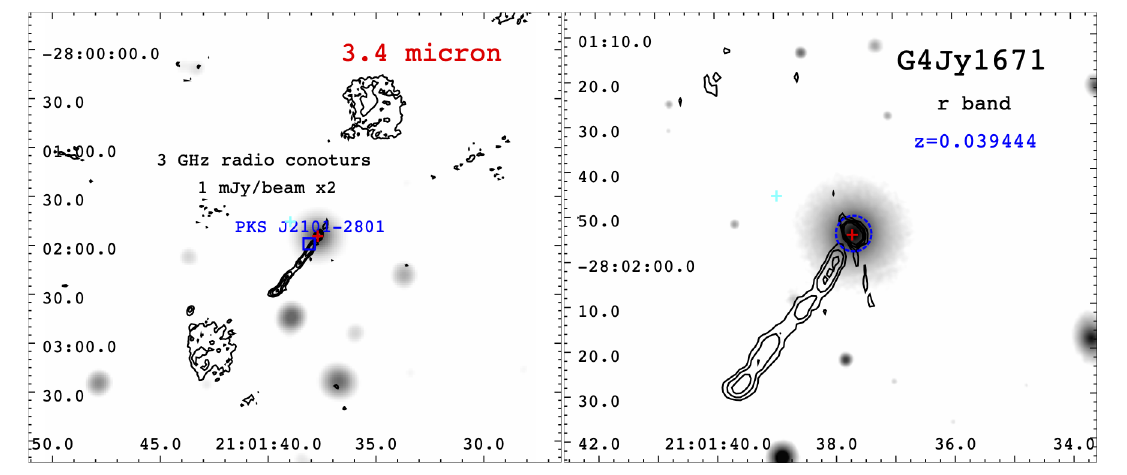}
\includegraphics[height=3.8cm,width=8.8cm,angle=0]{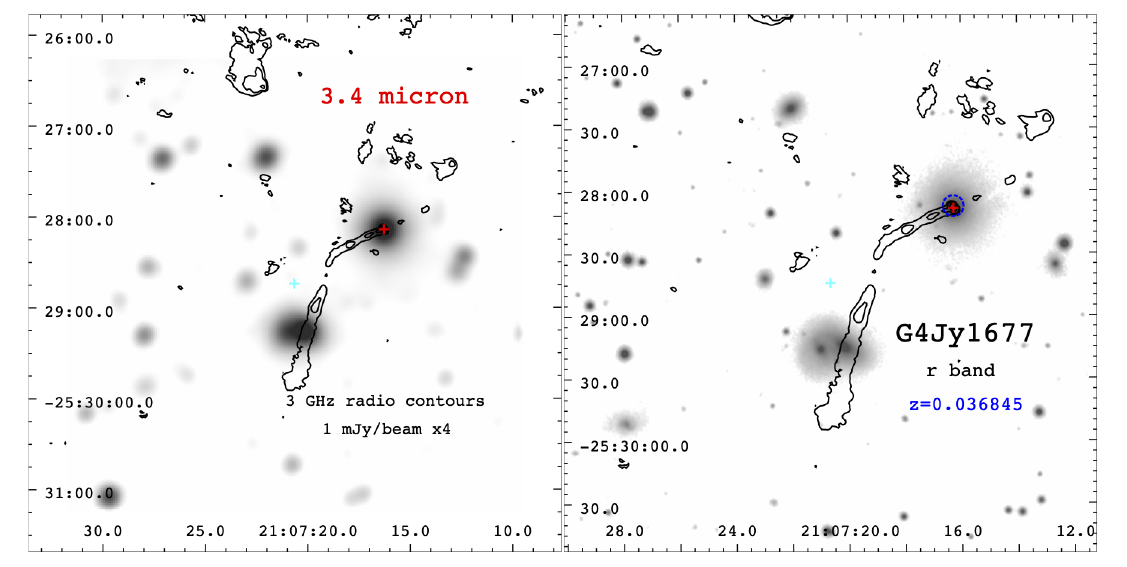}
\includegraphics[height=3.8cm,width=8.8cm,angle=0]{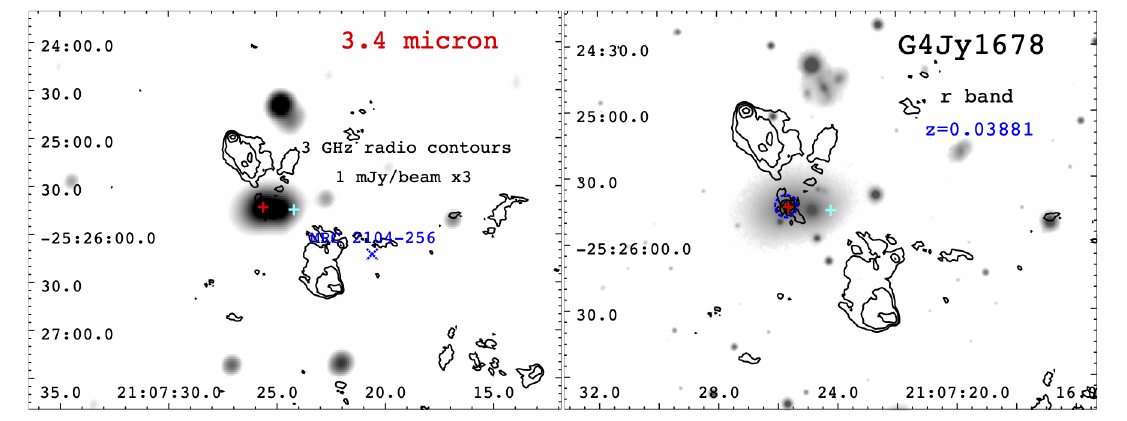}
\includegraphics[height=3.8cm,width=8.8cm,angle=0]{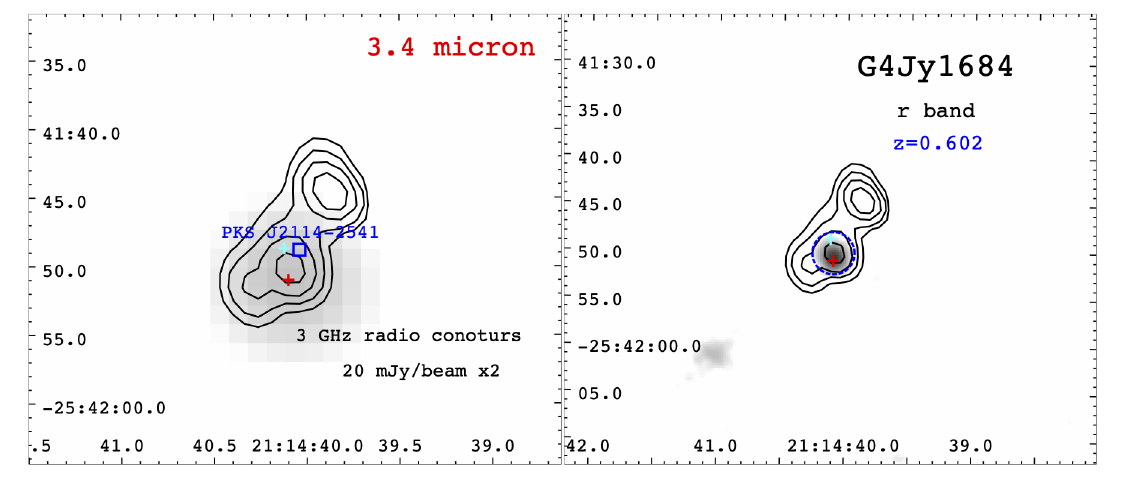}
\caption{Same as Figure~\ref{fig:example1} for the following \cs\ radio sources: \\ 
G4Jy\,1626, G4Jy\,1635, G4Jy\,1637, G4Jy\,1640, G4Jy\,1643, G4Jy\,1646, G4Jy\,1661, G4Jy\,1664, G4Jy\,1671, G4Jy\,1677, G4Jy\,1678, G4Jy\,1684.}
\end{center}
\end{figure*}

\begin{figure*}[!th]
\begin{center}
\includegraphics[height=3.8cm,width=8.8cm,angle=0]{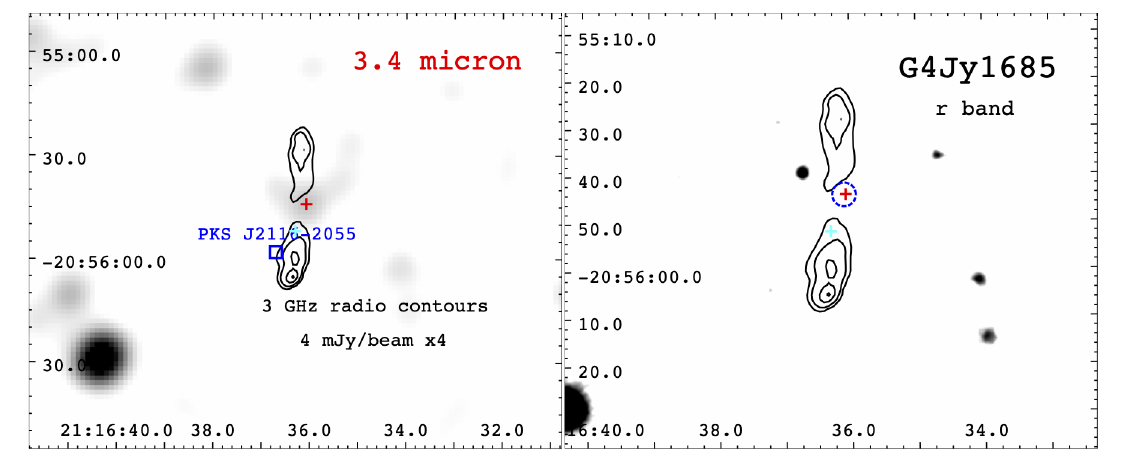}
\includegraphics[height=3.8cm,width=8.8cm,angle=0]{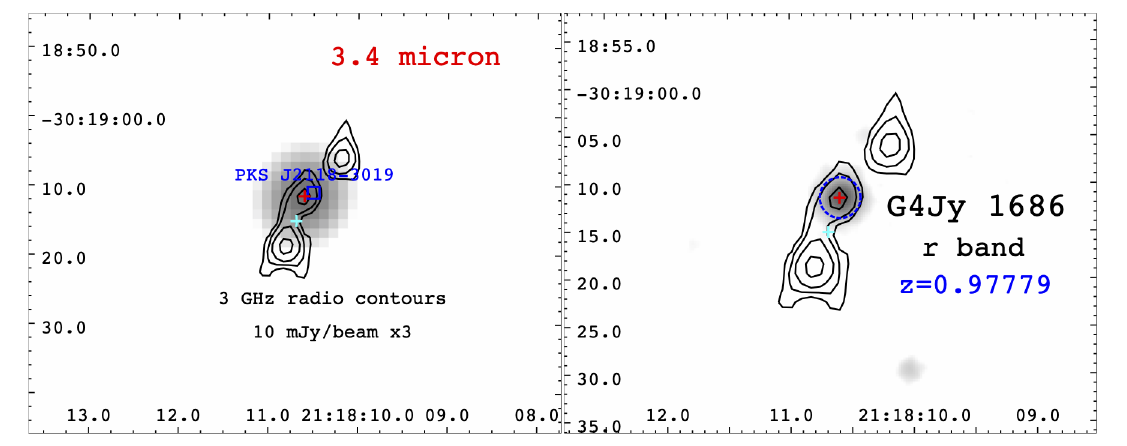}
\includegraphics[height=3.8cm,width=8.8cm,angle=0]{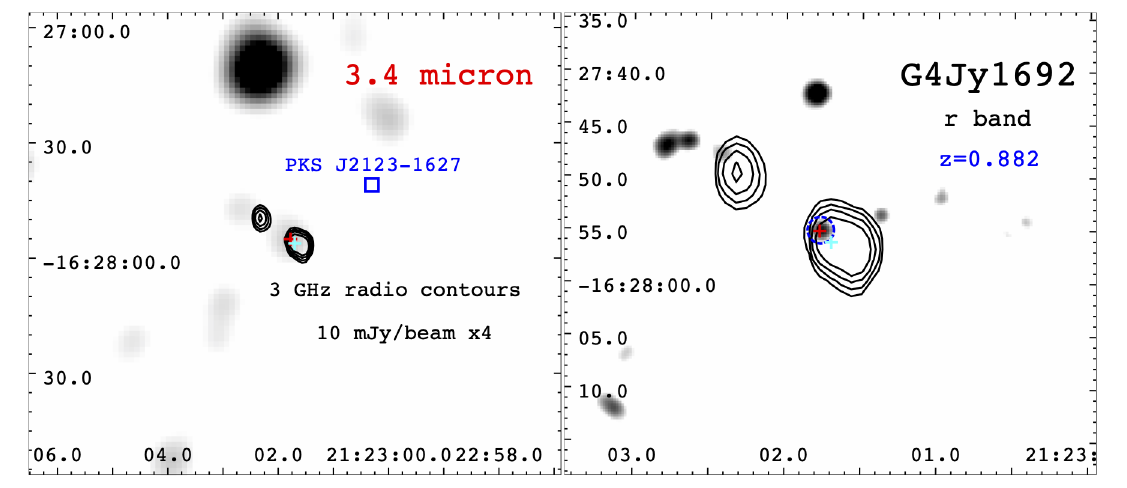}
\includegraphics[height=3.8cm,width=8.8cm,angle=0]{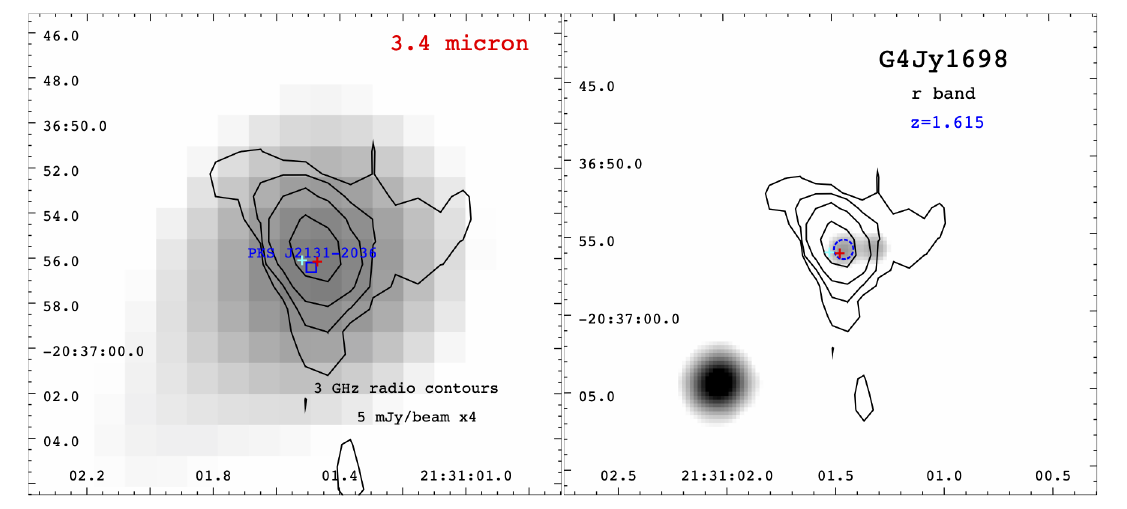}
\includegraphics[height=3.8cm,width=8.8cm,angle=0]{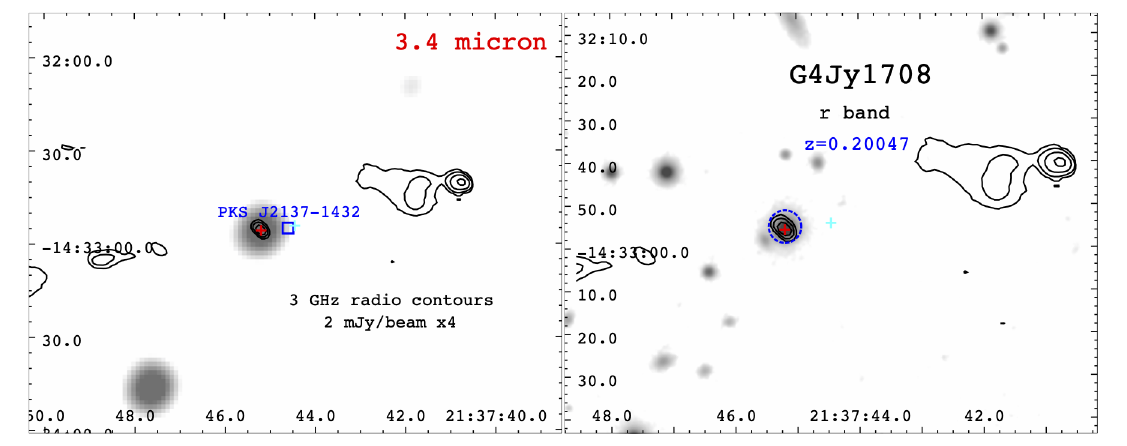}
\includegraphics[height=3.8cm,width=8.8cm,angle=0]{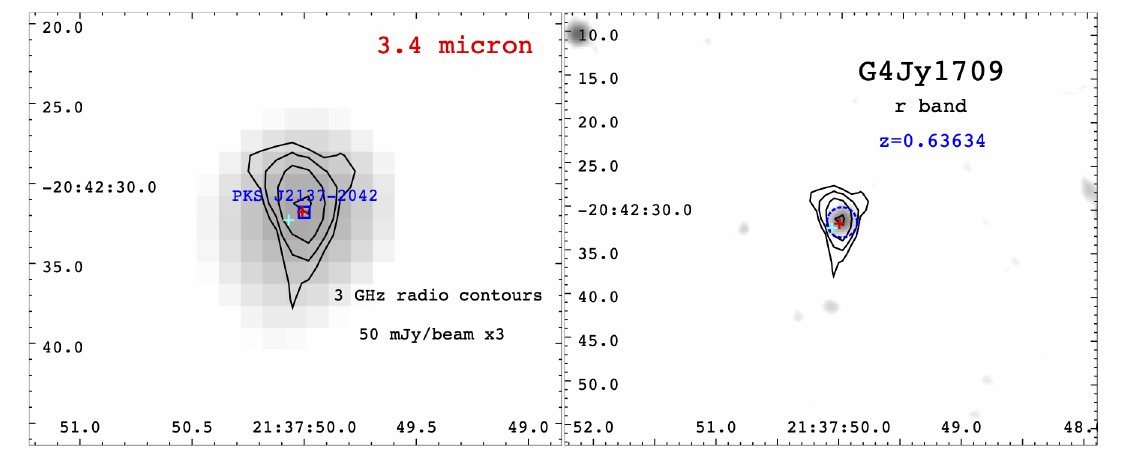}
\includegraphics[height=3.8cm,width=8.8cm,angle=0]{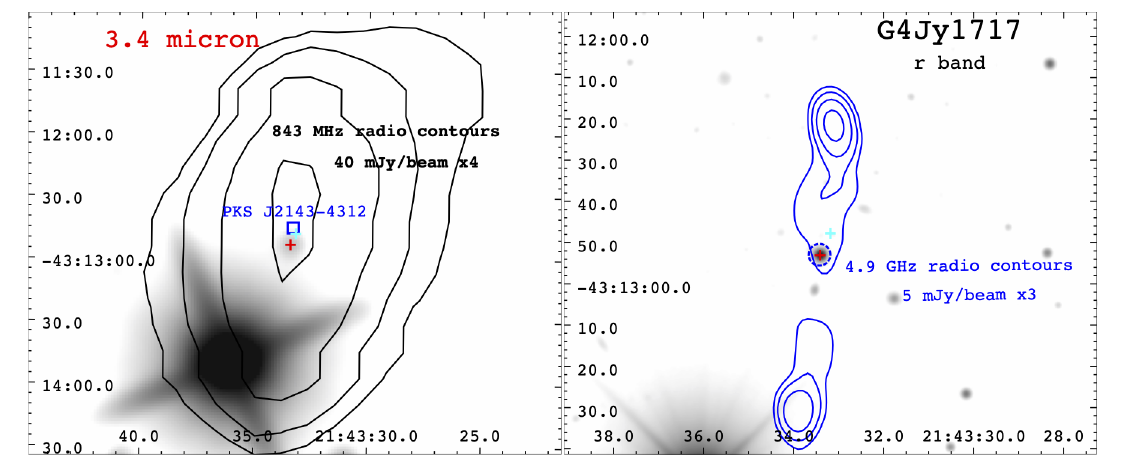}
\includegraphics[height=3.8cm,width=8.8cm,angle=0]{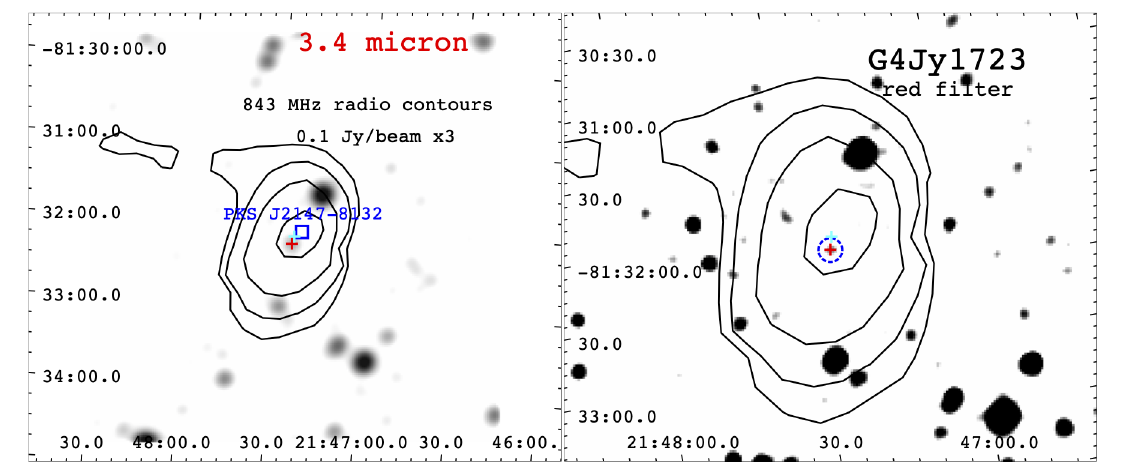}
\includegraphics[height=3.8cm,width=8.8cm,angle=0]{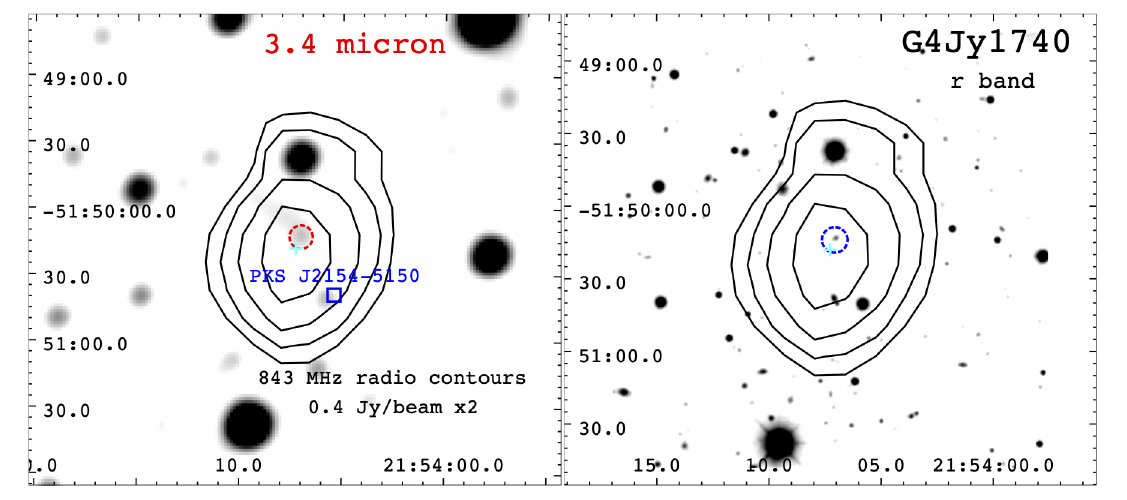}
\includegraphics[height=3.8cm,width=8.8cm,angle=0]{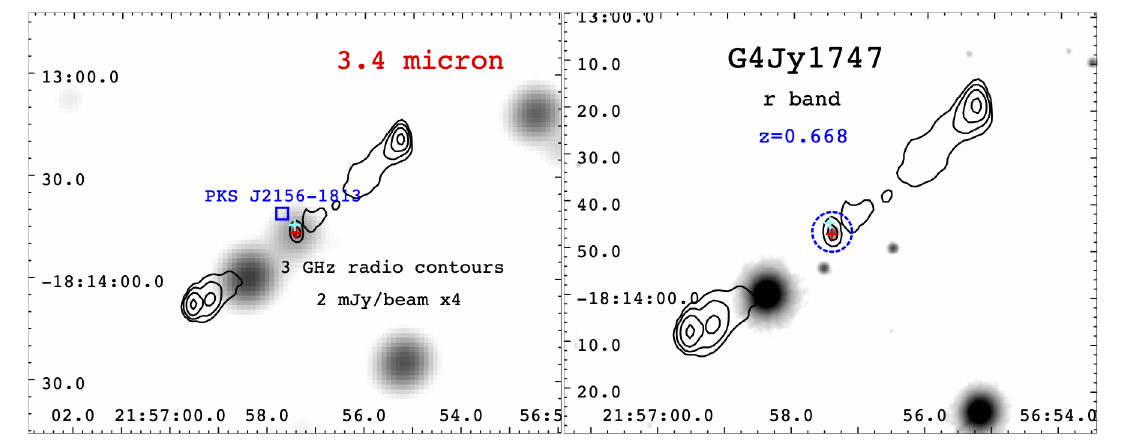}
\includegraphics[height=3.8cm,width=8.8cm,angle=0]{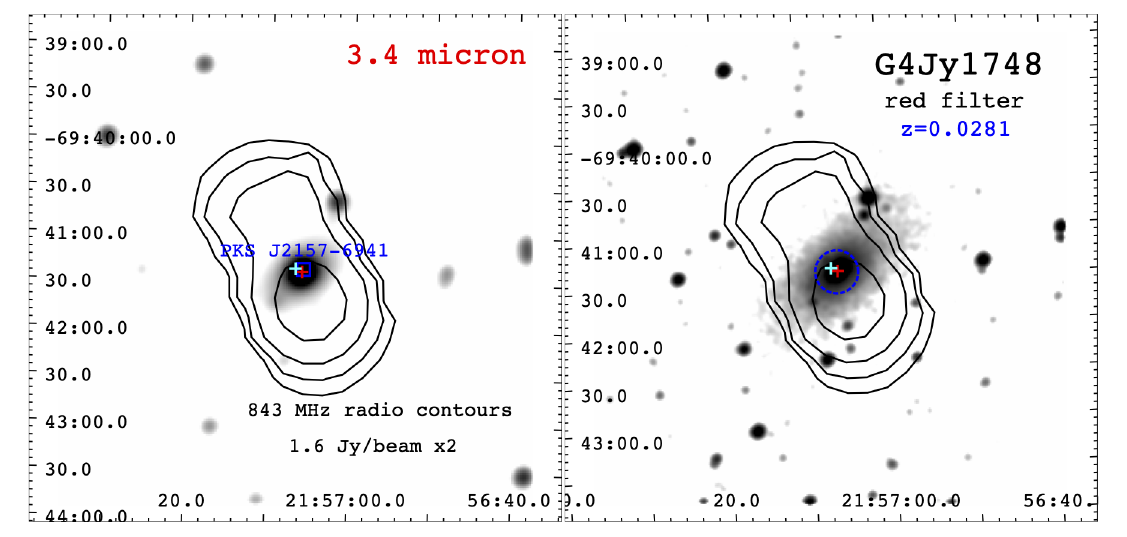}
\includegraphics[height=3.8cm,width=8.8cm,angle=0]{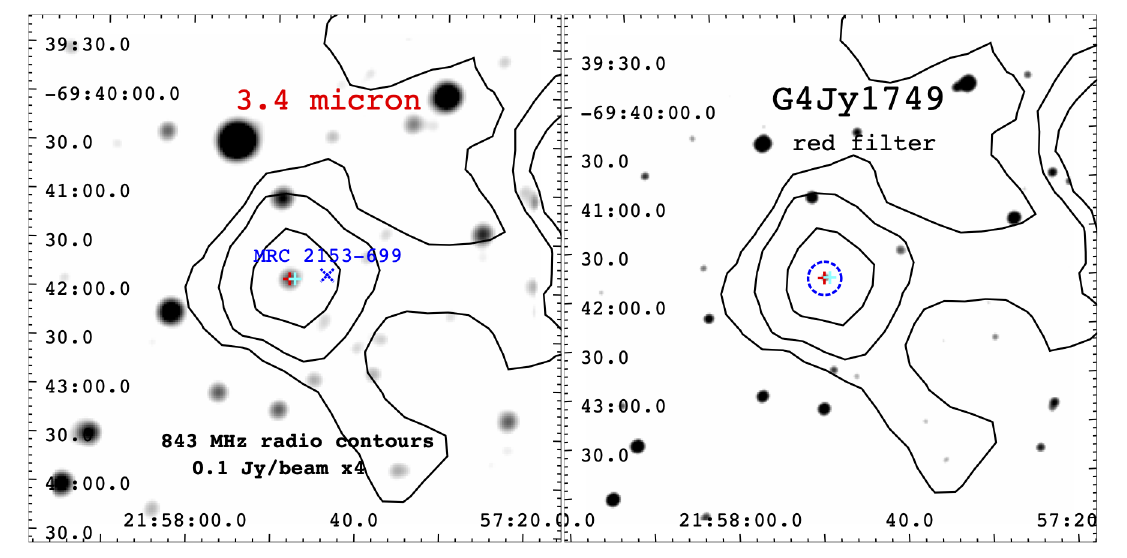}
\caption{Same as Figure~\ref{fig:example1} for the following \cs\ radio sources: \\ 
G4Jy\,1685, G4Jy\,1686, G4Jy\,1692, G4Jy\,1698, G4Jy\,1708, G4Jy\,1709, G4Jy\,1717, G4Jy\,1723, G4Jy\,1740, G4Jy\,1747, G4Jy\,1748, G4Jy\,1749.}
\end{center}
\end{figure*}

\begin{figure*}[!th]
\begin{center}
\includegraphics[height=3.8cm,width=8.8cm,angle=0]{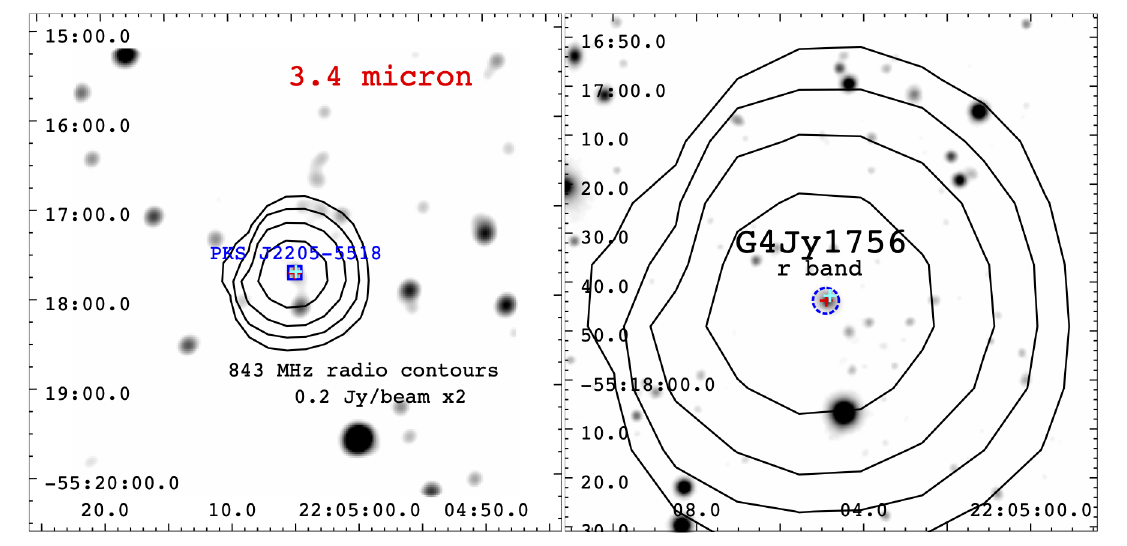}
\includegraphics[height=3.8cm,width=8.8cm,angle=0]{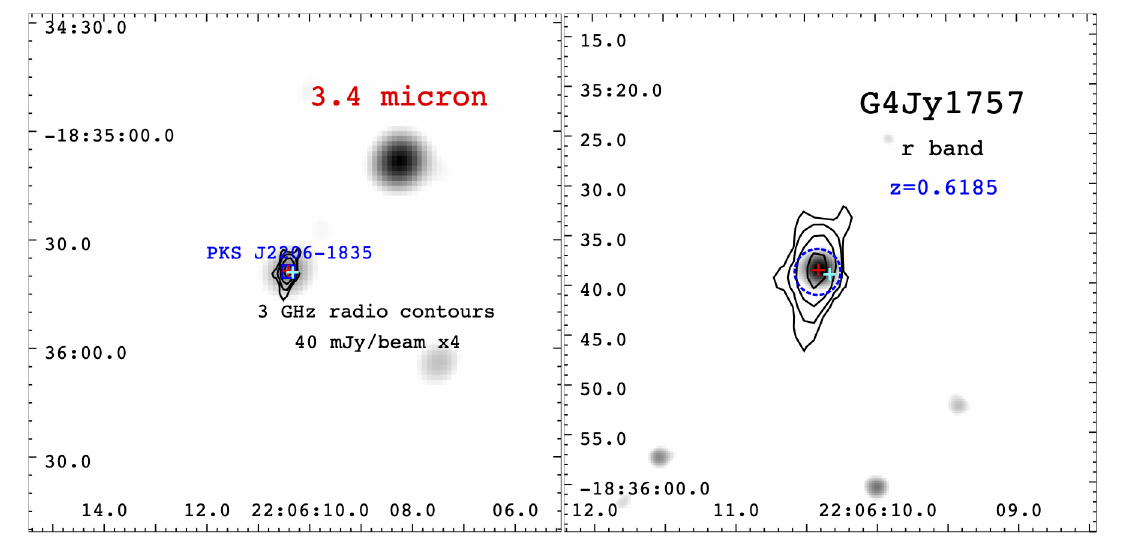}
\includegraphics[height=3.8cm,width=8.8cm,angle=0]{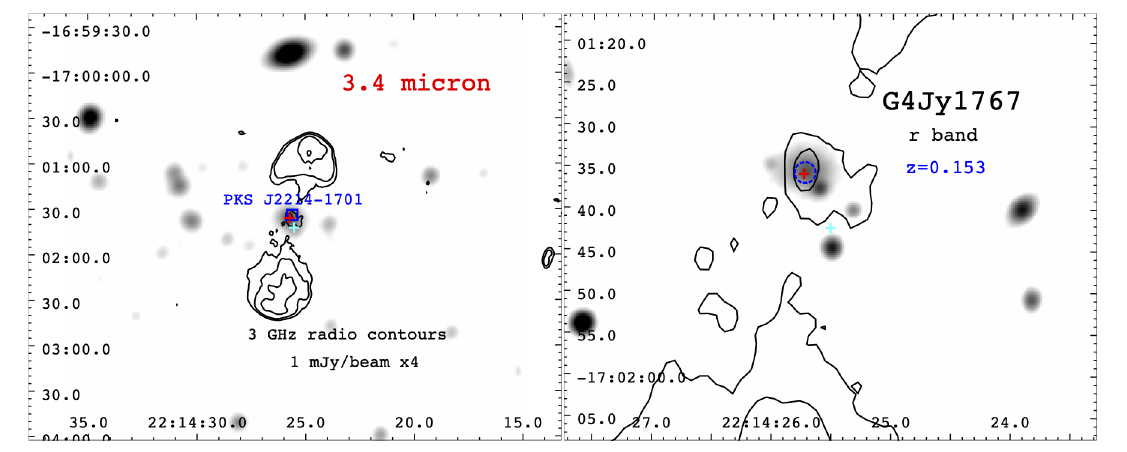}
\includegraphics[height=3.8cm,width=8.8cm,angle=0]{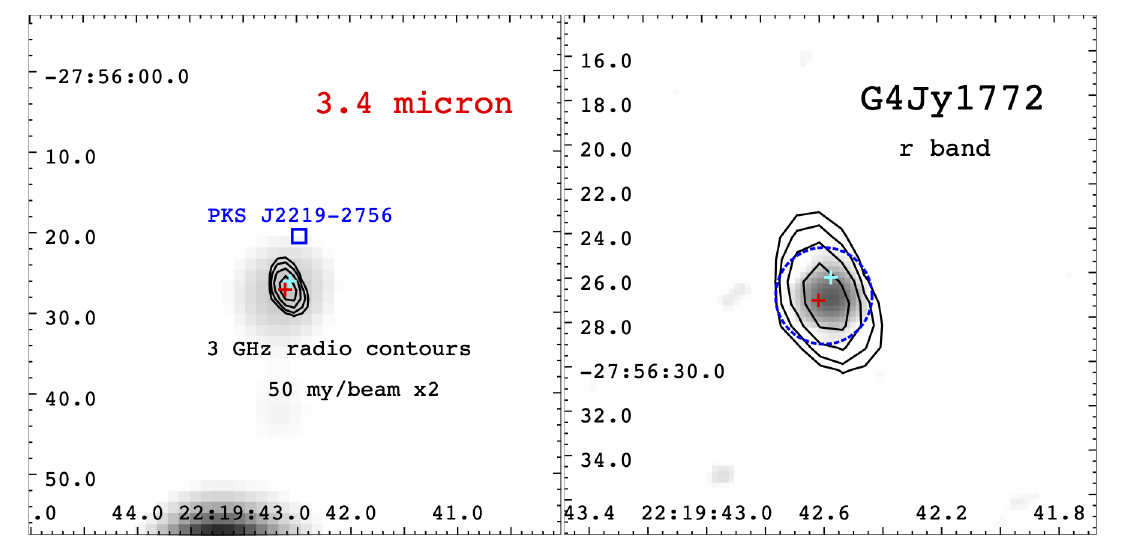}
\includegraphics[height=3.8cm,width=8.8cm,angle=0]{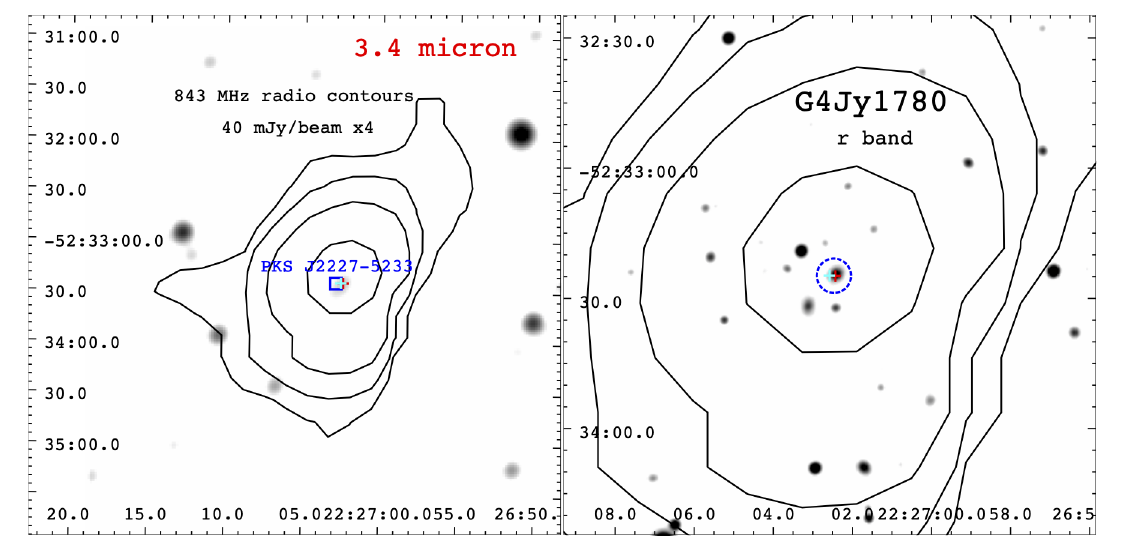}
\includegraphics[height=3.8cm,width=8.8cm,angle=0]{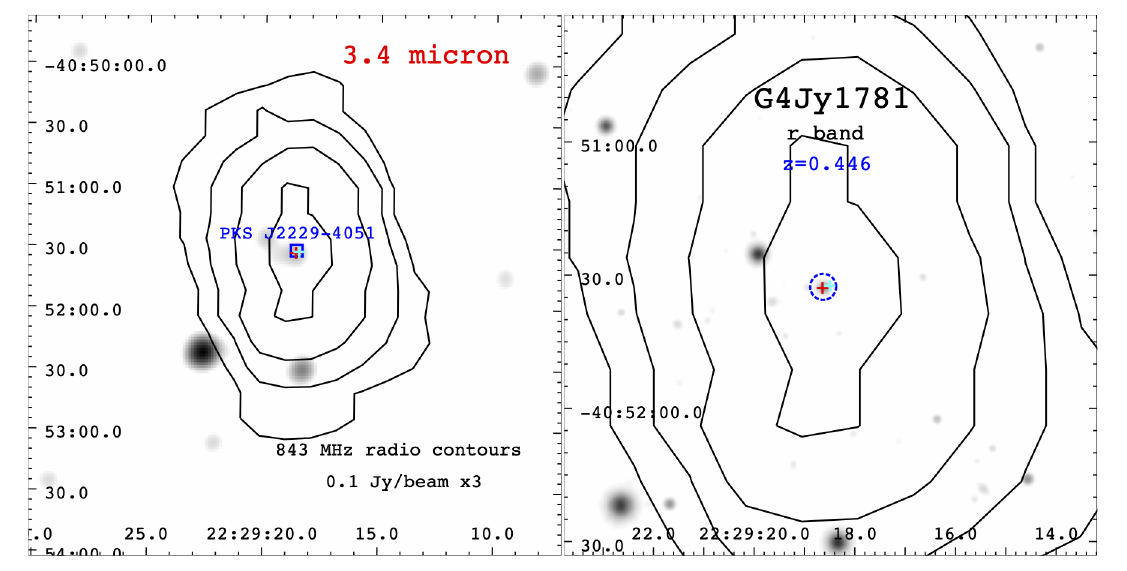}
\includegraphics[height=3.8cm,width=8.8cm,angle=0]{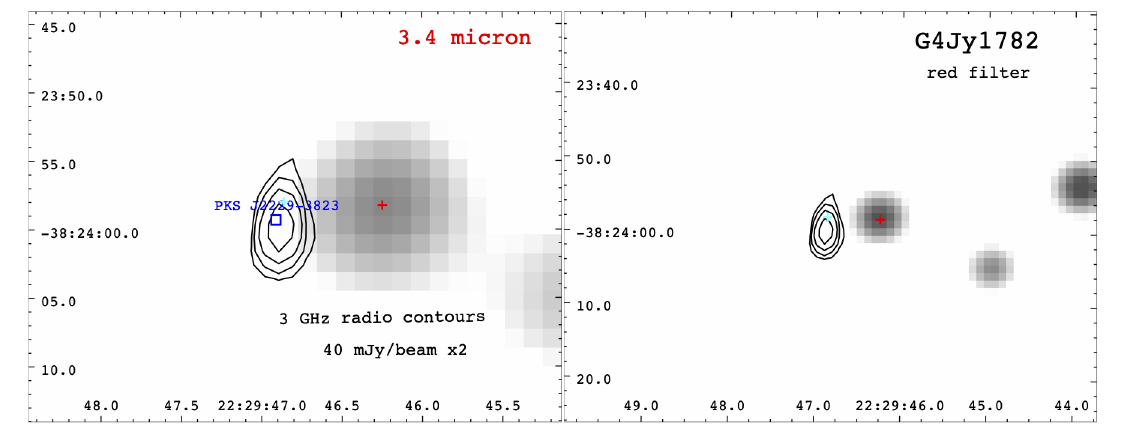}
\includegraphics[height=3.8cm,width=8.8cm,angle=0]{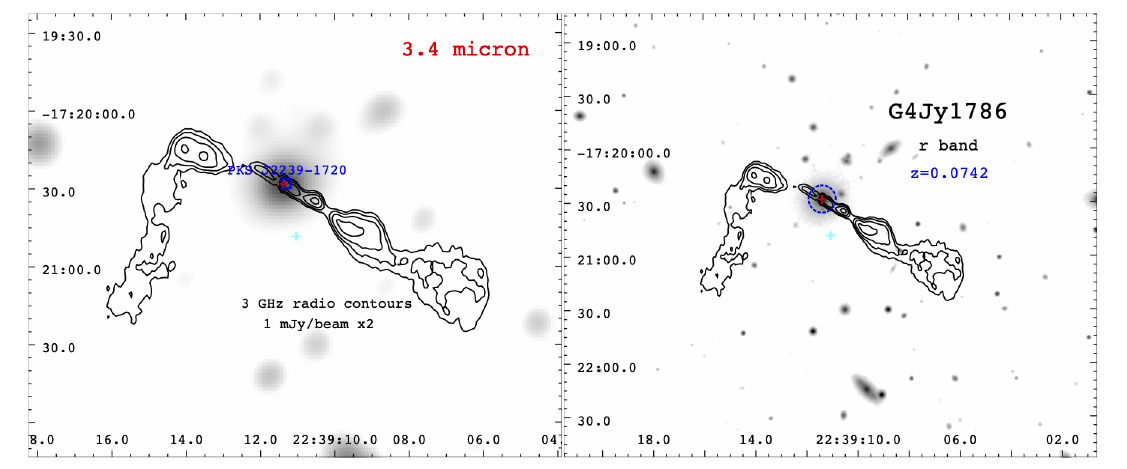}
\includegraphics[height=3.8cm,width=8.8cm,angle=0]{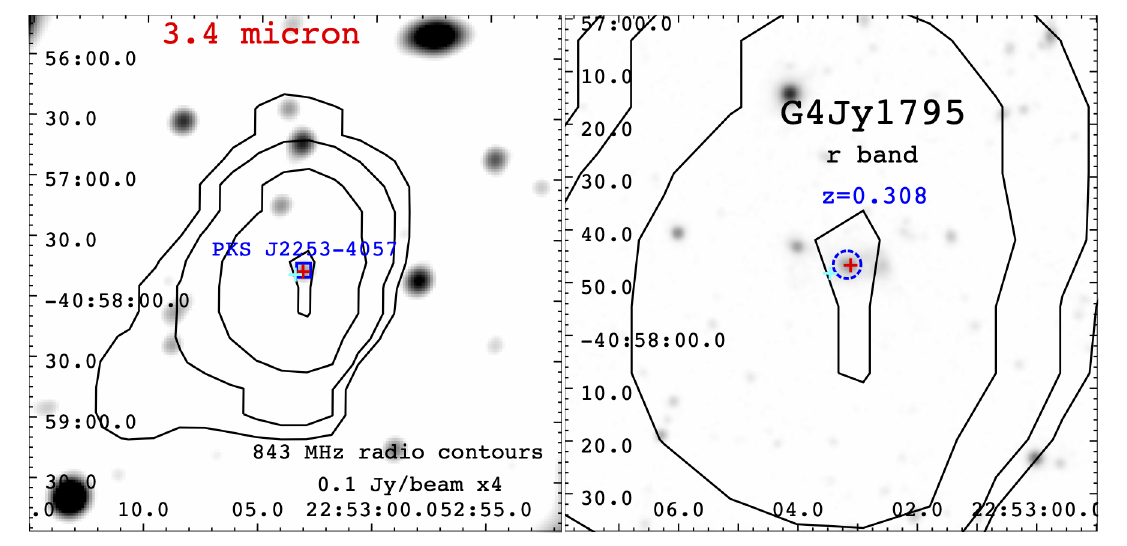}
\includegraphics[height=3.8cm,width=8.8cm,angle=0]{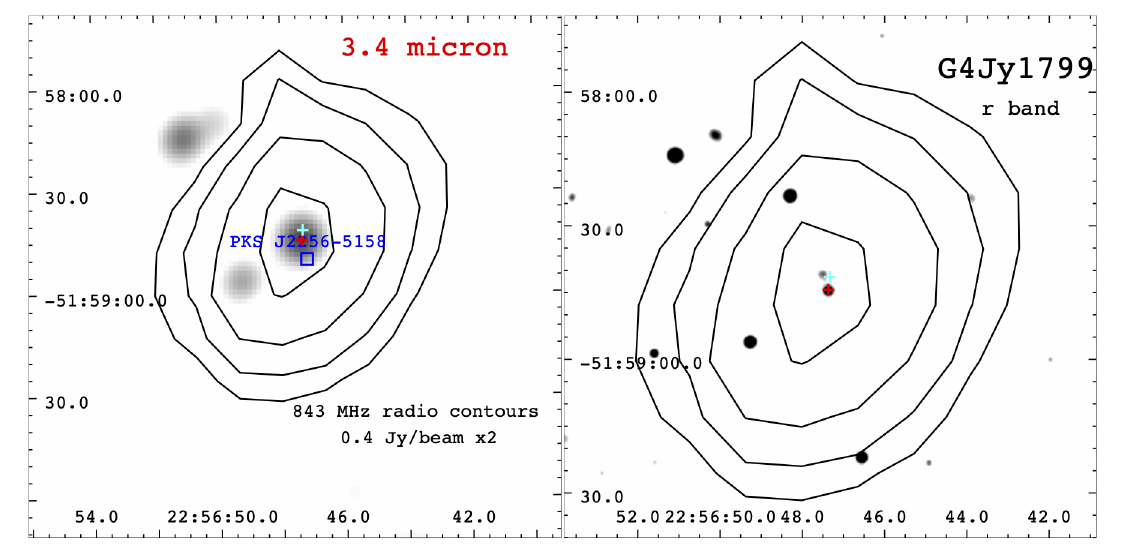}
\includegraphics[height=3.8cm,width=8.8cm,angle=0]{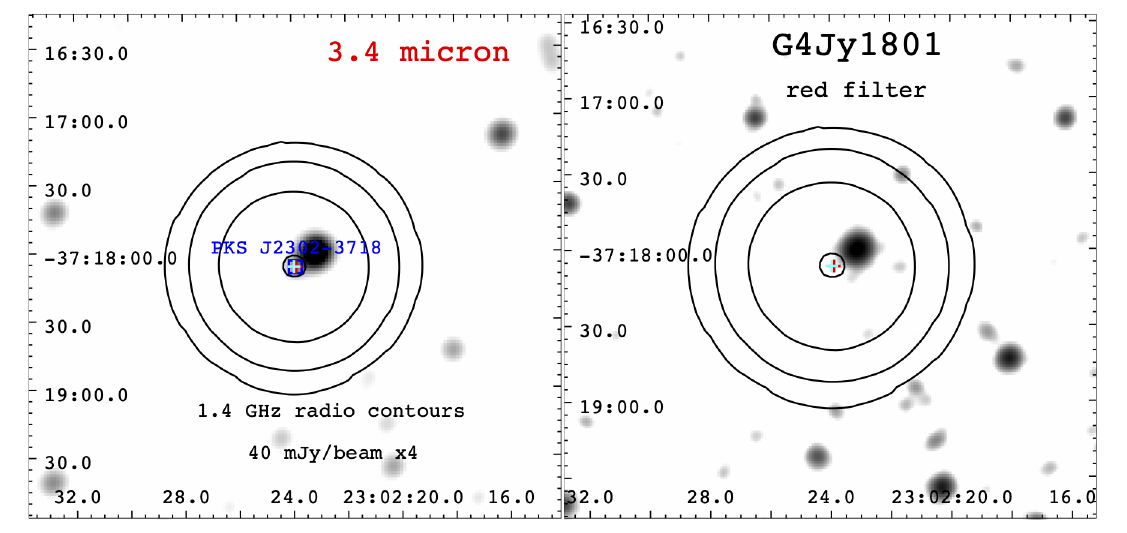}
\includegraphics[height=3.8cm,width=8.8cm,angle=0]{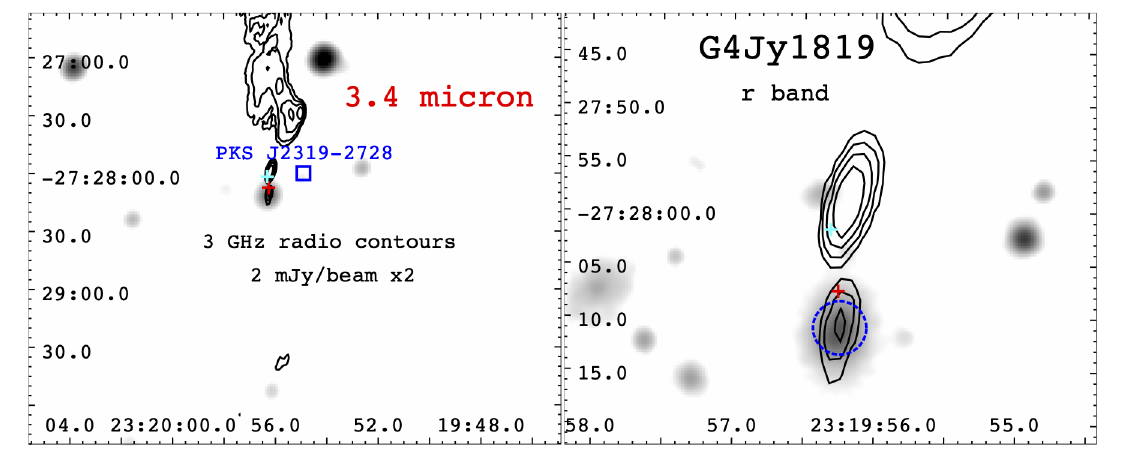}
\caption{Same as Figure~\ref{fig:example1} for the following \cs\ radio sources: \\ 
G4Jy\,1756, G4Jy\,1757, G4Jy\,1767, G4Jy\,1772, G4Jy\,1780, G4Jy\,1781, G4Jy\,1782, G4Jy\,1786, G4Jy\,1795, G4Jy\,1799, G4Jy\,1801, G4Jy\,1819.}
\end{center}
\end{figure*}

\begin{figure*}[!th]
\begin{center}
\includegraphics[height=3.8cm,width=8.8cm,angle=0]{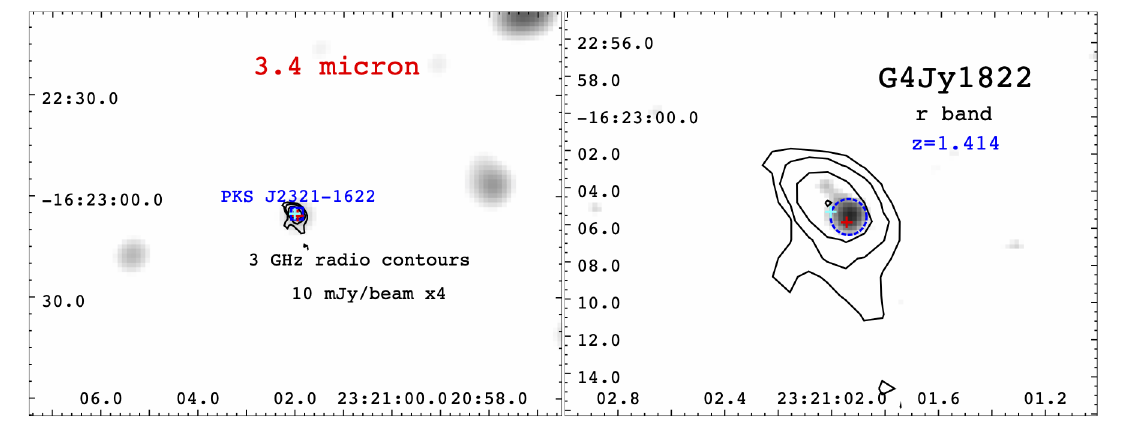}
\includegraphics[height=3.8cm,width=8.8cm,angle=0]{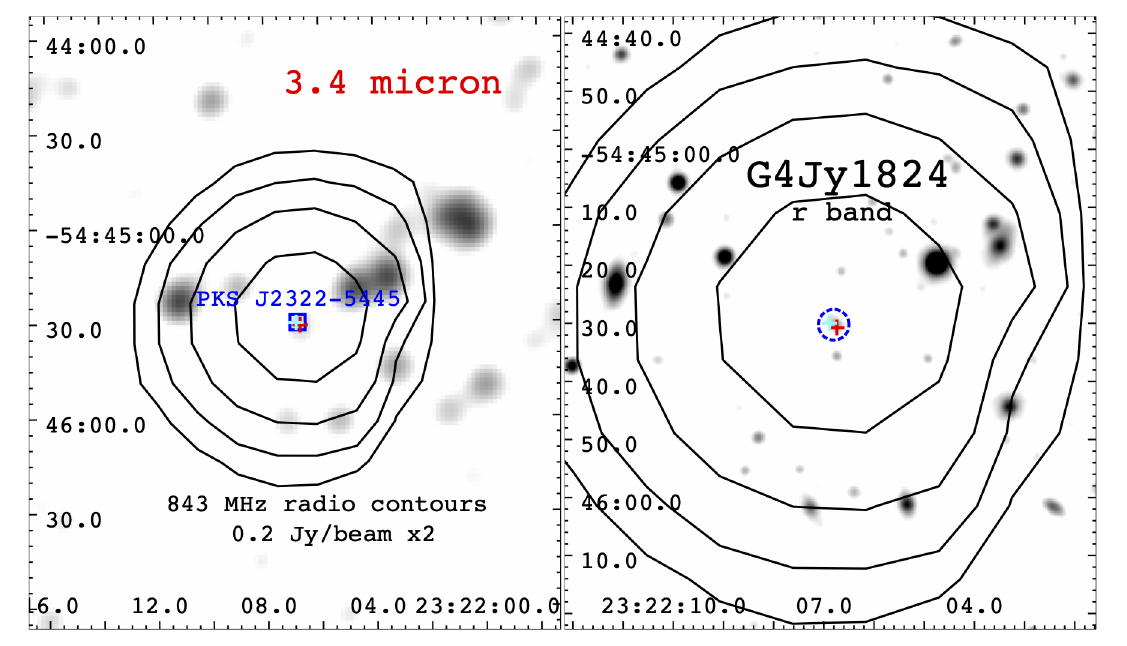}
\includegraphics[height=3.8cm,width=8.8cm,angle=0]{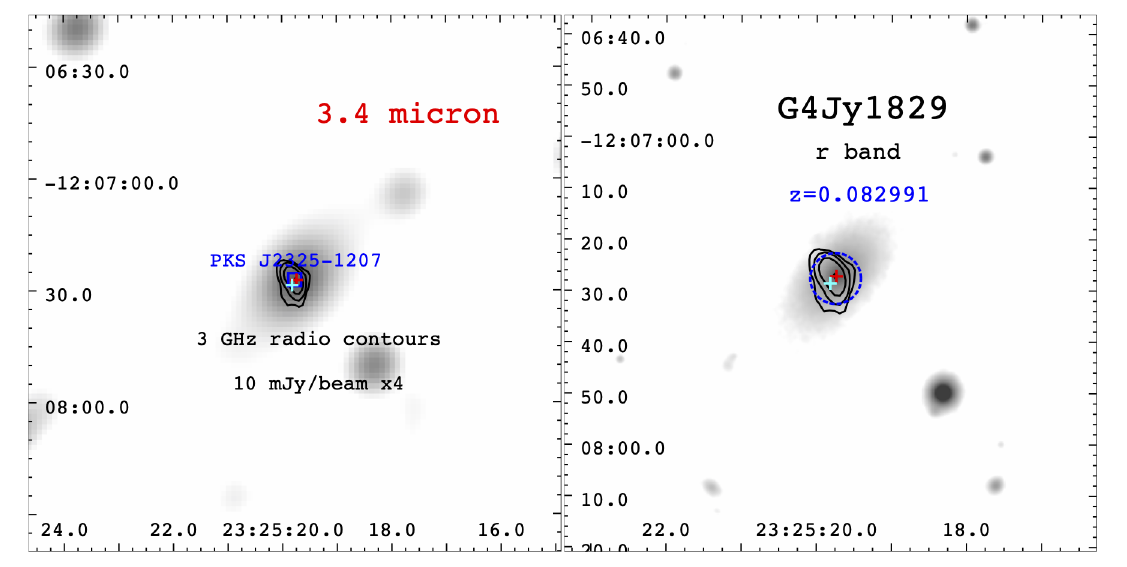}
\includegraphics[height=3.8cm,width=8.8cm,angle=0]{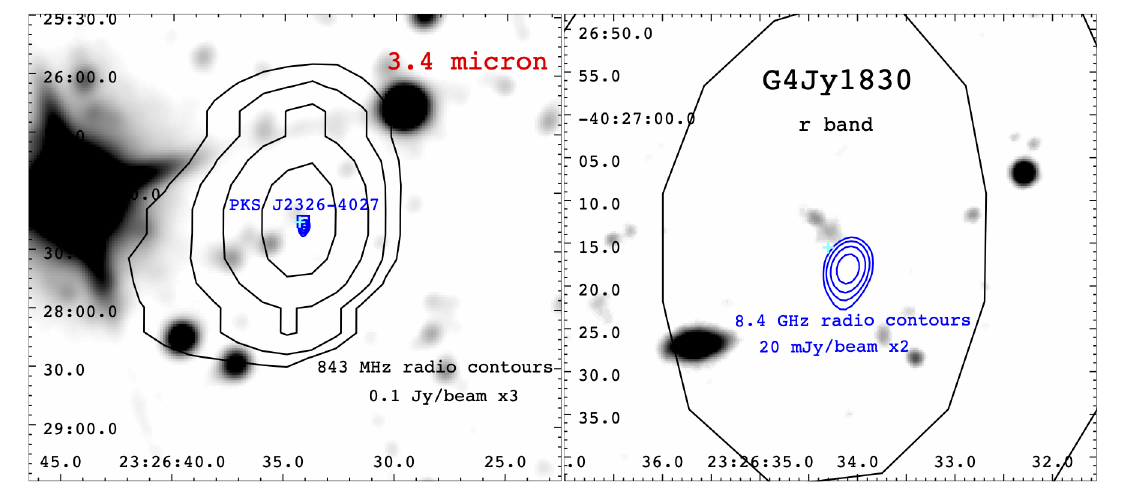}
\includegraphics[height=3.8cm,width=8.8cm,angle=0]{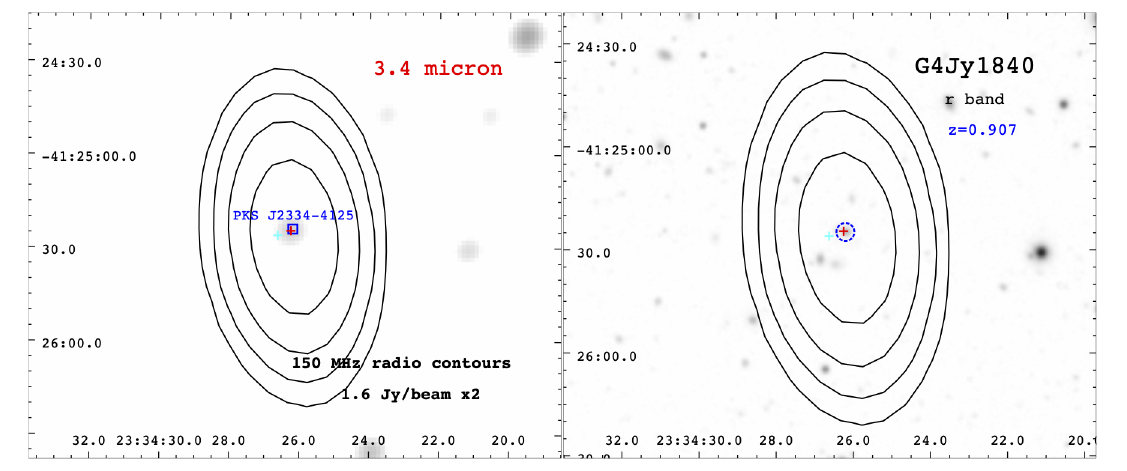}
\includegraphics[height=3.8cm,width=8.8cm,angle=0]{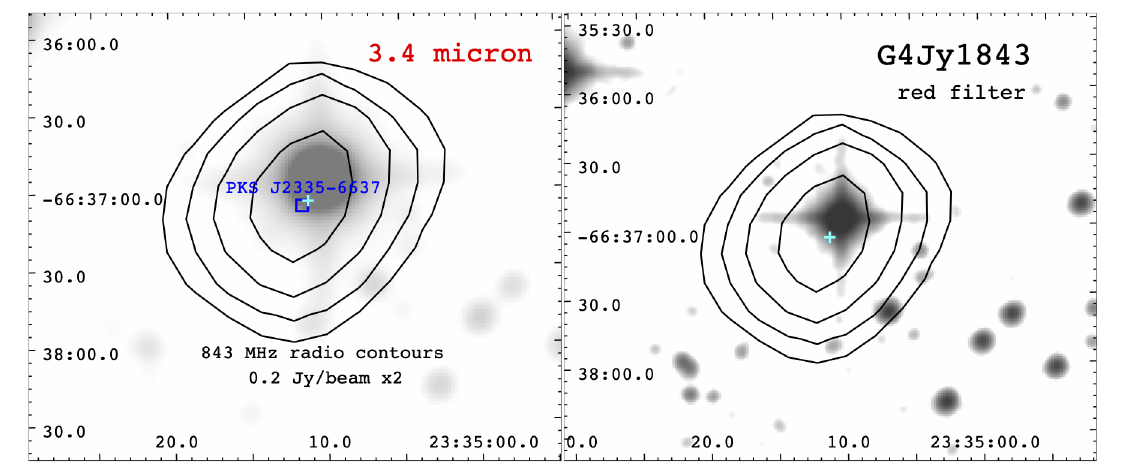}
\includegraphics[height=3.8cm,width=8.8cm,angle=0]{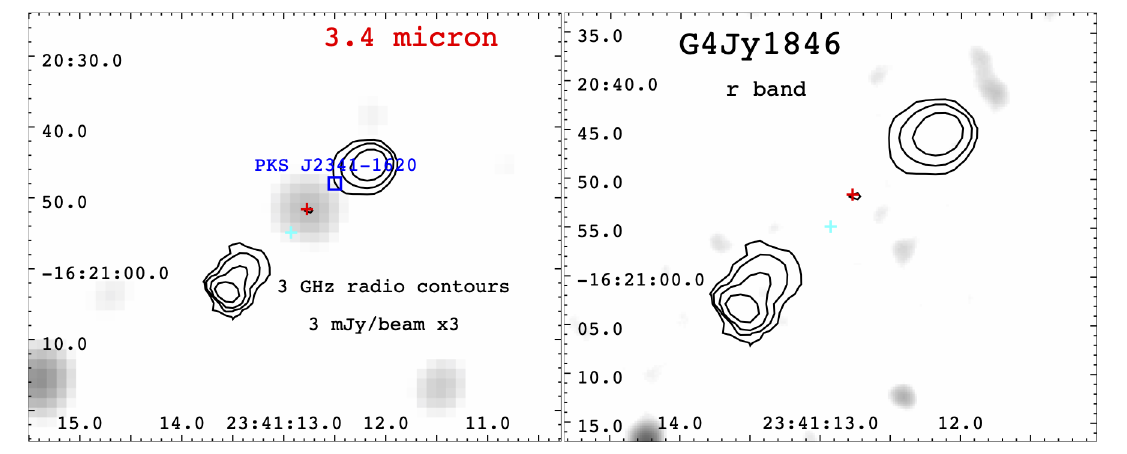}
\includegraphics[height=3.8cm,width=8.8cm,angle=0]{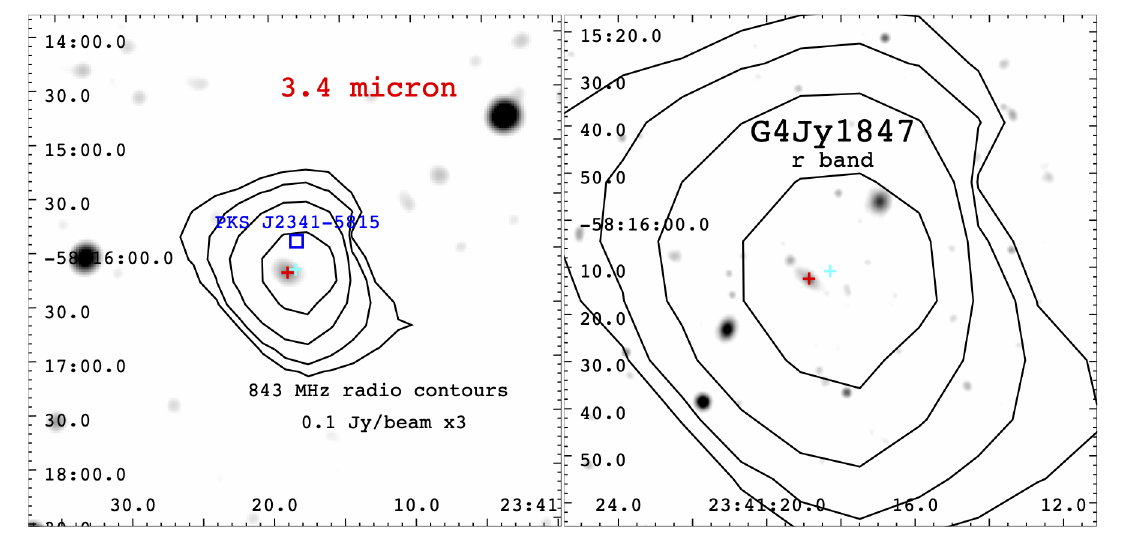}
\includegraphics[height=3.8cm,width=8.8cm,angle=0]{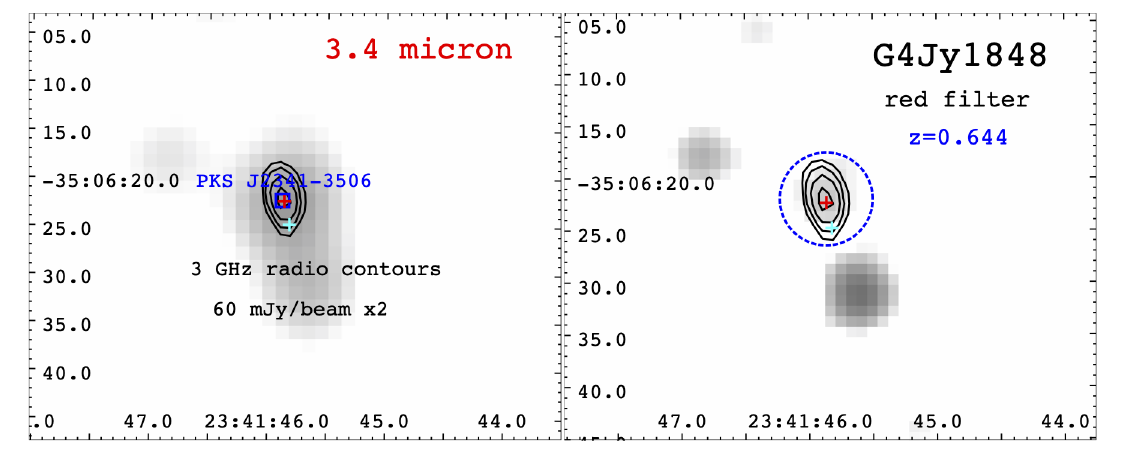}
\includegraphics[height=3.8cm,width=8.8cm,angle=0]{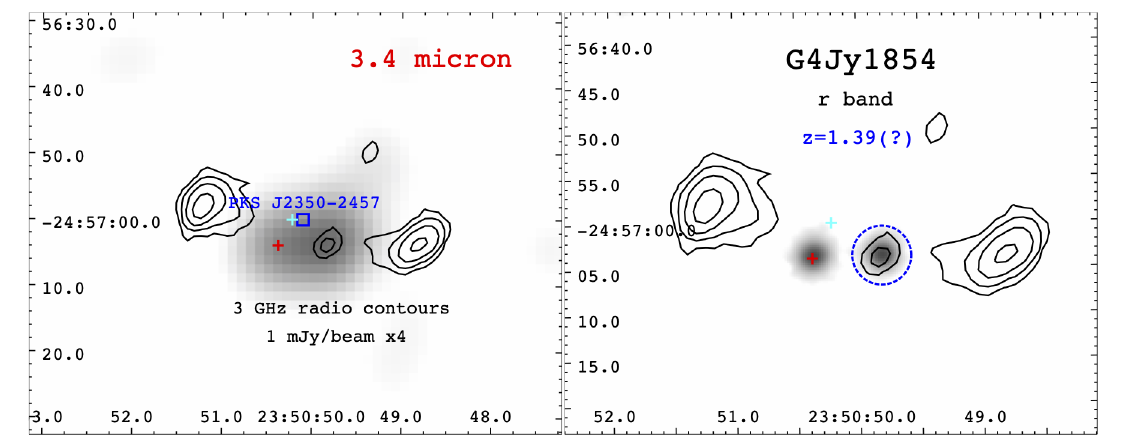}
\includegraphics[height=3.8cm,width=8.8cm,angle=0]{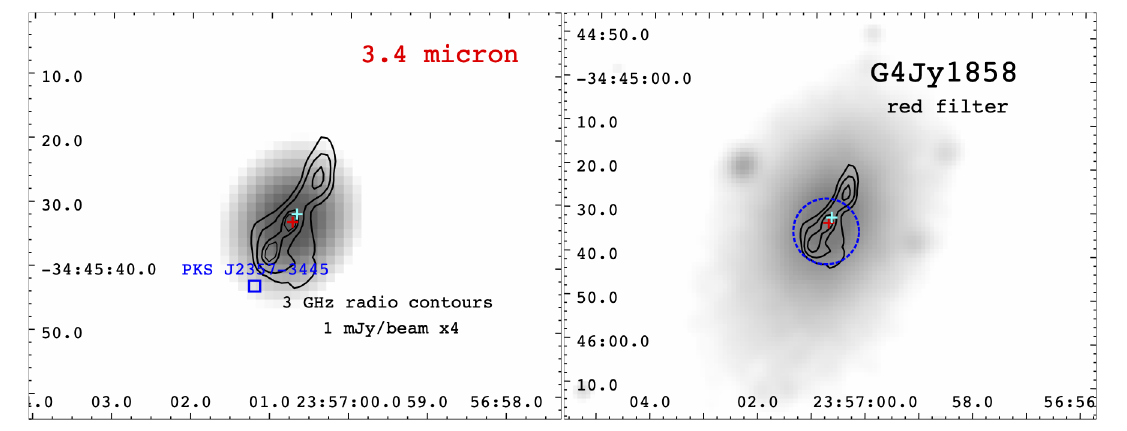}
\includegraphics[height=3.8cm,width=8.8cm,angle=0]{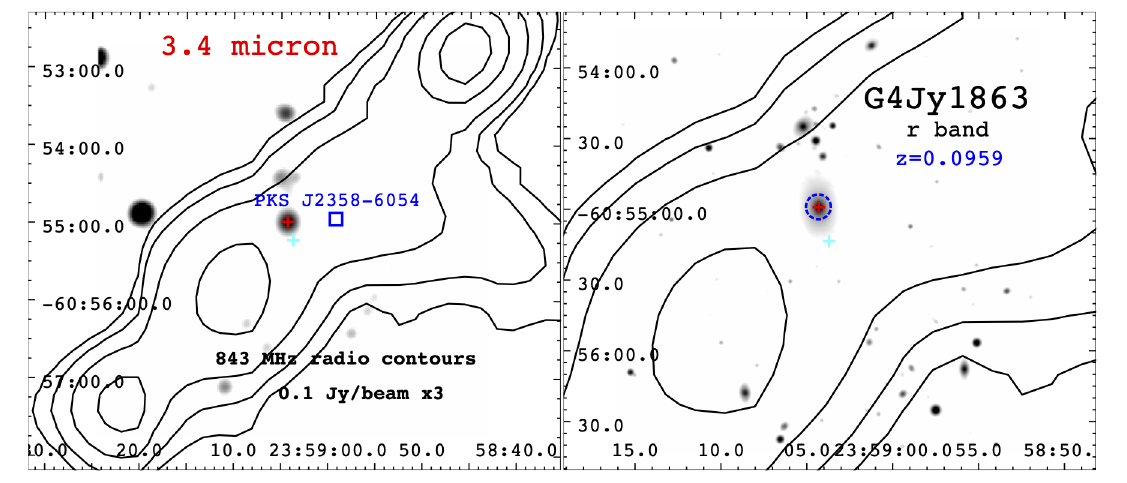}
\caption{Same as Figure~\ref{fig:example1} for the following \cs\ radio sources: \\ 
G4Jy\,1822, G4Jy\,1824, G4Jy\,1829, G4Jy\,1830, G4Jy\,1840, G4Jy\,1843, G4Jy\,1846, G4Jy\,1847, G4Jy\,1848, G4Jy\,1854, G4Jy\,1858, G4Jy\,1863.}
\end{center}
\end{figure*}

\section{Notes on individual sources}
\label{app:details}
Here we provide additional information, in addition to that retrievable from the main table. This list will be then updated in all forthcoming publications. Radio sources not listed below are those for which relevant information were not found in the literature. For the FR\,I and FR\,II radio classification of radio galaxies we mainly considered the information reported in the literature as well as the radio morphology observed in high resolution radio maps when available, adopting the same criteria of Capetti et al. (2017a,b). In several sources we also reported additional names provided in the literature, but full information about radio cross-identifications can be retrieved in Appendix B.
\vspace{0.2cm}

\underline{G4Jy\,4:} Hunstead et al (1978) reported the results of optical spectroscopic observations, collected with the 3.9-m Anglo-Australian Telescope, for 22 QSOs and emission-line galaxies associated with southern radio sources detected with Molonglo telescope at 408 MHz, and having flux densities above 0.95\,Jy. The first source in their sample is 0000-117, a QSO at $z$=1.465, having the optical position reported therein (00:03:22.12, -17:27:14.1 in J2000), that is a 3\arcsec.5 angular separation from the location of the counterpart assigned in our analysis with G4Jy\,4. Despite the one-side radio structure of G4Jy\,4, typical of core dominated QSOs, the source identified by Hunstead et al (1978) appears to be more consistent with the relatively brighter object located in the southern-eastern direction with respect to that coincident with the radio core of G4Jy\,4, thus having a $z$ estimate labelled with a question mark. It could be also associated to the radio source PKS\,0000-17, but the lack of optical information prevented us to claim this association.
\vspace{0.2cm}

\underline{G4Jy\,9:} a nearby radio source at $z$=0.2912 \citep{jones09} and correspondent to PKS\,0003-56. This source also belongs to the MS4 radio catalog \citep{burgess06a,burgess06b} and has a radio counterpart at 20 GHz \citep[a.k.a. AT20G\,J000558-562828;][]{murphy10}, optically identified as a normal galaxy \citep{mahony11} in agreement with the optical image reported in our analysis.
\vspace{0.2cm}

\underline{G4Jy\,12:} also known as PKS\,0003-83 and with a photometric redshift estimate of $z$=0.32 \citep{burgess06b}. It is a classical double source with unresolved components. The association provided by our analysis is consistent with that performed by Jones \& McAdam (1992) reporting an optical magnitude of 19.0 mag and a nearby fainter companion galaxy of 20.0 mag.
\vspace{0.2cm}

\underline{G4Jy\,20:} (a.k.a. PKS\,0008-44) has a photometric redshift estimate of $z$=1.0 reported in the MS4 optical identification analysis \citep{burgess06b}. In this case the lack of high resolution radio maps prevented us to claim that the optical counterpart is the one assigned in the MS4 catalog.
\vspace{0.2cm}

\underline{G4Jy\,26:} a celestial object belonging to several catalogs of southern radio sources and thus known also as PKS\,0012-38 and PMN\,J0015-3804 \citep{griffith93,gregory94}. It has a tentative photometric redshift estimate reported in the literature of $z$=0.57 \citep{burgess06b}, however having an IDF=3.0 if the optical counterpart associated in the MS4 sample is the correct one.
\vspace{0.2cm}

\underline{G4Jy\,33:} is also known as 3C\,8, PKS\,0016-12 and MRC\,0016-129. This source belongs to the original 3C catalog \citep{edge59} but not to its revised version \citep{spinrad85}. This radio source, classified as a high redshift radio galaxy, lies in a galaxy-rich large-scale environment \citep{wylezalek13} and it has a redshift estimate of $z$=1.589 \citep{best99}.
\vspace{0.2cm}

\underline{G4Jy\,43:} is a classical lobe dominated radio source, associated with PKS\,0020-25 since its large-scale radio structure resembles that visible at 4.8\,GHz \citep{kapahi98} and its optical identification \citep{mccarthy96b} corresponds to the WISE source associated in the G4Jy being a radio galaxy at redshift $z$=0.35.
\vspace{0.2cm}

\underline{G4Jy\,45:} is associated with PKS\,0021-29 (a.k.a. AT20G\,J002430-292853), optically identified as nearby QSO at $z$=0.40645 \citep{ho09}, having mid-IR colors similar to those of flat spectrum radio quasars \citep[FSRQs; see e.g.,][]{dabrusco12,dabrusco19}.
\vspace{0.2cm}

\underline{G4Jy\,48:} is associated with PKS\,0023-26, having both an [OIII] and a bolometric luminosity typical of type 2 QSOs \citep{dicken09a},  with a first redshift estimate of $z$=0.32162 \citep{holt08} more recently refined at $z$=0.32188 \citep{santoro20}. It is also a young radio source, showing two relatively symmetric lobes \citep{tzioumis02,morganti21a}, hosted by an early-type galaxy \citep{ramos11}. G4Jy\,48 was also observed during the X-ray survey of the 2\,Jy sample \citep[see][]{wall85}, showing an X-ray spectrum with a dominant jet component and a low intrinsic absorption \citep{mingo14}. Morganti et al. (2021a) recently observed this radio source using ALMA and discovered a very extended distribution of molecular gas revealing that, already on galaxy scales, the impact of the AGN is not limited to outflows. Given its high star formation rate, G4Jy\,48 lies in the region occupied by the star forming galaxies on the on the SFR-M$_*$ plane \citep[see e.g.,][]{bernhard21}.
\vspace{0.2cm}

\underline{G4Jy\,64:} has mid-IR colors of a FSRQs \citep{dabrusco14} and being at $z$=0.518 \citep{baker99}, this radio source is also known as PKS\,0032-20 and has a radio counterpart at 20 GHz (i.e., AT20G\,J003508-200359).
\vspace{0.2cm}

\underline{G4Jy\,70:} is identified with the QSO (a.k.a. PKS\,0035-39) at $z$=0.59313 \citep{thompson90,jones09} having a radio counterpart also listed in the MRC and in the MS4 catalogs \citep[see][respectively]{large81,burgess06a} as well as in the AT20G sample \citep[a.k.a. AT20G\,J003826-385948;][]{murphy10}.
\vspace{0.2cm}

\underline{G4Jy\,77:} is an extended radio source located at $\sim$6\,\arcmin\ from the X-ray position of the cool core galaxy cluster Abell\,85 \citep[see e.g.,][for details about its X-ray emission]{durret05,ichinohe15}. At the redshift of the galaxy cluster \citep[i.e., $z$=0.0557;][]{abell58,abell89,pislar97,oegerle01} the kpc scale is $\sim$1.1 kpc/arcsec, thus the distance between G4Jy\,77 and Abell\,85 is $\sim$360 kpc. It is a well-known radio phoenix \citep{kempner04}, example of fossil plasma present in galaxy clusters due to past AGN activity, with a very steep radio spectrum \citep{bagchi98,slee01,rahaman22}.
\vspace{0.2cm}

\underline{G4Jy\,78:} is associated with PKS\,0039-44, a radio galaxy that belongs to the 2\,Jy sample \citep{wall85} at $z$=0.346 \citep{tadhunter93,diserego94}.
\vspace{0.2cm}

\underline{G4Jy\,85:}, { belonging to the 2\,Jy sample \citep[a.k.a. PKS\,J0043-42; see e.g.,][]{wall85,morganti97}, is LERG \citep[][]{hine79} at $z$=0.0526 \citep{whiteoak72,tadhunter93} with a very extended radio structure \citep{morganti99} and a classical FR\,II radio morphology. It appears} to be located at the centre of a group/cluster of galaxies being also surrounded by a diffuse halo \citep{ramos11}. It also shows the presence of a bridge detected at infrared and optical frequencies related to the interaction with a nearby companion galaxy \citep[see also][]{inskip10}. Both hotspots are detected in the soft X-rays together with a relative faint extended emission of the ICM \citep{mingo17}.
\vspace{0.2cm}

\underline{G4Jy\,86:} is the nearby star forming galaxy NGC\,253 \citep[a.k.a. Sculptor Galaxy; see e.g.,][]{hoopes96} with a $z$=0.00081 \citep{springob05}. More details about this association are reported in the G4Jy catalog.
\vspace{0.2cm}

\underline{G4Jy\,90:} associated with PKS\,0048-44, has only a photometric redshift estimate of $z$=0.67 \citep{burgess06b}, not spectroscopically confirmed. 
\vspace{0.2cm}

\underline{G4Jy\,93:} (a.k.a. PKS\,0049-43) is a radio source with a tentative photometric redshift estimate of $z$=0.39 \citep{burgess06b} also listed in the catalog of $\gamma$-ray blazar candidates  \citep{dabrusco14,dabrusco19} with mid-IR colors similar to FSRQs.
\vspace{0.2cm}

\underline{G4Jy\,108:} is a radio galaxy with a typical FR\,II radio structure at $z$=1.019 \citep{best99}, also associated with PKS\,0056-17.
\vspace{0.2cm}

\underline{G4Jy\,113:} is a ``retired'' radio source \citep{coba20} with a $z$=0.0564, and located in the galaxy cluster Abell\,133 \citep{owen95}. McDonald et al. (2010) observed the presence of a thin H$\alpha$ filament toward the northeast, extending $\sim$25 kpc, co-spatial with an X-ray filament, and not consistent with a buoyant radio bubble. Radio and X-ray observations of the radio relic revealed that the relic lobe is energized by the central cD galaxy associated with G4Jy\,113, rather than by a shock generated in the radio relic \citep{fujita02}.
\vspace{0.2cm}

\underline{G4Jy\,120:} is a lobe dominated radio source. The SUMSS radio position marks the location of G4Jy\,120 at an angular separation of $\sim$17\,\arcsec from PKS\,0103-45, a value too large to claim this association.
\vspace{0.2cm}

\underline{G4Jy\,122:} is a classical double radio galaxy at $z$=0.4 \citep{tadhunter93}, optically classified as a narrow line radio galaxy \citep[see also][]{ramos11} and showing a prominent Fe K$\alpha$ emission line in its \xmm\ observation \citep{mingo14}. This radio source is known as PKS\,0105-16 and it is listed in the original 3C catalog as 3C\,32 \citep{edge59} but not in the 3CR revised version \citep{spinrad85}. Archival optical images seems to indicate that the host galaxy of G4Jy\,122 appears to be connected with an early-type galaxy of similar brightness located at $\sim$70 kpc in the north-western direction \citep{ramos11}.
\vspace{0.2cm}

\underline{G4Jy\,129:} is potentially associated with the radio source PKS\,0110-69 \citep[a.k.a. AT20G\,J011143-690016;][]{murphy10} due to the proximity to the intensity peak of the SUMSS radio map at 843 MHz.
\vspace{0.2cm}

\underline{G4Jy\,133:} according to our analysis is labelled with an IDF=3.0, however if follow up observations will confirm its WISE association reported in the G4Jy the radio source could be associated with PKS\,0114-47 \citep{white20b}, a giant radio galaxy \citep{jones92} with a $z$=0.146 \citep{jones09}.
\vspace{0.2cm}

\underline{G4Jy\,136:} (a.k.a. PKS\,0114-21) is a Compact Steep Spectrum (CSS) radio source with two hotspots embedded in a dense gaseous environment \citep{mantovani94} with a $z$=1.4153 \citep{debreuck10,seymour07}, with mid-IR emission dominated by a stellar continuum and several rest-frame optical emission lines detected in its spectrum \citep{nesvadba17}.
\vspace{0.2cm}

\underline{G4Jy\,143:} is associated with 3C\,38 (a.k.a. PKS\,0117-15), again not listed in the revised release 3CR \citep{spinrad85}, and it is classified as a radio galaxy at $z$=0.565 \citep{tadhunter93}, having a clear FR\,II morphology and showing high ionization emission lines detected in the optical spectrum. There is a nearby optical source, potentially a companion galaxy, however, due to the lack of spectroscopic information, we cannot claim that this is the case of a galaxy pair. 
\vspace{0.2cm}

\underline{G4Jy\,145:} is a QSO also at mid-IR frequencies \citep{dabrusco19}, originally located at $z$=0.837 \citep{hunstead78}. More recent ultraviolet observations give a redshift measurement of $z$=0.834 \citep{monroe16}.
\vspace{0.2cm}

\underline{G4Jy\,157:} belongs to the sample of Best et al. (1999) and is classified as a radio galaxy at $z$=0.372 showing a typical FR\,II radio morphology. 
\vspace{0.2cm}

\underline{G4Jy\,162:} has a clear optical counterpart, while the lack of a mid-IR associated source is mainly due to the artifacts present in the WISE image created by a nearby bright star located in the south-western direction. In the literature G4Jy\,162 is classified as a radio galaxy with a $z$=2.34665 \citep{best99,nesvadba17} being the most distant source listed in the \cs\ sample to date. G4Jy\,162 also shows a fairly regular emission line region morphology, albeit an increasing of the [OIII] surface brightness towards the northern side, and roughly aligned with the radio axis in the north-western side \citep{nesvadba17}.
\vspace{0.2cm}

\underline{G4Jy\,168:} is a classical double lobed radio source associated with PKS\,0129-073 (a.k.a. MRC\,0129-073 and PMN\,J0131-0703). There are three sources detected in the optical image that could be its potential counterpart, but in the WISE image only the brightest one is detected. This prevents us to label this source with an IDF different from 3.0.
\vspace{0.2cm}

\underline{G4Jy\,171:} is the nearby lenticular galaxy NGC\,612 \citep{ekers78} located at $z$=0.02977 \citep[see e.g.,][]{burbidge72,menzies89,dacosta91}, associated with the powerful radio source PKS\,0131-36 and showing a wide double radio structure, with the axis perpendicular to the disk of the galaxy. G4Jy\,171 also shows an unusual absorption in its optical spectrum mainly due to its host galaxy structure with gas and dust in its disk \citep{tadhunter93}.
\vspace{0.2cm}

\underline{G4Jy\,192:} is a double radio galaxy with a redshift $z$=0.41, evaluated comparing our finding chart with that reported in the literature \citep{mccarthy96b}. Three nearby companion optical galaxies appear to lie close to the host galaxy of G4Jy\,192.
\vspace{0.2cm}

\underline{G4Jy\,208:} (a.k.a. PKS\,0155-10) is a radio QSO at $z$=0.61847 \citep{burbidge68,best99,jones09}.
\vspace{0.2cm}

\underline{G4Jy\,213:} is a radio source at $z$=0.67680 \citep{jones04,croom04} with a mid-IR counterpart showing WISE colors similar to those of FSRQs \citep{dabrusco14}, and also tentatively classified as a QSO on the basis of its optical properties \citep{diserego94}.
\vspace{0.2cm}

\underline{G4Jy\,217:} was optically identified by Jauncey et al. (1978) and is a.k.a. PKS\,0202-76. It was classified as a QSO at $z$=0.38939 \citep[see also][]{jauncey78,danziger83,ho09}.
\vspace{0.2cm}

\underline{G4Jy\,219:} is classified as a radio galaxy with a photometric redshift estimate of $z$=0.45 belonging to the MS4 sample \citep{burgess06b}. In our analysis we used the radio map at 14.9 GHz to clearly identify the optical counterpart.
\vspace{0.2cm}

\underline{G4Jy\,227:} comparing the VLASS observation and the optical finding chart used in our analysis to those available in the literature at nominal frequency of 5 GHz \citep{reid99}, we confirmed the location of the host galaxy. G4Jy\,227 shows mid-IR colors typical of FSRQs \citep{dabrusco19} but does not have spectroscopic information.
\vspace{0.2cm}

\underline{G4Jy\,238:} also known as 3C\,62, is a HERG \citep[][]{hine79} with a classical FR\,II radio morphology with a $z$=0.13784 \citep{tadhunter93}, and also, a hard X-ray counterpart \citep[see e.g.,][]{koss17,kosiba22}.
\vspace{0.2cm}

\underline{G4Jy\,241:} shows a flat radio spectrum since it was selected as part of the Combined Radio All-Sky Targeted Eight GHz Survey \citep[a.k.a. CRATES\,J021645.12-474908.9;][]{healey07}. According to the literature this is a giant radio galaxy \citep[$\sim$1\,Mpc size;][]{christiansen77} hosted in an elliptical galaxy at $z$=0.06427 that is harbored in a galaxy-rich large-scale environment \citep{danziger83,zirbel97,jones09} and located in the direction of the Abell\,S\,239 galaxy cluster \citep{robertson90} that lies at similar distance,  being at $z$=0.0635 \citep{abell58,abell89}.
\vspace{0.2cm}

\underline{G4Jy\,247:} with no associated counterparts in the MRC, PKSCAT and PMN samples, has a counterpart at 20 GHz \citep[AT20G\,J021902-362607;][]{murphy10} and it also appears to be hosted in an elliptical galaxy \citep[see also][]{mahony11} clearly visible in both the mid-IR and in the optical images with a $z$=0.48881 \citep{jones09}.
\vspace{0.2cm}

\underline{G4Jy\,249:} is associated with PKS\,0219-706 (a.k.a. MRC\,0219-706 and PMN\,J0220-7022). This source was tentatively associated with an elliptical galaxy having a photometric redshift estimate of $z$=0.4 belonging to the MS4 sample \citep{burgess06b}, but according to our analysis, it has an IDF=3.0 since there are too many sources close to radio intensity peak that prevent us to claim the location of the host galaxy. 
\vspace{0.2cm}

\underline{G4Jy\,257:} as the previous source G4Jy\,249, is a radio galaxy with several counterparts at radio frequencies and a photometric redshift estimate of $z=$1.27 belonging to the MS4 sample \citep{burgess06b}.
\vspace{0.2cm}

\underline{G4Jy\,260:} is a lobe dominated radio QSO with a $z$=0.23224 \citep{best99,jones09} with mid-IR colors similar to those of $\gamma$-ray emitting QSOs \citep{dabrusco19}, and with a hard X-ray counterpart \citep{cusumano10,baumgartner13,koss17}.
\vspace{0.2cm}

\underline{G4Jy\,280:} is also associated with PKS\,0235-19 and is classified as a broad line radio galaxy with a classical FR\,II radio structure \citep{ramos11} at $z$=0.62 \citep{tadhunter93,best99}, with mid-IR colors of $\gamma$-ray blazars \citep{dabrusco14,dabrusco19}. Despite the fact that the radio core is not detected in the high resolution JVLA radio map at 5 GHz \citep{reid99}, its radio morphology confirms our assigned optical counterparts and association.
\vspace{0.2cm}

\underline{G4Jy\,290:} has only a photometric redshift estimate of $z$=0.72 \citep{burgess06b} but no clear optical identification.
\vspace{0.2cm}

\underline{G4Jy\,293:} is also associated with the MS4 radio galaxy MRC\,0245-558 with a photometric redshift estimate of $z$=0.82 \citep{burgess06b}, however the presence of several nearby companions, coupled with the lack of spectroscopic information, prevents us to positively assign an optical counterpart.
\vspace{0.2cm}

\underline{G4Jy\,301:} is a QSO located at $z$=1.002 \citep{murdoch84} with mid-IR colors similar to those of $\gamma$-ray detected FSRQs \citep{dabrusco19}.
\vspace{0.2cm}

\underline{G4Jy\,304:} is a CSS radio galaxy located at $z$=0.56288 \citep{tadhunter93,holt08} belonging to the 2Jy sample \citep{wall85,morganti93} and also known as PKS\,0252-71.
\vspace{0.2cm}

\underline{G4Jy\,312:} is associated with PKS\,0254-23 and it is classified as a radio galaxy located at $z=$0.509 \citep{mccarthy96b} showing a classical FR\,II radio morphology \citep[see also][]{kapahi98,best99}.
\vspace{0.2cm}

\underline{G4Jy\,326:} is a typical FR\,II radio galaxy with a $z$=0.268 \citep{mccarthy96b}.
\vspace{0.2cm}

\underline{G4Jy\,335:} is a high redshift radio galaxy at $z$=1.769 \citep{best99}.
\vspace{0.2cm}

\underline{G4Jy\,347:} is a 2.5\,Mpc giant radio galaxy \citep[a.k.a.MRC\,0319-454, PMN\,J0321-4510 and MSH\,03-43; see e.g.,][for a recent analysis]{burgess06b,malarecki15} located within a galaxy filament of the Horologium-Reticulum supercluster \citep{fleenor05}. The host galaxy \citep{bryant00} is located close to the north-eastern radio lobe, the only part of its radio structure reported in our finding chart, and it is associated with the optical source ESO\,248-G-10 having a weak counterpart at 20 GHz \citep{saripalli94}. The redshift of the host galaxy is $z=$0.0622 \citep{safouris09} and its radio structure in the north-eastern side goes through an environment having higher galaxy density than the southern radio structure, indicating the presence of a surrounding group of galaxies. There are also several galaxy clusters in its vicinity \citep{fleenor06}, namely: S0345, Abell 3111, Abell 3112 and APMCC369 at $z$=0.071, 0.078, 0.075 and 0.075, respectively.
\vspace{0.2cm}

\underline{G4Jy\,350:} is a large radio galaxy with the south-eastern radio lobe located at the same position of PKS\,0352-88 (a.k.a. MRC\,0352-884 and SUMSS\,J032359-881618), at $\sim$25\,\arcsec from the G4Jy position, as marked in the finding chart. A recent MeerKat observation revealed that the position of the mid-IR counterpart is consistent with that of WISE J032259.32-881600.4 \citep[][]{sejake22}.
\vspace{0.2cm}

\underline{G4Jy\,373:} is a radio galaxy with a typical FR\,II radio structure with a $z$=0.1126 \citep{carter83} behind the Fornax Cluster at an angular separation of $\sim$5\,\arcmin from NGC\,1399 (a.k.a. PKS\,0336-35) the Fornax BCG \citep[see e.g.,][]{carter83,killeen88a,hilker99}. 
\vspace{0.2cm}

\underline{G4Jy\,381:} is a FR\,II radio galaxy with a redshift estimate of $z$=0.0535 \citep{scarpa96,drinkwater01}, in the foreground of the galaxy cluster Abell\,3165 \citep{abell58,abell89,robertson90} but not related to it.
\vspace{0.2cm}

\underline{G4Jy\,386:} is also known as PKS\,0349-27 and PMN\,J0351-2744 and it was already detected as extended at 408 MHz \citep{schilizzi75a,schilizzi75b} as well as at higher frequencies \citep{reid99}. It was then optically identified with the same counterpart associated in both the G4Jy at mid-IR frequencies and in our analysis being \citep{bolton65} classified as a radio galaxy at $z=$0.6569 \citep{searle68,jones09}. This radio source shows extended emission in the optical band with remarkable features, including an extended narrow line region and bridges connecting G4Jy\,386 to two companion galaxies \citep{inskip10,ramos13}. These bridges are interpreted as due to tidal interaction with neighbor galaxies and/or mergers \citep{danziger84,tadhunter89}. X-ray emission was also detected between the lobes and, with an offset of a few arcsec, on the location of the northern hotspot \citep{mingo17}.
\vspace{0.2cm}

\underline{G4Jy\,387:} is a lobe dominated QSO at $z$=0.3245 \citep{lanzetta95,marziani96}, also known as 3C\,95 and PMN\,J0351-1429, inhabiting a galaxy-rich large-scale environment \citep[see e.g.,][]{hutchings96}. 
\vspace{0.2cm}

\underline{G4Jy\,392:} is a QSO (a.k.a. 3C\,94) showing a double radio structure and with a $z$=0.96354 \citep{lynds67,best99,ahn12}, recently included in the list of giant radio quasars having the size of the extended radio emission larger than 0.7 Mpc \citep{kuzmicz21}.
\vspace{0.2cm}

\underline{G4Jy\,404:} a radio galaxy at $z=$0.584 \citep{best99}.
\vspace{0.2cm}

\underline{G4Jy\,411:} is a well known $\gamma$-ray emitting blazar \citep[see e.g.,][and references therein]{acero15} known to emit also in the hard X-rays \citep[see e.g.,][]{cusumano10,koss17}, belonging to the class of FSRQs \citep[a.k.a. BZQ\,J0405-1308 as listed in the Roma-BZCAT][]{massaro09,massaro15b}, at $z$=0.57055 \citep{lynds67,marziani96}.
\vspace{0.2cm}

\underline{G4Jy\,415:} is one of the most luminous QSO at $z<1$ \citep[see e.g.,][]{punsly16}, optically identified by Hunstead (1971b) with a measured $z=$0.5731 \citep{kinman67,bechtold02,decarli10,johnson18}. G4Jy\,415 (a.k.a. PKS\,0405-12) shows X-ray emission arising from the northern hotspot \citep{sambruna04}, thus being included in the XJET database \citep{massaro11a}, and it is also listed in the Roma-BZCAT \citep{massaro15b} classified as blazar of uncertain type (i.e., BZU\,J0407-1211). Recent MUSE observations revealed the presence of six spatially extended line-emitting nebulae in the galaxy group where it is harbored, suggesting a connection between large-scale gas streams and the nuclear activity \citep{johnson18}. It also shows the detection of a narrow filament extending toward the QSO consistent with a cool intragroup medium filament similar to those occurring in cool-core galaxy clusters \citep[see e.g.,][]{mcdonald10}.
\vspace{0.2cm}

\underline{G4Jy\,416:} was originally identified with an optical counterpart by Hunstead (1971b) and then classified as a Gigahertz Peaked-Spectrum (GPS) radio sources \citep[see e.g.,][]{odea91,callingham17}. It has a redshift measurement of $z$=0.962 \citep{labiano07} { and it is also known as PKS\,0408-65 \citep{bolton79}}.
\vspace{0.2cm}

\underline{G4Jy\,417:} does not have an optical counterpart in the DSS image we retrieved from the archive, but it was identified with the same mid-IR counterpart listed in the G4Jy catalog in the literature \citep{hunstead71b,alvarez93}. However, we did not report any optical position and/or magnitude and we assigned it an IDF=4.1. G4Jy\,417 (a.k.a. PKS\,0409-752) also belongs to the MS4 catalog and it is classified as a narrow line radio galaxy with a typical FR\,II radio morphology at $z=$0.694 \citep{alvarez93,tadhunter93,diserego94}. It is harbored in a galaxy-rich environment \citep{ramos11} and shows evidence for a young stellar population \citep{holt07} and a far infrared excess \citep{dicken09b}.
\vspace{0.2cm}

\underline{G4Jy\,427:} has the same optical counterpart associated in the MS4 catalog \citep{burgess06b} where a photometric redshift estimate of $z$=0.42 is also reported. In the optical image used in our analysis we found a relatively brighter star in the south-western direction and several nearby companion galaxies of similar intensity suggesting that G4Jy\,427 could lie in a galaxy-rich environment. 
\vspace{0.2cm}

\underline{G4Jy\,436:} is also known as PKS\,0413-21, a core dominated quasar at $z$=0.808 \citep{wilkes86,best99} showing mid-IR colors similar to those of $\gamma$-ray blazars \citep{dabrusco14,dabrusco19}, and having a radio jet detected in the X-rays \citep{marshall05,massaro11a}.
\vspace{0.2cm}

\underline{G4Jy\,446:} is optically identified with a galaxy \citep{savage76} and has a photometric redshift estimate obtained thanks to the MS4 analysis that places the radio source at $z$=0.81 \citep{burgess06b}. In our analysis we could not clearly identify the radio core position and its optical counterpart. The mid-IR counterpart was not found in the G4Jy catalog \citep{white20b}. G4Jy\,436 also shows a CSS radio core \citep{randall11}
\vspace{0.2cm}

\underline{G4Jy\,453:} is a giant radio source already known since the MRC release \citep{jones92}, but the lack of high resolution image prevented us to verify the position of the host galaxy and if the redshift estimate reported in the literature is correct  \citep{kuzmicz21}.
\vspace{0.2cm}

\underline{G4Jy\,462:} is one of the dumb-bell FR\,I radio galaxies in the southern hemisphere \citep{ekers69,mcadam88,morganti93}, also known as IC\,2082 and PKS\,0427-53, with a twin tail \citep{carter81,lilly87}, and lying in the nearby galaxy cluster Abell\,S\,463. It has a redshift measurement of $z$=0.03931 \citep[see e.g.,][]{raimann05} showing weak emission lines in its optical spectrum. It also belongs to both the sample of local radio galaxies detected at 20 GHz \citep{sadler14}. We adopted here the same optical identification provided in the literature \citep[see also][]{jones92,burgess06b,white20b}. However, the lack of high resolution radio maps prevented us to classify this radio source as a wide-angle tail (WAT) radio galaxy \citep{burns81,owen76,odonoghue90,odonoghue93,sakelliou00,missaglia19}.
\vspace{0.2cm}

\underline{G4Jy\,492:} is a FR\,II radio galaxy at $z$=0.147 \citep{tadhunter93,diserego94,best99} with several nearby companion galaxies \citep{ramos13} and surrounding diffuse X-ray emission. { G4Jy\,492, also belonging to the 2\,Jy sample \citep[see e.g.,][]{wall85,morganti93}, has been recently observed in the X-rays and both its radio core and the northern hotspot were detected \citep{mingo17}. It was also detected in hard X-ray band \citep[see e.g.,][]{cusumano10,oh18}.}
\vspace{0.2cm}

\underline{G4Jy\,506:} was associated with the same optical counterpart in the literature \citep{bolton68,hunstead71b}. It shows a classical FR\,II radio structure. It is listed in the MS4 catalog with a photometric redshift estimate of $z$=0.22 \citep{burgess06b}.
\vspace{0.2cm}

\underline{G4Jy\,507:} is also known as NGC\,1692 a radio galaxy at $z$=0.035364 \citep{allison14}, optically identified in the literature \citep{bolton65,burbidge72,wills73,mccarthy96b,best99}. According to the optical image used in our analysis its entire radio structure lies within the brightness profiles of its host galaxy and it has two nearby companion galaxies, probably harbored in a galaxy-rich environment \citep{miller99}.
\vspace{0.2cm}

\underline{G4Jy\,510:} is a nearby QSO at $z$=0.533 \citep{wright79,henriksen91,bechtold02} showing MgII absorption system \citep{tytler87} and giant optical nebulae surrounding it \citep{helton21}. It shows a lobe dominated radio structure \citep{reid99} also listed in the CRATES catalog \citep{healey07}.
\vspace{0.2cm}

\begin{figure*}[!th]
\begin{center}
\includegraphics[height=6.6cm,width=16.cm,angle=0]{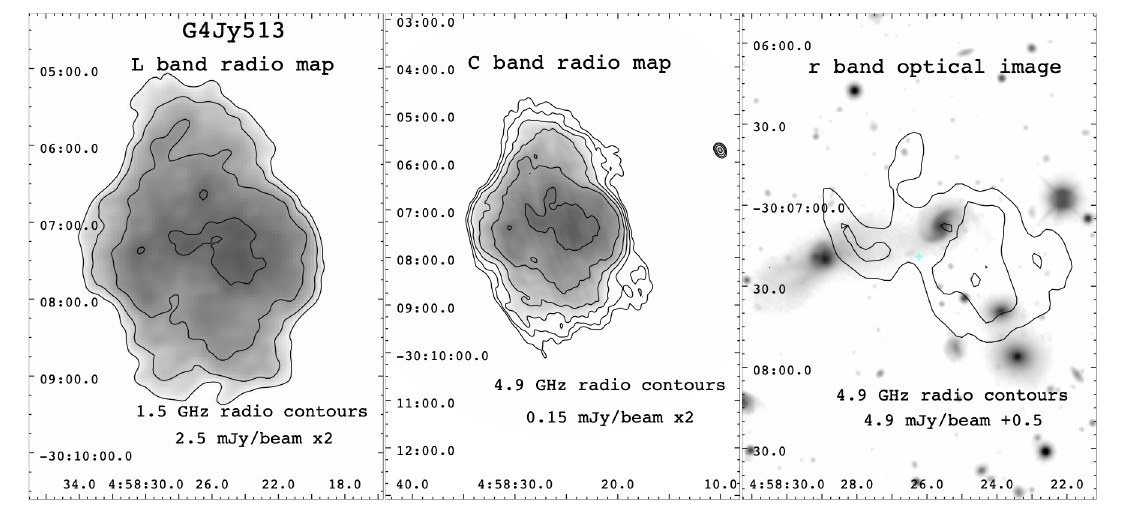}
\end{center}
\caption{Left and central panels) Radio maps of G4Jy\,513 at 1.4\,GHz and 4.9\,GHz, respectively, retrieved from the NVAS. The frequency of each radio map from which radio contours were drawn is reported together with the intensity of the first level and the binning factor indicating how they increase. Right panel) The $r$-band optical image collected from one of the surveys used in our analysis with three levels of radio contours drawn from the radio map at 4.9\,GHz starting at 4.9\,mJy and increasing by 0.5\,mJy. No clear detection of the radio core and of a potential optical counterpart is reported for this radio source showing diffuse radio emission.}
\label{fig:G4Jy513zoom}
\end{figure*}

\underline{G4Jy\,513:} is not optically identified but there is an optical interacting pair of galaxies within the highest radio contour drawn in our finding chart. We found two archival radio maps of G4Jy\,513, as reported in Figure~\ref{fig:G4Jy513zoom} in comparison with the archival $r$-band optical image, showing diffuse radio emission at both 1.4\,GHz and 4.9\,GHz but lacking a clear detection of its radio core. In this case the nearby radio source known as PKS\,0456-301 could be potentially associated with G4Jy\,513 thus suggesting the position of its optical counterpart as shown in the finding chart. G4Jy\,513 appear also associated with the galaxy cluster Abell\,3297 \citep[see e.g.,][and references therein]{robertson90}, even if spectroscopic confirmation is needed.
\vspace{0.2cm}

\underline{G4Jy\,517:} is a giant radio galaxy \citep[a.k.a. 0503-286 and MSH 05-22]{saripalli86,subrahmanya86,ishwara99,kuzmicz18} with a size of $\sim$1\,Mpc, located at $z$=0.0381 \citep{menzies89,fouque90,dacosta91,jones09}. We adopted the same association listed in the literature \citep{saripalli86,white20a} where 6dF\,J0505492-283519 is the host galaxy of the radio core, and inhabiting a small group \citep{subrahmanya08} with a  galaxy overdensity and a filamentary large-scale radio structure indicating a relatively low-density ICM environment.
\vspace{0.2cm}

\underline{G4Jy\,518:} is a FSRQ, also listed in the Roma-BZCAT \citep{massaro15b} as BZQ\,J0506-6109 with a $z$=1.093 \citep{wright77,bechtold02}. It also belongs to the MS4 catalog \citep{burgess06a} labelled as MRC\,0506-612 \citep{large81}, and it is a $\gamma$-ray emitting blazar \citep[see e.g.,][]{acero15,dabrusco19}.
\vspace{0.2cm}

\underline{G4Jy\,524:} is a radio galaxy (a.k.a. MRC\,0508-187 and TXS\,0508-187) with a lobe dominated radio structure. The optical image shows the presence of several nearby companion galaxies. G4Jy\,524 is also listed in a sample of ultra steep spectrum radio sources \citep{debreuck00}.
\vspace{0.2cm}

\underline{G4Jy\,530:} shows a large double-lobed radio morphology. The lacks of a mid-IR association and the resolution of radio maps used in our analysis did not permit us to locate the host galaxy in the optical image. However we tentatively associated this radio source with the narrow line radio galaxy PKS\,0511-48 (05:12:47.22, -48:24:16.4 in J2000) lying in the center of the radio structure and having a redshift estimate of $z$=0.30638 \citep{eracleous03,eracleous04}.
\vspace{0.2cm}

\underline{G4Jy\,531:} is listed in the MS4 sample \citep[a.k.a. MRC\,0511-305;][]{burgess06a} and it is classified as an asymmetric giant radio galaxy at $z=$0.05764 \citep{jones09}.
\vspace{0.2cm}

\underline{G4Jy\,538:} is a compact radio galaxy \citep{mccarthy96b} also known as PKS\,0519-20, with a $z=$1.086 \citep{best99} with a peaked spectrum determined thanks to the GLEAM observations \citep{callingham17}.
\vspace{0.2cm}

\underline{G4Jy\,540:} is a well known $\gamma$-ray emitting blazar \citep[a.k.a. PKS\,0521-36][]{bolton65,hunstead71b,danziger79a,acero15} at $z$=0.05655 \citep{sbarufatti06} listed as BZU\,J0522-3627 in the Roma-BZCAT \citep{massaro15b} with an X-ray emitting jet also detected by \chn\ \citep{birkinshaw02,massaro11a}.
\vspace{0.2cm}

\underline{G4Jy\,563:} is optically identified with a galaxy \citep{hunstead71a,bolton77} being also listed in the MS4 sample \citep{burgess06a}, with a photometric redshift estimate of $z$=0.184 \citep{burgess06b}.
\vspace{0.2cm}

\underline{G4Jy\,580:} has a mid-IR counterpart associated in the G4Jy that is the same listed in the MS4 sample \citep{burgess06a,burgess06b} as discussed in White et al. (2020b). However the presence of several optical sources around the radio intensity peak, marked in the finding chart,  coupled with relatively poor angular resolution of the radio map available, prevented us to claim a firm identification.
\vspace{0.2cm}

\underline{G4Jy\,590:} could be associated with PKS\,0601-34 (06:03:11.64, -34:26:45.1s in J2000) since, as reported in our finding chart, its position could be ``confused'' with that of the relatively brighter southern hotspot. In the MS4 sample is classified as a radio galaxy having a photometric $z$=0.58 \citep{burgess06b}.
\vspace{0.2cm}

\underline{G4Jy\,605:} is a double lobed radio galaxy optically identified in the literature \citep{bolton65,burbidge72}, and it has a close companion galaxy (i.e., being classifiable as a galaxy pair) clearly visible in the optical image.
\vspace{0.2cm}

\underline{G4Jy\,607:} is a radio galaxy at $z$=0.051 \citep{tritton72,wills04} having a flat radio spectrum \citep{healey07}.
\vspace{0.2cm}

\underline{G4Jy\,611:} is a LERG hosted in a dumb-bell galaxy \citep{frank13,ramos13,ineson15} located at the center of the Abell\,3391 galaxy cluster \citep{abell58,abell89} at $z$=0.054 \citep{danziger83,landt02} showing a classical WAT radio structure \citep{morganti99}. Its optical images show signatures of galaxy interactions with nearby companions \citep{ramos11}, while in the X-rays the ICM emission decreases in intensity in the region occupied by the northern radio lobe, tentatively indicating the possible presence of an X-ray cavity \citep{mingo17}.
\vspace{0.2cm}

\underline{G4Jy\,613:} is the BCG of the galaxy cluster Abell\,3395 \citep{abell58,abell89,brown91} with a $z$=0.051976 \citep{cava09}, being also detected in the X-rays \citep[see e.g.,][]{sun09} 
\vspace{0.2cm}

\underline{G4Jy\,614:} lies close to the galaxy cluster Abell\,3392 \citep{abell58,abell89,quintana95a,ebeling96}, originally detected in the X-rays by Trussoni et al. (1999) and then confirmed by Ramos Almeida et al. (2013), Ineson et al. (2015) and Mingo et al. (2017). The radio source is optically classified as a LERG, and it has a redshift estimate of $z$=0.054855 \citep{tadhunter93,wills04,jones09}. There are also several companion galaxies clearly detected in the optical image used to carry out our analysis. It is listed in both the MS4 sample \citep{burgess06a} and in the CRATES catalog \citep{healey07} as well as in the 2\,Jy one \citep[see e.g.,][]{wall85,morganti93,mingo14}. 
\vspace{0.2cm}

\underline{G4Jy\,618:} is the $\gamma$-ray FSRQ known as BZQ\,J0635-7516 \citep{massaro15b} with a $z$=0.653 \citep{hunstead78,bechtold02,danforth16} and with a relatively compact radio core \citep{jauncey89,morganti93}.
\vspace{0.2cm}

\underline{G4Jy\,619:} is a radio galaxy at $z$=0.055198 \citep{storchi96,jones09} also detected in the hard X-rays \citep{cusumano10,oh18}. It is listed in the literature as a giant radio galaxy \citep{danziger78,kronberg04,kuzmicz21} and, at larger scales, it shows a classical FR\,II radio morphology.
\vspace{0.2cm}

\underline{G4Jy\,639:} is also known as PKS\,0700-47 and it is listed in the MS4 sample classified as a radio galaxy and having a photometric redshift estimate of $z$=0.86 \citep{burgess06b}.
\vspace{0.2cm}

\underline{G4Jy\,642:} lies in a crowded optical field and, as for the previous radio source, it is listed in the MS4 sample \citep[a.k.a. MRC\,0704-427][]{burgess06b} with a photometric redshift estimate of $z=$1.33. 
\vspace{0.2cm}

\underline{G4Jy\,644:} can be considered a giant radio galaxy even if its extension is below the 1\,Mpc threshold \citep{ishwara99,malarecki15,proctor16,kuzmicz21}. It is also a double-double radio galaxy candidate, where the inner radio structure is misaligned with respect to the outer one \citep{jones92,bruni20}. G4Jy\,644 emits in the hard X-rays \citep[a.k.a. PBC\,J0709.2-3601][]{cusumano10,koss17,bruni21}, and it lies at $z$=0.11 \citep{parisi14}.
\vspace{0.2cm}

\underline{G4Jy\,651:} is a radio galaxy with a $z$=0.031358 \citep{jones09,allison14}, and its radio morphology on larger scales than those visible in our finding chart reveals a WAT structure. 
\vspace{0.2cm}

\underline{G4Jy\,653:} is a radio source (a.k.a. PKS\,0719-55) inhabiting a moderately rich environment \citep{zirbel96} and being listed in the MS4 sample with a photometric redshift estimate of $z$=0.22531 \citep{burgess06b}.
\vspace{0.2cm}

\underline{G4Jy\,672:} is a flat spectrum radio quasar, showing an ultraluminous accretion disk and a high kinetic luminosity jet \citep{punsly05}. It is listed in the Roma-BZCAT \citep[a.k.a. BZQ\,J0743-6726]{massaro09} as well as in the CRATES catalog \citep{healey07}, and it lies at $z=$1.512 \citep{begeron84,diserego94,bechtold02,jones09}. 
\vspace{0.2cm}

\underline{G4Jy\,680:} was optically identified by Schilizzi (1975) and it is a radio galaxy with a $z$=0.0699 \citep{danziger83}. G4Jy\,680 appears as a FR\,I radio source but at larger scales and at lower frequencies (e.g., 74\,MHz and 150 MHz) with respect to those we used to build radio contours over the optical image, its radio structure appears more extended.
\vspace{0.2cm}

\underline{G4Jy\,685:} is known as 3C\,195 and PKS\,0806-10, a FR\,II radio galaxy at $z$=0.10898 \citep{tadhunter93,jones09} listed in the 2Jy sample \citep{wall85,morganti93}. The detection of extended X-ray emission around the radio core \citep{ineson15,mingo17} coupled with optical signatures of galaxy interaction \citep{inskip10,ramos11} clearly indicates that G4Jy\,685 is harbored in a galaxy-rich large-scale environment. The southern radio hotspot is also detected in the X-rays \citep{mingo17}.
\vspace{0.2cm}

\underline{G4Jy\,706:} is a lobe dominated radio quasar at $z$=0.822 \citep{stickel93} with mid-IR colors similar to $\gamma$-ray blazars \citep{dabrusco14}
\vspace{0.2cm}

\underline{G4Jy\,718:} belongs to the MS4 sample (a.k.a. MRC\,0842-835) with a photometric redshift estimate of $z$=0.82 \citep{burgess06b}. The lack of optical detection in the finding chart at the location of the mid-IR associated source prevented us to assign this source an IDF=1.0. 
\vspace{0.2cm}

\underline{G4Jy\,721:} is the FR\,II radio galaxy 3C\,206, not listed in the revised 3CR catalog \citep{spinrad85}, with a $z$=0.19787 \citep{ho09}, and harbored in a galaxy cluster \citep{yee83,ellingson87,ellingson89,yates89}. It is also detected in the hard X-rays \citep[see e.g.,][]{ajello08a,ajello08b,oh18,kang20}.
\vspace{0.2cm}

\underline{G4Jy\,723:} is a radio quasar at $z$=0.521 \citep{browne77,hunstead78} listed in the 2Jy sample \citep[see e.g.,][]{morganti93,burgess06a} with mid-IR colors of $\gamma$-ray blazars \citep{dabrusco19}.
\vspace{0.2cm}

\underline{G4Jy\,730:} is a high redshift radio galaxy \citep[a.k.a. MRC\,0850-206][]{large81,debreuck00}, with a FR\,II radio morphology, with a $z$=1.337 \citep{best99}, for which the presence of a bright star in the north-eastern direction prevent the detection of its mid-IR counterpart. 
\vspace{0.2cm}

\underline{G4Jy\,734:} is a high redshift radio galaxy at $z$=1.665 optically identified in the literature \citep{best99,debreuck00} appearing to inhabit a large-scale environment with high galaxy density \citep{wylezalek13}. The lack of an optical detection in our finding chart prevented us to assign this source an IDF=1.0.
\vspace{0.2cm}

\underline{G4Jy\,747:} is a FR\,II radio galaxy, optically classified as Narrow Line Radio Galaxy \citep[NLRG;][]{ramos11}, at $z$=0.305 \citep{tadhunter93}, and listed in the 2\,Jy sample \citep{wall85,morganti97} with mid-IR colors of $\gamma$-ray blazars \citep{dabrusco19}. G4Jy\,747 lies in a relatively crowded field with several nearby companion galaxies, and appears as a double system, including the radio galaxy nucleus and a faint component a few kpc in the south-western direction \citep{ramos11}. Its morphological classification is the same attributed from the near infrared investigation presented by Inskip et al. (2010).
\vspace{0.2cm}

\underline{G4Jy\,796:} is a giant, double-lobed, radio galaxy \citep[a.k.a MRC\,0947-249][]{ishwara01} for which diffuse X-ray emission has been detected, arising from its extended radio structure \citep{laskar10}.  A redshift estimate of $z$=0.854 is reported in the literature \citep{mccarthy96b,kapahi98} but unconfirmed.
\vspace{0.2cm}

\underline{G4Jy\,818:} is a candidate double-double radio galaxy (a.k.a. MRC\,1002-215) with an unconfirmed redshift estimate of $z$=0.59 reported in the literature \citep{mccarthy96b}. Despite the presence of nearby galaxies on the western side, the host galaxy was not detected in our optical image.
\vspace{0.2cm}

\underline{G4Jy\,835:} is a CSS radio source \citep{prestage83} with a $z$=1.346 \citep{diserego94}
\vspace{0.2cm}

\underline{G4Jy\,836:} is a FR\,II radio galaxy listed in the MS4 sample with a photometric redshift estimate of $z$=0.7 \citep{burgess06b}. The presence of several optical sources prevented us from associating the radio core with its optical counterpart.
\vspace{0.2cm}

\underline{G4Jy\,837:} is an extremely powerful (i.e., with a luminosity higher than all 3C radio sources at the same redshift with the exception of 3C\,196) FR\,II radio galaxy \citep{punsly06} at $z$=1.28 \citep{murdoch84,stickel93,decarli10} with mid-IR colors of $\gamma$-ray blazars \citep{dabrusco19}.
\vspace{0.2cm}

\underline{G4Jy\,854:} is listed in the MS4 catalog as a radio galaxy with a photometric redshift estimate $z$=0.5 \citep{burgess06b}. It has an ultra steep radio spectrum \citep{debreuck00}. Both the mid-IR and the optical counterparts are not detected according to our analysis.
\vspace{0.2cm}

\underline{G4Jy\,876:} is a lobe dominated radio QSO for which the high resolution radio map used in our investigation allowed us to precisely locate the position of its host galaxy, being also coincident with that at other radio frequencies \citep{bolton79,large81,griffith93}. G4Jy\,876 seems to be associated with 3C\,246 \citep{edge59} with a $z$=0.345296 \citep{kinman67,veron88,lanzetta93,jones09,shi14}, and it resides in a galaxy-rich large-scale environment \citep{hintzen83,ellingson88,ellingson91,hutchings96}.
\vspace{0.2cm}

\underline{G4Jy\,894:} is a radio galaxy with an unconfirmed redshift estimate of $z$=0.59 reported in the literature \citep{mccarthy96b,best99}. 
\vspace{0.2cm}

\underline{G4Jy\,917:} is an unusual radio galaxy showing a circularly bent tail. It lies at $z$=0.033753 \citep[see e.g.,][]{burbidge72,tritton72,sandage78,allison14}, and it is harbored in the galaxy cluster Abell\,S\,665 \citep{abell58,abell89}.
\vspace{0.2cm}

\underline{G4Jy\,926:} is a classical lobe dominated radio source being associated with the nearby object PKS\,1131-19. We could not detect any counterpart in the optical image used to draw our finding chart at this location. However, according to Wills et al. (1974) there is an optical source showing Ca H and K broad and shallow absorption lines, but the finding chart they refer to, given by Moseley, Brooks, and Douglas (1970), seems to point a different target at more than 20\,\arcsec angular separation in the eastern direction from the radio position. 
\vspace{0.2cm}

\underline{G4Jy\,927:} is a lobe dominated QSO (a.k.a. PKS\,1131-17) at $z$=1.618 \citep{best99}.
\vspace{0.2cm}

\underline{G4Jy\,933:} is listed in the sample of Parkes half-Jansky flat-spectrum radio sources \citep{drinkwater97}, being a QSO with a radio jet extending up to 60 kpc in the north-west direction detected in the optical band \citep{uchiyama07} as well as in the X-rays \citep{sambruna02}, and with a $z$=0.55646 \citep{tadhunter93,drinkwater97,jones09}. A shorter tidal tail pointing to the west of the QSO is also detected in the optical band \citep{ramos11}. G4Jy\,933 is also classified as blazar of uncertain type in the Roma-BZCAT \citep[a.k.a. BZU\,J1139-1350][]{massaro15b}.
\vspace{0.2cm}

\underline{G4Jy\,934:} a FR\,II radio galaxy listed in the MS4 sample with a photometric redshift estimate of $z$=0.67 \citep{burgess06b}. The mid-IR associated source listed in the G4Jy catalog is the one marked in the finding chart and in agreement with the MS4 identification \citep{burgess06a}, which is based on the possible radio core detection proposed by Duncan \& Sproats (1992) as reported in White et al. (2020b). The MRC\,1136-320 radio source position is also cospatial with the associated mid-IR counterpart. The lack of any optical spectroscopic information prevented us to  associate it with its optical counterpart, also because the closest source to the radio centroid is the brightest one, as visible in our finding chart.   
\vspace{0.2cm}

\underline{G4Jy\,937:} is the well known Spiderweb Galaxy \citep[see e.g.,][]{pentericci97,pentericci98,pentericci02,carilli02,miley06} with a $z$=2.1585 \citep[see e.g.,][]{kuiper11}. 
\vspace{0.2cm}

\underline{G4Jy\,939:} is the brightest cluster galaxy (BGC) being located at the center of a galaxy cluster \citep{chapman00}, photometrically identified using the red sequence.
\vspace{0.2cm}

\underline{G4Jy\,952:} is a FR\,II radio galaxy listed in the MS4 sample with a photometric redshift estimate of $z$=1.35 \citep{burgess06b} having a mid-IR counterpart associated in the G4Jy catalog but lacking an optical correspondence.
\vspace{0.2cm}

\underline{G4Jy\,957:} is a radio galaxy with a $z$=0.117 \citep{danziger83} with several nearby optical and mid-IR sources.
\vspace{0.2cm}

\underline{G4Jy\,965:} is a core dominated QSO at $z$=0.258 \citep{jauncey78} with a compact steep spectrum radio core \citep{morganti97,reid99}.
\vspace{0.2cm}

\underline{G4Jy\,977:} is an Ultra Steep Spectrum radio source \citep[USS;][]{debreuck00,broderick07} for which optical follow up observations provided a redshift estimate of $z$=0.12 \citep{bryant09}, however the lack of the optical counterpart in our finding chart did not allow us to confirm this association.
\vspace{0.2cm}

\underline{G4Jy\,987:} is a candidate Hybrid Morphology Radio Source \citep[HyMoR: having a FR\,I radio morphology on one side and FR\,II on the other side of the radio core;][]{gopal00,cheung09} but in our finding chart we only focused on the northern side to highlight the host galaxy position. It lies at $z$=0.0874 \citep{danziger83,best99}.
\vspace{0.2cm}

\underline{G4Jy\,1009:} is a lobe dominated QSO at $z$=0.355209 \citep{wright79,eracleous03,jones09}.
\vspace{0.2cm}

\underline{G4Jy\,1034:} is the famous WAT hosted in the dumb-bell massive group of interacting elliptical galaxies NGC\,4782 and NGC\,4783 \citep[a.k.a. 3C\,278][]{borne84,borne88,desouza90,madejsky91,colina95} at $z$=0.015464 \citep{fairall92,quintana96}.
\vspace{0.2cm}

\underline{G4Jy\,1035:} is a radio galaxy at $z$=0.057353 \citep{melnick81,kaldare03} located in the central region of the Shapley Supercluster in the double system Abell\,S\,3528 and Abell\,3528A, with the BCG lying $\sim$4 kpc away from the the X-ray center, and having a tailed radio morphology coupled with steep radio spectra mainly due to a dense ICM of a pre-merging environment \citep{slee94,quintana95b,reid98,lopes18}.
\vspace{0.2cm}

\underline{G4Jy\,1038:} is one of the most famous $\gamma$-ray emitting QSO/blazar 3C\,279 \citep[see e.g.,][]{knight71,whitney71} with a $z$=0.5362 \citep[see e.g.,][]{marziani96}.
\vspace{0.2cm}

\underline{G4Jy\,1052:} is a NLRG with a CSS radio core (a.k.a. 1306-09) showing an inverted radio spectrum \citep{mukul20}. In the optical band G4Jy\,1052 shows a secondary nucleus being also harbored in a group of galaxies and it is undergoing interactions with other nearby sources hosted therein \citep{inskip10,ramos11,ramos13}, and with a $z=$0.46685 \citep{tadhunter93,holt08}.
\vspace{0.2cm}

\underline{G4Jy\,1071:} is a lobe dominated QSO at $z$=0.528 \citep{burbidge66,best99} with mid-IR colors of $\gamma$-ray blazars \citep{dabrusco19} and an MgII absorption system \citep{aldcroft94}.
\vspace{0.2cm}

\underline{G4Jy\,1080:} is the radio galaxy IC\,4296 \citep[see e.g.,][and references therein]{younis85,killeen86,killeen88b,smith00,wegner03,grossova19,condon21,grossova22}, the BCG of the Abell\,3565 galaxy cluster \citep{abell58,abell89}, a radio galaxy with a $z$=0.0125 \citep[see e.g.,][]{sandage78,efstathiou80}. 
\vspace{0.2cm}

\underline{G4Jy\,1083:} is a QSO with a WAT radio morphology \citep{hintzen84} with a $z$=0.625 \citep{burbidge66} and showing mid-IR colors of $\gamma$-ray blazars \cite{dabrusco14,dabrusco19}.
\vspace{0.2cm}

\underline{G4Jy\,1093:} is a radio galaxy at $z$=0.384 \citep{best99}.
\vspace{0.2cm}

\underline{G4Jy\,1129:} is a radio galaxy at $z$=1.094 \citep{best99}.
\vspace{0.2cm}

\underline{G4Jy\,1135:} is a radio galaxy (a.k.a. PKS\,1413-36) with a classical FR\,II morphology at large scales and with a $z$=0.07470 \citep{simpson93}, also emitting in the hard X-rays \citep{oh18}, and listed in the MS4 sample \citep{burgess06a}.
\vspace{0.2cm}

\underline{G4Jy\,1136:} is a FR\,II radio galaxy (a.k.a. PMN\,J1416-2146) at $z$=1.116 \citep{best00} for which the lack of an optical counterpart detected in our finding chart did not allow us to assign an IDF=1.0.
\vspace{0.2cm}

\underline{G4Jy\,1145:} is a radio galaxy showing a FR\,II radio structure, lying $z$=0.1195 \citep{burbidge67,grandi78,hunstead78,eracleous04},  and also emitting in the hard X-rays \citep{oh18,kosiba22}.
\vspace{0.2cm}

\underline{G4Jy\,1148:} is known as MRC\,1416-493 \citep{large81} and PKS\,1416-49 in the MS4 sample \citep{burgess06a}, classified as a radio galaxy with an intermediate FR\,I/II radio morphology at $z$=0.09142 \citep{simpson93,jones09}, and harbored in a non cool-core galaxy cluster \citep{worrall17} visible in the X-rays. G4Jy\,1148 also shows an X-ray cavity spatially associated with the north-eastern radio lobe and an X-ray excess associated with the south-western lobe, probably due to inverse-Compton emission.
\vspace{0.2cm}

\underline{G4Jy\,1152:} is a lobe dominated QSO at $z$=0.985 \citep{goncalves98} with mid-IR colors of $\gamma$-ray blazars \citep{dabrusco19}.
\vspace{0.2cm}

\underline{G4Jy\,1157:} is a lobe dominated QSO (a.k.a. PKS\,1421-38) for which the original spectroscopic observation identified only a single emission line, assuming this is due to MgII, it yields to a source redshift $z$=0.41 \citep{tritton71}, in agreement with subsequent analyses that refined the measurement of $z$=0.4068 \citep{veron88,eracleous03}.
\vspace{0.2cm}

\underline{G4Jy\,1158:} is the FSRQ PKS\,B1421-490 (a.k.a. PMN\,J1424-4913) with a one-sided jet and a unique knot detected in both the optical band and the X-rays \citep{gelbord05}. Marshall et al. (2011) reported a note claiming that there is an optical spectrum with a redshift estimate of $z$=0.662 but we could not retrieve this information from the literature and thus we marked it as uncertain.
\vspace{0.2cm}

\underline{G4Jy\,1161:} is listed in the equatorial sample of powerful radio sources \citep{best99} as a radio galaxy at $z$=1.632, however the lack of detection in the optical image we used in our analysis prevented us to confirm this association.
\vspace{0.2cm}

\underline{G4Jy\,1166:} is a lobe dominated QSO with a $z=$0.8295 \citep{bechtold02,jones09}
\vspace{0.2cm}

\underline{G4Jy\,1209:} a radio lobe dominated QSO with a $z$=0.942 \citep{schmidt66}.
\vspace{0.2cm}

\underline{G4Jy\,1225:} is a FSRQ listed in the Roma-BZCAT \citep[a.k.a. BZQ\,J1510-0543][]{massaro15b} with mid-IR colors of $\gamma$-ray blazars. In Peterson \& Bolton (1972) there is a reported redshift measurement of $z$=1.191, while the most recent optical spectrum place it at $z$=0.36 \citep{torrealba12}, thus we labelled it as uncertain in our analysis.
\vspace{0.2cm}

\underline{G4Jy\,1262:} is a radio galaxy listed in the MS4 sample (a.k.a. PMN\,J1530-4231) with a photometric redshift estimate of $z$=0.5 \citep{burgess06b}. 
\vspace{0.2cm}

\underline{G4Jy\,1276:} is a radio galaxy listed in the MS4 sample (a.k.a. PKS\,1540-73) with a photometric redshift estimate of $z$=0.68 \citep{burgess06b}.
\vspace{0.2cm}

\underline{G4Jy\,1279:} is listed in the literature as a double-double giant radio galaxy \citep{ishwara99,saripalli03,proctor16} with a $z$=0.1082 \citep{simpson93} with recurrent activity \citep{machalski10,kuzmicz17}, and a radio structure interpreted as due to the interaction of restarted jets with pre-existing relic cocoons \citep{safouris08}.
\vspace{0.2cm}

\underline{G4Jy\,1284:} is a HERG with a $z$=0.483 \citep{tadhunter93,hernan16}, and showing a double nucleus with a fainter companion $\sim$9 kpc to the south of the radio core \citep{inskip10,ramos11}.
\vspace{0.2cm}

\underline{G4Jy\,1301:} is a FR\,II radio galaxy at $z$=2.043 \citep{best99,venemans07}.
\vspace{0.2cm}

\underline{G4Jy\,1302:} is a radio galaxy at $z$=0.482 \citep{best99}, with a FR\,II radio structure. We adopted the same association proposed in the literature \citep{best99} and thus different from the mid-IR counterpart listed in the G4Jy catalog \citep{white20b}.
\vspace{0.2cm}

\underline{G4Jy\,1303:} is a FR\,II radio galaxy at $z$=0.109 \citep{best99}. 
\vspace{0.2cm}

\underline{G4Jy\,1330:} is a radio galaxy (a.k.a. PKS\,1621-11) at $z$=0.375 \citep{best99} having a bright star in the north-eastern direction visible in the mid-IR image that prevented us from assigning the mid-IR counterpart. 
\vspace{0.2cm}

\underline{G4Jy\,1336:} is a QSO with a compact radio structure with a $z$=1.124 \citep{jauncey84} and having mid-IR colors similar to $\gamma$-ray blazars \citep{dabrusco19}. It is also listed in the MS4 sample \citep[a.k.a. PKS\,1622-310][]{burgess06a}.
\vspace{0.2cm}

\underline{G4Jy\,1343:} is a nearby radio galaxy with a FR\,II morphology at $z$=0.166 \citep{best99}.
\vspace{0.2cm}

\underline{G4Jy\,1360:} is a LERG at $z$=0.0427 \citep{simpson93} in agreement with previous estimates present in the literature \citep{burbidge72,whiteoak72}, but not with the value of 0.024 reported in Danziger \& Goss (1983). It shows a FR\,II radio morphology in a high resolution map \citep{morganti93}.
\vspace{0.2cm}

\underline{G4Jy\,1365:} is a QSO with a $z$=0.799 having the same associated optical counterpart as in Best et al. (1999) but lacking a mid-IR detection probably due to the presence of a nearby bright source visible in our finding chart.
\vspace{0.2cm}

\underline{G4Jy\,1370:} a FR\,II LERG at $z$=0.236 \citep{best99}.
\vspace{0.2cm}

\underline{G4Jy\,1423:} is a FR\,II radio galaxy located at $z$=0.09846 \citep{tadhunter93} associated in our analysis thanks to the optical identification reported in the literature \citep{jauncey89,morganti93} being listed in the 2\,Jy sample \citep{wall85}.
\vspace{0.2cm}

\underline{G4Jy\,1432:} is a hard X-ray radio source \citep{baumgartner13,oh18} listed in the MS4 sample with a photometric redshift estimate of $z$=0.41 \citep{burgess06b}.
\vspace{0.2cm}

\underline{G4Jy\,1453:} is a radio galaxy listed in the MS4 sample (a.k.a. PKS\,1754-59 and AT20G\,J175906-594702) at the photometric redshift $z$=0.8 \citep{burgess06b} and optically identified in the literature \citep{hunstead71b}. The lack of a high resolution radio map prevented us confirming the host galaxy association.
\vspace{0.2cm}

\underline{G4Jy\,1472:} is the rare case of a a powerful radio-loud AGN in disk dominated galaxy with radio luminosity similar to powerful FR\,II radio galaxies \citep{morganti11}. G4Jy\,1472 (a.k.a. PKS\,1814-63 and PMN\,J1819-6345) is classified as a CSS radio source \citep[see e.g.,][and references therein]{morganti97} with a $z$=0.06412 \citep{thompson90,tadhunter93,holt08,morganti11}, a similar redshift estimate to that reported by Danziger \& Goss (1979) and Grandi (1983). It is also a hard X-ray source \citep{maselli10} listed in the Roma-BZCAT \citep[a.k.a. BZU\,J1819-6345 ][]{massaro15b}.
\vspace{0.2cm}

\underline{G4Jy\,1504:} is classified as a WAT (a.k.a. PKS\,1839-48) given the radio structure shown in Morganti et al. (1993) and Mingo et al. (2017) but classified, more recently, as a head-tailed radio galaxy by White et al. (2020b). The host galaxy identification adopted here is the same found in the literature and corresponding to an early-type galaxy at $z$=0.1108 \citep{simpson93,jones09,white20b} lacking strong optical emission lines \citep{tadhunter93}. G4Jy\,1504 is interacting with the two nearby companion galaxies, and it shows an apparent secondary nucleus located at $\sim$6 kpc in the south-eastern direction \citep{ramos11}. It also shows bright arc-like shell of $\sim$7 kpc in size, potentially due to gravitational lensing, in the north-western direction \citep{ramos11,ramos13}. The ICM emission around G4Jy\,1504 is clearly detected in the soft X-rays \citep{mingo17} and the radio source is also associated with a Fermi $\gamma$-ray object \citep{abdollahi20}.
\vspace{0.2cm}

\underline{G4Jy\,1513:} is a radio galaxy (a.k.a PKS\,1859-23) at $z$=1.430 \citep{best00}, and it resides in a region with several nearby companion galaxies. However, the lack of spectroscopic information about the nearby sources coupled with the lack of the counterpart associated with the radio emission at both mid-IR and optical frequencies did not allow us to confirm this association.
\vspace{0.2cm}

\underline{G4Jy\,1518:} also belongs to the equatorial sample of Best et al. (1999) where it is classified as a radio galaxy at $z$=0.226, as well as in the Molonglo Reference Catalogue of radio sources \citep[MRC\,1912-269][]{large81}.
\vspace{0.2cm}

\underline{G4Jy\,1532:} is a radio galaxy (a.k.a MRC\,1920-077 and TXS\,1920-077) listed in the sample of Best et al. (1999) with the optical counterpart correspondent to our association, but lacking a mid-IR correspondence. It has a redshift estimate of $z$=0.648.
\vspace{0.2cm}

\underline{G4Jy\,1555:} is a double-lobed radio galaxy \citep[see also][]{jones92} with a $z$=0.07507 \citep[a.k.a. 6dF\,J1933250-394021][]{scarpa96,jones09}, harbored in the Abell\,S\,820 galaxy cluster \citep{abell58,abell89,quintana95a}.
\vspace{0.2cm}

\underline{G4Jy\,1558:} is a Broad Line Radio Galaxy (BLRG; a.k.a. PKS\,1932-464), with a relatively broad and strong H$\alpha$ emission line, with a $z$=0.2307 \citep{tadhunter93,villar98,hernan16}, and, a FR\,II radio morphology \citep{morganti93}. G4Jy\,1558 shows a complex gas distribution resulting from the interaction with a nearby companion galaxy, and there is a knotty extended emission line nebula extending beyond the radio structure and the ionization cones, one of the largest ever detected around a radio galaxy at any redshift \citep{villar05}. The origin of the nebula is due to the presence of a star forming halo associated with the debris of the merger that triggered the activity. G4Jy\,1558 has sufficient luminosity at mid- to far-IR wavelengths to be classified as a luminous infrared galaxy \citep{inskip07}. The star formation structure can extend on the scale of a galaxy group, beyond the old stellar halo of the host galaxy \citep{villar05}. The gas in the emission line nebula is predominantly ionized by a mixture of AGN photoionization and emission from young stars \citep{inskip07,tadhunter11}. G4Jy\,1558 is a member of an interacting galaxy group which includes a highly disturbed starburst galaxy at a similar redshift, located at $\sim$100 kpc in the north-eastern direction \citep{inskip07}, and connected with G4Jy\,1558 by a series of arc-like irregular features up to $\sim$70 kpc distance from the galaxy centre \citep{ramos11,ramos13}.
\vspace{0.2cm}

\underline{G4Jy\,1562:} a radio galaxy listed in the MS4 sample with a photometric redshift estimate of $z$=1.92 \citep{burgess06b}.
\vspace{0.2cm}

\underline{G4Jy\,1565:} a QSO at $z$=0.452 \citep{tadhunter93,best99,hernan16} with mid-IR color of $\gamma$-ray blazars \citep{massaro12b,dabrusco14} and belonging to the 2\,Jy catalog \citep[a.k.a. PKS\,1938-15; see e.g.,][and references therein]{morganti93}.
\vspace{0.2cm}

\underline{G4Jy\,1569:} is a double-lobed radio galaxy belonging to the MS4 sample with a photometric redshift estimate of $z$=0.18 \citep{burgess06b} and showing an ultra steep radio spectrum \citep{debreuck00}. The radio source, a.k.a MRC\,1940-406, appear to be associated with the galaxy cluster Abell\,3646 \citep[see e.g.][]{abell89,robertson90}, even if this information requires optical spectroscopic observation to be confirmed.
\vspace{0.2cm}

\underline{G4Jy\,1587:} is also known as 3C\,404 \citep{edge59,bennett62}, with a counterpart at 20\,GHz \citep[a.k.a. AT20G\,J195611-073655;][]{murphy10} being classified as a radio galaxy at $z$=1.338 \citep{best00}. However the lack of mid-IR and optical counterparts in the images used in our analysis did not allow us to confirm the association found in the literature.
\vspace{0.2cm}

\underline{G4Jy\,1590:} is an LERG with a $z$=0.0581 \citep{simpson93} with a FR\,I radio morphology \citep{morganti93} harbored in a rich galaxy cluster at $z$=0.05845 \citep{stein96,ramos11,ineson15}, having the ICM clearly detected in the X-rays \citep{mingo17}.
\vspace{0.2cm}

\underline{G4Jy\,1605:} is the galaxy cluster relic of Abell\,3367 \citep{abell58,abell89} at $z$=0.055 \citep[see e.g.,][and references therein]{johnston08,owers09} located in the north-western direction from the head-tail radio-galaxy at its center, namely G4Jy\,1606, not belonging to our \cs\ catalog \citep[see][for additional details]{white20b}. 
\vspace{0.2cm}

\underline{G4Jy\,1613:} is the X-shaped radio source a.k.a. PKS\,2013-557 with a $z$=0.060629 with an optical spectrum showing the following emission lines: [OII]$\lambda$3727, [Ne III]$\lambda$3868, [OIII]$\lambda\lambda$4959,5007 \citep{tritton72,jones09}. It is a restarted giant radio galaxy \citep{malarecki15,kuzmicz18}, belonging to the sample selected as hard X-ray emitters \citep{cusumano10,oh18,ursini18,bruni19,bruni20}, and its X-shaped structure could be a signature of jet reorientation \citep{saripalli07}. Recent MeerKAT spectro-polarimetric observations revealed a double boomerang' radio structure at 1.28\,GHz mainly due to hydrodynamical backflows from the straight radio jets deflected by the large and oblique hot-gas halo of its host galaxy \citep{cotton20}. The radio source is also embedded in relatively faint cocoons with uniform brightness temperature and sharp edges due to subsonic expansion into the ambient intragroup medium \citep{cotton20}. The source is also listed in the MS4 and in the CRATES catalogs \citep[see][respectively]{burgess06a,healey07}.
\vspace{0.2cm}

\underline{G4Jy\,1618:} is an extended radio source \citep{jones92} located at $z$=0.352 \citep{buchanan06} listed in both the MS4 sample \citep{burgess06a} and in the catalogs of $\gamma$-ray blazar candidates \citep{dabrusco14,dabrusco19}.
\vspace{0.2cm}

\underline{G4Jy\,1626:} is a lobe dominated QSO at $z$=1.5 \citep{best99}. We did not detect its optical counterpart in the image collected from the surveys used in our analysis, thus in the finding chart we also report the radio position (i.e., 20:28:07.75, -15:21:21.7 in J2000) 
\vspace{0.2cm}

\underline{G4Jy\,1635:} with a $z$=0.13149 \citep[a.k.a 6dF\,J2033166-225317][]{hunstead78,jones09} is a radio galaxy living in a small group \citep{allington93}.
\vspace{0.2cm}

\underline{G4Jy\,1643:} is a FR\,I radio galaxy at $z$=0.0406 \citep{tritton72,whiteoak72} belonging to the galaxy cluster Abell\,S\,894 \citep{abell58,green88,abell89}.
\vspace{0.2cm}

\underline{G4Jy\,1646:} is an USS radio source located at $z$=1.464 \citep[a.k.a. MP\,J2045-6018]{debreuck01}, but lacking an optical counterpart in our finding chart, thus being labelled with a question mark.
\vspace{0.2cm}

\underline{G4Jy\,1664:} is a radio galaxy lying $z$=0.15662 \citep[a.k.a. 6dF\,J2056043-195635][]{stickel94,jones09} originally derived from stellar absorption features. It shows a GPS radio core \citep{odea91}. 
\vspace{0.2cm}

\underline{G4Jy\,1671:} (a.k.a. NGC\,6998 and PMN\,J2101-2802) is a winged radio galaxy in the low radio frequency images, showing a one-sided jet feeding the southern lobe. It lies at $z$=0.039444 \citep{tritton72,tadhunter93,jones09}, and it belongs to the galaxy cluster Abell\,3733 \citep{abell58,abell89,stein96,katgert98,smith00,kaya19}.
\vspace{0.2cm}

\begin{figure*}[!th]
\begin{center}
\includegraphics[height=12.cm,width=16.cm,angle=0]{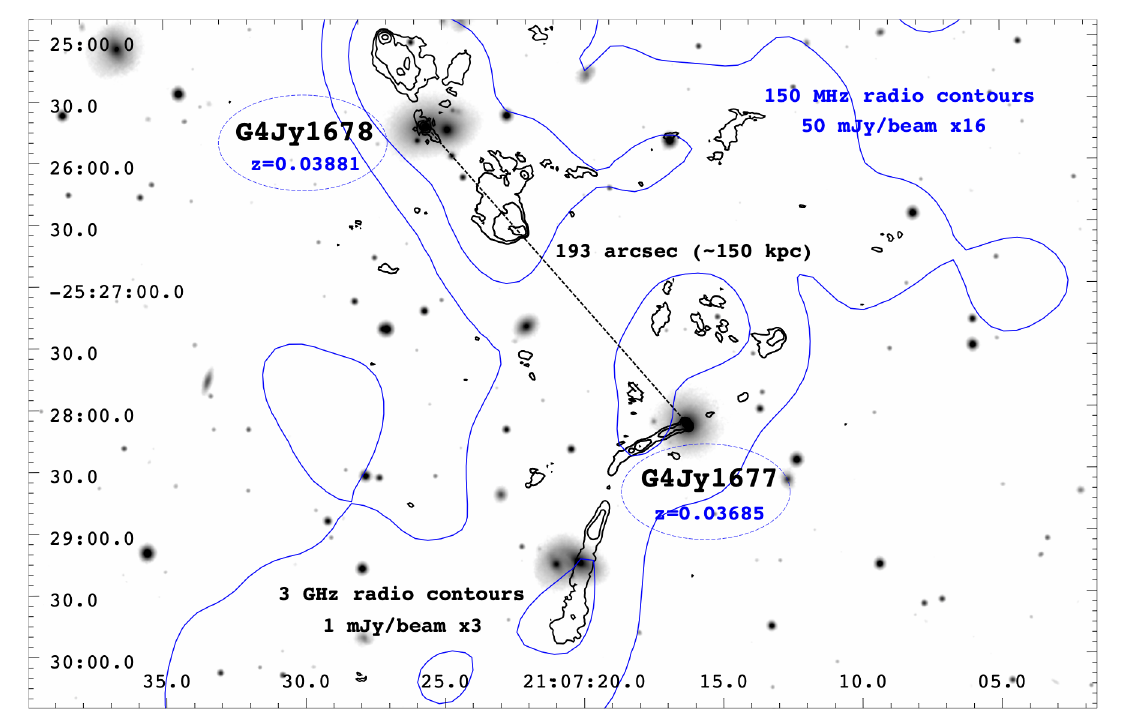}
\end{center}
\caption{The two radio galaxies G4Jy\,1677 and G4Jy\,1678 (a.k.a., NGC\,7016 \& NGC\,7018) harbored in the galaxy cluster Abell\,374 and lying at an angular separation of $\sim$190\,\arcsec (i.e., corresponding to $\sim$150 kpc at the galaxy cluster redshift). Radio contours from the TGSS and the VLASS archival images at 150\,MHz and 3\,GHz are overlaid, in blue and in black respectively, on the optical image in $r$ band available in the Pan-STARRS database.}
\label{fig:G4Jy1678-G4Jy1677}
\end{figure*}

\underline{G4Jy\,1677 \& G4Jy\,1678:} (a.k.a., NGC\,7016 \& NGC\,7018) both belong to the galaxy cluster Abell\,3744 \citep{abell58,abell89,dacosta98,katgert98,postman95} and lie at $z$=0.036845 and $z$=0.03881, respectively \citep{garilli93,tadhunter93,smith00}. More details about these interacting galaxies are given in the literature \citep[see e.g.,][and references therein]{worrall14} while the optical image with a larger field of view retrieved from the Pan-STARRS archive and with radio contours overlaid is shown in Figure~\ref{fig:G4Jy1678-G4Jy1677}.
\vspace{0.2cm}

\underline{G4Jy\,1684:} is a QSO with a steep radio spectrum \citep{baker95} with a $z$=0.602 \citep{baker99} 
\vspace{0.2cm}

\underline{G4Jy\,1686:} is a lobe dominated QSO at $z$=0.97779 \citep{searle68,aldcroft94,jones09} also listed in the MS4 sample and in the catalogs of $\gamma$-ray blazar candidates \citep[][respectively]{burgess06a,dabrusco19}.
\vspace{0.2cm}

\underline{G4Jy\,1692:} a FR\,II radio galaxy at $z$=0.882 \citep{best99}.
\vspace{0.2cm}

\underline{G4Jy\,1698:} also known as PKS\,2128-20 is a QSO at $z$=1.615 \citep{best99} with a nearby companion galaxy with no emission lines in its optical spectrum. G4Jy\,1698 does not show evidence for any net overdensity of red galaxies on either small ($\sim$150\,kpc) and/or large scales up to Mpc \citep{best03}
\vspace{0.2cm}

\underline{G4Jy\,1708:} is a lobe dominated QSO \citep{macdonald71,morganti93} with a $z$=0.20047 \citep{baldwin75,ho09} in the direction of the galaxy cluster Abell\,2361 \citep{abell58,abell89,bahcall69,mclure01}, and detected in the hard X-rays \citep{baumgartner13,koss17} with extended soft X-ray emission \citep{mingo17}. Optical images reveal a disturbed morphology with a shell in the western side embedded in an amorphous halo, and a faint tidal tail pointing to the south-eastern direction \citep{ramos11}.
\vspace{0.2cm}

\underline{G4Jy\,1709:} (a.k.a. PKS\,2135-20) is a BLRG at $z$=0.63634 \citep{tadhunter93,holt08,hernan16} with a CSS radio core \citep{morganti93}. It shows broad fan on the northern side, interpreted as due to a past interaction, and in agreement with the detection of a young stellar population in the nuclear region \citep{holt07}, and a far infrared excess \citep{dicken09b} being also extremely luminous at mid-IR frequencies \citep{ramos11}.
\vspace{0.2cm}

\underline{G4Jy\,1747:} is a radio QSO at $z$=0.668 \citep{dunlop89}.
\vspace{0.2cm}

\begin{figure*}[!th]
\begin{center}
\includegraphics[height=12.cm,width=16.cm,angle=0]{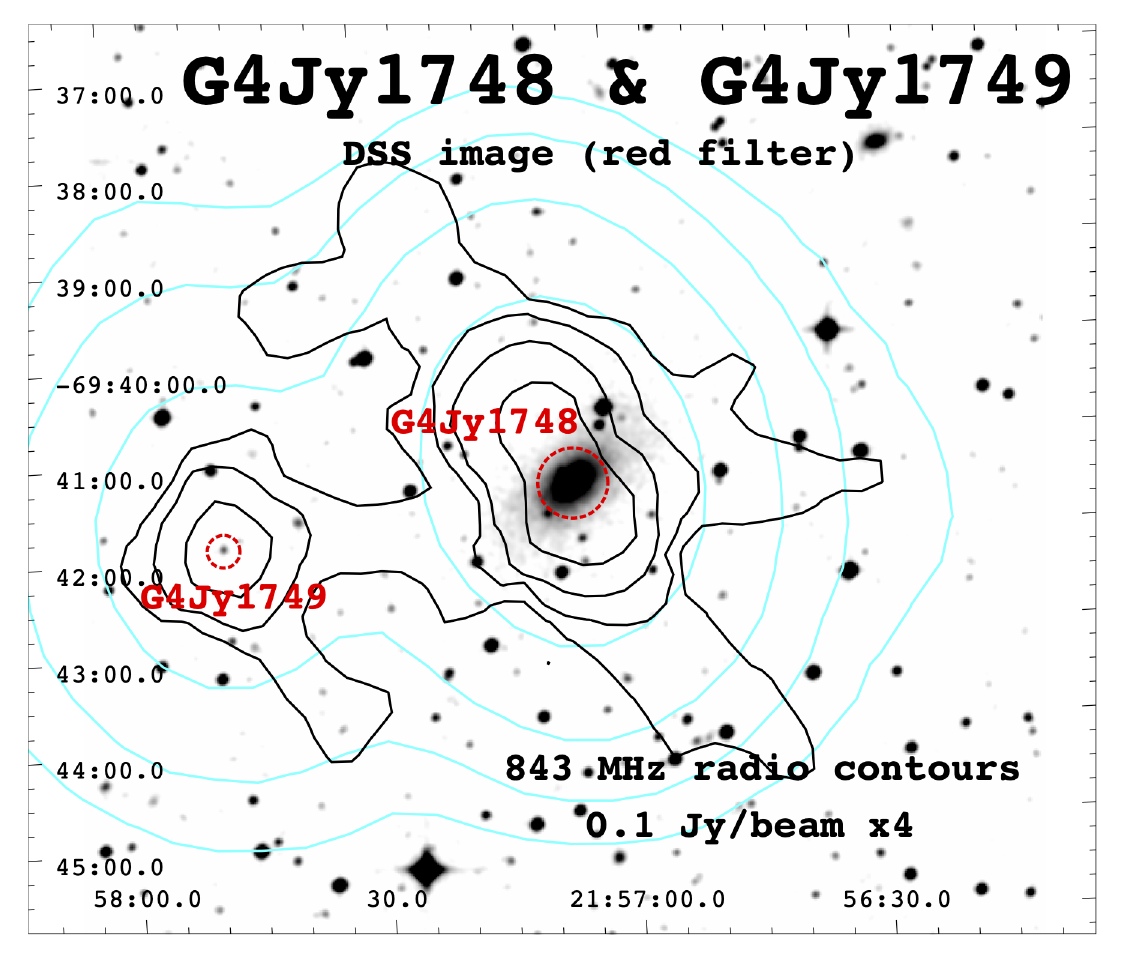}
\end{center}
\caption{The DSS optical image, in the red filter, of the field including the two radio sources G4Jy\,1748 and G4Jy\,1749, the former one harbored in a galaxy cluster at $z$=0.0281 and lying at an angular separation of $\sim$3.6\,\arcsec (i.e., corresponding to $\sim$120 kpc at its redshift) from G4Jy\,1749. Radio contours from the SUMSS at 843\,MHz are overlaid in black while those obtained from the GLEAM radio maps between 171\,MHz and 230\,MHz are reported in cyan, the former ones start at level of 0.1\,Jy/beam while the latter ones at 0.4\,Jy/beam and both increase by a factor of 4.}
\label{fig:G4Jy1748-G4Jy1749}
\end{figure*}

\underline{G4Jy\,1748 \& G4Jy\,1749:} the former one is a lobe dominated QSO \citep[a.k.a. PKS\,2152-69][]{jones92,morganti93} with a $z$=0.0281 \citep{marenbach82,tadhunter93} and showing a wide range of features associated with radio-galaxy/gas interactions typical of sources where radio mode feedback processes are occurring \citep{worrall12}. High resolution radio observations reveal a radio component at $\sim$10 arcsec in the north-eastern direction from the core, close to an optical highly ionized cloud. At larger scale, G4Jy\,1748 shows a FR\,II morphology with the northern lobe having a ``relaxed'' structure, while the southern lobe shows an edge-brightened, arc-like structure \citep{fosbury98}. The X-ray surface brightness has two depressions spatially associated with the radio lobes thus suggesting the presence of X-ray cavities inflated with radio plasma \citep{young05}. Both radio lobes have their hotspot detected in the X-rays \citep[see also][]{ly05,massaro11a}. G4Jy\,1748 has been also detected in both the hard X-ray and the $\gamma$-ray bands \citep[see e.g.,][respectively]{cusumano10,baumgartner13,abdollahi20}. On the other hand, at an angular separation of $\sim$3.6\,\arcmin\ from G4Jy\,1748, in the eastern direction, there is the nearby radio source G4Jy\,1749 as shown in Figure~\ref{fig:G4Jy1748-G4Jy1749}. In this case the lack of spectroscopic information on G4Jy\,1749 prevent us to claim that they belong to the same galaxy cluster as for G4Jy\,1677 \& G4Jy\,1678.
\vspace{0.2cm}

\underline{G4Jy\,1757:} is a QSO (a.k.a. PKS\,2203-18) with a $z$=0.61850 in a group of four QSOs \citep{morton82,stickel89}, also classified as FSRQ in the Roma-BZCAT \citep[a.k.a. BZQ\,J2206-1835;][]{drinkwater97,massaro09} and in the catalogs of $\gamma$-ray blazar candidates \citep{dabrusco14}.
\vspace{0.2cm}

\underline{G4Jy\,1767:} is a FR\,II radio galaxy (a.k.a. 3C\,444) with a $z$=0.153 \citep{lazareff75,dunlop89,tadhunter93} and being the BCG of the galaxy cluster Abell\,3847 \citep{abell58,heckman86,abell89,ramos13,ineson15}. It is embedded in a relatively dense ICM with clear X-ray cavities spatially associated with its radio lobes, driven by shocks \citep{croston11,vagshette17,mingo17}.
\vspace{0.2cm}

\underline{G4Jy\,1781:} is a QSO (a.k.a. PKS\,2226-41) at $z$=0.446 \citep{hewitt89}.
\vspace{0.2cm}

\underline{G4Jy\,1786:} is a WAT with a bent radio jet \citep{loken95} in the north-eastern direction, with a $z$=0.0742. G4Jy\,1786 is the BCG of the galaxy cluster Abell\,2462 \citep{owen95,quintana95a,dalton94} and also close to Abell\,3897 \citep{katgert98}, with the ICM clearly detected in the soft X-rays \citep{sun09}.
\vspace{0.2cm}

\underline{G4Jy\,1795:} (a.k.a. PKS\,2250-41) is a FR\,II radio galaxy hosted by an elliptical galaxy at a redshift of $z$=0.308 \citep{tadhunter93,inskip08} also listed in the MS4 and in the 2\,Jy samples \citep[][respectively]{burgess06a,morganti93}. It has an emission line region showing clear evidence of jet–cloud interactions \citep{tadhunter94,clark97,villar99,tilak05}, shaped as an arc around the western radio lobe with shock signatures \citep{inskip08}, probably due to star formation triggered by a faint companion. Recent ALMA observations revealed molecular gas with the broadest velocity profiles in the central, sub-kiloparsec region due to a direct interaction of the jet with dense clouds and outflowing molecular gas. On larger, kiloparsec-scales, the molecular gas appears to avoid the radio lobes, affected by the expanding cocoon around the radio source, likely dispersing and heating preexisting molecular clouds \citep{morganti21b}.
\vspace{0.2cm}

\underline{G4Jy\,1822:} is a radio galaxy at $z$=1.414 \citep{best99}. 
\vspace{0.2cm}

\underline{G4Jy\,1824:} This radio source, a.k.a PKS\,J2322-5445 and MRC\,2319-550, appears to lie in the direction of the galaxy cluster Abell\,1115 \citep[see e.g.][]{abell89,robertson90}. Optical spectroscopic observations are necessary to confirm this claim.
\vspace{0.2cm}

\underline{G4Jy\,1829:} is a young radio source with molecular gas ``hugging" its radio lobes \citep{morganti21a}. It lies
 at $z$=0.082991 \citep{schmidt65,costero77,owen95,jones09,fernandez16} in the central region of the massive cooling flow galaxy cluster Abell\,2597 \citep{abell58,abell89,crawford89}. At radio frequencies it shows straight and symmetric jets emerging from both sides, and an inverted spectrum of its radio core \citep{taylor99}, all confined by the ambient X-ray gas having the southern lobe deflected from its original south-western direction to the south \citep{pollack05}. Both ionized gas and cold molecular gas components wrap around the radio jet and the X-ray cavities \citep{tremblay16}, and the cospatial and comoving warm ionized and cold molecular components are consistent with a scenario of chaotic cold accretion, precipitation, and stimulated feedback \citep{tremblay18}. A ghost cavity at 330 MHz was also detected \citep{kokotanekov17}.
\vspace{0.2cm}

\underline{G4Jy\,1840:} is radio source with a $z$=0.907 \citep{diserego94} and listed also in the MS4 sample \citep{burgess06a}.
\vspace{0.2cm}

\underline{G4Jy\,1848:} is an USS radio galaxy with a $z$=0.644 \citep{debreuck06}.
\vspace{0.2cm}

\underline{G4Jy\,1854:} is a lobe dominated radio QSO having a steep radio spectrum and emission line, in its optical spectrum that, if interpreted as MgII, suggests a redshift estimate of $z$=1.39 \citep{thompson90}.
\vspace{0.2cm}

\underline{G4Jy\,1858:} is the FR\,I BCG (a.k.a. PKS\,2354-35) of the galaxy cluster Abell\,4059 \citep{abell58,abell89,green90} with a $z$=0.049021 \citep{green88,dacosta89,dacosta98,jones09}. The radio galaxy is classified as ``quiescent'' since it lacks any signature, in H$\alpha$ emission, of recent major events that could have disturbed the gas \citep{hamer16} and also shows a thick filament of ionized gas, with low velocity dispersion, toward the eastern side with LINER-like line ratios \citep{pagotto21}. The galaxy cluster shows a relatively strong cooling-flow \citep{schwartz91}, and there is evidence that radio emission is inflating X-ray cavities located in the ICM \citep{heinz02,dunn06}, that could be ghosts of a previous burst of G4Jy\,1854 due to the lack of any shock-heated ICM \citep{choi04}. There is also a central asymmetric ridge of X-ray emission extending for $\sim$30 kpc in the south-western direction from the galaxy cluster center \citep[see e.g.,][and references therein]{choi04}, colder and denser than the surrounding ICM, with super-solar metallicity and significant ICM asymmetry, possibly be due to a starburst/stripping event \citep{reynolds08}. 
\vspace{0.2cm}

\underline{G4Jy\,1863:} is a narrow-line hard X-ray selected giant radio galaxy \citep{cusumano10,malarecki15,oh18,ursini18,bruni20} with a $z$=0.0959 \citep{tritton72,danziger83} with several signatures of a past merger/interaction in its optical spectrum, as irregular shells and two faint arcs \citep{ramos11}. G4Jy\,1863 lies in the direction of the galaxy cluster Abell\,4067 \citep{abell58,abell89,teague90}.

\newpage
\section{Radio cross-identifications}
\label{app:crossid}
The first comparison was carried out with latest release (i.e., v1.01) of the Parkes radio catalog (PKSCAT90)\footnote{https://heasarc.gsfc.nasa.gov/W3Browse/all/pkscat90.html} \citep{bolton79} listing radio and optical data for $\sim$8000 radio sources and covering essentially all the sky south of declination +27$^\circ$ but largely excluding the Galactic Plane and the Magellanic Cloud regions as for the G4Jy. The original catalog included observations performed at frequencies of 408\,MHz and 2700\,MHz.

Then we compared the \cs\ sample with the Molonglo Reference Catalog of Radio Sources \citep[MRC,][]{large81}. The MRC is one of the largest homogeneous catalogs of radio sources observed at 408\,MHz, containing $\sim$12000 discrete sources with flux densities greater than 0.7 Jy in the declination range between +18.5$^\circ$ to -85$^\circ$ (in B1950 equinox) and excluding regions within 3$^\circ$ of the Galactic equator.

We also used the Texas Survey of 66841 discrete radio sources \citep[TXS;][]{douglas96} detected in the declination range between -35.5$^\circ$ and +71.5$^\circ$ (in B1950 equinox), which was performed at 365\,MHz. The Survey lists accurate positions with positional uncertainty of the order of arcseconds and flux densities of a few percent. The TXS Survey is 90\% complete at 0.4\,Jy and 80\% complete at 0.25\,Jy, being nearly free from spurious sources and has a low level of lobe-shift incidence.

In addition the Australia Telescope 20-GHz Survey \citep[AT20G][]{murphy10} was also compared with the \cs\ sample. The AT20G is a blind radio survey carried out at 20 GHz with the Australia Telescope Compact Array (ATCA) from 2004 to 2008, and covers the whole sky south of declination 0$^\circ$. The latest release of the AT20G source catalog lists 5890 sources above a 20-GHz flux-density limit of 40 mJy. All AT20G sources have total intensity and polarization measured at 20 GHz, and most sources south of declination -15$^\circ$ also have near-simultaneous flux-density measurements at 5 and 8 GHz with a completeness level of 91\% above 100 mJy/beam.

Then we crossmatched the \cs\ sample with the Parkes-MIT-NRAO catalog \citep[PMN][]{griffith93,wright94} in several regions of the sky: Southern, Zenith, Tropical and Equatorial surveys. These surveys were made using the Parkes 64-m radio telescope at a frequency of 4850 Hz with the NRAO multibeam receiver mounted at the prime focus. These surveys had a spatial resolution of $\sim$4.2 arcminutes. This survey covers 2.50 sr listing 23,277 radio sources to a flux limit ranging as a function of declination between $\sim$20 mJy at the southern survey limit and $\sim$50 mJy at the northern limit.

We found that 237 sources out of those 264 included in the \cs\ catalog have a radio counterpart in the PKSCAT90. In particular 171 of them have IDF=1.0 and this optical identification was augmented by a literature search for G4Jy\,538, G4Jy\,939, G4Jy\,1401, G4Jy\,1854 \citep[see e.g.,][]{bolton65,bolton66a,bolton66b,bolton68,hunstead71a,peterson73,peterson76,white87}. Then 249 out of 264 show a radio counterpart in the MRC catalog being detected at 408\,MHz, in particular 84 are also selected in the equatorial sample of Best et al. (1999), while 125 were selected to create the MS4 catalog, both built using MRC observations. There are also 235 \cs\ radio sources with a counterpart listed in the PMN catalog while only 126 and 118 in the TXS and in the AT20G catalogs, respectively.

In the following tables, we report the G4Jy name together with those available in several radio catalogs based on surveys carried out at different frequencies and mainly covering the Southern Hemisphere, namely: TXS, MRC, PKS, PMN catalogs. These cross-identifications can be used to retrieve observations out of different databases. These associations were mainly based on the NED and SIMBAD databases. The last two columns are dedicated to common names and to highlight those objects that are associated to Abell galaxy clusters as found in our literature search (see Appendix A).

This search for radio counterparts, as well as that on information regarding classifications and redshifts, was also augmented by the results achieved for the 2\,Jy catalog\footnote{https://2jy.extragalactic.info/The\_2Jy\_Sample.html} \citep[see e.g.,][]{wall85}, a southern sample of radio galaxies defined as having flux densities above 2 Jy at 2.7 GHz, declination below +10$^\circ$ and redshifts up to 0.7 and its full subsample \citep[see e.g.,][]{tadhunter93,morganti97} for which a large suite of multifrequency observations is already available \citep[see e.g.,][]{morganti93,morganti99,tadhunter02,ramos11,mingo14}. However, crossmatching the \cs\ sample with the 2\,Jy catalog we found that only 45 out of 264 radio sources are also listed therein. This analysis will be also updated in all forthcoming papers. 

\begin{table*}[]
\caption{Radio cross-identifications. Table 5 is published in its entirety in the machine-readable format. A portion is shown here for guidance regarding its form and content. }
\begin{center}
\begin{tabular}{rllllll}
\hline
G4Jy & TXS & MRC & PKS & PMN & Common & Abell \\
name & name & name & name & name & name & cluster \\
\noalign{\smallskip}
\hline 
\noalign{\smallskip}
4 & 0000-177 & 0000-177 & J0003-1727 & J0003-1727 & --& -- \\
9 & --& 0003-567 & J0005-5628 & J0005-5628 & --& -- \\
12 & --& 0003-833 & J0006-8306 & J0006-8305 & --& -- \\
20 & --& 0007-446 & J0010-4422 & J0010-4422 & --& -- \\
26 & --& 0012-383 & J0015-3804 & J0015-3804 & --& -- \\
27 & --& 0013-634 & J0016-6310 & J0016-6310 & --& -- \\
33 & 0016-129 & 0016-129 & J0018-1242 & J0018-1242 & 3C8 & -- \\
43 & 0020-253 & 0020-253 & J0023-2502 & J0023-2502 & --& -- \\
45 & 0022-297 & 0022-297 & J0024-2928 & J0024-2928 & --& -- \\
48 & 0023-263 & 0023-263 & J0025-2602 & J0025-2602 & --& -- \\
\noalign{\smallskip}
\hline
\end{tabular}\\
\end{center}
\end{table*}

\newpage
\section{Crossmatching GLEAM and WISE surveys: a statistical test}
\label{app:test}
Assuming those radio sources for which (i) the optical counterpart is coincident with the mid-IR one (i.e., IDF=1.0) plus those (ii) lacking an optical counterpart but having a mid-IR one associated with their radio core (i.e., IDF=4.1) as correct mid-IR associations, we count 203 out of 264 objects, being $\sim$77\% of the whole \cs\ sample. This fraction can be compared with the expected number of spurious associations that can arise when matching the G4Jy catalog with the AllWISE potential counterparts. We computed the chance probability of associations between mid-IR sources and those listed in the full G4Jy catalog since this was originally used to assign counterparts while the \cs\ sample was extracted out of it later.

We adopted here the same procedure described in \citep{massaro12b,dabrusco13,massaro14a,dabrusco14} to compute the probability of having spurious associations between those sources listed in the G4Jy catalog, using their brightness-weighted radio centroids as positions, and their AllWISE potential counterparts. Here we report just a brief overview of the method used. All crossmatches computed in the following analysis are based on their positions reported in the catalogs.

We started counting the total number of mid-IR counterparts $N(R)$ within circular regions of radius $R$ in the range between 0\,\arcsec\ and 10\,\arcsec, for each G4Jy source. Next, we generated 1000 mock catalogs, based on the distribution of the mid-IR sources around the brightness-weighted radio centroid reported in the G4Jy catalog, shifted it by a random value uniformly distributed between 10\,\arcsec\ and 20\,\arcsec\ in a random direction of the sky. The shifts used to create the mock catalogs were chosen to be not too distant from the original position reported in the G4Jy to guarantee that fake catalogs have the same sky distribution as the original G4Jy. This allowed us to crossmatch mock sources with real G4Jy objects taking into account the local density distribution of mid-IR sources \citep[see][for additional information]{massaro14a}. The total number of G4Jy sources in all mock catalogs is also preserved.

For each mock realization of the G4Jy catalog, we counted the number of fake associations with the AllWISE catalog occurring at angular separations $R$ smaller than 10\,\arcsec. Then we computed the mean number $\lambda(R)$ of these fake associations, averaged over the 1000 mock catalogs, verifying that $\lambda(R)$ has a Poissonian distribution. Increasing the radius by $\Delta\,R=$0\arcsec.1, we also calculated the difference $\Delta\,\lambda(R)$ between the number of mock sources within a radius of $R+\Delta(R)$ and those within $R$, defined as: $\Delta\,\lambda(R) = \lambda(R+\Delta\,{R}) - \lambda(R)$. 

\begin{figure*}[!ht]
\includegraphics[height=6.cm,width=9.2cm,angle=0]{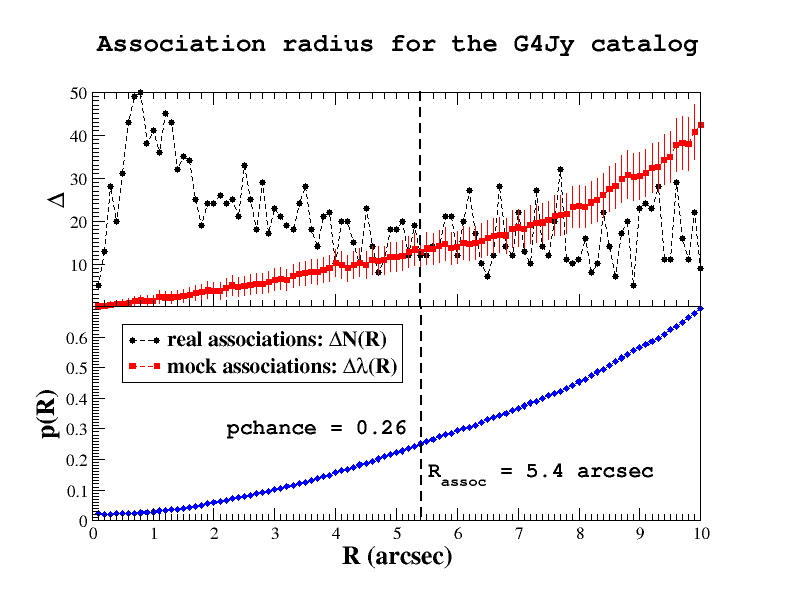}
\includegraphics[height=6.cm,width=9.2cm,angle=0]{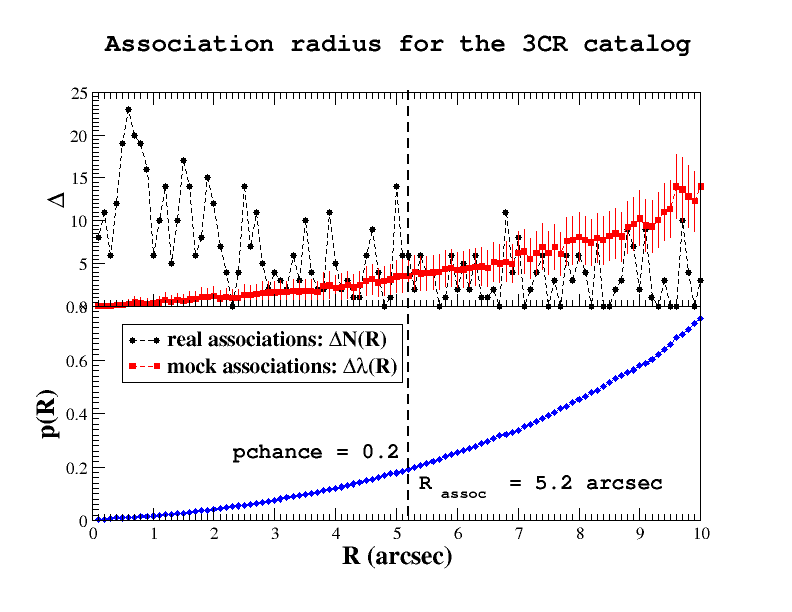}
\caption{The values of $\Delta\,\lambda(R)$ (red squares) and $\Delta\,N(R)$ (black circles) as a function of the angular separation $R$. Our choice of $R_{assoc}$ is marked by the vertical dashed line. It occurs when $\Delta\,\lambda(R)\simeq\Delta\,N(R)$. Uncertainties on the average $\lambda(R)$ values obtained by the crossmatches with mock catalogs were computed from their distributions at each $R$. The correspondent chance probability of having spurious associations at $R=R_{assoc}$ is then reported in the lower panel. Left panels correspond to the G4Jy catalog, used to extract the \cs\ sample while, for comparison, in the right panels we show the results of the same procedure applied to the 3CR sample. In the latter case we used NVSS coordinates similar to those adopted in the G4Jy catalog to estimate the brightness-weighted radio centroids at declinations above -39.5$^\circ$. Considering that in the Southern Hemisphere the source sky density (i.e., number of sources per square degree) of mid-IR potential counterparts is $\sim$10\% larger than that correspondent to the 3CR, results on these two different catalogs are in agreement.}
\label{fig:assocrad}
\end{figure*}

Finally, in Figure~\ref{fig:assocrad} we show the comparison between $\Delta\,N(R)$ (i.e., the difference between the number of real matches within a radius of $R+\Delta(R)$ and those within $R$) and $\Delta\,\lambda(R)$. For angular separations larger than $R_{assoc}=$5\arcsec.4 the $\Delta\,\lambda(R)$ curve begins to match that of $\Delta\,N(R)$. Thus we choose 5\arcsec.4 as the maximum angular separation at which we could consider the mid-IR source a reliable counterpart of the G4Jy radio object. An association between a G4Jy source and its potential mid-IR counterpart, occurring at angular separation above 5\arcsec.4 has almost the same probability of being either correct or random. Positional uncertainties of the NVSS and SUMSS radio surveys typically ranges between 1\,\arcsec to 5\,\arcsec and 2\,\arcsec to 10\,\arcsec, respectively \citep[see e.g.,][and references therein for a recent analysis]{vollmer05}. These were the radio surveys used in the G4Jy analysis to compute brightness-weighted radio centroids and associate radio sources with their mid-IR counterparts. Thus our statistical result is also in agreement with previous analyses.

As shown in Figure~\ref{fig:assocrad} the chance probability of spurious associations $p(R_{assoc})$ was computed as the ratio between the number of real associations $N(R_{assoc})$ and the average of those found in the mock realizations $\lambda(R_{assoc})$, corresponding to a value of $\sim$26\% \citep[see also][for additional details on $p(R_{assoc})$]{massaro11b, dabrusco13}. Thus the choice of $R_{assoc}$ is based on the comparison between differential distributions of real and average mock matches while $p(R_{assoc})$ on their cumulative ones. Adopting the same statistical procedure used here to search for mid-IR counterparts of blazars listed in the Roma-BZCAT \citep{massaro09,massaro15b}, we found $p(R_{assoc})$ below 1\% at radius of $\sim$3\,\arcsec \citep[see also][]{massaro13b,massaro14b,dabrusco19,demenezes20}. This was mainly due to the blazar nature (i) being core-dominated radio sources, thus mostly point-like objects at GHz frequencies, and (ii) having more  precise positions reported in the comparison catalog, as well as a combination of these two effects. 

The probability of having spurious associations (i.e., $\sim$26\%) is certainly in good agreement with our refined analysis on optical counterparts, for which the number of incorrect associations is expected to be $\sim$23\%. Then, angular separations $\theta_{ow}$ between the mid-IR counterpart, assigned in the G4Jy catalog, and the optical one with IDF=1.0 are all below 4\arcsec.8 with only one exception having 5\arcsec.6, as shown in Figure~\ref{fig:angsep}, thus consistent with being correct. In Figure~\ref{fig:assocrad} we also report the same plots but computed for the 3CR catalog \citep{spinrad85}. In this case the value of the association radius is 5\arcsec.2, estimated according to the same method previously described. This was computed using the NVSS coordinates of the 3CR radio sources that have a similar precision of those used in the G4Jy catalog to determine the brightness-weighted radio centroids. This association radius corresponds to a chance probability of spurious associations of $\sim$19\%. However the source sky density (i.e., number of sources per square degree) of mid-IR potential counterparts around 3CR sources is $\sim$10\% smaller than the one measured for the \cs\ catalog thus, taking into account of this, the chance probability computed for the \cs\ catalog is also in agreement with the expectations based on the 3CR catalog.
 
\begin{figure*}[]
\includegraphics[height=5.4cm,width=9.8cm,angle=0]{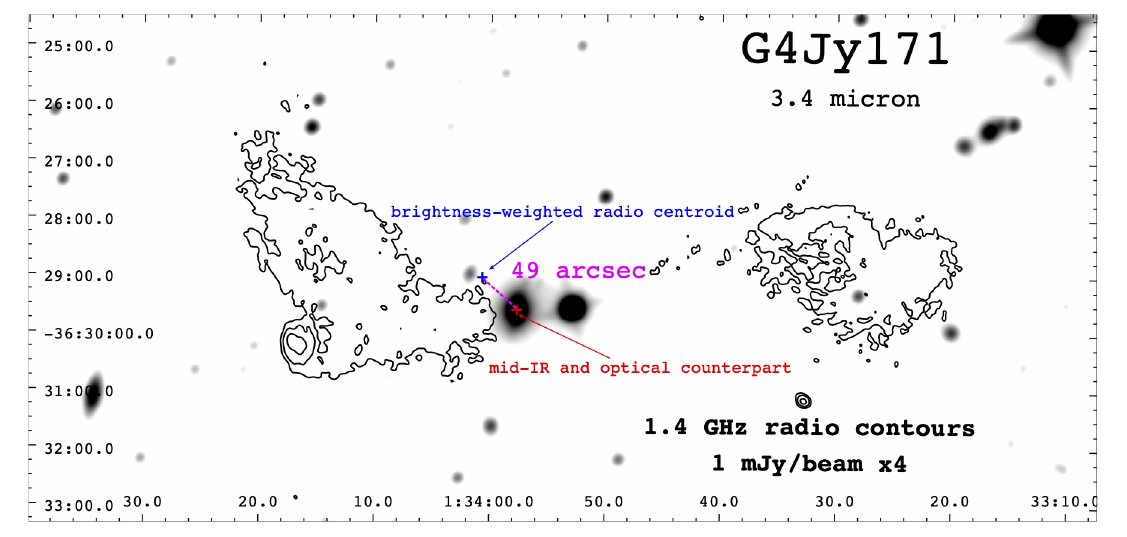}
\caption{Same as left panel of Figure~\ref{fig:example2} for G4Jy\,171. The blue cross marks the position of the brightness-weighted radio centroid while the red one that of the mid-IR and optical counterpart, associated with the radio core. The location of the radio core was identified thanks to the NVAS archival radio map at 1.4 GHz used also to draw radio contours overlaid to the mid-IR image. The radio core is not clearly detected in archival images of the TGSS, NVSS and SUMSS. the angular separation between the two crosses is $\sim$49\,\arcsec.}
\label{fig:example6}
\end{figure*}

In the \cs\ catalog 225 sources out of 264 have a mid-IR counterpart assigned by the original G4Jy catalog \citep{white20a,white20b}, and 136 out of 225 sources lie below the threshold of 5\arcsec.4 angular separation between the brightness-weighted radio centroid and the position of the assigned mid-IR counterpart. A similar situation occurs when comparing the brightness-weighted radio centroid with the optical position of the counterpart assigned thanks to our optical analysis. We have been able to find 211 optical counterparts out of 264 examined sources and only 109 have angular separation between radio and optical position below 5\arcsec.4, in agreement with the expectations of the statistical analysis. The main reason underlying the relatively high probability of getting spurious associations is the use of a brightness-weighted radio centroid since, as previously stated, does not always provide a reliable position of the host galaxy for sources having an extended radio morphology that is unresolved in radio maps used to compute it and/or for those clearly asymmetric. This motivated our analysis based on higher resolution radio images thus allowing us to determine the precise location of the host galaxies for the \cs\ catalog presented here. An extreme example on how the radio centroid can provide a misleading information about the position of the host galaxy is shown in Figure~\ref{fig:example6} where the high resolution radio map at 1.4 GHz available in the NVAS archive allowed us to confirm the position of the host galaxy, being the same assigned at mid-IR frequencies in the G4Jy catalog, but lying at $\sim$49\,\arcsec\ angular separation from the location of the brightness-weighted radio centroid. Diffuse radio emission arising from lobes could bias the location of the brightness-weighted radio centroid as shown in the case of G4Jy\,171 (Figure~\ref{fig:example6}).

\end{document}